\RequirePackage{fix-cm}
\documentclass[smallextended]{svjour3}       
\smartqed  
\usepackage{graphicx}
\usepackage{epsfig}
\usepackage{amsmath,amssymb}
\usepackage[table]{xcolor}
\usepackage{caption}
\usepackage{xcolor}
\usepackage{url}
\usepackage{hyperref}
\usepackage{orcidlink}
  \def\intl{\int\limits}
  
  \def\parder#1#2{\frac{\partial #1}{\partial #2}}
 \def\dedet#1#2{\left(\frac{\delta #1}{\delta t}\right)_{\rm #2}}
 \def\div#1{\frac{1}{r^2}\parder{}{r}\left(r^2 #1 \right)}
 
 \def\s{\sigma}
 \def\sr{\sigma_r}
 \def\tree{\textsc{Tree}}
 \def\co{{\cal O}}
\usepackage[acronym,nomain,nonumberlist]{glossaries}
\setacronymstyle{long-sc-short}
\makenoidxglossaries

\newacronym{SSE}{SSE}{Single Stellar Evolution}
\newacronym{BSE}{BSE}{Binary Stellar Evolution}
\newacronym{GRAPE}{GRAPE}{Gravity Pipeline}
\newacronym{NSC}{NSC}{Nuclear Star Cluster}
\newacronym{GC}{GC}{Globular Star Cluster}
\newacronym{HST}{HST}{Hubble Space Telescope}
\newacronym{JWST}{JWST}{James Webb Space Telescope}
\newacronym{ELT}{ELT}{Extremely Large Telescope}
\newacronym{M87}{M87}{Galaxy Messier Catalogue Nr. 87}
\newacronym{LIGO}{LIGO}{Laser Interferometer Gravitational Wave Observatory}
\newacronym{KAGRA}{KAGRA}{Kamioka Gravitational Wave Detector}
\newacronym{MOCCA}{MOCCA}{Monte Carlo Code}
\newacronym{CMC}{CMC}{Cluster Monte Carlo Code}
\newacronym{BBGKY}{BBGKY}{Bogoliubov Born Green Kirkwood Yvon}
\newacronym{AC}{AC}{Ahmad-Cohen Neighbour Scheme}
\newacronym{KS}{KS}{Kustaanheimo-Stiefel Regularization}
\newacronym{HARP}{HARP}{Hermite Accelerator Pipeline}
\newacronym{SPMD}{SPMD}{Single Program Multiple Data}
\newacronym{MPI}{MPI}{Message Passing Interface}
\newacronym{PN}{PN}{Post-Newtonian}
\newacronym{GPU}{GPU}{Graphical Processing Unit}
\newacronym{SSE2}{SSE}{Streaming SIMD Extension}
\newacronym{SIMD}{SIMD}{Single Instruction Multiple Data}
\newacronym{AVX}{AVX}{Advanced Vector Extension}
\newacronym{AR}{AR}{Algorithmic Regularization}
\newacronym{P3T}{P3T}{Particle-Particle Particle-Tree}
\newacronym{SDAR}{SDAR}{Slow Down Algorithmic Regularization}
\newacronym{MSTAR}{MSTAR}{Minimum Spanning Tree Algorithmic Regularization}
\newacronym{FROST}{FROST}{Fourth Order Forward Symplectic Integrator}
\newacronym{BiFROST}{BiFROST}{Binaries in FROST code}
\newacronym{CUDA}{CUDA}{Compute Unified Device Architecture}
\newacronym{MOBSE}{MOBSE}{Massive Objects in Binary Stellar Evolution}
\newacronym{NVLink}{NVlink}{NVIDIA High Speed GPU Interconnect}
\newacronym{CPU}{CPU}{Central Processing Unit}
\newacronym{FP}{FP}{Fokker-Planck}
\newacronym{FSI}{FSI}{Forward Symplectic Integrator}
\newacronym{MLT}{MLT}{Mixing Length Theory}
\newacronym{MESA}{MESA}{Modules for Experiments in Stellar Astrophysics}
\newacronym{HOSHI}{HOSHI}{Hongo Stellar Hydrodynamics Investigator}
\newacronym{MS}{MS}{Main Sequence (Stellar Astrophysics)}
\newacronym{HB}{HB}{Horizontal Branch (Stellar Astrophysics}
\newacronym{WD}{WD}{White Dwarf (Stellar Astrophysics)}
\newacronym{RG}{RG}{Red Giant (Stellar Astrophysics)}
\newacronym{KH}{KH}{Kelvin-Helmholtz (Stellar Astrophysics)}
\newacronym{CNO}{CNO}{Carbon Nitrogen Oxygen Burning Cycle (Stellar Astrophysics)}
\newacronym{HRD}{HRD}{Hertzsprung-Russell Diagram}
\newacronym{ZAMS}{ZAMS}{Zero Age Main Sequence (Stellar Astrophysics)}
\newacronym{pp}{pp}{Proton Proton Burning Cycle (Stellar Astrophysics)}
\newacronym{CMS}{CMS}{Cool Luminous Stars (Stellar Astrophysics)}
\newacronym{AGB}{AGB}{Asymptotic Giant Branch (Stellar Astrophysics)}
\newacronym{WR}{WR}{Wolf-Rayet (Stars)}
\newacronym{RSG}{RSG}{Red Supergiants (Stellar Astrophysics)}
\newacronym{LMS}{LMS}{Low Mass Stars}
\newacronym{HMS}{HMS}{High Mass Stars}
\newacronym{IMS}{IMS}{Intermediate Mass Stars}
\newacronym{NS}{NS}{Neutron Star}
\newacronym{BH}{BH}{Black Hole}
\newacronym{HeWD}{HeWD}{Helium White Dwarf}
\newacronym{COWD}{COWD}{Carbon Oxygen White Dwarf}
\newacronym{ONeWD}{ONeWD}{Oxygen Neon White Dwarf}
\newacronym{SN}{SN}{Supernova}
\newacronym{ECSN}{ECSN}{Electron Capture (EC) Supernova}
\newacronym{PISN}{PISN}{Pair Instability Supernova}
\newacronym{PPISN}{PPISN}{Pulsational Pair Instability Supernova}
\newacronym{IFMR}{IFMR}{Initial Final Mass Relation}
\newacronym{AIC}{AIC}{Accretion Induced Collapse}
\newacronym{MIC}{MIC}{Merger Induced Collapse}
\newacronym{RLOF}{RLOF}{Roche Lobe (RL) Overflow}
\newacronym{CEE}{CEE}{Common Envelope (CE) Evolution}
\newacronym{CV}{CV}{Cataclysmic Variable}
\newacronym{BPS}{BPS}{Binary Population Synthesis}
\newacronym{GR}{GR}{General Relativity, Relativistic}
\newacronym{LISA}{LISA}{Laser Interferometer Space Antenna}
\newacronym{SEVN}{SEVN}{Stellar Evolution for Nbody}
\newacronym{ASPS}{ASPS}{Advanced Stellar Population Synthesis}
\newacronym{IMF}{IMF}{Initial Mass Function}
\newacronym{IBP}{IBP}{Initial Binary Population}
\newacronym{MSP}{MSP}{Multiple Stellar Populations}
\newacronym{CMD}{CMD}{Colour-Magnitude Diagram}
\newacronym{IMBH}{IMBH}{Intermediate Mass Black Hole}
\newacronym{SMBH}{SMBH}{Supermassive Black Hole}
\newacronym{AMUSE}{AMUSE}{Astrophysical Multi-Purpose Software Environment}
\newacronym{BEANS}{BEANS}{Software Package for Big Data Analysis}
\newacronym{FITS}{FITS}{Flexible Image Transport System}
\newacronym{HDF5}{HDF5}{Hierarchical Data Format V.5}
\newacronym{COCOA}{COCOA}{Cluster Simulation Comparison with Observations}
\newacronym{NGC}{NGC}{New General Catalogue of Nebulae and Stars}
\newacronym{GALEV}{GALEV}{Galaxy Evolutionary Synthesis}
\newacronym{FSPS}{FSPS}{Flexible Stellar Population Synthesis}
\newacronym{IFU}{IFU}{Integral Field Unit}
\newacronym{SISCO}{SISCO}{Simulating IFU Star Cluster Observations}

\begin{document}

\def\aj{{AJ }}                   
\def\actaa{{Acta Astron. }}      
\def\araa{{ARA\&A }}             
\def\apj{{ApJ }}                 
\def\apjl{{ApJ }}                
\def\apjs{{ApJS }}               
\def\ao{{Appl.~Opt. }}           
\def\apss{{Ap\&SS }}             
\def\aap{{A\&A }}                
\def\aapr{{A\&A~Rev. }}          
\def\aaps{{A\&AS }}              
\def\azh{{AZh }}                 
\def\baas{{BAAS }}               
\def\bac{{Bull. astr. Inst. Czechosl. }}
\def\caa{{Chinese Astron. Astrophys. }}
\def\cjaa{{Chinese J. Astron. Astrophys. }}
\def\icarus{{Icarus }}           
\def\jcap{{J. Cosmology Astropart. Phys. }}
\def\jrasc{{JRASC }}             
\def\memras{{MmRAS }}            
\def\mnras{{MNRAS }}             
\def\na{{New A }}                
\def\nar{{New A Rev. }}          
\def\pra{{Phys.~Rev.~A }}        
\def\prb{{Phys.~Rev.~B }}        
\def\prc{{Phys.~Rev.~C }}        
\def\prd{{Phys.~Rev.~D }}        
\def\pre{{Phys.~Rev.~E }}        
\def\prl{{Phys.~Rev.~Lett. }}    
\def\pasa{{PASA }}               
\def\pasp{{PASP }}               
\def\pasj{{PASJ }}               
\def\rmxaa{{Rev. Mexicana Astron. Astrofis. }}%
\def\qjras{{QJRAS }}             
\def\skytel{{S\&T }}             
\def\solphys{{Sol.~Phys. }}      
\def\sovast{{Soviet~Ast. }}      
\def\ssr{{Space~Sci.~Rev. }}     
\def\zap{{Zs. f. Astroph. }}                 
\def\nat{{Nature }}              
\def\iaucirc{{IAU~Circ. }}       
\def\aplett{{Astrophys.~Lett. }} 
\def\apspr{{Astrophys.~Space~Phys.~Res. }}
\def\bain{{Bull.~Astron.~Inst.~Netherlands }} 
\def\fcp{{Fund.~Cosmic~Phys. }}  
\def\gca{{Geochim.~Cosmochim.~Acta }}   
\def\grl{{Geophys.~Res.~Lett. }} 
\def\jcp{{J.~Chem.~Phys. }}      
\def\jgr{{J.~Geophys.~Res. }}    
\def\jqsrt{{J.~Quant.~Spec.~Radiat.~Transf. }}
\def\memsai{{Mem.~Soc.~Astron.~Italiana }}
\def\nphysa{{Nucl.~Phys.~A }}   
\def\physrep{{Phys.~Rep. }}   
\def\physscr{{Phys.~Scr }}   
\def\planss{{Planet.~Space~Sci. }}   
\def\procspie{{Proc.~SPIE }}   

\let\astap=\aap
\let\apjlett=\apjl
\let\apjsupp=\apjs
\let\applopt=\ao

\title{Computational Methods for Collisional Stellar Systems}



\author{Rainer Spurzem \orcidlink{0000-0003-2264-7203} 
\and Albrecht Kamlah \orcidlink{0000-0001-8768-4510} }


\institute{R. Spurzem \at  \\[0.2cm]
              Astronomisches Rechen-Institut, Zentrum f\"ur Astronomie, \\ M\"onchhofstr. 12-14, 
                 69120 Heidelberg, Germany \\[0.2cm]
                   National Astronomical Observatories and Key Laboratory of Computational Astrophysics,
                                         Chinese Academy of Sciences, 20A Datun Rd., Chaoyang District,
                                         100101 Beijing, China \\[0.2cm]
              Kavli Institute for Astronomy and Astrophysics, Peking University,\\
Yiheyuan Lu 5, Haidian Qu, 100871, Beijing, China \\[0.2cm]
              \email{spurzem@ari.uni-heidelberg.de,\quad spurzem@nao.cas.cn}
              \and  \\[0.5cm]
              A. Kamlah \at \\[0.2cm]
              Max-Planck-Institut f\"ur Astronomie, K\"onigstuhl 17, 69117 Heidelberg, Germany \\[0.2cm]
              Astronomisches Rechen-Institut, Zentrum f\"ur Astronomie, \\ M\"onchhofstr. 12-14, 
                 69120 Heidelberg, Germany 
}

\date{Received: date / Accepted: date}

\maketitle
\newpage
\setcounter{tocdepth}{3}
\tableofcontents
\newpage

\begin{abstract}
Dense star clusters are spectacular self-gravitating stellar systems in our Galaxy and across the Universe - in many respects. They populate disks and spheroids of galaxies as well as almost every galactic center. In massive elliptical galaxies nuclear clusters harbor supermassive black holes, which might influence the evolution of their host galaxies as a whole. The evolution of dense star clusters is not only governed by the aging of their stellar populations and simple Newtonian dynamics. For increasing particle number, unique gravitational effects of collisional many-body systems begin to dominate the early cluster evolution. As a result, stellar densities become so high that stars can interact and collide, stellar evolution and binary stars change the dynamical evolution, black holes can accumulate in their centers and merge with relativistic effects becoming important. Recent high-resolution imaging has revealed even more complex structural properties with respect to stellar populations, binary fractions and compact objects as well as - the still controversial - existence of intermediate mass black holes in clusters of intermediate mass. Dense star clusters therefore are the ideal laboratory for the concomitant study of stellar evolution and Newtonian as well as relativistic dynamics. Not only the formation and disruption of dense star clusters has to be considered but also their galactic environments in terms of initial conditions as well as their impact on galactic evolution. This review deals with the specific computational challenges for modelling dense, gravothermal star clusters.


\end{abstract}

\section*{Preface}
\section{Astrophysical Introduction}
Stars play a fundamental role in astronomy - a large piece of information available to us about the Universe we inhabit comes from
stars. Coeval associations of stars, also called {\em star clusters}, are the birth place of most if not all stars. Star clusters form, evolve and ``die'' by dissolution all across the cosmic time, which is covered in the excellent review ``Star Clusters across Cosmic Time'' that focuses on the interplay between observations and theoretical knowledge~\cite{Krumholzetal2019}; our review, on the other hand, focuses more on the computational challenges to model a special class of dense star clusters. The term ``dense'' is not very well defined. We use it here in the sense that both the system should be ``gravothermal'' and that during at least some phases of the evolution close encounters or even direct collisions between stars or binaries occur. This definition constrains us to globular and young dense star clusters, as well as nuclear star clusters (NSCs). On nuclear star clusters there is another nice review~\cite{Neumayeretal2020}; nuclear star clusters are not in the focus of this work; however, if special computational issues have to be taken into account for nuclear star clusters we may elaborate on them here.\\
Globular star clusters (GCs) are thought to be the oldest objects in our Galaxy, their age covering a large fraction of the age of the Universe, and they are considered as fossil records of the time of early galaxy formation. GCs of variable age are found near all galaxies (except for the smallest dwarfs) and their specific frequency (number of clusters per galaxy mass unit, see e.g.~\cite{Harris1996}) differs as a function of galaxy type, highlighting the close relation between cluster and galaxy formation. The approximately 150 globular clusters of our own Milky Way have been studied in much more detail for their proximity. Today, star-by-star observations with the Hubble Space Telescope (HST), and proper motion studies using Gaia with high resolution spectroscopy to determine their stellar velocity dispersions~\cite{Bianchinietal2013} are possible. Small and big galaxies in the Local Group have systems of GCs, e.g. the Andromeda galaxy and the Magellanic clouds. Globular clusters in huge quantities have been detected around massive galaxies like M87 or other bright central cluster galaxies~\cite{Harrisetal2017} or at sites of star formation near the Antenna galaxies. Still this is – in cosmological scales – our neighborhood. Do clusters form normally following the cosmic star formation history, which peaks at redshifts of around 2~\cite{Reina-Camposetal2019}? Or do massive clusters form preferentially as special objects at much higher redshifts (e.g. $z\sim 6$~\cite{Boylan-Kolchin2018})? \\
Computer simulations of structure formation in the Universe begin to resolve GC scales~\cite{Ramos-Almendaresetal2020}, but they cannot compensate the current lack of deep observations. Only recently gravitational lensing from galaxy clusters has helped to identify candidates for proto-GCs at redshifts of $z>3$~\cite{Vanzellaetal2017}, and more recently even out to $z=6$~\cite{Vanzellaetal2019,Vanzellaetal2020,Vanzellaetal2021}. Future instruments such as Extremely Large Telescope (ELT) and others, will improve the situation significantly, while novel instruments such as James Webb Space Telescope (JWST) are already producing ground-breaking results here \cite{Claeyssensetal2023,Charbonneletal2023}. \\
We return to the ``dense'' and ``gravothermal'' nature of such star clusters. In gravothermal star clusters it is essential to consider the mutual gravitational interactions between many if not all of their stars. The cumulative effect of small angle gravitational deflections (encounters) between distant stars generates transport of heat and angular momentum (relaxation processes connected to these encounters are analogous to heat conduction and viscosity in a gaseous system), and this effect is dominant during certain stages of the star cluster evolution. In order to get the proper time scales connected to such relaxation processes in $N$-body simulations of star clusters typically a very large number of pairwise distant encounters needs to be followed (which asymptotically results in the computational complexity of all force calculations at a point of time $\approx N^2$, assuming a simple approach without parallelization and hybrid hardware). This physical constraint has led to the use of thermodynamics and statistical mechanics to model gravothermal star clusters~\cite{LyndenBellWood1968,Hachisuetal1978,Hachisu1979,Spurzem1991}. The star cluster could be modelled using computer codes for the gas dynamical evolution of stars. \textcolor{red}{The relaxation time in a star cluster is {\em long} compared to the dynamical timescale, while in stars the opposite is true (considering the photon diffusion time as relaxation time). This occurs because the mean free path in stellar systems is much longer than in stellar gaseous matter.
Therefore, conductivity and viscosity need to be defined in a different functional form than for the interior of stars (cf. the more detailed 
discussions in \cite{LyndenBellEggleton1980,LouisSpurzem1991}).}\\
An area where observations have picked up greatly through increased angular resolution and sensitivity of spectrographs is the identification of stellar binaries (e.g.~\cite{Giesersetal2018,Giesersetal2019,Kamannetal2020b}). Binary stars are an extremely important component of star clusters, because they form a dynamically active population which has a dramatic impact on the evolution of the host cluster (e.g.~\cite{Henon1961,Heggie1975,Elsonetal1987a}): for instance stellar exotica observed in clusters (blue stragglers, fast rotating stars, and X-ray binaries) all originate from binary star evolution. The special role that binary stars play in the life cycle of a cluster requires that we pin down as accurately as possible what fraction of stars form in binaries if we are to make progress when predicting the statistics of stellar populations at the later stages of a cluster's evolution. Close stellar encounters, including direct collisions, become a reality in the densest region of dense star clusters. How often these take place, and what the outcomes of such events are remain a puzzle that is only now beginning to be solved. Last, but not least compact objects form in binaries and take part in few-body interactions and stellar evolution of binaries; ultimately, binaries consisting of compact objects only are possible sources of gravitational waves to be detected by ground-based gravitational wave dectectors that are operational today, such as (Advanced) Laser Interferometer Gravitational-Wave Observatory ((a)LIGO)~\cite{Aasietal2015,Abbottetal2018d,Abbottetal2019b}, (Advanced) Virgo Interferometer ((a)Virgo)~\cite{Acerneseetal2015,Abbottetal2018d,Abbottetal2019b} and Kamioka Gravitational Wave Detector (KAGRA) (e.g.~\cite{Abbottetal2018d,Abbottetal2020e,Akutsuetal2019}). 
\textcolor{violet}{Note also the next (third) generation space based detectors in planning (Einstein Telescope\cite{Branchesietal2023} and Cosmic Explorer\cite{Reitzeetal2019}).} \\
The numerical and computational tools to model such dense, massive star clusters are based on either approximate models from statistical physics, or the direct $N$-body simulation approach. Currently, the latter approach is very dominant since it allows to include details of astrophysics (binaries, stellar evolution, tidal fields) more easily. However, in order to establish the degree of reliability of $N$-body simulations approximate models have been very important. These are mostly based on the Fokker-Planck approximation in statistical mechanics, and the numerical solution of resulting kinetic equations by using, e.g., Monte Carlo techniques, direct numerical solution, or gaseous and moment models). Since they are important and fundamental to understand results $N$-body simulations this review contains a basic description of them, too. \\
GCs and NSCs (dense and gravothermal) are ideal laboratories to examine the theoretical physical processes (heat conduction, angular momentum transport through viscosity) and their influence on the formation and evolution of extreme stellar populations like X-ray binaries or blue stragglers, compact objects (neutron stars, black holes), and are ideal test beds for stellar population synthesis models and stellar evolution.

\section{Theoretical Foundations}
Computational modelling of star clusters requires to follow the complex interplay of thermodynamic processes such as heat conduction and relaxation with the physics of self-gravitating systems, the stochastic nature of star clusters having finite particle number $N$, and the astrophysical knowledge and models for the evolution of single and binary stars and of external tidal forces. GCs are a very good laboratory for this, because their dynamical and relaxation timescales are well separated from each other and from the lifetime of the cluster and the Universe in its entirety. This article deals with ``direct'' $N$-body simulations, which are suitable for systems where the interaction between dynamics and relaxation is important (sometimes also denoted as ``gravothermal'' systems~\cite{LyndenBellWood1968}). Other kinds of $N$-body simulations are useful for example for hydrodynamics (``smoothed particle hydrodynamics''), galaxy dynamics (``collisionless systems'') or cosmological $N$-body simulations of structure formation in the Universe, and are not covered here. The main distinction of those from the models presented here, is that the dynamics of systems dominated by two-body relaxation (``collisional systems'') require typically very high accuracy (typical energy error per crossing time $\Delta E/E < 10^{-5}$ or smaller) over very long physical integration times (thousands of crossing times). The term ``collisional'' refers here to elastic gravitational encounters (relaxation encounters), which drive the ``thermal'' cluster evolution. Other processes, such as close encounters, encounters involving one or more binaries, and direct collisions also happen in the system. \\
Let us begin with the definition of some useful time scales. A typical particle crossing time $t_{\rm cr}$ in a star cluster is
\begin{equation}
t_{\rm cr} = \frac{r_{\mathrm{h}}}{\sigma_{\mathrm{h}}}\ ,
\label{1.1}
\end{equation}
where $r_{\mathrm{h}}$ is the radius containing 50 \% of the (current) total mass and
$\sigma_{\mathrm{h}}$ is a typical velocity associated with the root mean square random motion (velocity dispersion) taken
at $r_{\mathrm{h}}$. If virial equilibrium prevails, we have
$\sigma_{\mathrm{h}}^2 \approx GM_{\mathrm{h}}/r_{\mathrm{h}}$ (where the sign $\approx$ here and
henceforth means ``approximately equal'' or ``equal within an order
of magnitude''), thus
\begin{equation}t_{\rm cr} \approx \sqrt{\frac{r_{\mathrm{h}^3}}{G M_{\mathrm{h}}}} \ .
\label{1.2}
\end{equation}
This relation is equal to the dynamical timescale, which is also used for example
in the theory of stellar structure and evolution. Global dynamical
adjustments of the system, like oscillations, are connected
with this timescale. Taking the square of Eq.~\ref{1.2} yields
$t_{\rm cr}^2 \approx r_{\mathrm{h}}^3/(GM_{\mathrm{h}})$ which is related to Kepler's third law,
because the orbital velocity in a Keplerian point mass potential has the
same order of magnitude as the velocity dispersion in virial equilibrium. 
Unlike most laboratory gases stellar systems are not usually in
thermodynamic equilibrium, neither locally nor globally.
Radii of stars are usually extremely small
relative to the average inter-particle distances of stellar
systems (e.g. the radius of the
sun is $r_\odot\approx 10^{11}$ cm, a typical distance between
stars in our galactic neighbourhood is of the order of $10^{18} $cm).
Only under rather special conditions in the centres of galactic nuclei
and during the short high-density core collapse phase of a
globular cluster, stellar densities might become large enough that
stars come close enough to each other to collide, merge or disrupt
each other. Therefore it is extremely unlikely under normal conditions that two stars
touch each other during an encounter; encounters
or collisions usually are elastic gravitational scatterings.
The mean inter-particle distance is large
compared to $p_0=2Gm/\s^2$, which is the impact parameter for a $90^o$ deflection in a typical encounter of two stars
of equal mass $m$, where the relative velocity
at infinity is $\sqrt{2}\s$,
with local 1D velocity
dispersion $\s$. Thus most encounters are small-angle deflections.
The relaxation time $t_{\rm rx}$
is defined as the time after which the root mean square velocity increment due
to such small angle gravitational deflections is of the same order as
the initial velocity dispersion of the system. We use
the local relaxation time as defined by~\cite{Chandrasekhar1942}:
\begin{equation}
t_{\rm rx} = \frac{9}{16 \sqrt{\pi}} \frac{\sigma^3}{G^2 m \rho
    \ln(\gamma N)} \ .
\label{1.3}
\end{equation}
$G$ is the gravitational constant, $\rho$ the mean stellar mass density, $\sigma$ the 3D velocity dispersion, $N$ the total particle number; this definition was used by~\cite{Larson1970a,BettwieserSpurzem1986}, because it naturally occurs when computing collisional terms (as in Eq.~\ref{2.1.23}), if the velocity distribution function is written as a series of Legendre polynomials~\cite{SpurzemTakahashi1995}, with numerical factors being unity (for equipartition terms of lowest order\cite{SpurzemTakahashi1995}) or only little different from unity, such as 9/10 for the collisional decay of anisotropy~\cite{BettwieserSpurzem1986}). Other definitions of relaxation can be found frequently, for example in~\cite{Spitzer1987}. They differ only by numerical factors, except for the so-called Coulomb logarithm $\ln(\gamma N)$, which may take different functional forms. For common forms of the Coulomb logarithms only $\gamma$ is of order unity, but may take different values (e.g. 0.11~\cite{GierszHeggie1994a}, or 0.4~\cite{Spitzer1987}).\\
Assuming virial equilibrium a fundamental proportionality turns out:
\begin{equation}
\frac{t_{\rm rx}}{t_{\rm dyn}} \propto  \frac{N}{\ln(\gamma N)} \ .
\label{1.4}
\end{equation}
(cf. e.g.~\cite{Spitzer1987}). As a result, for very large $N$, dynamical equilibrium is attained much faster than thermodynamic equilibrium. Therefore, even if treated them as ``gaseous'' spheres, stellar systems evolve qualitatively different from stars; in stars the thermal timescale is short compared to the dynamical timescale~\cite{BettwieserSugimoto1984}. Another interesting consequence of the long thermal timescale in star clusters is that anisotropy can prevail for many dynamical times. If one assumes a purely kinetic temperature definition, it ensues that in star clusters the temperatures (or velocity dispersions) can remain different for different coordinate directions over many dynamical times. For example, in a spherical system (using polar coordinates) the radial velocity dispersion of stars (``temperature'') $\s_{r}^2$ could be different from the tangential one $\s_{t}^2$. For the relaxation time above the 3D velocity dispersion $\s^2 = \s_{r}^2 + 2\s_{t}^2$ is used. If axisymmetric or triaxial the tangential velocity dispersion can be decomposed into two different dispersions $2\s_{\mathrm{t}}^2 = \s_\theta^2+\s_\phi^2$. \\
A full account on the relevance of anisotropy for star clusters is beyond the scope of this paper; exemplary we mention here that interest in anisotropy was recently sparked by anisotropic mass segregation in rotating star clusters, both globular and nuclear~\cite{Szolgyenetal2018,Szolgyenetal2019,Szolgyenetal2021,Torniamentietal2019,Kamlahetal2022b}.

\section{Direct Fokker-Planck and moment models}
\begin{figure}[t]
\includegraphics[width=\textwidth]{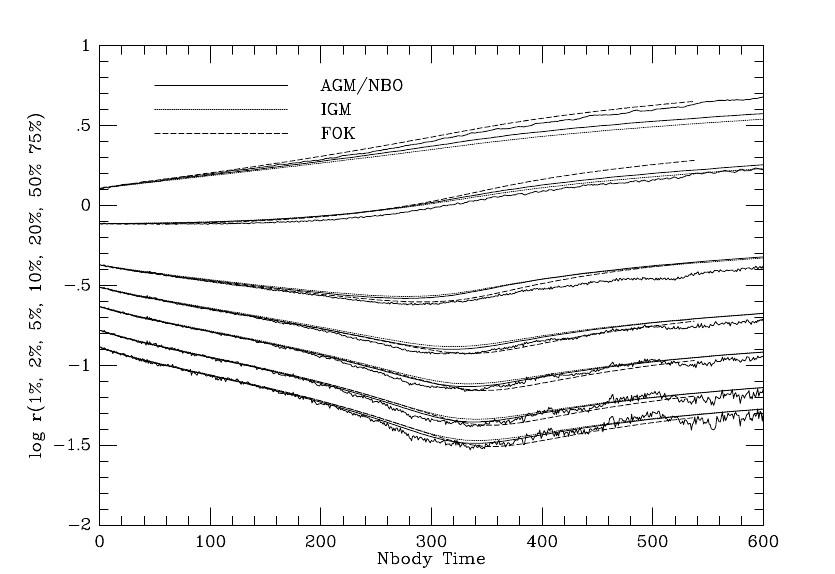}
 \caption{The time evolution of Lagrangian radii for star clusters modelled by an anisotropic gaseous model (AGM)~\cite{LouisSpurzem1991}, isotropic gas model~\cite{Heggie1984}, an isotropic Fokker-Planck model, and direct $N$-body simulation (from~\cite{GierszSpurzem1994}, the figure is also contained in Chapter 8 of~\cite{BinneyTremaine2008}.
 }
\label{FigGierszSp}
\end{figure}

Models based on the Fokker-Planck approximation (also denoted as approximate or statistical models) have been designed and implemented in times when it was very difficult to simulate large star clusters directly by $N$-body simulations. Dramatic development in hardware and software has made now possible direct $N$-body simulations of up to a million bodies with realistic astrophysics and binaries~\cite{Wangetal2015,Wangetal2016} (\texttt{Dragon} simulations). However, the use of the approximate models is very useful to understand the nature of physical processes in star clusters (such as heat conduction or viscosity); by comparison with $N$-body models they can mutually support each other. \\
\textcolor{violet}{The Fokker-Planck approximation to describe two-body relaxation in spherical star clusters is the foundation for all Monte Carlo models used nowadays (MOCCA and CMC, see Sect.~\ref{Monte Carlo Models} below). Therefore it is deemed useful to provide a deeper than usual insight into its theoretical foundations here.}\\
Statistical models have been employed to clarify the physical nature of relaxation processes in star clusters, such as core collapse~\cite{LyndenBellWood1968}, post-collapse evolution due to an energy source from binaries undergoing close encounters with single stars~\cite{InagakiLyndenBell1983}, gravothermal oscillations~\cite{SugimotoBettwieser1983,BettwieserSugimoto1984,CohnHutWise1989}. These methods have also been important for the study of anisotropy, mass segregation, and rotation later on.\\
Comparison and mutual adjustment of parameters in order to get agreement between statistical models and direct $N$-body simulations was started by~\cite{Aarsethetal1974} for a first pre-collapse comparison, followed by an extended study using statistical averages of $N$-body simulations to match gaseous models~\cite{GierszHeggie1994a,GierszHeggie1994b,GierszSpurzem1994}, see also Fig.~\ref{FigGierszSp}; in~\cite{SpurzemAarseth1996} it was shown that relaxation processes consistent with theory dominate core collapse in star clusters, and~\cite{Makino1996} showed for the first time a signature of gravothermal oscillations in a $N$-body simulation. 
Classical models based on the Fokker-Planck approximation use quite strong approximations, like spherical symmetry (in general, with some exceptions allowing axisymmetry), dominance of relaxation encounters, modelling all few-body effects (binary-single and binary-binary close encounters) in a statistical way. A mass spectrum would be modelled by discrete dynamical components with different masses (except Monte Carlo models, see below). With the increasing need to have star cluster models matching detailed observations of star clusters, the use of Fokker-Planck type models was no more practical. That hardware and software developments made more and more realistic particle numbers possible in direct $N$-body simulations has been another reason for the decline of statistical models. \\
There is one remarkable exception, the Monte Carlo technique - while it is also based on Fokker-Planck theory it uses a quasi $N$-body realization and allows state-of-the-art models up to the current time (see Sect.~\ref{Monte Carlo Models}). In the following subsections, nevertheless, we will outline the basic theory beneath the statistical models, because in some areas (rotating, flattened star clusters, NSCs, anisotropy) they are still important today in order to analyse results of $N$-body simulations.

\subsection{Fokker-Planck approximation}
\label{Fokker-Planck}
The Fokker-Planck approximation truncates the so--called BBGKY hierarchy {\color{red}(named after Bogoliubov–Born–Green–Kirkwood–Yvon)} of kinetic equations at lowest order assuming that for most of the time all particles are uncorrelated with each
other and only coupled via the smooth global gravitational potential 
\textcolor{red}{(see Chapter 8.1, Sects. 2 and 3 of~\cite{BinneyTremaine2008}; the following paragraph is a summary from their text). We start with most general $N$-particle distribution function $f^{(N)}$, which depends on positions and velocities $\vec{r_i},\vec{v_i}$ of a set of $N$ particles and time $t$:
\begin{equation}
f^{(N)} (\vec{r_i},\vec{v_i},i=1,\ldots,N,t)
\end{equation}
It provides a probability to find all the particles at their given positions and velocities. In $6N$ dimensional phase space the particles are an incompressible fluid following Liouville's ``big'' equation
\begin{equation}
\frac{D}{Dt}f^{(N)} = 0
\label{eq:Liouville}
\end{equation}
where the derivative is the Lagrangian derivative. 
If all particles are uncorrelated $f^{(N)} = (f^{(1)})^N$, i.e. the $N$-particle distribution function is just the $N$-fold product of a single particle distribution function. That is the case for example in a collisionless stellar system, where all particles just follow their trajectories determined by a global smooth gravitational potential and any direct interaction between two or few particles (stars) is negligible. For collisional stellar systems, however, gravitational encounters (two-body relaxation) change the phase space distribution, particles are not uncorrelated anymore.
The theoretical ansatz in that case would be to define a two-body correlation function $g$ by
\begin{equation}
f^{(2)} (\vec{r_i},\vec{v_i},i\!=\!1...2,t)\!=\!f^{(1)} (\vec{r_1},\vec{v_1},t)\cdot f^{(1)} (\vec{r_2},\vec{v_2},t) + g(\vec{r_i},\vec{v_i},i\!=\!1...2,t)
\end{equation}
So, from knowing $f^{(1)}$ and $g$ we get $f^{(2)}$; by using higher order correlation functions one can get from $f^{(n)}$ to $f^{(n+1)}$. Integration of Eq.~\ref{eq:Liouville} step-by-step over single particles provides a sequence of equations for $f^{(N-1)}$ to $f^{(1)}$, which is the BBGKY hierarchy. However it is usually not very helpful, because all the correlation functions for $2,...N\!-\!1$ need to be known. For practical purposes in collisional stellar systems, where two-body relaxation is important (e.g. open, globular, and nuclear star clusters), it is sufficient to deal with the two-body correlation, which is done phenomenologically in two different ways for distant and close correlations (encounters and binaries), as described below.\\
Higher than two-body correlations are rarely important. There could be a relation to Sundmann's famous theorems~\cite{Sundman1907,Sundman1909}, which state that in the three-body problem direct three-body collisions occur only with a negligibly small probability; it means whenever three particles get close to each other, there will be always a sequence of separate close two-body encounters, practically never\footnote{meaning the set of initial conditions, which lead to a triple collision is mathematically of measure zero compared to the general space of initial conditions; in physical language: what is needed is that three bodies have in total zero angular momentum with respect to their center of mass.}
the three bodies will simultaneously get extremely close together. Burrau's three body problem is a nice demonstration of that behaviour~\cite{Szebehely1967}; whether the fundamental assumption of dominance of two-body correlations is in fact realized or not can only be checked computationally by comparison of models based on the Fokker-Planck approximation (such as also Monte Carlo models) with direct $N$-body simulations. An example for a situation of a very high density in a collapsing core of a star cluster, where higher correlations become important, can be found in \cite{Tanikawaetal2012}.}\\
Instead of determining a general correlation function one resorts to a phenomenological description of the effects of collisions by computing local diffusion coefficients directly from the known solution of the two-body problems. Diffusion coefficients $D(\Delta v_i)$ and $D(\Delta v_iv_j)$ denote the average rate of change of $v_i$ and $v_iv_j$ due to the cumulative effect of many small angle deflections during two-body encounters, at a given radius $r$ (from here assuming spherical symmetry). Let $m$, $\vec{v}$ and $m_f$, $\vec{v_f}$ be the mass and velocity of a star from a test and field star distribution, respectively (both distributions can but need not to be the same). In Cartesian geometry (for the local velocities) the diffusion coefficients are defined by
\begin{equation}
 D(\Delta v_i)  = 4\pi G^2 m_f \ln\Lambda \parder{h(\vec{v})}{v_i}
 \ \ ;\ \ D(\Delta v_i\Delta v_j)  = 4\pi G^2 m_f \ln\Lambda\cdot
    \frac{\partial^2 g(\vec{v})}{\partial v_i \partial v_j} \ \ ;
\label{2.1.20}
\end{equation}
Local means here that we do not explicitly consider the dependence of $f$ on the spatial coordinate. $g$, $h$ are
the Rosenbluth potentials defined in~\cite{Rosenbluthetal1957}
\begin{equation}
 h(\vec{v})  = (m+m_f)
 \int \frac{f(\vec{v}_f)}{\vert\vec{v}-\vec{v}_f\vert}
 d^3\!\vec{v}_f \ \ ;\ \
 g(\vec{v})  = m_f \int f(\vec{v}_f) \vert\vec{v}-\vec{v}_f\vert
 d^3\!\vec{v}_f \ .
\label{2.1.21}
\end{equation}
Note that provided the distribution function $f$ is given in terms
of a convenient polynomial series as in Legendre polynomials the
Rosenbluth potentials can be evaluated analytically to arbitrary
order, as was seen already by~\cite{Rosenbluthetal1957}, see for a modern re-derivation and its use for star cluster dynamics~\cite{GierszSpurzem1994,SpurzemTakahashi1995,Schneideretal2011}.
With these results we can finally write down the local Fokker-Planck
equation in its standard form
for the Cartesian coordinate system of the $v_i$:
\begin{eqnarray}
\label{2.1.22}
\parder{f}{t} &+& \vec{v}_i\parder{f}{\vec{r}_i}+\vec{{\dot v}}_i
\parder{f}{\vec{v}_i} = \dedet{f}{enc} \ ; \\
\dedet{f}{enc} &=&
   - \sum_{i=1}^3 \parder{}{v_i}\Bigl[
        f(\vec{v}) D(\Delta v_i) \Bigr]
    + \frac{1}{2} \sum_{i,j=1}^3
       \frac{\partial^2}{\partial v_i\partial v_j}\Bigl[
        f(\vec{v})
         D(\Delta v_i\Delta v_j)
       \Bigr] \ .
\label{2.1.23}
\end{eqnarray}
The subscript ``enc'' should refer to encounters, which are the driving
force of two-body relaxation.
Still Eq.~\ref{2.1.22} is a six-dimensional integro-differential equation; its direct numerical simulation in stellar dynamics can presently only be done by further simplification. If the encounter term is zero, Eq.~\ref{2.1.22} is transformed into Liouville's equation for a collisionless system. For a self-gravitating system Eqs.~\ref{2.1.22} and \ref{2.1.23} are not sufficient, since the knowledge of the gravitational potential $\Phi$ is necessary. This can be seen above from the $\vec{\dot{v}}_i$ term - its computation requires to know the gravitational force. In moment or gas models described below (for spherical symmetry) Poisson's equation takes the simple form Eq.~\ref{2.3.7}; for orbit averaged Fokker-Planck or Monte Carlo models (see Sect.~\ref{OrbitAverage} and \ref{MOCCA}) the gravitational potential enters directly into the energy as constant of motion (cf. Eq.~\ref{radialvelocity}).

\subsection{Moment or Gas Models}
The local Fokker-Planck equation (Eq.~\ref{2.1.22}) is utilized in another
way for gaseous or conducting gas sphere models of star clusters. Integrating
it over velocity space with varying powers of the velocity coordinates
yields a system of equations in the spatial coordinates; the local
approximation is used in the sense that the orbit structure of the
system is not taken into account, diffusion coefficients and all
other quantities are assumed to be well defined just as a function of
the local quantities (density, velocity dispersions and so on). The system
of moment equations is truncated in third order by a phenomenological equation of heat transfer. Such an approach has been suggested by
\cite{LyndenBellEggleton1980,Heggie1984} and generalized to anisotropic systems by~\cite{Bettwieser1983,BettwieserSpurzem1986,LouisSpurzem1991}. In the following the derivation of the model
equations is described.

\subsubsection{The ``Left Hand Sides''}
\label{sect2.3.1}
In spherical symmetry, polar coordinates $r$
$\theta $, $\phi$
are used and $t$ denotes the time. The vector
$\vec{v} = (v_i), i=r,\theta,\phi$, denotes the velocity
in a local Cartesian coordinate system at the spatial point
$r,\theta,\phi$. In the interest of brevity
$u=v_r$, $v=v_\theta$, $w=v_\phi$ is used. The distribution function $f$,
which due to spherical symmetry is a function of $r$, $t$,
$u$, $v^2+w^2$ only, is normalized according to
\begin{equation}
\rho(r,t) = \int f(r,u,v^2+w^2,t) du\,dv\,dw,
\label{2.2.1}
\end{equation}
where $\rho(r,t)$ is the mass density; if $m$ denotes
the stellar mass, we get the particle density $n=\rho/m$. Then
\begin{equation}
{\bar u} = \int u f(r,u,v^2+w^2,t) du\,dv\,dw,
\label{2.2.2}
\end{equation}
is the bulk radial velocity of the stars.
Note that for the analogously defined quantities ${\bar v}$ and
${\bar w}$ we have in spherical systems
${\bar v} = {\bar w} = 0$ (rotating, axisymmetric systems: ${\bar w} \ne 0$).
In order to go ahead to the anisotropic gaseous model equations
we now turn back to the left hand side of the Fokker-Planck
equation Eq.~\ref{2.1.22}, which is the collisionless Boltzmann or Vlasov operator.
For practical reasons we prefer for the left hand side
local Cartesian velocity coordinates, whose
axes are oriented towards the $r$, $\theta$, $\phi$ coordinate space
directions. With the Lagrange function
\begin{equation}
{\cal L} = \frac{1}{2}\bigl({\dot r}^2 + r^2{\dot\theta}^2 +
         r^2 \sin^2\!\!\theta\, {\dot\phi}^2\bigr) - \Phi(r,t)
\label{2.3.1}
\end{equation}
the Euler-Lagrange equations of motion for a star moving in
the cluster potential $\Phi$ become:
\begin{equation}
  {\dot u}  = - \parder{\Phi}{r} + \frac{v^2\!+\!w^2}{r} \ \ ;\ \
  {\dot v}  = - \frac{uv}{r} + \frac{w^2}{\tan\theta} \ \ ;\ \
  {\dot w}  = - \frac{uw}{r} - \frac{vw}{r\tan\theta} \ \ .
 \label{2.3.2}
\end{equation}
The complete local Fokker-Planck equation, derived from Eq.~\ref{2.1.22},
attains the form
\begin{equation}
\parder{f}{t} + u\parder{f}{r} + {\dot u}\parder{f}{u} +
     {\dot v}\parder{f}{v} + {\dot w}\parder{f}{w} = \dedet{f}{enc}\ ,
\label{2.3.3}
\end{equation}
where the term subscribed by ``enc'' denotes the terms
involving diffusion coefficients as in Eq.~\ref{2.1.23}.
Moments $\langle i,j,k \rangle $ of $f$ are defined in the
following way (all integrations range from $-\infty $ to $\infty $):
\begin{eqnarray}
 \langle 0,0,0\rangle &:=& \rho = \int f dudvdw  \ \ ; \ \
 \langle 1,0,0\rangle  :=  \bar{u} = \int uf dudvdw \\
 \langle 2,0,0\rangle &:=& p_r + \rho\bar{u}^2 = \int u^2 f dudvdw \\
 \langle 0,2,0\rangle &:=& p_\theta = \int v^2 f dudvdw \ \ ; \ \
 \langle 0,0,2\rangle  :=  p_\phi = \int w^2 f dudvdw \\
 \langle 3,0,0\rangle &:=&  F_r + 3\bar{u}p_r + \bar{u}^3 =
                            \int u^3 f dudvdw \\
 \langle 1,2,0\rangle &:=&  F_\theta + \bar{u}p_\theta = \int uv^2 f dudvdw \\
 \langle 1,0,2\rangle &:=&  F_\phi + \bar{u}p_\phi = \int uw^2 f dudvdw\ .
 \label{2.3.4}
\end{eqnarray}
Note that the definitions of $p_i$ and $F_i$ are such that they
are proportional to the random motion of the stars. Due to spherical
symmetry we have $p_\theta = p_\phi =: p_t$ and
$F_\theta = F_\phi =: F_t/2$. By $p_r = \rho\sr^2$ and $p_t = \rho\s_t^2$
the random velocity dispersions are given, which are closely related
to observables in GCs and galaxies. It is convenient to define velocities of energy transport by
\begin{equation}
 v_r  = \frac{F_r}{3 p_r} + u \ \ ; \ \
 v_t  = \frac{F_t}{2 p_t} + u \ .
\label{2.3.5}
\end{equation}
By multiplication of the Fokker-Planck equation \ref{2.3.3} with
various powers of $u$, $v$, $w$ we get up to second order the
following set of moment equations for $\bar{u}$ dropped in the following):
 \begin{eqnarray}
 \parder{\rho}{t} + \div{u\rho} &=& 0 \\
 \label{2.3.6a}
 \parder{u}{t}+u\parder{u}{r} + \frac{GM_r}{r^2} +
     \frac{1}{\rho}\parder{p_r}{r} + 2\frac{p_r - p_t}{\rho r} &=& 0 \\
 \label{2.3.6b}
 \parder{p_r}{t} + \div{u p_r} +
 2 p_r \parder{u}{r} + \div{F_r} - \frac{2F_t}{r} &=&
  \dedet{p_r}{enc,bin3} \\
 \label{2.3.6c}
  \parder{p_t}{t} + \div{u p_t} +
 2 \frac{p_t u}{r} + \frac{1}{2}\div{F_t} + \frac{F_t}{r} &=&
  \dedet{p_t}{enc,bin3} \ .
 \label{2.3.6d}
\end{eqnarray}
The terms labeled with ``enc'' and ``bin3'' symbolically denote the collisional terms resulting from the moments of the right
hand side of the Fokker-Planck equation (Eq. \ref{2.1.23}) and an energy generation by formation and hardening of three body encounters. Both will be discussed below.
With the definition of the mass $M_r$ contained in a sphere of radius $r$
 \begin{equation}
\parder{M_r}{r} = 4 \pi r^2 \rho
\label{2.3.7}
\end{equation}
the set of Eqs. \ref{2.3.6a}--\ref{2.3.6d} is equivalent to gas-dynamical equations coupled with Poisson's equation. Since moment equations of order $n$ contain moments of order
$n\!+\!1$, it is necessary to close the system of the above equations by an independent closure relation. Here we choose the heat conduction closure, which consists of a phenomenological
Ansatz in analogy to gas dynamics. It was first used (restricted to isotropy) by~\cite{LyndenBellEggleton1980}. It is assumed that heat transport is proportional to the temperature gradient,
where we use for the temperature gradient an average velocity dispersion $\sigma^2 = (\sr^2 + 2\s_t^2)/3$ and assume $v_r = v_t$ (this latter closure was first introduced by~\cite{BettwieserSpurzem1986}). Therefore, the last two equations to close our model are
\begin{equation}
  v_r - u + \frac{\lambda}{4\pi G\rho t_{\rm rx}} \parder{\s^2}{r} = 0
  \ \ ; \ \ v_r = v_t  \ .
\label{2.3.11}
\end{equation}
With Eqs. \ref{2.3.6a}--\ref{2.3.6d}, \ref{2.3.7}, and \ref{2.3.11} we have now seven equations for our seven dependent variables $M_r$, $\rho$, $u$, $p_r$, $p_t$, $v_r$, $v_t$.
\subsubsection{The ``Right Hand Sides''}
\label{sect2.3.3}
All right hand sides of the moment equations \ref{2.3.6a}--\ref{2.3.6d} are calculated by multiplying the right hand side (the encounter term) of the Fokker-Planck equation as it occurs
in Eq.~\ref{2.1.23} with the appropriate powers of $u$, $v$ and $w$ and integrating over velocity space. Since the diffusion coefficients in Eq.~\ref{2.1.20} also contain the distribution function $f$, the equation depends non-linearly on it. That has led in the early papers to a simplification by using an isotropized background distribution function $f_b$ inside the diffusion coefficients, different from the actual one~\cite{Larson1970a,Cohn1980}. In~\cite{BettwieserSpurzem1986,EinselSpurzem1999,Schneideretal2011} there is always full consistency between the background and the actual distribution function.

\subsection{Orbit averaged Fokker-Planck models and rotation}
\label{OrbitAverage}
The direct solution of the six-dimensional integro-differential equations (Eq.~\ref{2.1.22} and Eq.~\ref{2.1.23}) is generally not possible. To have numerical solutions of the Fokker-Planck equation directly one applies Jeans's theorem and transforms $f$ into a function of the classical integrals of motion of a particle in a potential under the given symmetry, as e.g. energy $E$ and modulus of the angular momentum $J^2$ in a spherical potential or $E$ and $z$-component of angular momentum $J_z$ in axisymmetric coordinates. Thereafter, the Fokker-Planck equation can be integrated over the accessible coordinate space for any given combination of constants of motion and the orbit-averaged Fokker-Planck equation ensues. By transformation from $v_i$ to $E$ and $J$ and via the limits of the orbital integral the potential enters both implicitly and explicitly. In a two-step scheme alternatively solving the Poisson- and Fokker-Planck equation a direct numerical solution is obtained for spherical systems in 1D (using $E$ only~\cite{Cohn1980}), or in 2D (using both $E$ and $J^2$~\cite{Cohn1979,Takahashi1996a,Takahashi1996b,Takahashi1997}).

\begin{table}
\centering
\begin{tabular}{lcccccc}
\hline \hline $\omega_0$ & $T_{\text {rot }} / T_{\text {kin }}$ (\%) & $e_{\text {dyn }}(0)$ & $r_{\text {tid }} / r_{\mathrm{c}}(0)$ & $r_{\mathrm{h}} / r_{\mathrm{c}}(0)$ & $\tau_{\mathrm{rc}}(0)$ & $\tau_{\mathrm{rh}}(0)$ \\
\hline 0.00 & 0.00 & -0.001 & 18.72 & 2.70 & 19.24 & 91.88 \\
0.05 & 0.23 & 0.002 & 18.61 & 2.70 & 19.23 & 91.77 \\
0.10 & 0.89 & 0.013 & 18.25 & 2.68 & 19.22 & 90.80 \\
0.20 & 3.38 & 0.051 & 16.83 & 2.66 & 19.20 & 89.71 \\
0.30 & 7.00 & 0.105 & 14.99 & 2.62 & 19.21 & 87.73 \\
0.40 & 11.23 & 0.165 & 13.08 & 2.55 & 19.22 & 84.12 \\
0.50 & 15.61 & 0.224 & 11.46 & 2.48 & 19.27 & 80.49 \\
0.60 & 19.81 & 0.278 & 9.94 & 2.39 & 19.40 & 76.32 \\
0.70 & 23.71 & 0.327 & 8.77 & 2.30 & 19.50 & 71.78 \\
0.80 & 27.18 & 0.368 & 7.69 & 2.20 & 19.71 & 67.37 \\
0.90 & 30.25 & 0.403 & 6.88 & 2.12 & 19.86 & 63.24 \\
1.00 & 32.99 & 0.433 & 6.22 & 2.04 & 20.02 & 59.63 \\
\hline
\end{tabular}
\caption{\textcolor{red}{Initial conditions of rotating King models from \cite{EinselSpurzem1999} with $W_0=6$. $T_{\mathrm{rot }} / T_{\mathrm{kin }}$ is the ratio of bulk rotational energy to total kinetic energy in percent, $e_{\text {dyn }}(0)$ is the dynamical ellipticity, $r_{\mathrm{tid }} / r_{\mathrm{c}}(0)$ is the ratio of the tidal radius to core radius, $r_{\mathrm{h}} / r_{\mathrm{c}}(0)$ is the ratio of the half-mass radius to core radius, $\tau_{\mathrm{rc}}(0)$ is the central relaxation timescale and $\tau_{\mathrm{rh}}(0)$ is the half-mass relaxation timescale. All of these quantities are shown for $t=0$ of system time units. Table data are from \cite{EinselSpurzem1999}, but note that in this paper the per cent sign (\%) in the header for $T_{\mathrm{rot }} / T_{\mathrm{kin }}$ had been omitted erroneously.}}
\label{einsel-table1}
\end{table}

One of the main uncertainties in this method is that for non-spherical mass distributions the orbit structure in the system may depend on unknown non-classical third integrals of motion which are neglected.
First 2D models of axisymmetric, rotating globular star clusters~\cite{EinselSpurzem1999} used initial models obtained published earlier~\cite{Goodman1983,LuptonGunn1987,LongarettiLagoute1996}, which are generalizations of the standard King models~\cite{King1966a}, using its dimensionless central potential value $W_0$ and a new dimensionless rotation parameter $\omega_0$. These models have a rigid rotation in the core, maximum of the rotation curve around the half-mass radius and a differentially decreasing rotation in the halo. They are still mainly supported by ``thermal'' pressure (velocity dispersions), the rotational energy provides a smaller part of the energy. Typical initial data for $W_0=6$ and different rotation parameters are seen in Table~\ref{einsel-table1}: the ratio of total rotational to total kinetic energy,
dynamical ellipticity, ratios of tidal and half-mass radii to initial core radii, the central relaxation time and finally the half-mass relaxation time in system units are given (for definitions see~\cite{EinselSpurzem1999}). 
\begin{figure}[ht]
\includegraphics[width=\textwidth]{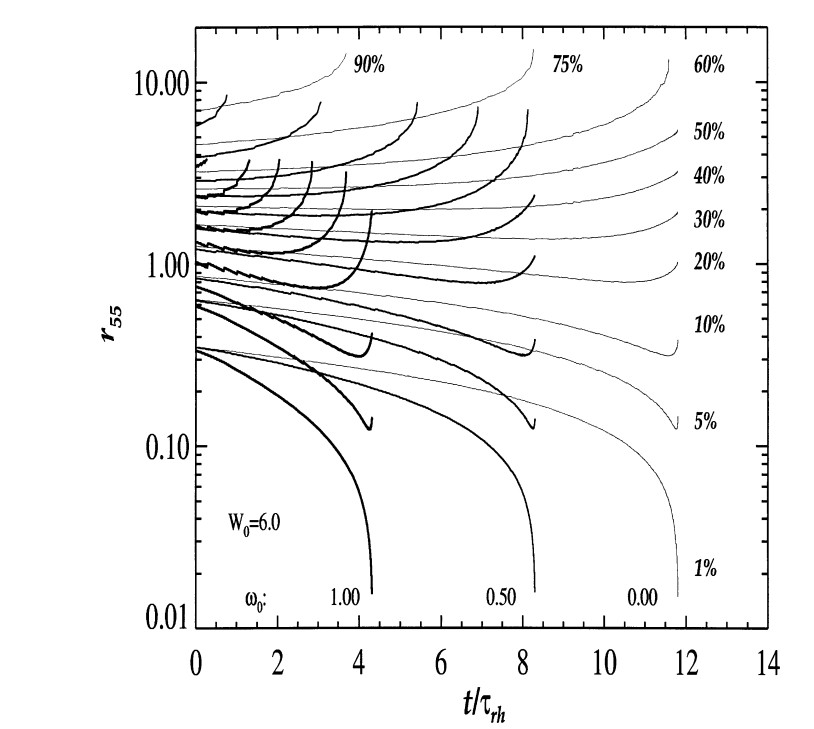}
 \caption{Evolution of mass shells (Lagrange radii $r_{55}$) for models with King central potential parameter $W_0=6$, and three different dimensionless rotation parameters (see Table~\ref{einsel-table1}) $\omega_0 = 0.0, 0.3, 0.6$ as indicated. Shown are the radii for mass columns containing the indicated percentage
of total initial mass in the direction of the $\theta = 54^{\circ}.74$ angle. From~\cite{EinselSpurzem1999}.}
\label{einselspfig1}
\end{figure}

Fokker-Planck models showed that in presence of rotation there is an effective viscosity transporting angular momentum outwards and accelerating cluster evolution significantly as compared to a spherical cluster (see Fig.~\ref{einselspfig1} and~\cite{EinselSpurzem1999}). A series of follow-up papers include post-collapse and multi-mass models~\cite{Kimetal2002,Kimetal2004,Kimetal2008} and found an accelerated rotation in the core for heavy masses sinking to the core - as it was predicted by the combined gravogyro and gravothermal ``catastrophes'' predicted by~\cite{Hachisu1979,AkiyamaSugimoto1989}. One rotating model included in the \texttt{Dragon} simulations~\cite{Wangetal2016}, however, did not show accelerated evolution. Whether this is due to heavy mass loss by stellar evolution (not included in earlier papers) or due to a small deviation from the proper initial model is not clear. There is an urgent need for more coverage of rotating stellar clusters by direct $N$-body simulations, see some first progress~\cite{Tiongcoetal2022,Livernoisetal2022,Kamlahetal2023}.
The initial models of Table~\ref{einsel-table1} are still in frequent use, in particular if realized as $N$-body configurations for $N$-body models~\cite{Hongetal2013,Tiongcoetal2017,Tiongcoetal2022,Livernoisetal2022,Kamlahetal2023}. Notice also the alternative rotating models of Varri~\cite{VarriBertin2012,Varrietal2018}, which are more suitable with regard to the outer cluster zones under influence of tidal fields.

\section{Monte Carlo Models}
\label{Monte Carlo Models}
Monte Carlo models of star clusters are the only ones which are still intensively used up to the present time, even though they are based on the Fokker-Planck approximation, in the same way as Fokker-Planck or gaseous/moment models. Sometimes this may not be clear to every reader of current papers using Monte Carlo models, because they provide data equivalent to $N$-body simulations - particles with masses, positions and velocities at certain times. Astrophysics (stellar single and binary evolution, stellar collisions, relativistic binaries...) has been included very much like in $N$-body models. This review is not about Monte Carlo models, but a brief summary of their history and entry points to the current literature should be given.

\subsection{H{\'e}non and Spitzer type method}
\label{HenonSpitzer}
As the name suggests Monte Carlo models are based on the principle that stars have an orbit in a known self-consistent potential; random perturbations are applied, which model the effect of relaxation by distant gravitational encounters. Spitzer's method follows the orbits of stars in the global potential of the cluster and randomly applies kicks in velocity to the stars; at the end of a long series of papers they included binaries and a mass spectrum~\cite{Spitzeretal1971a,Spitzeretal1971b,Spitzeretal1972a,Spitzeretal1972b,Spitzeretal1973,Spitzeretal1975a,Spitzeretal1975b,Spitzeretal1980}.\\
H{\'e}non's method is using the phase space of constants of motion of a star in a spherically symmetric potential, energy and angular momentum. Deflections are selected randomly, and their effect on angular momentum and energy computed and applied~\cite{Henon1971}. The method was extended to include astrophysical effects, including binaries and stellar evolution~\cite{Stodolkiewicz1982,Stodolkiewicz1986}. These models still allowed for ``superstars'', i.e. one particle in the Monte Carlo model could represent many real stars.\\
Current Monte Carlo models are based on H{\'e}non's method, but restricted to star-by-star modelling (much like $N$-body), where every star is a particle in the Monte Carlo simulation. This only made it possible to include all astrophysical effects in the same way than it is done in $N$-body simulations. This new line of Monte Carlo models was initiated by Giersz~\cite{Giersz1998} (code name \textsc{MOCCA}) and the Northwestern team~\cite{Joshietal2000} (code name \textsc{CMC}). 

\subsection{\textsc{MOCCA} and \textsc{CMC}}
\label{MOCCA}
The Monte Carlo codes based on the H{\'e}non scheme use constants of motion (specific energy $E$, specific angular momentum $L$) as basic variables, properties of stars in the simulation. If the spherically symmetric gravitational potential $\Phi(r)$ is known, the pericenter $r_{\rm min}$ and apocenter $r_{\rm max}$ of the orbit are known. At every point of the orbit $r$ the radial velocity is known from
\begin{equation}
v_r = \pm\sqrt{2\Bigl(E - \Phi(r)\Bigr) - \frac{L^2}{r^2}}.
\label{radialvelocity}
\end{equation}
The orbital integral defines the orbital time $\tau$ by
\begin{equation}
\frac{\tau}{2} = \intl_{r_{\rm min}}^{r_{\rm max}} \frac{dr}{v_r}.
\end{equation}
With $p(r) = (2/\tau)\cdot (dr/v_r)$ one gets a probability distribution function, used to randomly pick a radial position $r_i$ for the star on its orbit (which should be distributed according to $p(r)$).
Let $m_i$ be the stellar mass of stars ($i=1,\ldots,n$), then the spherically symmetric gravitational potential can be computed according to~\cite{Henon1971}
\begin{equation}
\Phi(r) = G \left(-\frac{1}{r}\sum_{i=1}^k m_i - \sum_{i=k+1}^n\frac{m_i}{r_i}   \right).
\end{equation}
In addition to that two angles $\theta$ and $\phi$ are randomly picked, so as to have a three dimensional position of the star. Velocities are obtained from $E$, $L$, and $U(r_i)$ (one more random number needed). In that way a model star cluster is produced whose data structure is three dimensional - equivalent to that of an $N$-body simulation. To model the relaxation effect, two neighbouring stars are selected and a mean squared deflection angle chosen, which is proportional to the time-step over the relaxation time. Using this angle changes in $E$ and $L$ are computed. Binaries and close encounters between them have been first modelled completely stochastically as well (using random impact parameters, and using random realization of known cross sections). More recently a few-body integration is done in both \textsc{MOCCA} and \textsc{CMC} codes. This is a very rough account of Monte Carlo principles, the reader interested in more details is referred to the papers cited in the next paragraphs.\\
An account of the state of the \textsc{MOCCA} code and comparison with $N$-body simulations is published here~\cite{Gierszetal2013,Gierszetal2015}. Recently, it has been used for a large number of simulations of Galactic and extragalactic clusters, the \textsc{MOCCA} Survey Database has been published~\cite{Hongetal2020b,Levequeetal2021,Levequeetal2022b,Levequeetal2023}. The \textsc{CMC} code\cite{Rodriguezetal2021b} has been developed in parallel, with matching models to observations~\cite{Ruietal2021}, and an overview of the current state of the code~\cite{Rodriguezetal2022}. Examples of current use of this code focus on compact remnants and their gravitational wave emission, such as e.g.~\cite{Rodriguezetal2021a,Kremeretal2021,Yeetal2022}.\\
Both Monte Carlo codes have been very successful in terms of generating a large amount of cluster simulations to be compared with observational data and also to follow the evolution of special objects. However, one should not forget their serious limitations:
\begin{itemize}
    \item if we have a number of massive objects in a central high density region - the assumption of a smooth spherically symmetric potential breaks down;
    \item at high densities and if many binaries are present, the assumption that there are uncorrelated two-body relaxation encounters and close few-body encounters, which can be clearly separated, breaks down. 
    \item Taking into account external tidal fields is quite difficult, though in simple cases not impossible, due to the strictly spherical cluster centered gravitational potential. 
\end{itemize}
The bottom line is that Monte Carlo models have to be used in order to get an overview of large parameter ranges of star cluster evolution, but in many cases a check by comparison with direct $N$-body simulations is desirable. They do not suffer from all the problems mentioned above; however, also direct numerical solutions of the $N$-body problem have certain issues, see Sect.~\ref{reliable}. A nice overview of current Monte Carlo models is in \cite{Vasiliev2015}, who also present a somewhat restricted Monte Carlo code for rotating systems (see Sect.~\ref{Rotation}). 

\section{Direct $N$-body Simulations -- Methods and Algorithms}
To integrate the orbits of particles in time under their mutual
gravitational interaction the total gravitational potential at each
particle's position is required. Poisson's equation in integral
form gives the potential $\Phi$ generated at a point in coordinate
space $\vec{r}$ due to a smooth mass distribution $\rho(\vec{r})$
\begin{equation}
\Phi(\vec{r}) = - G \int \frac{\rho(\vec{r}^\prime)}
      {\vert\vec{r}^\prime - \vec{r}\vert} d^3\vec{r}^\prime  \ \ .
\label{4.1}
\end{equation}
A discrete particle distribution in $N$-body simulations is given by
\begin{equation}
\rho(\vec{r}) = \sum_{i=1}^N\,m_i\,\delta(\vec{r}-\vec{r}_j)
\end{equation}
with $N$ particles of mass $m_i$ distributed at positions $\vec{r}_j$. 
Putting this into the integral Poisson equation (\ref{4.1}) we get Newton's law for point masses:
\begin{equation}
\Phi(\vec{r}) = - G \sum_{j=1}^N \frac{m_j}{\vert\vec{r}-\vec{r}_j\vert}.
\end{equation}

\subsection{\textsc{Nbody} - the growth of an industry}

\begin{table}
	\centering
	\resizebox{\textwidth}{!}{%
\begin{tabular}{>{\color{violet}}c>{\color{violet}}l>{\color{violet}}l}
\hline \hline Year & \multicolumn{1}{>{\color{violet}}c}{ Keyword } & \multicolumn{1}{>{\color{violet}}c}{ Reference } \\
\hline $1961 \ldots \ldots$ & Force polynomial & \cite{Aarseth1963} \\
& Individual time steps & \cite{Aarseth1963} \\
& Gravitational softening & \cite{Aarseth1963} \\
$1966 \ldots \ldots$ & Spherical harmonics & \cite{Aarseth1967} \\
$1969 \ldots \ldots$. & Two-body KS regularization & \cite{KustaanheimoStiefel1965} \\
$1972 \ldots \ldots$ & Three-body regularization & \cite{AarsethZare1974} \\
$1973 \ldots \ldots$. & Global regularization & \cite{Heggie1974} \\
& AC neighbor scheme & \cite{AhmadCohen1973} \\
$1978 \ldots \ldots$. & Co-moving coordinates & \cite{Aarseth1979} \\
$1979 \ldots \ldots$ & Regularized AC & \cite{Aarseth1985a} \\
$1980 \ldots \ldots$. & Planetary formation & \cite{LecarAarseth1986}\\
$1986 \ldots \ldots$. & Hierarchical block-time steps & \cite{McMillan1986} \\
$1989 \ldots \ldots$. & Chain regularization & \cite{MikkolaAarseth1990} \\
$1990 \ldots \ldots$. & Particle in box scheme & \cite{Aarsethetal1993} \\
$1991 \ldots \ldots$ & Collisional tree code & \cite{McMillanAarseth1993} \\
$1992 \ldots \ldots$. & Chain $N$-body interface & \cite{Aarseth1994} \\
$1993 \ldots \ldots$ & Hermite integration & \cite{Makino1991a,MakinoAarseth1992} \\
$1995 \ldots \ldots$. & Synthetic stellar evolution & \cite{Toutetal1997} \\
& Tidal circularization & \cite{Mardlign1995a,Mardling1995b} \\
& Slow chain regularization & \cite{MikkolaAarseth1998} \\
$1996 \ldots \ldots$ & Hierarchical stability & \cite{MardlingAarseth1999} \\
$1998 \ldots \ldots$. & Evolution of hierarchies & \cite{MardlingAarseth1999} \\
& Stumpff KS method & \cite{MikkolaAarseth1998} \\
$1999 \ldots \ldots$ & HARP-6 procedures & \cite{Aarseth1999b} \\
& Symplectic integrators & \cite{MikkolaTanikawa1999a,MikkolaTanikawa1999b} \\
& \textsc{Nbody6++} SPMD / MPI acceleration & \cite{Spurzem1999} \\
$2000 \ldots \ldots$ & Single stellar evolution (\textsc{SSE}) & \cite{Hurleyetal2000} \\
$2002 \ldots \ldots$ & Binary stellar evolution (\textsc{BSE}) & \cite{Hurleyetal2002b} \\
$2003 \ldots \ldots$ & GRAPE-6 procedures & \cite{Makinoetal2003} \\
$2006 \ldots \ldots$ & 2.5PN in \textsc{Nbody5} & \cite{Kupietal2006} \\
$2007 \ldots \ldots$ & Direct $N$-body GPU acceleration & \cite{PortegiesZwartetal2007} \\
$2008 \ldots \ldots$ & AR with PN terms & \cite{MikkolaMerritt2008} \\
$2010 \ldots \ldots$ & Updated AR for few-body problems & \cite{HellstroemMikkola2010} \\
$2012 \ldots \ldots$ & \textsc{Nbody} codes GPU acceleration & \cite{NitadoriAarseth2012} \\
$2013 \ldots \ldots$ & MPI acceleration on GPU clusters /  \textsc{ phiGPU} & \cite{Bercziketal2013} \\
& 3.5PN in \textsc{Nbody6} & \cite{Bremetal2013} \\
$2015 \ldots \ldots$ & SSE/AVX acceleration on GPU clusters & \cite{Wangetal2015} \\
$2017 \ldots \ldots$ & Forward symplectic integrators (FSI)  & \cite{DehnenHernandez2017} \\
$2020 \ldots \ldots$ & P${}^3$T with SDAR in \textsc{PeTar} & \cite{Wangetal2020c} \\
& & \cite{Wangetal2020d} \\
$2021 \ldots \ldots$ & Minimum spanning tree \textsc{MSTAR}/\textsc{BiFrost} & \cite{Rantalaetal2021}\\
\hline
\end{tabular}}
\caption{\color{violet} Table showing important algorithmic, hardware and software development stepping stones in the development of direct $N$-body codes. The table is adapted from \cite{Aarseth1999b}, corrected in some places, but expanded to more recent developments. The abbreviations used are as follows:}
\begin{itemize}
\color{violet}
    \item KS: Kustaanheimo-Stiefel 
    \item AC: Ahmad-Cohen
    \item HARP-6 / GRAPE-6: special-purpose computers named Hermite AcceleratoR Pipeline-6 / GRAvity piPE-6
    \item PN: Post-Newtonian
    \item SPMD / MPI: Single Program Multiple Data scheme/Message Passing Interface.
    \item AR: Algorithmic chain 
    \item GPU: graphics processing unit
    \item SSE / AVX : Advanced Vector Extension/Streaming SIMD (Single Instruction, Multiple Data) Extension for vectorization in the CPU (central processing unit).
    \item P${}^3$T / SDAR: particle-particle particle-tree/Slow-Down Agolrithmic chain
    \item MSTAR / BiFrost: Minimum spanning tree + algorithmic chain / Binaries in Frost 
\end{itemize} 
\label{Algorithmic stepping stones}
\end{table}
It was already discovered by Sebastian von Hoerner in the earliest published $N$-body simulations that the relaxation time~\cite{Chandrasekhar1942} is relevant for star cluster evolution and that the formation of close and eccentric binaries occurs as the rule rather than as an exception. It was particularly difficult to accurately integrate them, effectively the simulation would be stopped if close binaries demanded too small time-steps~\cite{vHoerner1960,vHoerner1963}.\\
About at the same time a young postdoc - Sverre Aarseth - in Cambridge developed a direct $N$-body integrator for galaxy clusters with gravitational softening, thereby avoiding von Hoerner's problems with tight binaries~\cite{Aarseth1963}. His code was based on Taylor series evaluation of the gravitational force up to its second derivative. Eight years later regularization methods~\cite{KustaanheimoStiefel1965} were implemented in Aarseth's direct $N$-body code~\cite{Aarseth1971}. This allowed to proceed past the binary deadlock detected in von Hoerner's models.\\
Another direct $N$-body code by Roland Wielen appeared on the market, and in a seminal paper~\cite{Aarsethetal1974} fair agreement was shown between Aarseth's and Wielen's codes and a Monte Carlo code by Lyman Spitzer (see above Sect.~\ref{HenonSpitzer}). However, only at the turn of the century Aarseth and von Hoerner could compare their codes, and von Hoerner published a remarkable account on ``how it all started''~\cite{vHoerner2001}.\\
Already in 1985 the code \textsc{Nbody5}~\cite{Aarseth1985a} had become a kind of ``industry standard'', attaining world wide use. It employed Taylor series using up to the third derivative of the gravitational force, in a divided difference scheme based on four time points, with individual particle time-steps. Also there were regularizations for more than two bodies, such as the classical chain regularization~\cite{MikkolaAarseth1990}, and the Ahmad-Cohen~\cite{AhmadCohen1973} neighbour scheme already in \textsc{Nbody5}. The advent of vector and parallel computers demanded an optimization towards hierarchically blocked time-steps and the Hermite scheme (Sect.~\ref{Hermite})~\cite{McMillan1986,MakinoAarseth1992}, because it used only two time points, which made memory management easier. This became known as \textsc{Nbody6}.\\
The growth of the ``industry''\cite{Aarseth1999b} included further improvements in the regularization techniques~\cite{MikkolaAarseth1996,MikkolaAarseth1998,Aarseth1999a} and a comprehensive book summary~\cite{Aarseth2003b}. Table~\ref{Algorithmic stepping stones} summarizes the main algorithmic, hardware and software development stepping stones in the direct $N$-body community up until today. 

\subsection{The \textsc{Nbody6} scheme}
In the following, the \textsc{Nbody6} integrator is described in some more detail (note that \textsc{Nbody7} already contains parallelization through GPU acceleration and will be treated in the next section). \textsc{Nbody6} and its parallelized and GPU accelerated offspring (\textsc{Nbody6++, Nbody6GPU, Nbody6++GPU, Nbody7}, see Table~\ref{versions}) is still the most widely used method for direct $N$-body simulations, but recently also new approaches have been published (cf. Sect.~\ref{newapproaches}).

\subsubsection{The Hermite scheme}
\label{Hermite}
The Hermite scheme and \textsc{Nbody6} go back to~\cite{MakinoAarseth1992}; in conjunction with a hierarchically blocked time-step scheme (see below and~\cite{McMillan1986}) it improved the performance on vector computers and turned out to be efficient for all of recent parallel and innovative hardware (general and special purpose parallel computers, GRAPE and GPU). Assume a set of $N$ particles with positions $\vec{r}_i(t_0)$ and
velocities $\vec{v}_i(t_0)$ ($i=1,\ldots, N$) is given at time $t=t_0$, and let us look at a selected test particle at $\vec{r} = \vec{r}_0=\vec{r}(t_0)$ and $\vec{v} = \vec{v}_0 = \vec{v}(t_0)$. Note that here and in the following the index $i$ for the test particle $i$ and also occasionally the index $0$ indicating the time $t_0$ will be dropped for brevity; sums over $j$ are to be understood to include all $j$ with $j\ne i$, since there
should be no self-interaction. Accelerations $\vec{a}_0$ and their time derivatives ${\bf{\dot a}}_0$ are calculated explicitly:
\begin{equation}
 \vec{a}_0 = \sum_j Gm_j \frac{\vec{R}_j}{R_j^3} \ \ ; \ \
 \vec{{\dot a}}_0 = \sum_j Gm_j \left(
  \frac{\vec{V}_j}{R_j^3} - 3 (\vec{V}_j\!\cdot\!\vec{R}_j)
   \frac{\vec{R}_j}{R_j^5} \right) \ ,
\label{4.1.1}
\end{equation}
where $\vec{R}_j:=\vec{r}\!-\!\vec{r}_j$, $\vec{V}_j := \vec{v}\!-\!\vec{v}_j$, $R_j:=\vert\vec{R}_j\vert$, $V_j:=\vert\vec{V}_j\vert $. By low order predictions,
\begin{eqnarray}
 \vec{x}_p(t) &=& \frac{1}{6}(t-t_0)^3\vec{{\dot a}}_0
          +\frac{1}{2}(t-t_0)^2\vec{a}_0 + (t-t_0)\vec{v} + \vec{x} \ ,\\
 \vec{v}_p(t) &=& \frac{1}{2}(t-t_0)^2\vec{{\dot a}}_0
          + (t-t_0)\vec{a}_0 + \vec{v} \ ,
\label{4.1.2}
\end{eqnarray}
new positions and velocities for all particles at $t>t_0$ are calculated and used to determine a new acceleration and its derivative directly
according to Eq. \ref{4.1.1} at $t=t_1$, denoted by $\vec{a}_1$ and $\vec{{\dot a}}_1$. On the other hand $\vec{a}_1$ and $\vec{{\dot a}}_1$ can also be obtained from a Taylor series using higher derivatives of $\vec{a}$ at $t=t_0$:
\begin{eqnarray}
 \vec{a}_1 &=& \frac{1}{6}(t-t_0)^3 \vec{a}_0^{(3)} +
              \frac{1}{2}(t-t_0)^2 \vec{a}_0^{(2)} +
              (t-t_0)\vec{{\dot a}}_0 + \vec{a_0} \ ,\\
 \vec{{\dot a}}_1  &=& \frac{1}{2}(t-t_0)^2 \vec{a}_0^{(3)} +
               (t-t_0) \vec{a}_0^{(2)} + \vec{{\dot a}}_0 \ .
\label{4.1.3}
\end{eqnarray}
If $\vec{a}_1$ and $\vec{{\dot a}}_1$ are known from direct summation (from Eq. \ref{4.1.1} using the predicted positions and velocities) one can invert the equations above to determine the unknown higher order derivatives of the acceleration at $t=t_0$ for the test particle:
\begin{eqnarray}
 \frac{1}{2} \vec{a}^{(2)} &=& -3 \frac{\vec{a}_0 - \vec{a}_1}{(t-t_0)^2}
   - \frac{2\vec{{\dot a}}_0 + \vec{{\dot a}}_1}{(t-t_0)} \\
 \frac{1}{6} \vec{a}^{(3)} &=& 2 \frac{\vec{a}_0 - \vec{a}_1}{(t-t_0)^3}
   - \frac{\vec{{\dot a}}_0 + \vec{{\dot a}}_1}{(t-t_0)^2} \ ,
\label{4.1.4}
\end{eqnarray}
This is the Hermite interpolation, which finally allows to correct positions and velocities at $t_1$ to high order from
\begin{eqnarray}
 \vec{x}(t) = \vec{x}_p(t) + \frac{1}{24}(t-t_0)^4 \vec{a}_0^{(2)}
                                 +\frac{1}{120}(t-t_0)^5 \vec{a}^{(3)} \ ,\\
 \vec{v}(t) = \vec{v}_p(t) + \frac{1}{6}(t-t_0)^3 \vec{a}_0^{(2)}
                                 +\frac{1}{24}(t-t_0)^4 \vec{a}_0^{(3)} \ .
\label{4.1.5}
\end{eqnarray}

\begin{figure}
\includegraphics[width=\textwidth]{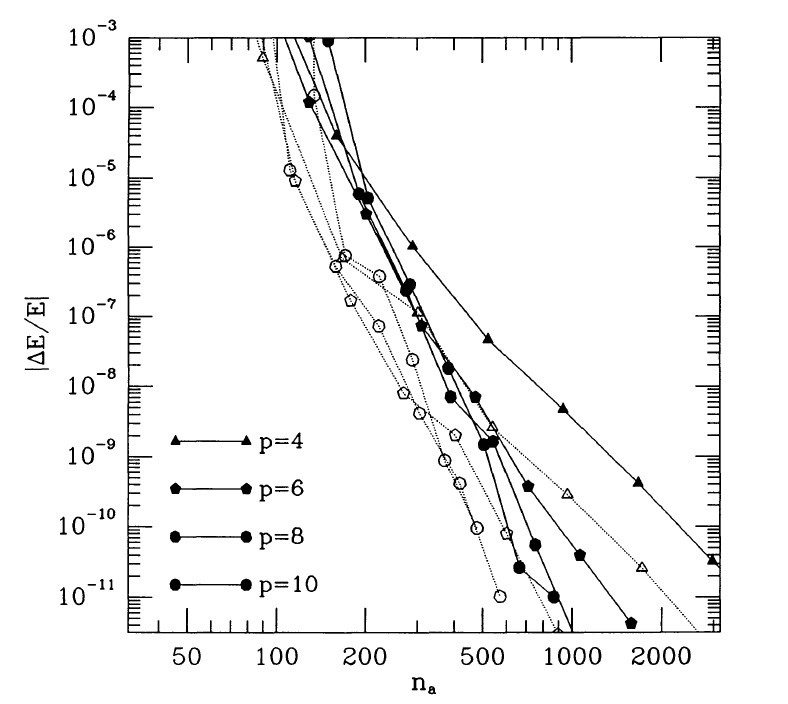}
 \caption{The relative energy error as the function of the number of steps.
 A time-step criterion using differences between predicted and corrected
 values is used, different from Eq. \ref{4.1.6}. Dotted curves are for
 Hermite schemes, solid curves for Aarseth schemes. The step number $p$
 denotes the order of the integrator. From~\cite{Makino1991a}.
 }
\label{Fig3Mak91}
\end{figure}
Taking the time derivative of Eq. \ref{4.1.5} it turns out that the
error in the force calculation for this method is $\co (\Delta t^4)$,
as opposed to standard leap-frog scheme, which has a force
error of $\co (\Delta t^2)$ (but see new developments in Sect.~\ref{newapproaches}). Additional errors induced by approximate
potential calculations (particle mesh or \tree ) create potentially
even larger errors than that. In Fig. \ref{Fig3Mak91}, however,
it is shown that the above Hermite method used for a real $N$-body integration
sustains generally an error of $\co(\Delta t^4)$ for the entire calculation.

\subsubsection{Time-step Choice}
Aarseth~\cite{Aarseth1985a} provides an empirical time-step criterion
\begin{equation}
\Delta t = \sqrt{\eta \frac{ \vert\vec{a}\vert \vert\vec{a}^{(2)}\vert
                          + \vert\vec{{\dot a}}\vert^2 }{
                          \vert\vec{{\dot a}}\vert \vert\vec{a}^{(3)}\vert
                          + \vert\vec{a}^{(2)}\vert^2 }} \ .
\label{4.1.6}
\end{equation}
The error is governed by the choice
of $\eta$, which in most practical applications is taken to be
$\eta = 0.01 - 0.04$.
It is instructive to compare this with the inverse square of the
curvature $\kappa$ of the curve $\vec{a}(t)$ in coordinate space
\begin{equation}
\frac{1}{\kappa^2} =
 \frac{1+\vert\vec{\dot a}\vert^2}{\vert\vec{a}^{(2)}\vert^2} \ .
\end{equation}
Clearly under certain conditions the time-step choice of Eq. \ref{4.1.6}
becomes similar to choosing the time-step according to the curvature of
the acceleration curve; since it was determined just empirically, however,
it cannot generally be related
to the curvature expression above. In~\cite{Makino1991a} a different time
step criterion has been suggested, which appears simpler and more
straightforwardly defined, and couples the time-step to the difference
between predicted and corrected coordinates.
The standard Aarseth time-step criterion from Eq. \ref{4.1.6} has been used in
most $N$-body simulations so far, because it seems to achieve an optimal step better than (on average) the mathematically more sound Makino step (see the time-step related discussion in~\cite{Sweatman1994}).\\
Since the position of all field particles can be
determined at any time by the low-order prediction from Eq. \ref{4.1.2}, the
time-step of each particle (which determines the time at which the
corrector of Eq. \ref{4.1.5} is applied) can be freely chosen according to
the local requirements of the test particle; the additional error induced
due to the use of only predicted data for the full $N$ sums of Eq. \ref{4.1.1}
is negligibly small, for the benefit of not being forced to keep all
particles in lockstep. Such an individual time-step scheme is in particular
for non-homogeneous systems very advantageous, as was quantitatively
pointed out by~\cite{MakinoHut1988}.
Particles in the high density core of a star clusters need to be updated
much more often than particles on orbits very far from the centre.
They show that the gain in computational speed due to the individual time-step scheme (as compared
to a lockstep scheme where all particles share the minimum required time-step)
is of the order $N^{1/3}$ for homogeneous and $N^1$ for strongly spatially
structured systems~\cite{MakinoHut1988}.\\
For the purpose of vectorization and parallelization it is better not
to have the particles continuously distributed on a time axis. Consequently,
\cite{Makino1991b} uses a hierarchical scheme, still on the basis of
Eq. \ref{4.1.6}; but a change of the time-step is considered only
if that equation yields a variation of $\Delta t$ compared
to the last step by more than a factor of 2 (increase or decrease).
If this is the case a variation by $2$ is applied only. Thus
in model units all time-steps are selected from the set
$\{2^{-i}\vert i=0,...i_{\rm max}\}$ with $k = i_{\rm max}$ determined by the
condition that $\Delta t_{\rm min} > 2^{-i_{\rm max}}$ for
the minimum time-step $\Delta t_{\rm min}$ determined from Eq. \ref{4.1.6}.
For core collapse simulations of star clusters of a few ten thousand
particles $i_{\rm max}$ goes up to about 20; empirically and
theoretically~\cite{MakinoHut1988} $\Delta t_{\rm min}\propto N^{-1/3}$,
so for large $N$ $i_{\rm max}$ becomes larger, however, on
the other hand, how large $i_{\rm max}$ grows for fixed $N$
depends on the selected criteria for so--called KS regularisation of
perturbed two--body motion (see below). The implementation of the
block step scheme indeed uses an even stronger condition than the above
described one, it is demanded that not only the time-steps, but also
the individual accumulated times of each particles are commensurate
with the time-step itself.
This ensures that for any particle $i$ and any time $T_i = t_i + \delta t_i $
{\it all} particles with $\delta t_j < \delta t_i $ have
for their own time $T_j = t_j + \delta t_j = T_i $, where the last
equality is the non--trivial one. Such procedure is important for
the parallelization of the algorithm. For example it has as a consequence
that at the big time-steps always huge groups of particles are due
for correction, sometimes even all particles (at the largest steps).

\subsubsection{Ahmad-Cohen neighbour scheme}
Another refinement of the Hermite or Aarseth ``brute force'' method is
the two-time-step scheme, denoted as neighbour or Ahmad-Cohen scheme
\cite{AhmadCohen1973}. For each particle a neighbour radius is defined,
and $\vec{a}$ and $\vec{\dot a}$ are computed due to neighbours and
non-neighbours separately. Similar to the Hermite scheme the
higher derivatives are computed separately for the neighbour
force (irregular force) and non-neighbour force (regular force).
Computing two time-steps, an irregular small $\Delta t_{\rm irr}$ and
a regular large $\Delta t_{\rm reg}$, from
these two force components by Eq. \ref{4.1.6} yields a time-step
ratio of $\gamma := \Delta t_{\rm reg}/\Delta t_{\rm irr}$ being in
a typical range of 5--20 for $N$ of the order $10^3$ to $10^4$. The
reason is that the regular force has much less fluctuations than
the irregular force.
The Ahmad-Cohen neighbour scheme is implemented in a self-regulated way,
where at each regular time-step a new neighbour list is determined using a given
neighbour radius $r_{si}$ for each particle. If the neighbour number found
is larger than the prescribed optimal neighbour number, the neighbour radius
is increased or vice versa. In
\cite{Aarseth1985a,MakinoHut1988} more complicated algorithms to adjust
the neighbour radius are described. On the
contrary to~\cite{MakinoHut1988}, who find an optimal
neighbour number of $N_{n,\rm opt} \propto N^{3/4}$
we find that adopting a constant neighbour number of the
order of $50-200$ is sufficient at least up to $N=10^6$. The reason is that
by using special purpose machines or parallelization for parts of
the code, an optimal neighbour number is not well defined, so the
neighbour number can be selected according to accuracy and efficiency
requirements. After each regular time-step the new
neighbour list is communicated along with the new particle positions
to all processors of the parallel machine, thus making it possible to
do the irregular time-step in parallel as well.\\
Using a two-time-step or neighbour scheme
again increases the computational
speed of the entire integration by a factor of at least proportional
to $N^{1/4}$~\cite{Makino1991a}. Both the regular and
irregular time-steps are arranged in the
hierarchical, commensurable way.

\subsubsection{Regularizations}
As the relative distance $r$ of the two bodies becomes small, their time-steps are reduced to prohibitively small values, and
truncation errors grow due to the singularity in the gravitational potential. Such a close encounter is characterised by an impact parameter $p$ and a relative velocity at ``infinity'' (in practice some distance inside the cluster) $v_\infty$. A close encounter is characterized by
\begin{equation}
p < p_{90} = 2 G(m_1 + m_2)/v_\infty^2,   \label{eq:p90}
\end{equation}
where $p_{90}$ is the impact parameter related to a 90 degree deflection in a two-body problem; $G$, $m_1$, $m_2$, $v_\infty$ are the gravitational constant, the masses of the two particles and their relative velocity at infinity. In the cluster centre, it is very likely that two stars come very close together in a hyperbolic encounter. So, if the separation of two particles gets smaller than $p_{90}$ they are candidates for regularization. To be actually regularized, the two particles have to fulfil two more sufficient criteria: that they are approaching each other, and that their mutual force is dominant. These sufficient criteria are defined as
\begin{eqnarray}
& {\mathbf R}\cdot {\mathbf V} > 0.1 \sqrt{(G(m_1+m_2)R} \nonumber\\
& \gamma := \frac{\vert {\mathbf a}_{\rm pert}\vert \cdot R^2}{G(m_1+m_2) } < 0.25 \nonumber
\label{eq:suff}
\end{eqnarray}
Here, ${\mathbf a}_{\rm pert}$ is the vectorial differential force exerted by other perturbing particles onto the two candidates, $R$, ${\mathbf R}$, ${\mathbf V}$ are scalar and vectorial distance and relative velocity vector between the two candidates, respectively. The factor 0.1 in the upper equation allows nearly circular orbits to be regularized; $\gamma < 0.25$ demands that the relative strength of the perturbing forces to the pairwise force is one quarter of the maximum. These conditions describe quantitatively that a two-body subsystem is dynamically separated from the rest of the system, but not necessarily unperturbed.\\
The idea is to take both stars out of the main integration cycle, replace them by their centre of mass (c.m.) and advance the usual integration with this composite particle instead of resolving the two components. The internal motion of the two members of the regularized pair (henceforth KS pair, for {\bf K}ustaanheimo and {\bf S}tiefel~\cite{KustaanheimoStiefel1965}) is done in a separate coordinate system. However, as was already noted by~\cite{Aarseth1971} there is no need for the perturbation of the KS pair from other stars to be small. \\
The internal motion of a KS pair is integrated in a 4D vector space obtained from a combined canonical and time transformation of relative Cartesian positions and velocities. The coordinate transformation goes back to Levi-Civita in 2D~\cite{LeviCivita1916}. A full generalization to higher dimensions is only possible over the mathematical object of a field, the next one to be quaternions in 4D. Kustaanheimo and Stiefel found a way to transform forward from 3D to 4D and back from 4D to 3D by working over a skewed field of quaternions (sacrificing some commutativity rules; their mathematical language was different though).
A modern theoretical approach to this subject can be found e.g. in~\cite{NeutschScherer1992}; the complete formalism including also the time transformation can be found in~\cite{Mikkola1997a}.
Aarseth uses this method to integrate the KS pairs in 4D space, and when using the back-transformation automatically returning to Cartesian 3D space~\cite{Aarseth1971}. The KS transformation converts the motion in a singular Newtonian gravitational potential into a harmonic oscillator in 4D space, which has no singularity. 
Since the harmonic potential is regular, numerical integration with
high accuracy can proceed with much better efficiency, and there
is no danger of truncation errors for arbitrarily small separations.
The internal time--step of such a KS--regularized pair is independent
of the eccentricity and of the order of some 50--100 steps per orbit.\\
While regularization can be used for any analytical two--body
solution even across a mathematical singularity (collision), it is practically applied to perturbed pairs only.
Once the perturbation $\gamma$ falls below a critical value
of $\approx 10^{-6}$,
a KS pair is considered unperturbed, and the analytical solution for
the Keplerian orbit is used instead of doing numerical integration.
The two-body KS regularization occurs in the code either for
short-lived hyperbolic encounters or for persistent binaries. \\
Close encounters between single particles and binary stars are
also a central feature of cluster dynamics. The chain regularization~\cite{MikkolaAarseth1998} is invoked if a KS pair has a  close encounter with another single star or another pair.
Especially, if systems start with a large number of primordial (initial) binaries, such encounters may lead to stable (or quasi-stable) hierarchical triples, quadruples, and higher multiples.
They are treated by using special stability criteria~\cite{MardlingAarseth2001}.\\
Every subsystem - KS pair, chain or hierarchical subsystem -
could be perturbed by other single stars. Perturbers are typically those objects that get closer to the object
than $R_{\rm sep} = R/\gamma_{\rm min}^{1/3}$, where $R$ is the
typical size of the subsystem; for perturbers, the components of the subsystem are resolved in their
own force computation as well. Algorithmic regularization is a relatively recent method based on a time transformed leap-frog method~\cite{MikkolaMerritt2008}; it does not employ the KS transformation. See for its use and application the next subsection. 

\subsection{Parallel and GPU computing and \textsc{Nbody}}
\label{parallel}
A fundamental problem was raised by Daiichiro Sugimoto about 30 years ago~\cite{Sugimotoetal1990} - direct numerical simulations of globular star clusters - with order of a million stars in direct $N$-body - could not be completed for decades if extrapolating the standard evolution of computational hardware at that time (Moore's law) for the future. Therefore astronomers in the Department of Astronomy at Tokyo University started wire-wrapping and designing a new integrated circuit, a special purpose computer chip named GRAPE (={\bf GRA}vity {\bf P}ip{\bf E}). The work was continued with great success by the team of Junichiro Makino, the GRAPE chips were finally assembled into GRAPE accelerator mainboards containing several chips (such as HARP, GRAPE-4, GRAPE-6)~\cite{Makinoetal1993,Makinoetal1997,MakinoTaiji1998,Makinoetal2003}.\\ 
The GRAPE chip is an application specific integrated circuit (ASIC), which could only compute gravitational forces between particles (it also computed the time derivative of the force, to be directly applicable to the Hermite scheme of \textsc{Nbody6}). A GRAPE board is a multi-core (multi-chip) parallel computing device (e.g. GRAPE-4 board contained 48 chips with shared memory, each chip contained one pipeline for force calculation; the GRAPE-6 chip contained 6 force pipelines~\cite{Makinoetal2003}).\\ 
Custom built computing clusters using GRAPE were built outside of Japan e.g. in Rochester and Heidelberg~\cite{Harfstetal2007}. In the following years, graphical processing units (GPU) widely replaced GRAPE; direct $N$-body implementations were done on GPU clusters~\cite{PortegiesZwartetal2007,Schiveetal2008}. Interfaces have been written such that GRAPE users could right away also use GPU with the newly invented programming language CUDA (Yebisu~\cite{NitadoriMakino2008}, Sapporo~\cite{Gaburovetal2009}, Kirin~\cite{Bellemanetal2008,Bellemanetal2014}). Still somewhat state of the art is \textsc{Nbody6GPU}, which includes GPU acceleration of \textsc{Nbody6} using CUDA kernels for single node servers~\cite{NitadoriAarseth2012}. Many of these kernels written by Keigo Nitadori are still in current use, even for the massively parallel programs such as \textsc{Nbody6++GPU}, see below.\\
At the same time another development started, parallelization of \textsc{Nbody6} with the (at that time) new standard MPI (message passing interface). \textsc{Nbody6++}~\cite{Spurzem1999} uses the
SPMD (Single Program Multiple Data) scheme to run many instances of the code in parallel, while distributing force computations for different particles to the processors of a massively parallel computer. From time to time data transfers using MPI communication routines are necessary, to make sure all processors are synchronous. Systems with hundreds of processing units were used at the time (e.g. CRAY T3E), which demanded efficient coding of the communication scheme. Copy and ring algorithms were developed, and asynchronous data transfer and computation implemented~\cite{Makino2002,Dorbandetal2003}.\\
\textcolor{red}{A copy algorithm keeps always a complete copy of all particle data on every parallel process; parallelization is over groups of particles due for the correction step; communication sends around all new particle positions and velocities in the Hermite scheme to all other processes. In contrast the ring algorithm uses a domain decomposition, every process has its specific set of particles (at least for some time), and instead of communicating particle positions and velocities partial gravitational forces and their time derivatives are communicated. A copy algorithm has been implemented by \cite{Spurzem1999,Hemsendorfetal2002} for \textsc{Nbody6++}, and a ring algorithm is used in \textsc{phiGRAPE}~\cite{Harfstetal2007}.
All these communication algorithms have been implemented long time ago using the \textsc{MPI\_SENDRECV} routine in a cyclic fashion - for $p$ processes $p\!-\!1$ communication steps are needed\footnote{$p$ is the number of processing elements running an MPI process, typically one multi-core CPU runs 1-4 MPI processes per node, across several nodes; each of them is connected to GPUs on the node}. Every process simultaneously sends data to its next neighbour and receives data from its other neighbour, in a ring structure. Therefore these algorithms are also denoted as systolic communication algorithms (both copy and ring). Nowadays \textsc{MPI\_ALLGATHER} or \textsc{MPI\_ALLREDUCE} may be used, but their implementations are not transparent and vary; the latter would normally use a \textsc{Tree}-based implementations (instead of systolic) - the number of communication steps is then only $\log_2(p)$ (while our systolic algorithm needs ${\cal O}(p)$ steps). It can be shown that asymptotically (for large data chunks and low latency) both algorithms are equivalent with regard to the total time required, because the \textsc{Tree} based algorithm uses increasingly large data packages, while in systolic algorithms every step communicates the same amount of data.\cite{Dorbandetal2003}.
Hence, currently still the systolic communication with a copy algorithm is used in \textsc{Nbody6++GPU}. If going to ten or hundred million bodies the copy algorithm may become too large for system memories, and should be updated to the ring algorithm with domain decomposition, which is not a fundamental problem (already used in the \textsc{phiGRAPE} code, which is a simple variant of the Hermite scheme with blocked hierarchical time steps). }
While both ring and copy algorithms scale linearly with $p$ a hypersystolic algorithm exists which scales only with $\sqrt{n_p}$~\cite{Lippertetal1996,Lippertetal1998}. For GRAPE a special implementation of a hypersystolic algorithm for 2D meshes of processing elements has been proposed~\cite{Makino2002}. 
\textcolor{red}{Hypersystolic and other \textsc{Tree} based communication algorithms can play out their strengths in case of a huge number of processes (as in case of \textsc{GRAPE} for example) with relatively modest computational load on every process. On the contrary the current \textsc{Nbody6++GPU} algorithm requires modest numbers of processes (order 10-100 with current particle numbers of up to around a million, may be more in the future), which have big computing loads and very large data chunks to communicate.}\\
\textsc{Nbody6++}~\cite{Spurzem1999} parallelizes both force loops with MPI, for the regular and the neighbour force in the Ahmad-Cohen scheme.\\ 
A ring communication algorithm with domain decomposition in the future would also help for situations when there are many small block time-steps with few particles to integrate. The current code \textsc{Nbody6++} (and its successors \textsc{Nbody6++GPU} using GPU) only invoke parallel MPI execution if the number of particles in a time block is large enough (like e.g. 50-100, best value has to be tested for every hardware). For smaller blocks all processors are redundantly computing everything without communication, to avoid the overhead connected with MPI. Since the special hierarchical time-step scheme of \textsc{Nbody6} favours time blocks with many particles this is for usual globular cluster simulations no bottleneck. However, in case of very high central density, like in nuclear star clusters with central supermassive black hole (see Sect.~\ref{nuclear}) the parallel performance gets degraded. \\
With the advent of clusters, where nodes would be running MPI, and each node having a GPU accelerator, \textsc{Nbody6++GPU} was created - on the top level MPI parallelization is done for the force loops (coarse grained parallelization) and at the bottom level each MPI process calls its own GPU to accelerate the force calculation~\cite{Bercziketal2013}, using Nitadori's Yebisu library~\cite{NitadoriMakino2008} for the regular force only.
Secondly, an AVX/SSE implementation accelerates prediction and neighbour (irregular) forces, and also a number of other features had been severely optimized and improved (such as particle selection at block times)~\cite{Wangetal2015}. This code currently keeps the record of the largest direct $N$-body simulation of a globular cluster with all required astrophysics (single and binary stellar evolution, stellar collisions, tidal field), simulated over 12 Gyrs~\cite{Wangetal2016}.\\
In recent years, inspired also by LIGO/Virgo/KAGRA gravitational wave detections~\cite{Abbottetal2016a}, numerous current updates have been made with regard to stellar evolution of massive stars in singles and binaries~\cite{Kamlahetal2022b}, and with regard to collisional build up of stars (mass loss at stellar collisions allowed) and intermediate mass black holes~\cite{Rizzutoetal2021,Rizzutoetal2022,ArcaSeddaetal2021a}. The current code is available via \textsc{Github}\footnote{\url{https://github.com/nbody6ppgpu}}. Note that a different service is provided by Long Wang, quoted here~\cite{Varrietal2018}. That alternative version of \textsc{Nbody6++GPU} has been recently used by the Padova team, replacing the older SSE/BSE stellar evolution by MOBSE (see e.g.~\cite{DiCarloetal2021}).\\
Star clusters with primordial (initial) binaries inevitably lead to binaries of black holes. If two black holes get close enough to each other, either during a hyperbolic encounter or due to close Newtonian three-body or four body interactions, Post-Newtonian corrections have to be taken into account. They take the form of an expansion of the relative acceleration between the two bodies in terms of $(v/c)^{2{\rm i}}$, denoted as PNi-terms. PN1, PN2, and PN3 are conservative, producing periastron shifts of orbits, while PN2.5 and PN3.5 provide the energy and angular momentum loss due to gravitational radiation. The first implementation was done in \textsc{ Nbody5} up to PN2.5~\cite{Kupietal2006}; and for \textsc{ Nbody7}~\cite{Aarseth2012}. Also relativistic spin-spin and spin-orbit interactions of orders PN1.5, PN2.0, PN2.5 have been recently included~\cite{Bremetal2013}. The most recent version of \textsc{ Nbody7}~\cite{Banerjeeetal2020} includes also the full PN terms by using the ARChain (algorithmic regularization chain) method~\cite{MikkolaMerritt2008}. \textsc{ Nbody7} is GPU accelerated, but has not yet the MPI parallelization of \textsc{ Nbody6++} and \textsc{ Nbody6++GPU}. Generally binary evolution governed by Post-Newtonian terms has been compared with full numerical solutions of general relativity; deviations between fully relativistic and Post-Newtonian evolution only occur during the final period of merger, in a time span usually negligible for astrophysical purposes. The reader interested in the original citations with regard to the derivation and justification of Post-Newtonian terms is referred to~\cite{Kupietal2006,MikkolaMerritt2008,Bremetal2013} for further references therein.\\
When black holes merge they experience a kick due to asymmetric gravitational wave emission, see e.g. the MOCCA implementation~\cite{Morawskietal2018}; a similar model is already used in \textsc{ Nbody7}~\cite{Banerjeeetal2020}, and this is a field where \textsc{ Nbody6++GPU} is currently lagging behind, current work is ongoing on it. The following Table~\ref{versions} gives a summary of the features of different variants of the \textsc{ Nbody} codes.
\begin{table}[t]
\caption{Comparison of the code versions}
\let\V\checkmark
    \begin{tabular}{|p{2cm}|c|c|c|c|>{\color{violet}}c|c|>{\color{violet}}c|>{\color{violet}}c|>{\color{violet}}c|} \hline
                        & ITS & ACS & KS & HITS & PN & AR & CC & MPI & GPU \\\hline
    \textsc{Nbody1}     & \V  &     &    &      &    &    &    &     &    \\
    \textsc{Nbody2}     &     & \V  &    & \V   &    &    &    &     &    \\
    \textsc{Nbody3}     & \V  &     & \V &      &    &    &    &     &    \\
    \textsc{Nbody4}     &     &     & \V & \V   &    &    & \V &     &    \\ 
    \textsc{Nbody5}     & \V  & \V  & \V &      &(\V)&    & \V &     &    \\
    \textsc{Nbody6}     &     & \V  & \V & \V   & r  &    & \V &     &    \\
    \textsc{Nbody6GPU}  &     & \V  & \V & \V   & \V &    & \V &     & \V \\
    \textsc{Nbody6++}   &     & \V  & \V & \V   &    &    & \V & \V  &    \\
    \textsc{Nbody6++GPU}&     & \V  & \V & \V   & r  &    & \V & \V  & \V \\
    \textsc{Nbody7}     &     & \V  & \V & \V   & \V & \V &    &     & \V \\ \hline
    \end{tabular}\\
\V:  Included in standard version of that level \\
ITS: Individual time--steps~\cite{Aarseth1985a} \\
ACS: Ahmad-Cohen neighbour scheme~\cite{AhmadCohen1973}\\
KS: KS--regularization of few-body subsystems~\cite{KustaanheimoStiefel1965} \\
HITS: Hermite scheme integration method combined with hierarchical
      block time-steps~\cite{MakinoAarseth1992}\\
PN: Post-Newtonian~\cite{Kupietal2006,MikkolaMerritt2008,Aarseth2012}\\
r: restricted PN, only orbit-averaged energy loss by gravitational radiation~\cite{Rizzutoetal2021,Rizzutoetal2022,ArcaSeddaetal2022}\\
(\V): only included in special version of the code~\cite{Kupietal2006}\\
AR: Algorithmic regularization~\cite{MikkolaMerritt2008} \\
CC: Classical chain regularization~\cite{MikkolaAarseth1998} \\
MPI: Message Passing Interface, multi-node multi-CPU parallelization~\cite{Spurzem1999} \\
GPU: use of GPU acceleration~\cite{NitadoriAarseth2012} (if also MPI: multi-node many GPU~\cite{Bercziketal2013})
\label{versions}
 \let\V\undefined
\end{table}
Table~\ref{timing} shows for \textsc{Nbody6++GPU} a model fit, obtained from a number of simulations using a range of particle numbers $N$ and MPI process number $N_p$, where each MPI process also uses a GPU\cite{Huangetal2016}. Eight different pieces of the code have been profiled as indicated. The fit shows the following key informations:
\begin{itemize}
\item[1] regular and irregular force computation are very well parallelized ($\propto N_p^{-1}$);
\item[2] regular force computation still scales with approximately $N^2$, but with a very small factor in front, due to the fast GPU processing.
\item[3] MPI communication and synchronization provide a bottleneck, no further speedup possible for more than 8-16 MPI processes.
\item[4] Also prediction and sequential parts on the host are bottlenecks if going for large $N$, because they scale approximately with $N^{1.5}$, and do not scale down with processor number.
\end{itemize}
The timing model is already a few years old, the current code version has made progress in MPI parallelization of prediction (pos. 3). To improve the communication scaling faster MPI or NVLink\footnote{NVIDIA High Speed GPU Interconnect} communication hardware will be beneficial (pos. 5, 6). Note that all numerical factors in the fit dependent on the specific hardware used - CPUs, GPUs, communication lines between CPU nodes and between CPU and GPU.

\renewcommand{\arraystretch}{1.5}
\begin{table}
	\centering
	\resizebox{\textwidth}{!}{%
\begin{tabular}{clclclclcl}
\hline \hline Description & Timing & Expected scaling & \\
 & variable & $N$ & $N_{\mathrm{p}}$ &  Fitting value [sec] \\
 \hline
 Regular force computation  & $T_{\text {reg }}$ & $\mathcal{O}\left(N_{\text {reg }} \cdot N\right)$ & $\mathcal{O}\left(N_p^{-1}\right)$ & $\left(2.2 \cdot 10^{-9} \cdot N^{2.11}+10.43\right) \cdot N_p^{-1}$ \\
Irregular force computation & $T_{\text {irr }}$ & $\mathcal{O}\left(N_{\text {irr }}\cdot\left\langle N_{n b}\right\rangle\right)$ & $\mathcal{O}\left(N_p^{-1}\right)$ & $\left(3.9 \cdot 10^{-7} \cdot N^{1.76}-16.47\right) \cdot N_p^{-1}$ \\
Prediction & $T_{\text {pre }}$ & $\mathcal{O}\left(N^{k n_p}\right)$ & $\mathcal{O}\left(N_p^{-k p_p}\right)$ & $\left(1.2 \cdot 10^{-6} \cdot N^{1.51}-3.58\right) \cdot N_p^{-0.5}$ \\
Data moving & $T_{\text {mov }}$ & $\mathcal{O}\left(N^{k n_m 1}\right)$ & $\mathcal{O}(1)$ & $2.5 \cdot 10^{-6} \cdot N^{1.29}-0.28$ \\
MPI communication (regular) & $T_{\text {mcr }}$ & $\mathcal{O}\left(N^{k n_{c r}}\right)$ & $\mathcal{O}\left(k p_{c r} \cdot \frac{N_p-1}{N_p}\right)$ & $\left(3.3 \cdot 10^{-6} \cdot N^{1.18}+0.12\right)\left(1.5 \cdot \frac{N_p-1}{N_p}\right)$ \\
MPI communication (irregular) & $T_{\text {mci }}$ & $\mathcal{O}\left(N^{k n_{c i}}\right)$ & $\mathcal{O}\left(k p_{c i} \cdot \frac{N_p-1}{N_p}\right)$ & $\left(3.6 \cdot 10^{-7} \cdot N^{1.40}+0.56\right)\left(1.5 \cdot \frac{N_p-1}{N_p}\right)$ \\
Synchronization  & $T_{\text {syn }}$ & $\mathcal{O}\left(N^{k n_s}\right)$ & $\mathcal{O}\left(N_p^{k p_s}\right)$ & $\left(4.1 \cdot 10^{-8} \cdot N^{1.34}+0.07\right) \cdot N_p$ \\
Sequential parts on host  & $T_{\text {host }}$ & $\mathcal{O}\left(N^{k n_h}\right)$ & $\mathcal{O}(1)$ & $4.4 \cdot 10^{-7} \cdot N^{1.49}+1.23$\\
\hline
\end{tabular}}
\caption{\textcolor{red}{Profiling model developed for \textsc{Nbody6++GPU}, details explained in main text; reproduced from~\cite{Huangetal2016}}}
\label{timing}
\end{table}
\renewcommand{\arraystretch}{1.0}

\begin{figure}[ht]
\includegraphics[width=0.59\textwidth]{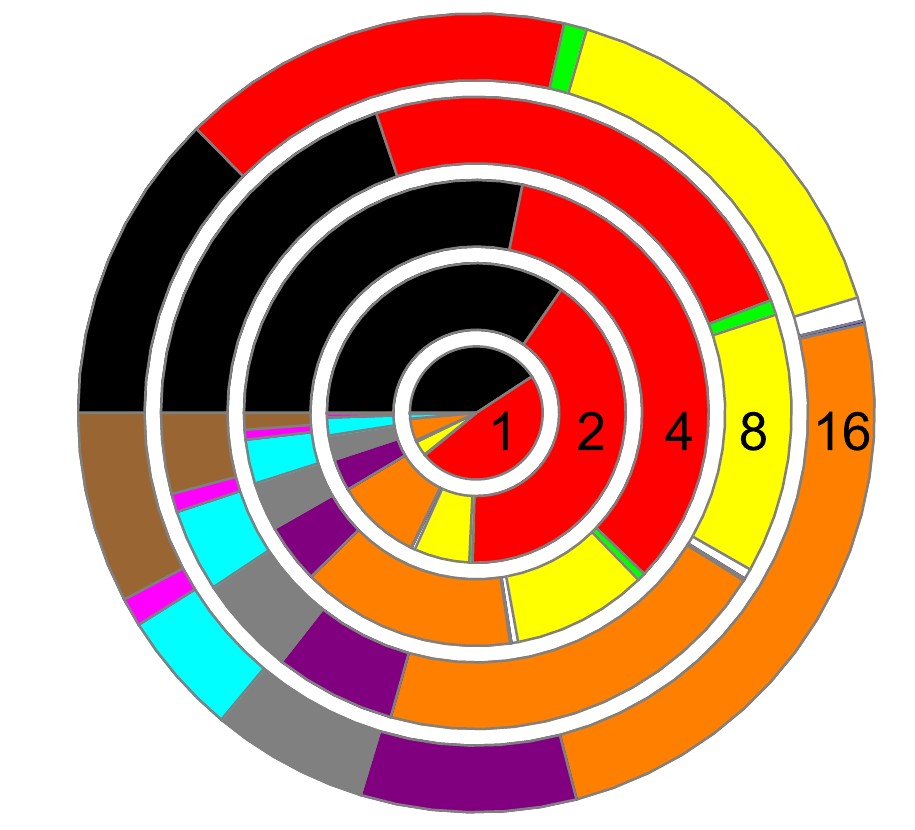}
\includegraphics[width=0.4\textwidth]{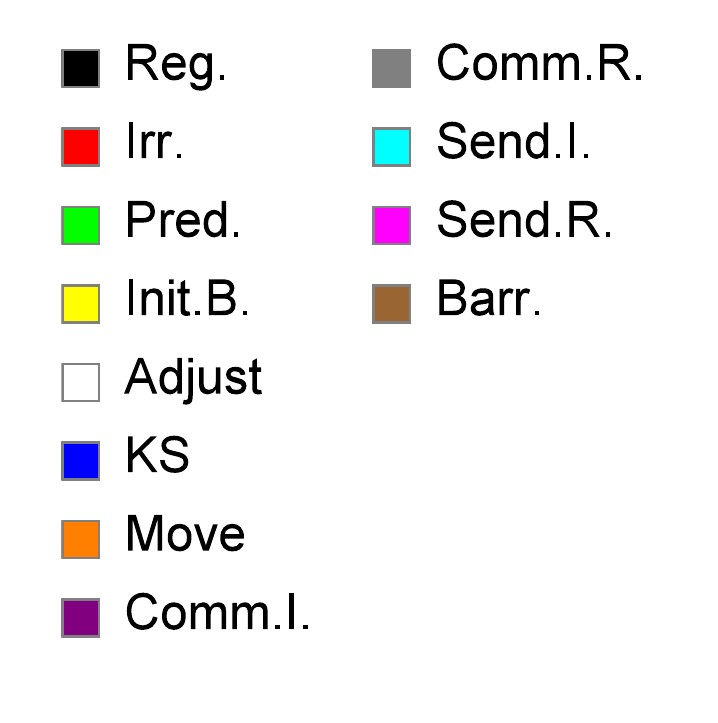}
 \caption{Pie chart showing the time fraction spent in different parts of the \textsc{Nbody6++GPU} code for a one million body simulation without initial (primordial) binaries. Different rings show different number of MPI processes $N_p$ (inside to outside
1, 2, 4, 8 and 16.). Colours explained in main text.}
\label{piechart}
\end{figure}

Fig.~\ref{piechart} shows a similar information in principle than Table~\ref{timing}, but here the eye should inspect the relative weight of the different components, when increasing the number of MPI processes. The coloured fields correspond to the code parts discussed above, but a little more segmented:
\begin{itemize}
\item[a] Reg. and Irr. correspond to regular and irregular force computation in Table~\ref{timing};
\item[b] Pred. is prediction;
\item[c] Move is data moving;
\item[d] Comm.R, Send.R., Comm.I. and Send.I is MPI communication (regular, irregular)
\item[e] Barr. is synchronization
\item[f] Init.B., Adjust, KS, refer to sequential parts on the host. 
\end{itemize}
The bottom line to stress from these results is that even for one million bodies the bottleneck of the parallel code is {\em NOT} the regular force (which would be extremely dominant in a sequential processing), so it is {\em NOT} the stumbling block for going to much higher particle number, these are prediction and communication. 
There are also \textsc{ phiGPU}~\cite{Bercziketal2013}, \textsc{ phiGRAPE}~\cite{Harfstetal2007}, and \textsc{ HiGPU}~\cite{CapuzzoDolcettaetal2008,Spera2014}. All of them are using the Hermite scheme with hierarchically blocked time-steps, and are fully parallelized and GPU accelerated. There is no Ahmad-Cohen neighbour scheme and no regularization, which means that on a serial computer they would be much slower than e.g. \textsc{ Nbody6++GPU}. But with a very efficient parallelization and GPU acceleration this is partly compensated; they have been used for astrophysical problems where star-by-star modelling could be neglected, such as e.g. galactic nuclei and galaxy mergers with supermassive black holes (cf. e.g~\cite{Zhongetal2014,Zhongetal2014,Lietal2017,Lietal2019,Bortolasetal2018}).\\
An interesting feature is that these codes implement higher order Hermite integration schemes (in \textsc{ phiGPU} $4^{\rm th}$, $6^{\rm th}$, or $8^{\rm th}$ order can be chosen,  \textsc{ HiGPU} uses $6^{\rm th}$ order. There is another $6^{\rm th}$ and $8^{\rm th}$ order Hermite integrator~\cite{NitadoriMakino2008}; so far these higher order integrators have seen relatively little use, consistent with the conclusion that the $4^{\rm th}$ order integrator is an optimal choice for performance and accuracy~\cite{Makino1991a}.

\subsection{Are N-Body simulations reliable?}
\label{reliable}
At this point the reader may expect that direct $N$-body simulation turn out to be the most reliable (although computationally most expensive) way to simulate the dynamical evolution of a gravitating system consisting of $N$ point masses. It does not involve any serious approximations and assumptions, as e.g. the Fokker-Planck approximation and the Monte Carlo codes. By reducing the $\eta$-values in the time-step (Eq.~\ref{4.1.6}) any accuracy can be achieved in principle, as long as machine accuracy permits it. Usually for accuracy and time-step choice globally conserved quantities are used, such as energy and angular momentum, and center of mass conservation (position and velocity).\\
However, for a system with $N$ particles phase space has $6N$ dimensions, and a check of say energy, angular momentum, and center of mass alone only checks whether the numerically calculated system remains within an allowed $6N-9$ dimensional hypervolume. There is no a priori information how ``exact'' the ``true'' individual trajectories are reproduced in the simulation within this hypervolume. It was early pointed out that, due to repeated close encounters occurring between particles, initial configurations that are very close to each other, quickly diverge in their evolution from each other~\cite{Miller1964}. In that work it was shown that the separation in phase space of two trajectories increases exponentially with time, or with other words, the evolution of the configuration is extremely sensitive to initial conditions (particle positions and velocities). The timescale of exponential instability is as short as a fraction of a crossing time, and the accurate integration of a system to core collapse would require of order $\co(N)$ decimal places~\cite{Goodmanetal1993,Kandrupetal1994}. These papers argue that the problem is caused by two-body encounters, but chaotic orbits in non-integrable potentials can be a source of exponential instability and thus cause unreliable numerical integrations as well. \\
However, the situation is not as bad as it seems. $N$-body simulations of star clusters or galactic nuclei do not always exploit the detailed configuration space of all particles. Quantities of interest are global or somehow averaged quantities, like Lagrangian radii or velocity dispersions averaged in certain volumes. As it was nicely demonstrated in a pioneering series of papers~\cite{GierszHeggie1994a,GierszHeggie1994b,GierszHeggie1996,GierszHeggie1997} such results are not sensitive to small variations of initial parameters. They took statistically independent initial models (positions and velocities at the beginning selected by different random number sets) and showed that the ensemble average of the dynamical evolution of the system always evolved predictably and in remarkable accord with results obtained from the Fokker-Planck approximation. The method was also partly and successfully used in~\cite{GierszSpurzem1994}, which focused on the evolution of anisotropy and comparisons with the anisotropic gaseous models of the author of this paper, or in more recent examples~\cite{Rizzutoetal2021,Rizzutoetal2022} where the formation of intermediate mass black holes was analyzed over a large set of $N$-body simulations, using statistically independent initial models.\\
As a consequence, it should be remembered, however, that great care has to be taken when interpreting results of $N$-body simulations on a particle by particle basis, for example determining rates of specific types of encounters, which could produce mergers in a large direct $N$-body model.\\
The long-term behaviour of dynamical systems as the solar system are
being studied by $N$-body simulations as well, but clearly there are
much higher requirements on the accuracy of the individual orbits in contrast to the star cluster problem. Therefore for the solar system dynamics symplectic methods, using a generalized leap-frog, like the widely used Wisdom-Holman symplectic mapping method~\cite{WisdomHolman1991} are the standard integration method. Symplectic mapping methods do not show secular errors in energy and angular momentum. However, in their standard implementation they require a constant time-step (but see recent new developments described in the following subsection). A generalization using a time transformation simultaneously with the generalized leap-frog has been suggested which can cope with variable time-steps~\cite{Mikkola1997b}. \\
It has been proposed to reduce secular errors in Hermite schemes and direct $N$-body simulations to a level comparable with symplectic methods by using a time-symmetric scheme. A small variation in the Hermite corrector is needed, which allows to iterate to convergence (few iterations usually enough) and individual time-steps made reversible through another iteration~\cite{Hutetal1995,Funatoetal1996,Makinoetal1997}. How well this generally works and its relation to symplectic schemes is presently not clear. But it has been well used for direct $N$-body simulations of planet formation and planetary systems~\cite{Kokuboetal1998,Makinoetal1997}. These codes though are still on the level of \textsc{Nbody4}, because they do not use the Ahmad-Cohen neighbour scheme - even in the smallest steps full force calculations over all $N$ particles are needed. \textsc{Nbody6++} has been similarly improved using an extended Hermite scheme to allow iteration, for a hybrid $N$-body and Fokker-Planck simulation of planetesimal growth in protoplanetary disks~\cite{Glaschkeetal2014,AmaroSeoaneetal2014} (no GPU implementation with \textsc{Nbody6++GPU} yet).\\
In~\cite{MikkolaAarseth1998} it is stressed that even with a newly applied classical method secular errors in the integration of close binaries can be strongly reduced. One should keep in mind though, that the $N$-body integration schemes discussed in this paper yield excellent results in the star cluster research (see Sect. 4) but are unsuitable for long-term solar system studies, because they generally have secular errors, although small. Due to the inherently physically chaotic nature of star clusters remaining small secular errors can usually be tolerated. It means that the solution found in the computer always stays near a permitted solution of the underlying Hamiltonian, even if it does not stay on the one trajectory which belongs to the initial conditions~\cite{QuinlanTremaine1992}. But a recent dynamical study has reiterated that it may not be sufficient just to check a few globally conserved quantities, because that could be dominated by a few high energy objects (binaries) and could cover up errors in other parts of the system~\cite{WangHernandez2021}.\\
As outlined above in star cluster simulations
the secular errors are being kept small relative
to typical values of energy and angular momentum and
an accurate reproduction of all individual stellar orbits is not
generally required.

\subsection{New Approaches}
\label{newapproaches}
\begin{figure}[t]
\includegraphics[width=\textwidth]{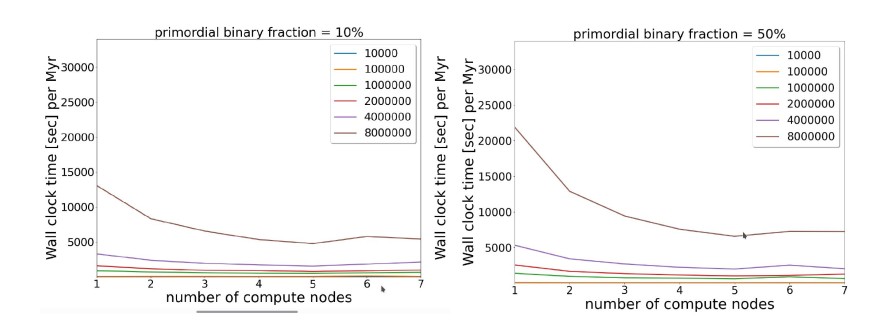}
 \caption{Strong scaling of the novel PeTar code\cite{Wangetal2020d}, showing wall clock times obtained on the Juwels-Booster as a function of number of compute nodes, for particle numbers up to eight million and binary fractions of up to 50\%. The benchmarks have been done by Qi Shu, and will be published in~\cite{Spurzemetal2022}.
 }
\label{petar-scaling}
\end{figure}

A completely new code, called \textsc{PeTaR} has been introduced~\cite{Wangetal2020d}. It is a hybrid $N$-body code, which combining the P${}^3$T (particle-particle particle-tree) method~\cite{Oshinoetal2011,Iwasawaetal2015,Iwasawaetal2016,Iwasawaetal2017} and a slow-down time transformed symplectic integrator (\textsc{SDAR})~\cite{Wangetal2020c}. The latter is mathematically similar to a KS~\cite{KustaanheimoStiefel1965} regularization, using a time transformation in a similar way (based on the Poincar\'e\ transform of the Hamiltonian~\cite{PretoTremaine1999,MikkolaMerritt2008}), but regarding the canonical coordinate transformation it uses an extended classical phase space rather than the 4D KS space. Both regularization methods also employ a slow-down procedure. 
\textsc{PeTar} uses the parallelization framework for developing particle simulation codes (\textsc{FDPS}~\cite{Iwasawaetal2016,Iwasawaetal2020}) to manage the particle-tree construction and long-range force calculation.
For single and binary stellar evolution the standard SSE and BSE packages are used as in e.g. MOCCA~\cite{Hurleyetal2005,Banerjeeetal2020,Kamlahetal2022b}.\\
The code is conceptually ahead of \textsc{Nbody6++GPU} in several respects; parallelization of a large number of hard binaries is included and a domain decomposition makes it easier to go to particle numbers much larger than $10^6$, as well as the use of the \textsc{Tree} scheme for distant groups of particles, rather than the Ahmad-Cohen neighbour scheme in \textsc{Nbody6++GPU}. We show in Fig.~\ref{petar-scaling} its excellent strong scaling obained on the Juwels Booster supercomputer in Germany~\cite{JUWELS2021}.
The paper presenting the \textsc{PeTar} code makes some much stronger statements about \textsc{Nbody6++GPU}, claiming it is impossible to implement binary parallelization and domain decomposition in it. We think the given arguments are not sound, a thorough discussion of these issues will follow here in the future depending on further work with \textsc{Nbody6++GPU}. Also the paper presents a couple of convincing comparisons between both codes, but a real long time model comparison with many binaries is still work in progress. The problem is, that the single and binary stellar evolution in \textsc{Nbody6++GPU} has a much stronger coupling to the dynamical evolution of binaries and close encounters - that is from a programmer's viewpoint a trouble (as stated in~\cite{Wangetal2020d}), but it is important when following very hard and relativistic binaries~\cite{Rizzutoetal2021,Rizzutoetal2022}. Even the \textsc{Tree} algorithm is not really essential - inherently it can be much faster than the direct Hermite scheme (even with Ahmad-Cohen neighbour scheme), but it depends on the accuracy required. We have shown by comparison of \textsc{Nbody6++GPU} with the \textsc{Bonsai Tree} code~\cite{Bedorfetal2012b} that if same accuracy is required, performance difference is not significant~\cite{Huangetal2016}. This can be also seen in Sect.~\ref{parallel} from the performance analysis of  \textsc{Nbody6++GPU} - even for more than a million bodies the regular distant forces are not the bottleneck, they have been effectively parallelized. \\
Another novel approach is based on forward symplectic integrators (FSI)~\cite{Chin1997,ChinChen2005,Chin2007,DehnenHernandez2017}. It is called \textsc{Frost}~\cite{Rantalaetal2021}, uses MPI parallelization and GPU acceleration and shows competitive benchmarks for very large $N$ (million and more). The authors note that the FSI approach has hardly been used in the community, even though it generates very accurate orbital integrations. However, as discussed above, for most star cluster simulations orbit integration with maximum machine precision is not really needed; however, for the innermost regions of nuclear star clusters, with stellar orbits around supermassive black holes, this will be important and \textsc{Frost} may turn out to be excellently suited for such environment. On the other hand the planetary system and protoplanetary disk simulation community regularly uses either symplectic schemes or the improved iterated Hermite schemes (see above Sect.~\ref{parallel}). A very significant innovation in \textsc{Frost} is \textsc{Mstar} - a fast parallelized algorithmically regularized integrator. Instead of classical and algorithmic chains, which assemble their particles linearly (in a chain), \textsc{Mstar} uses a minimum spanning \textsc{Tree}, so the chain can have branches. That leads to smaller average distances within chain particles as before, reduces computing time and errors. Such an algorithm could be as well used for the other chain regularizations.\\ 
Finally, hybrid codes have been constantly designed and used for a while to amend the direct $N$-body for a large number of distant particles. It started already with \textsc{Nbody5} which was coupled to a \textsc{Tree} scheme~\cite{McMillanAarseth1993}; \textsc{Nbody6++} has been hybridized with a series expansion code~\cite{Hemsendorfetal2002} (sometimes also called self-consistent field~\cite{HernquistOstriker1992}, SCF). Last, but not least there is a new hybrid code \textsc{Etics}~\cite{Meironetal2014}, which has been coupled with \textsc{phiGRAPE}, and applied to the loss cone problem in a nuclear star cluster around a binary supermassive black hole~\cite{Avramovetal2021}.

\section{Astrophysics in star clusters}
\label{Astrophysics in star clusters}
\subsection{Single Stellar Evolution}
\label{Single stellar evolution}
In realistic star cluster simulations all stars undergo stellar evolution as time proceeds, see e.g. \cite{Churchetal2009} and therefore, a large array of stellar evolutionary processes must be considered. We briefly outline the fundamentals of single stellar evolution (Sect.~\ref{Single stellar evolution}) because it is essential to understand the complexities that need to be modelled before we move on to an area, in which collisional $N$-body simulations find some of their strongest applications, which is binary stellar evolution (Sect.~\ref{Binary Stellar Evolution}) in dense star clusters. The discussion in this Sect.~\ref{Single stellar evolution} is primarily based on~\cite{Kippenhahnetal2012}, but a more recent review may also be found in~\cite{SalarisCassini2017}.\\
A star is a self-gravitating object of a hot plasma, which emits energy at the surface in form of photons (and from the inner regions in the form of neutrinos). Furthermore, it is spherically symmetric in the absence of rotation, magnetic fields and a sufficiently close companion or multiple companion stars that induce interior oscillations and bulges through tidal interaction or deform the star through mass transfer. These are typical assumptions in 1D modelling of single stars and they yield four fundamental structure equations that govern stellar evolution under the assumption of hydrostatic equilibrium. Any deviation from hydrostatic equilibrium will become increasingly important in harder binary stars.\\
Energy transport in a star is either radiative or convective (where convective transport can also include some conduction, which is not that important). Whether it is one or the other is given by the Schwarzschild stability criterion, which compares the temperature gradient in the radiative case with the temperature gradient by an adiabatic movement of matter elements: $\nabla_{\mathrm{rad}}< \nabla_{\mathrm{ad}}$. The less practical Ledoux criterion also takes into account a possible gradient in the density and chemical composition of a star. If some matter is unstable according to the Ledoux criterion, then convection will set in and will mix the material until stellar homogeneity. This process will diminish these gradients. Therefore, in practice the Schwarzschild criterion is more commonly used. \\
Radiative energy transport is commonly described using a diffusion equation. For convection, on the other hand, there is a convective gradient and since no complete theory of convection exists, the problem is approximated using mixing-length theory (MLT). MLT describes the convective temperature gradient $\nabla_{\mathrm{c}}$ surprisingly well despite a large number of unrealistic assumptions, e.g., all convective ``bubbles'' are assumed to travel the same distance due to buoyancy forces until they disappear leading to the dispersion of all their energy to the next level. MLT is parameterised globally by $\alpha_{\mathrm{MLT}}$, which is the ratio of the mixing length to the pressure scale height. $\alpha_{\mathrm{MLT}}$ is around $2$ for Solar models. Since in deep stellar interiors, convection is very efficient and thus the ``blobs" move adiabatically, it is $\nabla_{\mathrm{c}} \simeq \nabla_{\mathrm{ad}}$. In the outermost layers (low densities), however, convection is not so efficient, a lot of energy is lost by a blob moving up and the energy transport is super-adiabatic even approaching the radiative gradient in the extreme: $\nabla_{\mathrm{rad}} > \nabla_{\mathrm{c}} > \nabla_{\mathrm{ad}}$. \\
The chemical composition of a star changes with time due to nuclear reactions in its interior. It can also be subject to convective mixing, sedimentation, rotation (angular momentum transport) and hydrodynamical instabilities. The inclusion of all of these effects is difficult, because it requires 3D treatment; but most currently used stellar evolution codes, such as Modules for Experiments in Stellar Astrophysics (\textsc{MESA}) ~\cite{Paxtonetal2011,Paxtonetal2013,Paxtonetal2015,Paxtonetal2016,Paxtonetal2018,Paxtonetal2019} or HOngo Stellar Hydrodynamics Investigator (\textsc{HOSHI})~\cite{Takahashietal2016,Takahashietal2018,Takahashietal2019,Yoshidaetal2019} are 1D. 

\subsubsection{Two fundamental principles of stellar evolution}
\label{The two fundamental principles of stellar evolution}
The general evolution of a star following the assumptions above is governed by two fundamental principles: the first one is the virial theorem (gravitational energy $E_\mathrm{g} = -2 E_\mathrm{i}$ total internal energy), which follows from the assumption of hydrostatic equilibrium in the star that is represented by a self-gravitating sphere (Sect.~\ref{Single stellar evolution}). The virial theorem implies that on contraction of a star that is modelled as an ideal gas, half of the liberated energy is radiated away and the other half is used to increase the internal energy, which means that the star is heating up. In other words, if stars lose energy from the surface, the star must contract overall and heat up, which is a consequence of its negative heat capacity. That does not mean that some parts like the stellar envelope are not expanding over the star's evolution, but what is certain that the largest part of the star is contracting over the life-time and heating up. Interestingly, massive stars, which are radiation pressure dominated, approach the limit of an unbound structure, which is one of the reasons why they lose mass much more easily. \\
The second fundamental principle is the Coulomb repulsion, which determines the sequence of nuclear burning phases. Due to the virial theorem above that leads to a general increase in the interior stellar temperature, nuclear burning phases follow a sequence of light to heavier elements, i.e. they start with hydrogen (H) burning (the main sequence (MS) phase), followed by helium (He) burning (horizontal branch (HB) phase), the carbon (C) burning phase and so on. This burning sequence stops when an iron (Fe) core is reached, because any further nuclear fusion is endothermic. We reach an ``onion-like" stellar structure: in the outer layers original stellar material is still processing (H fusing to He), while at the centre an Fe and Nickel (Ni) core forms simultaneously if the stellar mass is large enough.

\newpage
\subsubsection{Timescales, energy conservation and homology}
\label{Timescales, energy conservation and homology in stellar evolution}
The following timescales are extremely useful in characterising the evolution of stars:
\begin{enumerate}
	\item hydrostatic timescale $\tau_{\mathrm{hydro}}$: let us assume that the internal stellar forces are not balanced and the star is not in hydrostatic equilibrium anymore. The timescale to return to hydrostatic equilibrium is: $\tau_{\mathrm{hydro}}\simeq \frac{1}{2}(G\bar{\rho})^{-1/2}$. It is extremely short, on the order of seconds for White Dwarfs (WDs), of minutes for the Sun and of the order of days for Red Giants (RGs).\\
    \item Kelvin-Helmholtz (thermal) timescale $\tau_{\mathrm{KH}}$: it is defined as the timescale during which the entire internal energy of star would be radiated away by its current luminosity.
    For the Sun it is on the order of 10 million years.\\
	\item nuclear timescale $\tau_{\mathrm{nuc}}$: let us assume that the whole luminosity comes only from the nuclear energy reservoir within the star and that the luminosity stays constant at the current state for the duration of the thought experiment. For the Sun this the emission of all nuclear energy as radiation is on the order of 70 billion years.
\end{enumerate}
In most phases of stellar evolution, we have  $\tau_{\mathrm{hydro}} \ll \tau_{\mathrm{KH}} \lessapprox \tau_{\mathrm{nuc}}$ and mostly also even $\tau_{\mathrm{KH}} \ll \tau_{\mathrm{nuc}}$ for MS and core He burning (CHeB) stars. In late stellar evolution phases $\tau_{\mathrm{KH}}$ can approach $\tau_{\mathrm{nuc}}$.\\
If we look at the global energy conservation in stellar evolution, we arrive at the homology (``similarity") relations for stars. From these we can derive a mass-luminosity relation that is very fundamental in stellar physics. For MS stars, the homology analysis yields $L\simeq \mu^4 M^3$, where $\mu$ is the mean molecular weight ($rT\sim \mu m$). This relation implies that the luminosity does not directly depend on energy generation; also the proportionality factor predominantly depends on the opacity of the stellar material, which in turn is determined by its chemical composition. If the energy generation in the star changes, it will adjust itself such that is has the same luminosity as before. \\
Furthermore, a mass-radius ($M\!-\!R$) relation is derived from the homology relations for stars. The relation now depends on the energy generation too in contrast with the mass-luminosity ($M\!-\!L$) relation. For the two main nuclear cycles on the MS, we get for the CNO-cycle $R\sim \mu^{0.61}M^{0.78}$ and for the pp-cycle we obtain $R\sim \mu^{0.125}M^{0.5}$. \\
The $M\!-\!L$ relation, the $M\!-\!R$ relation and the Stefan-Boltzmann law for black-body radiation leads to the equation for the MS in the Hertzsprung-Russell diagram (HRD): $\mathrm{log}(L)=8\times\mathrm{log}(T_{\mathrm{eff}})+\mathrm{const.}$ and also to lines of constant radius in the HRD, which follow $\mathrm{log}(L)=4\times \mathrm{log}(T_{\mathrm{eff}})+\mathrm{const.}$. \\
The lifetime of stars is derived from the $M\!-\!L$ relation and $\tau_{\mathrm{nuc}}\sim E_{\mathrm{nuc}}/L \sim M/L$ to get $\tau_{\mathrm{nuc}}\sim M^{-2}$. This means that more massive stars are brighter, but have shorter lifespans. In phases beyond the MS, the nuclear reaction energy release is smaller and the luminosities are generally larger, which leads to shorter lifetimes. Consequently, the total lifetime of a single star is dominated by its time spent on the MS. \\
Through the homology relations values of the central temperature $T_{\mathrm{c}}$, central pressure $P_{\mathrm{c}}$, and central density $\rho_{\mathrm{c}}$ of a star on the MS are obtained, which all depend on the stellar mass and the nuclear energy generation. Increasing stellar mass along the MS leads to: (1) increase of central temperature $T_{\mathrm{c}}$; (2) decrease of central density  $\rho_{\mathrm{c}}$ if the CNO-cycle ($1~\mathrm{M}_{\odot} \lessapprox M$) is the dominant nuclear burning mechanism, while $\rho_{\mathrm{c}}$ increases if the pp-cycle ($M\lessapprox 1~\mathrm{M}_{\odot}$) dominates; (3) decrease of the central pressure. Hence, with increasing mass, stars along the MS are hotter and radiation pressure becomes increasingly dominant until it dominates completely for very high mass stars. \\
Finally, we discuss the homologous contraction of a gaseous sphere. This analysis yields a relation between the central temperature and central density. For ideal gases, the contraction thereof leads to heating of the gas and for non-relativistic strongly degenerate gases, this contraction leads to cooling in a transition from non-degenerate to strongly degenerate region. This means that low mass stars will never ignite certain elements, because at some stage they become degenerate in the core and the central temperature drops upon further contraction. 

\subsubsection{Fundamental parameters - mass and composition}
\label{Fundamental parameters - Mass and composition}
While they are incredibly useful to understand fundamental relations in stellar astrophysics, the homology relations (see Sect.~\ref{Timescales, energy conservation and homology in stellar evolution}) cannot be applied over the full evolution of the star and are typically only applied to MS stars. We need other ways to describe the full evolution of a star. In general, the fundamental parameters of stellar evolution are the zero-age MS (ZAMS) mass and the (homogeneous) chemical composition. \\
Other very important parameters independent of mass and composition are rotation and magnetic fields. Rotation can lead to additional interior mixing, which changes the chemical composition of the star. Magnetic fields may influence the pressure balance and interact with convection and rotation, which is probably most important for massive stars. 

\subsubsection{Mass change of stars - stellar winds}
\label{Mass change of stars - Stellar winds}
The masses of all stars change throughout their lives through winds, parameterised by a stellar mass loss rate $\dot{M}$. Stellar winds are the outflows of matter leaving the stellar surface with an energy sufficient to escape from the star's gravity. The main question is what the nature of the force is that is powerful enough to overcome the star's gravity. Different types of stars have different winds. Recently, excellent reviews of the winds of lower mass stars were written by~\cite{Decin2020} and similarly of high mass stars by~\cite{Vink2021}. 
\begin{enumerate}
	\item Hot luminous stars (HMSs), such as massive MS or evolved stars ($R\sim 10~\mathrm{R}_{\odot}$), have strong and \textit{fast} (terminal wind velocities of $v_{\infty}\sim 2000-3000$~$\mathrm{km\,s}^{-1}$) stellar winds powered by radiative line driving (radiative forces that are exerted on atomic lines, such as ionized C,N,O or Fe-group elements; resonance lines in optically think regions just a couple of $R_{\odot}$ around the HMS). These have very high mass loss rates $\dot{M}$ of $10^{-8}-10^{-4}$~$\mathrm{M}_{\odot}/\mathrm{yr}$.\\
	\item Cool luminous stars (CMSs), such as AGB intermediate mass stars (IMS) ($R > 100~\mathrm{R}_{\odot}$) have strong and \textit{slow} ($v_{\infty}\leq 25$~~$\mathrm{km\,s}^{-1}$) stellar winds that are pulsation-driven. These two have very high mass loss rates $\dot{M}$ of $10^{-8}-10^{-4}$~$\mathrm{M}_{\odot}/\mathrm{yr}$. The fact the CMSs are cool, it is believed that close to the stellar atmosphere, these stars can form dust grains, because the pulsations from the star can form regions of large density just above the stellar atmosphere. The  dust grains absorb momentum and collide with surrounding gaseous species and thus you get a launch of a stellar wind.  \\
	\item Solar-type stars (LMSs) have hot surrounding coronae and have a \textit{weak} stellar wind 
	that is a pressure-driven coronal wind of intermediate speeds ($v_{\infty}\leq 400-800$~$\mathrm{km\,s}^{-1}$). They have very low mass loss rates $\dot{M}$ of $10^{-14}$~$\mathrm{M}_{\odot}/\mathrm{yr}$.
\end{enumerate} 
Many stellar evolution models used inside $N$-body codes express wind acceleration by a $\mathrm{\Gamma}$ factor. $\mathrm{\Gamma}$ is defined as the ratio of radiative over gravitational acceleration. Radiative acceleration is due to radiative pressure and introduces an extra force acting on a spherically symmetric, isothermal wind. It is related to electron scattering $\mathrm{\Gamma}_{\mathrm{e}}$ or dust scattering $\mathrm{\Gamma}_{\mathrm{d}}$, for example. These quantities are introduced into the momentum equation of an isothermal, spherically symmetric stellar wind, which leads to an effective gravitational acceleration $g_{\mathrm{eff}}(r)$. Using $g_{\mathrm{eff}}(r)$, we can calculate the escape velocities and these are lower by the introduction of the extra force. However, it depends very strongly on the distance to the stellar surface, where this additional force is introduced; the farther out it occurs, the less impactful it becomes on the overall stellar mass loss rate. Therefore, since dust grains form very \textit{close} to the star (e.g. in CMSs), these are very impactful on the mass loss rate. In red supergiants (RSGs), on the other hand, these grains form much farther out and therefore, dust-driven winds are generally not relevant here. \\  
Moreover, radiation transport and the chemistry in the wind are both essential to a full modelling of a stellar wind. It is important to state that in general, there is no full theory of stellar winds available~\cite{Decin2020}. Furthermore, the layperson is overwhelmed by the large number of mass loss rate prescriptions derived predominantly from observations, which differ enormously in magnitude and slope~\cite{Decin2020}. The choice of mass loss recipe has an enormous impact on the outcome of realistic $N$-body simulations and the dynamics of the star cluster as described in this review. As an astrophysical community, we are just at the beginning of unravelling the complexities of specific stellar winds, such as Wolf-Rayet (WR) stars~\cite{SanderVink2020} or the impact of pulsations and variability on winds in AGB and post-AGB stars~\cite{Trabucchietal2019} before a fully self-consistent theory can be envisioned.

\subsubsection{Formation of compact objects and their natal masses, kicks and spins}
\label{Formation of compact objects and their natal masses, kicks and spins}
\begin{center}
\begin{figure}
\includegraphics[scale=0.19]{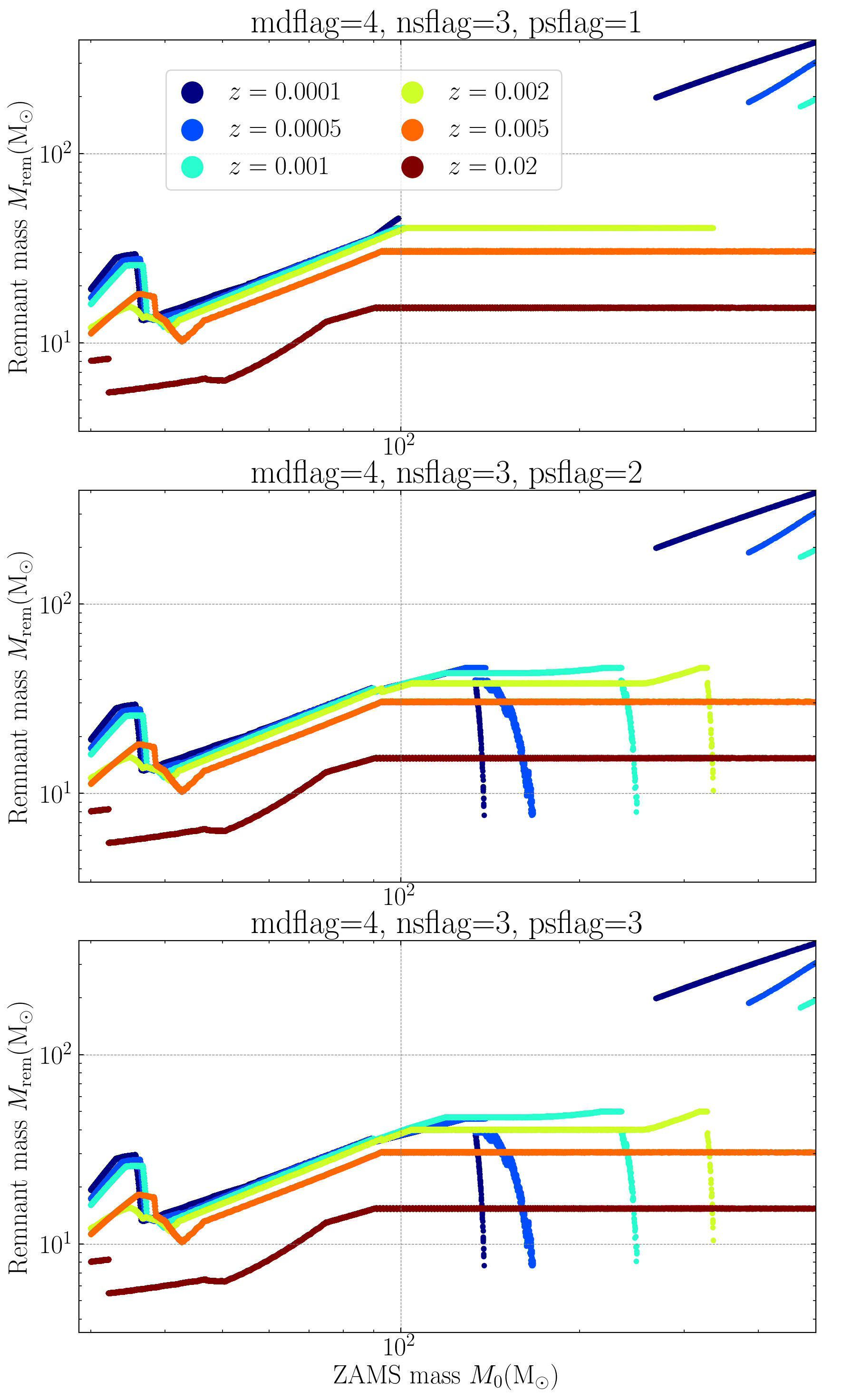}
 \caption{Initial-final-mass-relations (IFMRs) of the black holes (BHs) from \textsc{McLuster} samples ($N=2.5\times 10^4$ single ZAMS stars) depending on six different metallicities ranging from $Z$=0.0001 to Solar metallicity at $Z$=0.02. The \textsc{McLuster} version uses \texttt{level C} stellar evolution~\cite{Kamlahetal2022a}. Shown are the recipes for the ``strong" (\texttt{psflag}=1) on top~\cite{Belczynskietal2016}, ``weak" (\texttt{psflag}=2) in the middle~\cite{Leungetal2019c,Leungetal2020b} and the ``moderate" (\texttt{psflag}=3) (P)PISNe on the bottom~\cite{Leungetal2019c,Leungetal2020b}. The ZAMS stars suffer wind mass loss via metallicity-dependent winds (\texttt{mdflag}=4) (no bi-stability jump) from~\cite{Belczynskietal2010} and the core-collapse SNe are set to ``rapid"~\cite{Fryeretal2012} (Figure from~\cite{Kamlahetal2022a}).}
\label{IFMR_McLuster_Kamlah}
\end{figure}
\end{center}
Depending on the progenitor star core mass, a compact object such as a white dwarf (WD), neutron star (NS) or black hole (BH) may form. Oftentimes binary processes are involved~\cite{Willemsetal2005,Fragosetal2009,Wongetal2012,Wongetal2014}, but these are discussed in the next sub-chapter. The following processes apply to all single stars in the relevant mass ranges. The formation of a compact object is associated with a natal remnant mass, a natal kick and a natal spin, which are all subject to significant theoretical and observational uncertainty. Nevertheless, it is important to model these as accurately as possible, because the global dynamical evolution of a collisional stellar system critically depends on these. The natal mass depends on a number of factors and we will only focus now on the collapse mechanism and its associates fallback and not the mass loss in the progenitor star, although it is also instrumental. The impact of the mass loss has been discussed already in a previous section. Traditionally, the natal masses of the WDs (and their three main types HeWDs, COWDs, ONeWDs) and their dependence on the progenitor masses are modelled by~\cite{Hurleyetal2000,HurleyShara2003}. For NSs a maximum mass of around $2.5$~$\mathrm{M}_{\odot}$~\cite{Linares2018,Linares2020} and the relationship follows typically~\cite{Hurleyetal2000}, but the exact masses are unknown because of the large uncertainties mainly in the internal structure of a NS~\cite{LattimerPrakash2004,Lattimer2012}. In addition to~\cite{Hurleyetal2000}, the possibility of a so-called electron-capture SNe (ECSNe) that leads to the formation of a NS~\cite{Nomoto1984,Nomoto1987,Podsiadlowskietal2004b,Kieletal2008a,Ivanovaetal2008,Leungetal2020a}, which has very important properties that are discussed below, has been included in many codes~\cite{Belczynskietal2008,Banerjeeetal2020,Kamlahetal2022a}. 
Most attention has arguably been paid to the remnant BH masses~\cite{EldridgeTout2004b,Belczynskietal2008,Fryeretal2012} and a number of collapse mechanisms for certain mass ranges have been proposed. In simulations the most popular prescriptions are the rapid or delayed core-collapse SNe models by~\cite{Fryeretal2012} in combination with various (pulsational) pair instability SNe ((P)PISNe) stellar evolution recipes~\cite{Fryeretal2001,Yoshidaetal2016a,SperaMapelli2017,Woosley2017,WoosleyHeger2021,Belczynskietal2016,Leungetal2019c,Leungetal2020b}. Fig.~\ref{IFMR_McLuster_Kamlah} shows a suite of small simulations when \textsc{McLuster} ~\cite{Kuepperetal2011a,Kamlahetal2022a,Levequeetal2022a} is used as a population synthesis tool with \texttt{level C} stellar evolution \cite{Kamlahetal2022a}. It shows all relevant remnant mass phases, which can be sub-divided into a core-collapse SNe, PPISNe, PISNe and a direct collapse phase in increasing ZAMS mass (this is an extension of the core-collapse SNe models for ZAMS masses above which PISNe is ineffective; in our case an extension of the rapid SNe models by~\cite{Fryeretal2012}). Two interesting conclusions can immediately be drawn here: first, the metallicity is incredibly important for the production of high mass BHs, because progenitor stars with high metallicities will contain more metal lines for radiative wind mass loss. Secondly, the (P)PISNe prescriptions available from theory can have an enormous impact on the abundance of BHs. This might particularly important in Population III (Pop-III) star clusters, where intermediate mass black hole (IMBH) progenitor stars are postulated to have large enough masses and crucially also low enough metallicities from birth to evolve by (P)PISNe from interior evolution alone (e.g.~\cite{Kamlahetal2023,Wangetal2022} for recent $N$-body simulations of these clusters; see Sect.~\ref{Initial stellar mass function} for a more general discussion of Pop-III stars in the initialisation of star cluster simulations). 
\\
\begin{center}
\begin{figure}
\includegraphics[width=1.0\textwidth]{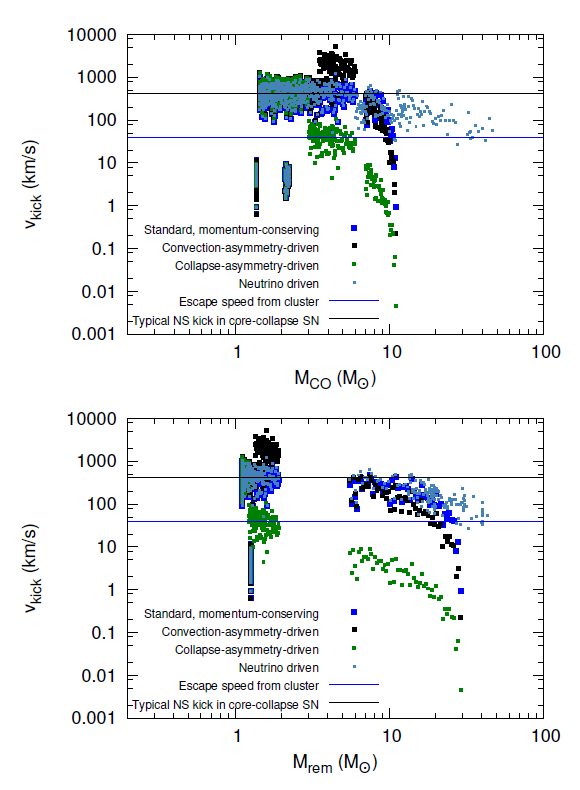}
 \caption{Plot showing natal kick prescriptions, $v_{\mathrm{kick}}$ (all of them are at least also available in \textsc{Nbody6++GPU, MOCCA, McLuster} and \textsc{PeTar}), as generated by \textsc{Nbody7} in~\cite{Banerjeeetal2020}. A metallicity of Z=0.0001 is assumed here. The models feature rapid core-collapse SNe from~\cite{Fryeretal2012} and strong (P)PISNe from~\cite{Belczynskietal2016}. Due to the logarithmic vertical axis, direct-collapse BHs with a fallback fraction, $f_{\mathrm{fb}}$ = 1 and $v_{\mathrm{kick}}$ = 0 are not shown in these panels. The sharp drop in $v_{\mathrm{kick}}$ with increasing $m_{\mathrm{CO}}$ or $m_{\mathrm{rem}}$ is the approach towards direct collapse. The typical $v_{\mathrm{esc}}$ for the $M_{\mathrm{cl}}(0)\simeq 5.0 \times 10^4~\mathrm{M}_{\odot}$ and $r_{\mathrm{h}}(0)\simeq2$~pc clusters considered here (blue, solid line). The velocity dispersion of the Maxwell distribution from all the kick models are scaled is $265.0~\mathrm{km\,s}^{-1}$ from~\cite{Hobbsetal2005}. It is apparent that for these settings the collapse asymmetry driven kicks will produce most (stellar mass) BHs below $v_{\mathrm{esc}}$ of the cluster (Figure adapted from~\cite{Banerjeeetal2020}).}
\label{Kicks_Banerjee.png}
\end{figure}
\end{center}
The magnitude of natal kicks results, broadly speaking, from an inherent asymmetry in the SNe process and generally their magnitude is rather uncertain~\cite{HansenPhinney1997,Hobbsetal2005}. They affects the dynamical stability of a binary (if one of the binary stars is forming a compact object) and are even able to disrupt a binary completely. This also implies that a large amount of gravitational binding energy in binaries may be removed from the cluster in this way and this will consequently impact the global cluster evolution. WDs are associated with low velocity kicks of the order of $10^0$~$\mathrm{km\,s}^{-1}$~\cite{Fellhaueretal2003}, while neutron stars may reach kicks above even $10^3$~$\mathrm{km\,s}^{-1}$ except in the case in which they form as a result of an electron-capture SNe (ECSNe). Here they receive then kicks of order of only $10^0$~$\mathrm{km\,s}^{-1}$~\cite{GessnerJanka2018} meaning that they can be retained in a star cluster (simulation)~\cite{Clark1975,Abbottetal2017a,Abbottetal2020a,Manchesteretal2005,Kamlahetal2022a}. Natal kicks for BHs and NSs, which do not undergo an ECSNe (or AIC or MIC, see next chapter), receive kicks traditionally scaled by fallback during the SNe in simulations~\cite{Belczynskietal2008,Fryeretal2012}. The more fallback of stellar material there is onto the proto-remnant core, the lower the resulting kick is. Furthermore, in simulations, it is typically assumed that the asymmetry is produced by a dominant process~\cite{Banerjeeetal2020,Banerjee2021a}: convection-asymmetry driven kicks~\cite{Schecketal2004,FryerYoung2007,Schecketal2008}, collapse-asymmetry driven kicks~\cite{BurrowsHayes1996,Fryer2004,MeakinArnett2006,MeakinArnett2007} or neutrino-driven natal kicks~\cite{Fulleretal2003,FryerKusenko2006,Banerjeeetal2020,Banerjee2021a}. These lead to different retention fractions of BHs in star cluster simulations~\cite{Banerjeeetal2020}, which can be seen in Fig.~\ref{Kicks_Banerjee.png} for a sample of \textsc{Nbody7} simulations from~\cite{Banerjeeetal2020}. It is apparent that for these settings the postulated collapse asymmetry driven kicks will produce most (stellar mass) BHs below $v_{\mathrm{esc}}$ of the cluster.\\
The natal spins of compact objects are important in general binary evolution and can also have significant impact on the mergers of compact objects, for example in a BH-BH merger~\cite{Morawskietal2018,Morawskietal2019}. In the following, we focus on BHs, but the same arguments can be extended to NSs and WDs and the discussion is largely taken from~\cite{Kamlahetal2022a}. In general, the spin angular momentum of the parent star does not necessarily translate directly into the natal spin angular momentum of the BH upon collapse. To quantify the spin,~\cite{Kerr1963} define a dimensionless parameter $a_{\mathrm{spin}}$ that accounts for the natal spin angular momentum.~\cite{Banerjee2021a} assumes that the magnitude of $a_{\mathrm{spin}}$ for the BHs is set directly at the moment of birth without any related mass accretion of GR coalescence processes. 
\begin{center}
\begin{figure}
\includegraphics[width=1.0\textwidth]{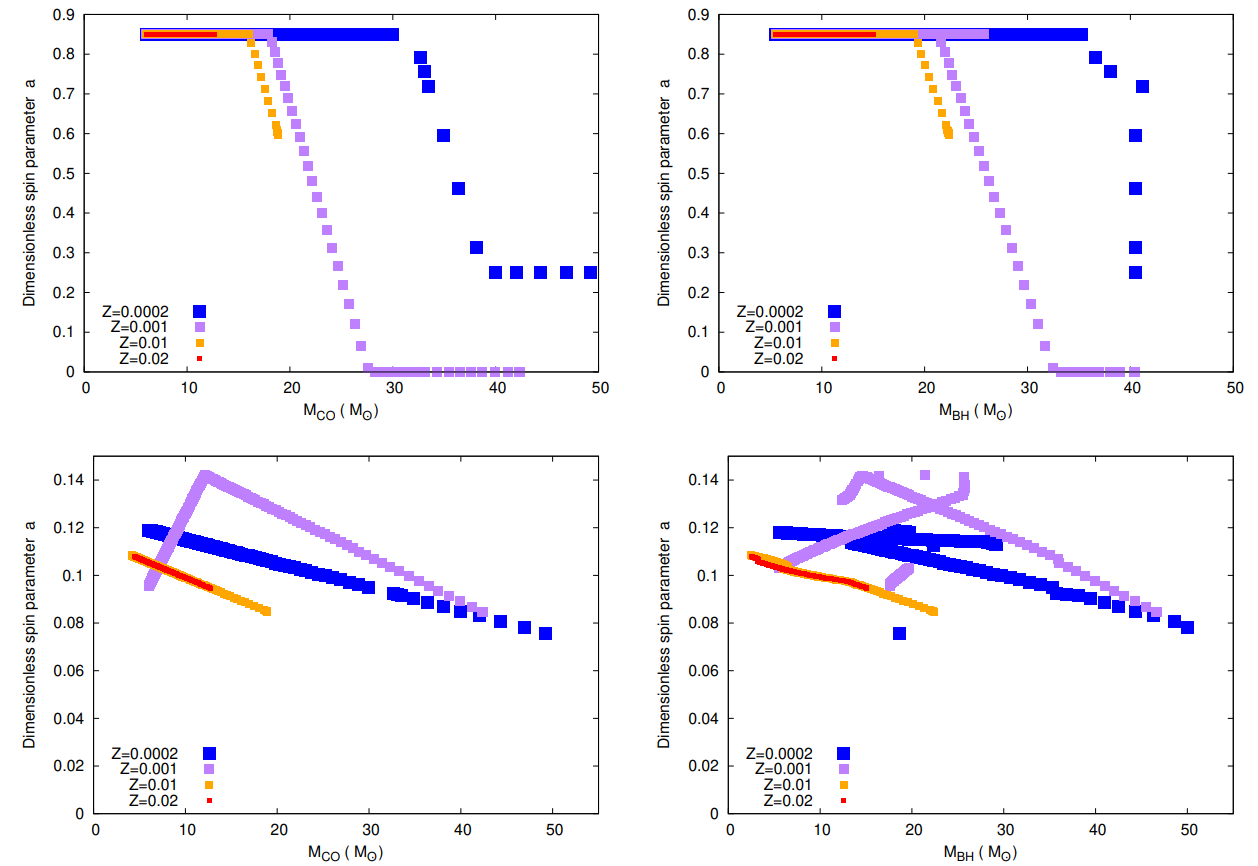}
 \caption{Plot showing the magnitude of dimensionless spin parameter, $a_{\mathrm{spin}}$, of stellar-remnant BHs at birth (i.e., of BHs that have not undergone any mass accretion or GR coalescence after their formation) as a function of the progenitor star’s carbon-oxygen core mass, $M_{\mathrm{CO}}$ (left column), and the BH mass, $M_{\mathrm{BH}}$ (right colummn; all of them are at least also available in \textsc{Nbody6++GPU, MOCCA, McLuster} and \textsc{PeTar}), as generated by \textsc{Nbody7} in~\cite{Banerjee2021a}.\\
 Top panels: the $N$-body models corresponding to these panels employ the “Geneva model” of~\cite{Belczynskietal2020} for BH spin and comprise only single stars initially, whose ZAMS masses range from $(0.08-150.0)~\mathrm{M}_{\odot}$ and which are distributed according to a standard Kroupa IMF~\cite{Kroupa2001}. The models feature rapid core-collapse SNe from~\cite{Fryeretal2012} and strong (P)PISNe from~\cite{Belczynskietal2016}. The models are shown for four metallicities, $Z$= 0.0002, 0.001, 0.01, and 0.02 as indicated in the legends. \\
 Bottom panels: the $N$-body models corresponding to these panels employ the “MESA model” of~\cite{Belczynskietal2020} for BH spin. The other model characteristics are the same as those in the top panels except that the “weak” PPISNe mass prescription~\cite{Leungetal2019c} is utilized (resulting in the non-monotonic behaviour with respect to $M_{\mathrm{BH}}$, which, here, extends up to
 $\simeq 50~\mathrm{M}_{\odot}$ as opposed to the models in the top panels, where $M_{\mathrm{BH}}$ is capped at $\simeq 40.5~\mathrm{M}_{\odot}$ due to the use of~\cite{Belczynskietal2016} (Figure adapted from~\cite{Banerjee2021a}).}
\label{NatalSpins_Banerjee.png}
\end{figure}
\end{center}
In the following we highlight three natal spin models that are available now in \textsc{Nbody7, Nbody6++GPU, McLuster, PeTar} and \textsc{MOCCA}, see also~\cite{Kamlahetal2022a}. The simplest model of BH natal spins, the \texttt{Fuller} model, produces zero natal spins~\cite{Banerjee2021a} as here the Tayler-Spruit magnetic dynamo can essentially extract all of the angular momentum of the proto-remnant core, leading to nearly non-spinning BHs~\cite{Spruit2002,FullerMa2019,Fulleretal2019}. The second spin model is the \texttt{Geneva} model~\cite{Eggenbergeretal2008,Ekstroem2012,Banerjee2021a}. The basis for this model is the transport of the angular momentum from the core to the envelope. This is only driven by convection, because the \texttt{Geneva} code does not have magnetic fields in the form of the Taylor-Spruit magnetic dynamo. This angular momentum transport is comparatively inefficient and leads to \textit{high} natal spins for low to medium mass parent O-type stars, whereas for high mass parent O-type stars, the angular momentum of the parent star may already have been transported away in stellar winds and outflows and thus the natal BH spins may be low. The third and last spin model is the \texttt{MESA} model, which also accounts for magnetically driven outflows and thus angular momentum transport~\cite{Spruit2002,Paxtonetal2011,Paxtonetal2015,Fulleretal2019,Banerjee2021a}. This generally produces BHs with much \textit{smaller} natal spins than the \texttt{Geneva} model described above. The \texttt{Geneva} and the \texttt{MESA} models and their metallicity dependence are shown in Fig.~\ref{NatalSpins_Banerjee.png}.

\subsection{Binary stellar evolution}
\label{Binary Stellar Evolution}
In addition to the astrophysical processes that affect all stars in isolation, the proximity (orbital period $P_{\mathrm{orb}} \leq 10^4$~days~\cite{Eggleton1996}) to another star or compact object through the frequent encounters in collisional stellar systems or through intrinsic binary evolution, can affect the individual stars or compact objects dramatically and we need to account for these in the simulations. A population synthesis code should include them all~\cite{Eggleton2006}. 

\subsubsection{Stellar Spin and orbital changes due to mass loss or gain}
\label{Stellar Spin and orbital changes due to mass loss or gain}
If two stars are in a binary, they can transfer mass via stellar winds and therefore also transfer angular momentum even if they are not yet undergoing Roche-lobe overflow (RLOF)~\cite{Hurleyetal2002b,Eggleton2006,Tout2008b}. If a secondary star accretes mass by passing through the wind of the primary star, it is spun up intrinsically by a fraction of the spin angular momentum that is lost by the donor star. The accretion rate is traditionally modelled by~\cite{BondiHoyle1944}, which depends on the wind velocity $v_{\mathrm{W}}$. This quantity is observationally difficult to determine~\cite{Decin2020} and should be proportional to the escape velocity from the stellar surface of the star~\cite{Hurleyetal2002b}. \\
The mass variations between companion stars also changes the orbital parameters of the binary star. In general, the eccentric orbit is circularised as a result of mass transfer being more effective at periastron than apastron. Additionally, the accretor star is slowed down by the drag induced by the wind it passes through and this dissipates angular momentum away from the system. The orbital circularisation timescale $\tau_{\mathrm{circ}}$ as result of mass transfer is orders of magnitudes larger than the equivalent timescale caused by tidal friction for the same binary star system (see Sect.~\ref{Orbit circularisation and stellar spin synchronisation by tidal damping}).

\subsubsection{Effects of tidal damping}
\label{Orbit circularisation and stellar spin synchronisation by tidal damping}
Observations show that the rotation of close binary stars is synchronised with the orbital motion without any dynamical mass transfer having taken place~\cite{Lurieetal2017,Mazeh2008,MeibomMathieu2005}. Therefore, there must exist a torque that transfers angular momentum between the stellar spin and the orbit in such a way that the binary approaches the observed equilibrium state that is characterised by corotation (spin-orbit synchronisation timescale $\tau_{\mathrm{sync}}$) and a circular orbit (circularisation timescale $\tau_{\mathrm{circ}}$)~\cite{Zahn1977,Hut1981,Hurleyetal2002b,Tout2008b}. Alternatively, dissipation of energy might also lead to an accelerated in-spiral of the binary stars~\cite{Hut1980,Rasioetal1996,Tout2008b}. \\
When two binary star members are detached but sufficiently close, tidal interaction between them becomes important. The mere presence of a companion star causes a tidal force that elongates a star along the line between the centres of mass, thereby resulting in tidal bulges (see e.g.~\cite{Hurleyetal2002b}). \\
When the binary component rotates uniformly with a circular orbital motion, then the tidal bulges on its stellar surfaces are steady and the stars are in hydrostatic equilibrium. In such a scenario, we also speak of equilibrium tides. However, when this condition no longer holds, the hydrostatic equilibrium is disrupted and the star undergoes forced stellar oscillations. This scenario is described by a combination of equilibrium and now also dynamical tides, the latter of which produce much smaller tidal bulges than the former and they can also take any orientation~\cite{Eggletonetal1998,Eggleton2006,Hurleyetal2002b,Zahn1970,Zahn1974,Zahn1975,Zahn1977,Siessetal2013}.
Dissipative processes within a star cause the tides to be misaligned with the line of centres. This results in a torque that transfers angular momentum between the stellar spin and the orbit~\cite{Hurleyetal2002b}. This dissipation is non-conservative and happens on relatively long timescales~\cite{Eggleton2006}. \\
The dissipative processes within a star depend on the stellar structure. Typically, a distinction is made between stars with appreciably deep convective envelopes and stars with radiative envelopes. The tides dissipate energy and the binary system approaches an equilibrium state that is characterised by a circular orbit and corotation~\cite{Zahn1977,Hut1981,Hurleyetal2002b,Toutetal2008}. \\
In stars with appreciably deep convective envelopes, turbulent viscosity that acts on equilibrium tides (the same effect on dynamical tides is negligible~\cite{Zahn1975,Zahn1977}) is the most efficient form of dissipation~\cite{Kopal1978,Hut1981,Hurleyetal2002b}. The dissipation takes shorter than the nuclear burning timescale $\tau_{\mathrm{nuc}}$ (see Sect.~\ref{Timescales, energy conservation and homology in stellar evolution})~\cite{Zahn1989,Zahn1991,Hurleyetal2002b}. \\
In stars with radiative envelopes, radiative dissipation near the surface of the star causes an asymmetry in the internal stellar oscillations induced by tides and the tidal field itself. This leads to a torque that is necessary for the binary system to approach the equilibrium state~\cite{Zahn1977,Zahn1989,Zahn1992,Hurleyetal2002b} and in sufficiently close binaries this happens on shorter timescales than the nuclear burning timescale $\tau_{\mathrm{nuc}}$~\cite{Zahn1975}. This radiative damping on the dynamical tides is the most efficient process to achieve the equilibrium state in binary stars with member stars that do not have an outer convective zone. However, if they do then the aforementioned turbulent friction on the equilibrium tides provides the primary torquing~\cite{Zahn1975,Zahn1977,Zahn1989}. \\
$\tau_{\mathrm{sync}}$ and $\tau_{\mathrm{circ}}$ in binary stars with convective envelopes are typically orders of magnitude smaller than those with radiative envelopes~\cite{Zahn1977,Hurleyetal2002b}. $\tau_{\mathrm{sync}}$ and $\tau_{\mathrm{circ}}$ are generally not equal except in a limiting case~\cite{Zahn1977}.
\\
If the stars are degenerate but have sufficient stellar structure, i.e. WDs and NSs, then the above two dissipative mechanisms cannot be used as the stellar structure is significantly different. WDs will have very low spins, because the progenitor AGB star has already spun down in its expansion. Furthermore, in WD-WD binaries, the orbit will already be circularised (in the absence of WD natal kicks~\cite{Fellhaueretal2003}) due to the stellar wind mass and thus angular momentum loss. For this reason only the synchronisation timescale $\tau_{\mathrm{sync}}$ due to degenerate damping is of importance here and it is only applicable for close systems. $\tau_{\mathrm{sync}}$ in WD-WD, WD-NS and NS-NS binaries could exceed the age of the Universe~\cite{Campbell1984}.  

\subsubsection{Dynamical mass transfer and its stability}
\label{Dynamical mass transfer and its stability}
Apart from mass transfer through stellar winds, mass transfer can also happen via RLOF. This happens when the primary star fills it RL as a result of stellar expansion or in-spiral. The subsequent mass transfer then happens through the innermost Lagrange point. Typically, this process depends strongly on the mass ratio of the binary~\cite{Eggleton1983a} and happens in corotating, circularised binaries but in some instances, it can also occur in highly eccentric binaries, that are a result of tidal capture. \\
In the RLOF mass transfer, also angular momentum is transferred. The stability of the mass transfer traditionally determined by
three logarithmic derivatives of radii with respect to the mass of the RL-filling star following~\cite{Webbink1985b,Webbink2003}: the derivative describing the rate of change of the RL radius $R_{\mathrm{L}}$ for conservative mass transfer (total mass and angular momentum conservation) $\zeta_{\mathrm{L}}$~\cite{Eggleton2006}, the derivative at constant entropy $s$ and composition of each isotope $X_{\mathrm{i}}$ throughout the star $\zeta_{\mathrm{ad}}$ and a third derivative that describes the rate of change of the radius of the primary with mass in equilibrium $\zeta_{\mathrm{eq}}$. The mass transfer rate $\dot{M}$ depends on the relative values of these derivatives~\cite{Tout2008b}:
\begin{enumerate}
	\item $\zeta_{\mathrm{ad}} < \zeta_{\mathrm{L}}$ $\rightarrow$ $\dot{M}$ increases rapidly, there is positive feedback and the mass transfer is unstable, the secondary star cannot accrete at such a high rate and it expands $\rightarrow$ formation of a common envelope (CE) around the two stars~\cite{Paczynski1976,Ivanovaetal2013,Ivanova2019}.
	\item $\zeta_{\mathrm{eq}} < \zeta_{\mathrm{L}} < \zeta_{\mathrm{ad}}$ $\rightarrow$ $\dot{M}$ decreases in its immediate response, but then expands on a thermal timescale.
	\item $\zeta_{\mathrm{L}} < \zeta_{\mathrm{ad}}$ \& $\zeta_{\mathrm{L}} < \zeta_{\mathrm{eq}}$ $\rightarrow$ $\dot{M}$ decreases initially, because the stellar radius decreases. RLOF happens again, when the primary fills it RL again.
\end{enumerate}
On these basis of these exponents alone, it is possible to make a number of arguments on the evolution of Cataclysmic Variables (CVs), Algols and other exotic binary stars. 

\subsubsection{Common-envelope evolution}
\label{Common-envelope evolution (CEE)}
The process of CE evolution (CEE) is instrumental in compact binary and close binary formation.~\cite{Ivanovaetal2013,Ivanova2016,Ivanova2018,Ivanova2019,Paczynski1976}. A CE is the outcome when $\zeta_{\mathrm{ad}} < \zeta_{\mathrm{L}}$ in RLOF or when two stars collide, where one of the stars has a dense core. Generally, CEE occurs when the primary star transfers more mass on dynamical timescales than secondary can accept. It strongly depends on the instabilities in the RLOF preceding the formation of a CE~\cite{Olejaketal2021}. The CE expands and thus rotates more slowly than the orbit of the secondary and primary star. This causes friction, the binary spirals in and transfers orbital energy to the envelope. Either so much energy in this process is transferred that the envelope is expelled completely leaving behind a close binary in corotation or in the process of in-spiral the binaries coalesce~\cite{Eggleton2006,Hurleyetal2002b,Toutetal1997}. \\
The CE is traditionally modelled with the ``$\alpha \lambda$" energy-formalism~\cite{Webbink1984,Toutetal1997,Hurleyetal2002b}, which assumes energy is conserved and where $\alpha$ ($\alpha < 1$ if no other energy sources other than the binding and orbital energy are present; it can be as high as $\alpha = 5$ otherwise~\cite{Fragos2019}) is the ``efficiency" of the energy re-use and $\lambda$ is a measure of the binding energy between the envelope and the core of the donor star and should depend on the type of the star, its mass and its luminosity~\cite{DewiTauris2000,Claeysetal2014,Ivanova2019,Olejaketal2021}. This picture is very simplistic and does not take into account the myriad of processes that go on during CEE, which are also not fully understood yet~\cite{Ivanovaetal2013,IvanovaNandez2016,Ivanova2019,Ivanovaetal2020}. On the other hand, the $\alpha \lambda$ energy-formalism is computationally very efficient and therefore it is widely used in population synthesis codes that require fast and robust stellar evolution computations~\cite{Hurleyetal2002b,Belczynskietal2008,Claeysetal2014,Eldrigeetal2017,Mapelli2018b,Breiviketal2020a,Kamlahetal2022a}. Some of these also allow for recombination energy of hydrogen in the cool outer layers of the CE being transferred back into the binding energy of the CE. \\
Recently, a new formalism has been developed by~\cite{Tranietal2022}, which solves a binary orbit under gas friction with numerical integration. This means that the authors do not approximate CE as an instantaneous process, unlike in many binary population synthesis (BPS) codes around. The new formalism, which can be easily implemented in BPS codes, provides a significant upgrade, which can explain observations of post-CE binaries which non-zero eccentricities~\cite{Kruckowetal2021}. \\
In a binary consisting of a NS or a BH and a giant star, after the CE has been ejected and if the binary survives this phase, the H-rich envelope of giant stars might be stripped completely off. Now, the binary consists of a BH or a NS orbiting a naked He star. There might now be subsequent mass transfer from the naked He star to the NS of BH. This post-CE RLOF mass transfer leaves behind a so-called ``ultra-stripped" He star that explodes in an ultra-stripped SNe~\cite{Taurisetal2013b,Tauris2015,Taurisetal2017}. This type of SNe is significantly different from the typical core-collapse SNe and the process of ultra-stripping leads to a significant decrease in BH-NS and BH-BH mergers and a slight increase in NS-NS mergers~\cite{Schneideretal2021}. 

\subsubsection{Mergers and general relativistic merger recoil kicks}
\label{Coalescence and collisions}
\begin{center}
\begin{figure}
\includegraphics[width=1.0\textwidth]{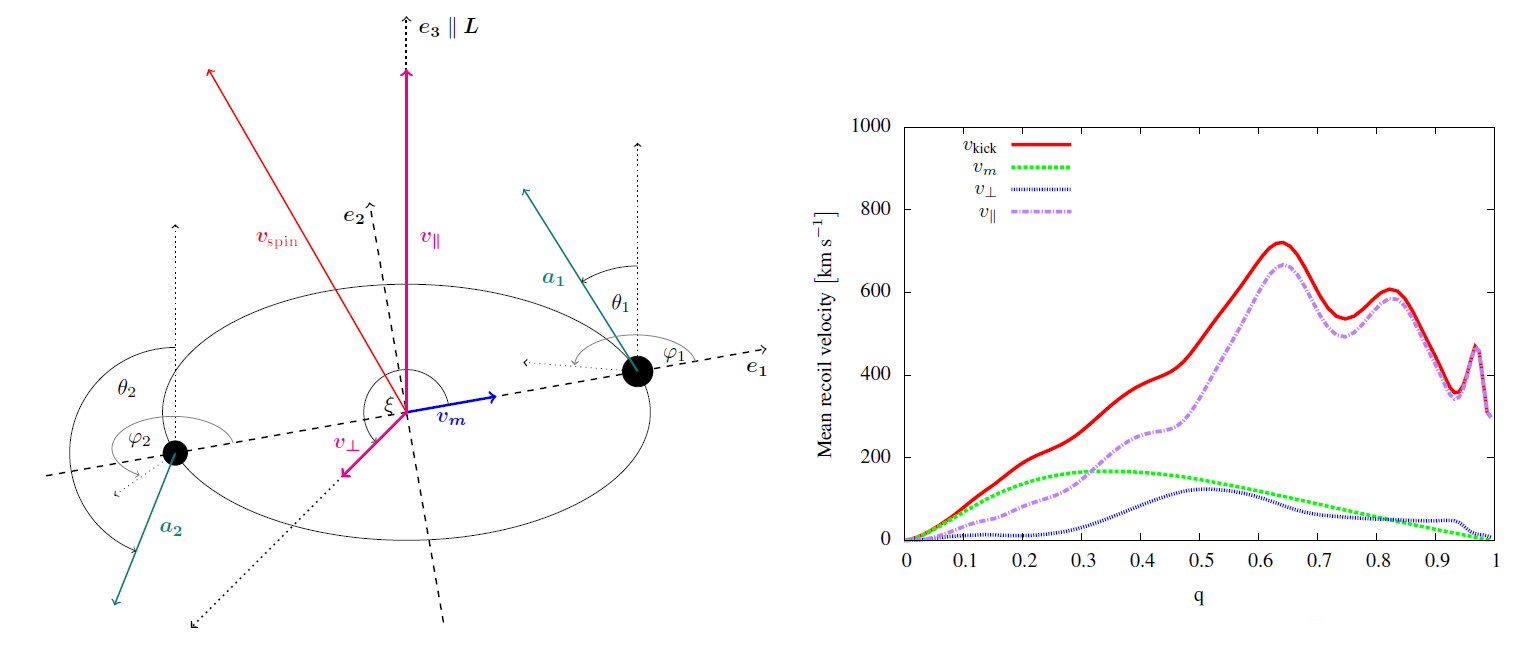}
 \caption{Left hand side: Conceptual picture presenting components of the recoil
velocity. Dashed lines represent a Cartesian coordinate system in
the orbital plane: $\mathbf{e_1}$ and  $\mathbf{e_3}$. A vertical dotted line is a line perpendicular
to orbital plane ($\mathbf{e_3}$, parallel to orbital angular momentum).
The red vector is the kick component related to spin asymmetry, and
magenta vectors are its projections on the plane and parallel to
$\mathbf{e_3}$. The blue vector represents the mass inequality contribution. The black
filled circles represent a pair of BHs, their spins and orientation
in a spherical coordinate system are illustrated. This drawing also
reflects typical proportions between recoil velocity components.\\
Right hand side: Example of how each recoil velocity component depends
on mass ratio $q$ for a metallicity dependent spin vector from~\cite{Belczynskietal2017}. $q$ is the only variable for the determination of $v_{\mathrm{m}}$. Other components and the overall kick velocity depend also on spin magnitudes and orientations, in this
case the mean value is plotted. As we can see, the major component
is almost always $v_{\parallel}$, others only play a role for low $q$.
(Figures and captions adapted from~\cite{Morawskietal2018} (combination of Fig.~2 and Fig.~3)).
 }
\label{Morawski_BHkick}
\end{figure}
\end{center}
An outcome of CEE may be the coalescence of the two stars. The subsequent merger product depends on the relative compactness of the two stars and thus it depends on the stellar evolutionary stage~\cite{Toutetal1997,Hurleyetal2002b}. If similar in stellar type, then the two stars mix completely. If one is much more compact than the other, then the more compact core sinks to the centre and the other mixes with the envelope. An unstable Thorne-{\.Z}ytkow object is created if the merger involves a NS or a BH~\cite{ThorneZytkow1977}. Detailed calculations on the merger outcomes following coalescence and collisions, which are less likely than coalescence, but still relevant in star clusters (e.g.~\cite{Rizzutoetal2021,Rizzutoetal2022}), depending on the initial stellar types have been tabulated in~\cite{Hurleyetal2002b}. \textcolor{red}{A coalescence for our purposes here means that at least one of the members is a star with a core and that the binary has a circular orbit before merging, while a collision means an actual physical collision, where none of the binary members is an evolved stellar type, but the member can also be a compact object.} Generally, the mixing and the final masses of the merger products are highly uncertain and only approximations can be made according to our current knowledge~\cite{Olejaketal2020a,Kamlahetal2022a}. There are recent attempts to unravel the masses and compositions of merger products of massive stars with hydrodynamical codes~\cite{Costaetal2022,Balloneetal2023} and they can be used to give approximate formulae for $N$-body or BPS codes in the future. \\
The merger of compact objects is associated with a general relativistic (GR) merger recoil kick due to the asymmetry in the GW (see also~\cite{Kamlahetal2022a} for a brief discussion with respect to \textsc{Nbody6++GPU} and \textsc{MOCCA}). The recoil velocity in this process depends on the mass ratio of the two compact objects and their spin vectors~\cite{Loustoetal2012} and can reach several hundreds $\mathrm{km\,s}^{-1}$ on average~\cite{Morawskietal2018,Morawskietal2019}, which is much larger than typical star cluster escape speeds. Fig.~\ref{Morawski_BHkick} (from~\cite{Morawskietal2018,Morawskietal2019}) shows the a conceptual picture of the geometry of a GR merger recoil kick in a BH-BH merger and the dependence of the mean recoil velocity on the mass ratio $q$ of the two BHs for a metallicity dependent spin model from~\cite{Belczynskietal2017}. It can be seen that $q$ has a huge impact on whether a GR merger recoil kick velocity exceeds the escape speed of the surrounding stellar (and gaseous) material or not. Equal mass mergers might be retained in nuclear star clusters~\cite{Schoedel2014b} and extreme mass ratio mergers might theoretically even be retained in open clusters (although IMBHs will probably not form there)~\cite{Bakeretal2007b,Bakeretal2008,PortegiesZwartetal2010,Baumgardtetal2018}. For (nearly) non-spinning BHs (\textsc{Fuller} model~\cite{FullerMa2019}), the kick velocity is smaller than for high spins. For non-aligned natal spins and small mass ratios, the asymmetry in the GW may produce GR merger recoils that reach thousands of $\mathrm{km\,s}^{-1}$~\cite{Bakeretal2008,vanMeteretal2010b}. The calculation of the mass ratio is straightforward and the spins may be calculated from e.g.~\cite{HoffmanLoeb2007} or~\cite{JimenezFortezaetal2017}. \\
Generally, the orbital angular momentum of the BH-BH dominates the angular momentum budget that contributes to the final spin vector of the post-merger BH and therefore, within limits, the final spin vector is mostly aligned with the orbital momentum vector~\cite{Banerjee2021a}. In the case of physical collisions and mergers during binary-single interactions, the orbital angular momentum is not dominating the momentum budget and thus the BH spin can still be low.~\cite{Banerjee2021a} also includes a treatment for random isotropic spin alignment of dynamically formed BHs. Additionally,~\cite{Banerjee2021a} assumes that the GR merger recoil kick velocity of NS-NS and BH-NS mergers~\cite{ArcaSedda2020,Chattopadhyayetal2021} to be zero but assigns merger recoil kick to BH-BH merger products from numerical-relativity fitting formulae of~\cite{vanMeteretal2010b} (which is updated in~\cite{Banerjee2022}). The final spin of the merger product is then evaluated in the same way as a BH-BH merger. \\
The inclusion of these kicks in direct $N$-body simulations is still unusual (e.g.~\cite{DiCarloetal2019,DiCarloetal2020a,DiCarloetal2020b,DiCarloetal2021,Rizzutoetal2021,Rizzutoetal2022,Kamlahetal2022a,Kamlahetal2022b} all do not include these in addition to missing PN terms), but it is worth mentioning~\cite{ArcaSeddaetal2021a} do include the GR merger recoil kicks by posterior analysis. \textsc{Nbody7}~\cite{Aarseth2012,Banerjeeetal2020,Banerjee2021a} on the other hand does include GR merger recoil kicks based on~\cite{Loustoetal2012,HoffmanLoeb2007}. In \textsc{MOCCA} numerical relativity (NR) models~\cite{Campanellietal2007,Rezollaetal2008,Hughes2009,vanMeteretal2010b,JimenezFortezaetal2017} have been used to formulate semi-analytic descriptions for \textsc{MOCCA} and \textsc{Nbody} codes~\cite{Morawskietal2018,Morawskietal2019,Banerjee2021a,BelczynskiBanerjee2020,ArcaSeddaetal2021a,Banerjee2022}. Recently, GR merger recoil kicks have also been added to \textsc{Nbody6++GPU}~\cite{ArcaSeddaetal2022} following~\cite{Campanellietal2007,JimenezFortezaetal2017} and with this code version, the whole kick process can be followed self-consistently.

\subsubsection{Accretion or merger induced collapse}
\label{Neutron star production by accretion or merger induced collapse}
In sufficiently close double degenerate COWD-COWD, ONeWD-ONeWD or COWD-ONeWD binary stars, sufficiently high and dynamically stable RLOF mass accretion of hot CO-rich matter may lead to a heating of the outer layers of the secondary, which will result in the ignition of nuclear burning~\cite{SaioNomoto2004}. If carbon burning is ignited in the COWD envelope, the heat will be transported the stellar core by conduction and then the secondary will evolve into an ONeWD~\cite{SaioNomoto1985,SaioNomoto1998}, which will eventually collapse into a NS if the critical mass of the ONe core is surpassed ($M_{\mathrm{ecs}}$=1.38~$\mathrm{M}_{\odot}$)~\cite{Nomoto1984,Nomoto1987,Belczynskietal2008}. This ONeWD collapse is referred to as accretion induced collapse (AIC). If, on the other hand, the ignition happens in the centre then the star will undergo a SN-Ia explosion, which leaves no remnant behind. \\
Double degenerate COWD binaries may also coalesce without undergoing dynamically stable mass transfer. During this process the less massive forms a thick, turbulent accretion disk and the more massive COWD will accrete matter close to the Eddington limit. Here the carbon will be ignited on the envelope of the secondary and thus the outcome will be a ONeWD and no SN-Ia will happen~\cite{SaioNomoto2004}. Again, if the ONe core mass surpasses $M_{\mathrm{esc}}$, then the ONeWD will collapse into a NS and this is known as a merger-induced collapse (MIC). Other pathways for MIC are mergers of a ONeWD with any type of WD companion if the resulting merger mass surpasses the critical mass for NS formation~\cite{SaioNomoto1998,Belczynskietal2008}. \\
The distinction between AIC and MIC is made, because the former may be observed already through their stable mass transfer phase or in low-mass X-ray binary stars and the latter may be observed through gravitational waves observed with LISA~(Laser Interferometer Space Antenna\cite{Ruiteretal2019}).

\subsubsection{Gravitational radiation and magnetic braking}
\label{More slow non-conservative processes: gravitational radiation and magnetic braking}
Gravitational radiation emitted from sufficiently close binary stars ($P\leq 0.6$~days) transports angular momentum away from the system and drives it to a mass transfer state that might result in coalescence~\cite{PetersMathews1963,Hurleyetal2002b,Eggleton2006}. The effect this radiation has on the orbit of the binary (excluding PN terms) may be obtained by averaging the rates of energy loss and angular momentum loss over an approximately Keplerian orbit~\cite{Peters1964,Eggleton2006}. Gravitational radiation will circularise the orbit on the same timescale as the orbit shrinks until coalescence. \\
In co-rotating and sufficiently close binary stars, magnetic braking slows down the rotation of the individual star with a convective envelope, but also drains angular momentum from the orbit of the binary star, because tidal friction between the stars may conserve co-rotation~\cite{Mestel1968a,Mestel1968b,MestelSpruit1987,Eggleton2006}. As a result, this process will force a close binary to a state of RLOF within Hubble time. In some situations, this process is dominating binary evolution, such as in CVs above the orbital period gap~\cite{Schreiberetal2016,Zorotovicetal2016,Bellonietal2018b}. In spin-spin period evolution ($P-\dot{P}$) of pulsars this process is also important (e.g.~\cite{KielHurley2006,KielHurley2009}). Both processes outlined above are non-conservative.

\subsection{Combining stellar evolution with collisional N-body codes}
\label{Combining stellar evolution with collisional $N$-body codes}
There are two main methods that stand out in practice concerning the integration of the complicated stellar evolution into $N$-body codes. Both of these, interpolation between tables or approximation of stellar evolution data by some interpolation (fitting) formulae as functions of mass, age and metallicity, has unique advantages and disadvantages that have been known for a long time~\cite{Eggleton1996}. As it stands now, the two approaches are not in competition, but rather complement one another~\cite{Hurleyetal2000}. 

\subsubsection{Interpolation between tables}
\label{Interpolation between tables}
This method calculates stellar parameters from detailed evolutionary tracks (e.g.~\cite{Polsetal1998}). These evolutionary tracks are derived from 1D stellar evolution codes and are in tabular format. They are necessarily rather large and therefore, this approach has historically been limited by memory availability on hardware~\cite{Eggleton1996,Hurleyetal2000,Agrawaletal2020}. Unlike fitting formulae, stellar parameters from the given set of detailed tracks are calculated in real time with this method. Hence, one just needs to change the input stellar tracks to generate a new set of stellar parameters. It has been claimed that this approach is the most flexible, robust and efficient today when combining detailed stellar evolution with stellar dynamics~\cite{Agrawaletal2020}. \\
\cite{MaederMeynet1989,Schalleretal1992,Alongietal1993,Bressanetal1993,Fagottoetal1994a,Fagottoetal1994b,Claret1995,ClaretGimenez1995} constructed such tables, which were later then expanded upon and refined by~\cite{Polsetal1998}. In the aforementioned works, the convective mixing or overshooting length $l_{\mathrm{OV}}$ presents another hurdle, which describes the average distance by which convective cells push into stable regions (or radiative regions from Schwarzschild condition~\cite{Biermann1932,Gabrieletal2014}) beyond the convective boundary~\cite{Schalleretal1992,Polsetal1998,JoyceChaboyer2018}. This treatment was modified by~\cite{Polsetal1998} and replaced with a ``$\nabla$ prescription", which is based on the stability criterion itself ($\delta_{OV}=0.12$ was found to best reproduce observations~\cite{Schroederetal1997,Polsetal1997,Polsetal1998,Hurleyetal2000}). This new criterion avoids physical discontinuities for disappearing classical convective cores. Further quantities that will influence the calibration of the luminosity $L$ of a stellar evolution model are the nuclear reaction rates and the core Helium abundance $Y$. Another source of large uncertainty was left largely unchanged by~\cite{Polsetal1998}. This uncertainty has been described by the~\cite{Polsetal1998} as the ``Achilles heel" in stellar evolution codes. This uncertainty is in the mixing length of $\alpha_{\mathrm{MLT}}$, which is derived from mixing-length theory~\cite{BoehmVitense1958} to describe heat transport in the convective regions of stars~\cite{JoyceChaboyer2018,Pasettoetal2018} (see also Sect.~\ref{Single stellar evolution}).~\cite{Polsetal1998} set $\alpha_{\mathrm{MLT}}$=2.0 (based on the Solar model). But not all stars with convective regions exhibit identical convective properties and $\alpha_{\mathrm{MLT}}$ can show large variations from star to star~\cite{JoyceChaboyer2018}. \\
Even today methods stellar evolution by interpolation between tables are being developed with increasing success as hardware memory capabilities also improve:
\begin{itemize}
    \item \textsc{SEVN}~\cite{Speraetal2015,SperaMapelli2017,Speraetal2019}, which has been completed for binary evolution~\cite{Siciliaetal2022} has been used extensively to study the evolution gravitational wave source progenitor stars. Additionally, it is not available as \textsc{SEVN2.0}, which is integrated in \textsc{PeTar} \cite{Wangetal2020d}.
    \item and \textsc{COMBINE}~\cite{Kruckowetal2018} codes, which also has binary evolution implemented~\cite{Kruckow2020,Kruckowetal2021} has also been used extensively to study the evolution gravitational wave source progenitor stars. 
    \item \textsc{METISSE} code~\cite{Agrawaletal2020}, which is based on the \textsc{STARS}~\cite{Eggleton1971,Eggleton1972,Eggletonetal1973a,Eggletonetal1973b,Pols1995,Polsetal1997,Schroederetal1997}, \textsc{MESA}~\cite{Paxtonetal2011,Paxtonetal2013,Paxtonetal2015,Paxtonetal2016,Paxtonetal2018,Paxtonetal2019} and \textsc{BEC}~\cite{Yoonetal2006,Yoonetal2012,Brottetal2011,Koehleretal2015,Szecsietal2015,Szecsietal2022}. Unlike \textsc{SEVN} or \textsc{COMBINE}, this code does not yet account for binary stars. In general, \textsc{METISSE} will be another promising candidate for combining full stellar dynamics with detailed stellar evolution.
\end{itemize}

\subsubsection{Interpolation/Fitting formulae}
\label{Interpolation/Fitting formulae}
\begin{center}
\begin{figure}
\includegraphics[width=0.95\textwidth]{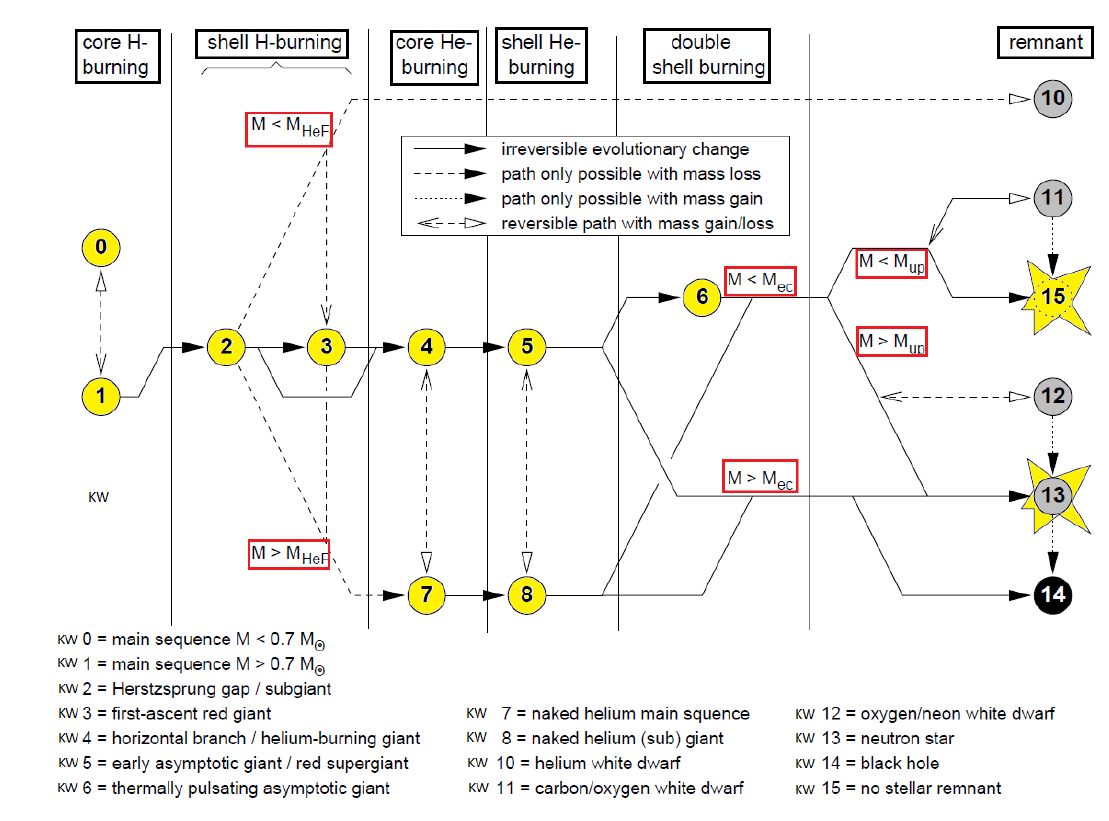}
 \caption{Diagram showing the complete stellar evolution for the \textsc{SSE} code. It shows the possible paths of evolution through the various single stellar evolution phases. The paths between the individual stellar types are marked as general \textit{irreversible} paths and \textit{irreversible} paths only possible with \textit{mass loss} or \textit{mass gain}. Furthermore, there are also \textit{reversible} paths with \textit{mass gain or mass loss}. The meaning of the masses is as follows: $m_{\mathrm{ HeF}}$ is the mass of the star to constitute the development of a degenerate He core on the GB and ignite He in a degenerate He flash at the top of the GB.  $m_{\mathrm{ ec}}$ and $m_{\mathrm{ up}}$ deal with the remnant masses and pathways of Supernovae (SNe), when the AGB evolution is terminated, which is after the CO-core mass reaches a limiting value and undergoes SNe. $m_{\mathrm{ up}} = 1.6~M_{\odot}$  and $m_{\mathrm{ ec}} = 2.25~M_{\odot}$ depend on metallicity of the star and refers to the mass-range, where C burning leads to the formation of a degenerate ONe-core. This might collapse due to electron-capture (EC) of $^{24}\mathrm{Mg}$ and result in NS production, see Sect.~\ref{Formation of compact objects and their natal masses, kicks and spins}. However, in almost all stars, mass loss in the TPAGB phase leads to a shedding of the envelope, so that the final remnant is a WD. If this mass is extreme, then we get might get a mass-less remnant. If the mass of the core $M_{\mathrm{ c,BAGB}} < 1.6~M_{\odot}$, then the result is a COWD. With $M_{\mathrm{ c,BAGB}} \geq 1.6$, then we will get a ONeWD. If $M_{\mathrm{ c,BAGB}} > 2.25~M_{\odot}$, then the star is massive enough to form Fe core, which result in SNes, which end up either in a NS or a BH. If mass loss comes into the equation, then we might get NS production from stars within $m_{\mathrm{ up}} < M_{\mathrm{ c,BAGB}} < m_{\mathrm{ ec}}$ (Figure adapted from~\cite{Hurleyetal2000}).}
\label{Hurley_SSE}
\end{figure}
\end{center}
A first attempt to incorporate simple stellar evolution fitting formulae in a direct $N$-body code was done by~\cite{Aarseth1996} on the basis of~\cite{Eggletonetal1989}. Later, as a successor to~\cite{Eggletonetal1989} was created using the method developed by~\cite{Polsetal1998}. They based their code on the original Cambridge \textsc{STARS} stellar evolution program by~\cite{Eggleton1971,Eggleton1972,Eggletonetal1973a,Eggletonetal1973b,Pols1995,Polsetal1997,Schroederetal1997}. The result are the famous \textsc{Single Stellar Evolution} (\textsc{SSE}) fitting formulae, which for the first time included metallicity as a free parameter~\cite{Hurleyetal2000,Hurley2008a,Hurleyetal2013a}. Fig.~\ref{Hurley_SSE} shows the complex discretization of stellar phases and the possible evolutionary pathways between them in the \textsc{SSE} package. The figure has been included, because this fundamental structure still remains in many stellar evolution production codes today (see below).\\
In general, such fitting formulae take much more care and thus time to set up than method of interpolating between tables~\cite{Churchetal2009}, because the movement of a star in the Hertzsprung-Russell-Diagram (HRD) is highly non-uniform and erratic. Furthermore, they are also less adaptable to changes in stellar tracks, for example, when they need to be adjusted due to some new development in astrophysics. On the other hand, the \textsc{SSE} provides us with rapid, robust and analytic formulae, which can be easily modified and integrated into an $N$-body code along the lines of~\cite{Aarseth1996} and give stellar luminosity, radius and core mass of the stars as functions of mass, metallicity and age for all stellar evolutionary phases~\cite{Hurleyetal2000,Railtonetal2014}. \\
However, these formulae necessarily also discard a lot of crucial stellar evolution information~\cite{Hurley2008a}.  \textcolor{
red}{For example, stellar mixing depends on several timescales and internal stellar structure parameters~\cite{Olejaketal2020a} and so these cannot be modelled directly by the fitting formulae. Only the outcomes can be parameterised for stellar types of the individual stars along the lines of~\cite{Hurleyetal2002b}. }\\
Despite these fundamental complications in stellar evolution modelling that persist to this day (see e.g.~\cite{JoyceChaboyer2018,Pasettoetal2018,TangJoyce2021,Agrawaletal2022}) and which translate directly into the continuous and differentiable fitting formulae (polynomial form from least square fitting~\cite{Hurleyetal2000}), the \textsc{SSE} code has successfully, for the first time, provided us with a method by which we can evolve stars from ZAMS masses (0.1-100)~$\mathrm{M}_{\odot}$ (the models from~\cite{Polsetal1998} only reach 50~$\mathrm{M}_{\odot}$, but the \textsc{SSE} formulae can be safely extrapolated to 100~$\mathrm{M}_{\odot}$~\cite{Hurley2008a}) rapidly and accurately (within 5\% of detailed stellar evolution models over all phases of the evolution~\cite{Hurleyetal2000}) in $N$-body simulations throughout all evolutionary phases taking into account all of the astrophysical processes outlined in Sect.~\ref{Single stellar evolution} and offering a metallicity range from 0.0001 to 0.03 with $Z_{\odot}\simeq 0.02$ being Solar metallicity as an input parameter. \\
However, for a complete picture we also need to model the binary evolution processes outlined in Sect.~\ref{Binary Stellar Evolution}. For the fitting formulae this is provided by the \textsc{Binary Stellar Evolution} (\textsc{BSE}) code~\cite{Hurleyetal2002b,Hurley2008b,Hurleyetal2013b}, which is an add-on of the \textsc{SSE} package. This has been a huge success story and many full dynamical cluster simulations have utilised \textsc{SSE \& BSE} to evolve the stars, e.g.~\cite{Wangetal2016,Askaretal2017a,DiCarloetal2019,DiCarloetal2020a,DiCarloetal2020b,DiCarloetal2021,Rizzutoetal2021,Rizzutoetal2022,Kamlahetal2022a}.
The \textsc{SSE \& BSE} codes have been the foundation for many other BPS codes:
\begin{itemize}
    \item \textsc{COMPAS}~\cite{TeamCompas2021}
    \item \textsc{MSE}~\cite{Hamersetal2020}
    \item \textsc{MOBSE}~\cite{GiacobboMapelli2018,GiacobboMapelli2019b,Mapellietal2020a} and related code called \textsc{ASPS} also used in~\cite{Lietal2023}, 
    \item \textsc{StarTrack}~\cite{Belczynskietal2002,Belczynskietal2008,Belczynskietal2020}
    \item \textsc{COSMIC}~\cite{Breiviketal2020a} and its implementations in \textsc{CMC}~\cite{Kremeretal2019,Rodriguezetal2022}
    \item \textsc{BSE-LevelC}~\cite{Kamlahetal2022a} and its implementation in \textsc{McLuster} \cite{Kuepperetal2011a,Kamlahetal2022a}.
\end{itemize}
The fitting formulae from the \textsc{SSE} code are also implemented in BPS code \textsc{binary\_c}~\cite{Izzardetal2004,Izzardetal2006,Izzardetal2009}. \\
New fitting formulae have recently been constructed, which are derived from fitting to 1D \textsc{HOSHI} stellar evolution models~\cite{Takahashietal2016,Takahashietal2018,Takahashietal2019,Yoshidaetal2019} to extremely massive low metallicity (EMP; Pop-III) stars~\cite{Tanikawaetal2020,Tanikawaetal2021a,Tanikawaetal2021c,Hijikawaetal2021}. These are constructed such that they can be implemented into any of the \textsc{BSE} variants mentioned above in a straightforward fashion and therefore also into stellar dynamics codes such as \textsc{Nbody6++GPU}~\cite{Wangetal2015}. 

\subsection{Initial conditions for star cluster simulations}
\label{Initial conditions for star cluster simulations}
\subsubsection{Global star cluster initial conditions}
\label{Global star cluster initial conditions}
Defining appropriate global initial conditions for star cluster simulations is highly non-trivial as the formation of a star cluster and the stars within it depend on a large number of parameters that are very uncertain due to a lack of better theoretical understanding and or observations. In the following, we give an overview of the most important parameters in this context for $N$-body simulation of star clusters.

\subsubsection{Initial 6D phase space distribution}
\label{Initial 6D phase space distribution}
In order to initialise an $N$-body star cluster simulation, we need to distribute the $N$ particles in 6D phase space.
A statistical approach as described in Sect.~\ref{Fokker-Planck} is taken to realize a star cluster, which follows the probability density distribution $f(\vec{r},\vec{v},t)$. The full 6D distribution function is rarely known explicitly; under the assumption of steady state, such that $f$ does not depend on time, the Jeans's theorem~\cite{BinneyTremaine2008} allows us to express $f$ as a function of integrals of motion of a single star moving in the gravitational potential $\Phi(r)$. For now we assume spherical symmetry, so we have for example specific energy and specific angular momentum: $f=f(\vec{r},\vec{v}) = f(E,L)$, which are defined as $E = v^2/2 + \Phi(r)$ and $ L = \vert\vec{r}\times\vec{v}\vert$ (cf. Sect.~\ref{MOCCA}). Deviations from spherical symmetry can be taken into account as well, see for example Sect.~\ref{Rotation} for the importance of initial bulk rotation. \\
Examples of such self-consistent distribution functions are given by 
\begin{equation}
    f(E) = F_n E^{n-3/2},
\end{equation}
where $n$ is an integer index and $F_n$ a normalization factor to make sure that $f(E)$ is properly normalized as a probability density function. For $n=5$ this results in the famous Plummer model~\cite{Plummer1915}, and for $n=7/4$ another famous solution, a density cusp~\cite{BahcallWolf1976,FrankRees1976} around supermassive black holes is found~\cite{Pretoetal2004}. In ~\cite{BinneyTremaine2008} these models are also called stellar polytropes, because their density distribution is the same as a gaseous polytrope~\cite{Chandrasekhar1939} of the same index $n$. Analytical density distributions exist for $n=0$, $n=1$, and $n=5$~\cite{Kippenhahnetal2012}, but for stellar systems only $n=5$ is physically useful. \\
The theory of gaseous spheres also knows the isothermal solution, which is obtained for $n=\infty$; in stellar dynamics the equivalent is the isothermal sphere
\begin{equation}
    f(E) = F_\infty \exp(-E/\sigma^2 ).
\end{equation}
Here $\sigma^2$ is the r.m.s. stellar velocity dispersion, analogous to the temperature in a gaseous sphere. These models have some problem, because their radial extent is unlimited (Plummer and isothermal) or even their mass is infinite (isothermal). Therefore, and since real star clusters are often subject to a tidal cutoff due to the host galaxy, a cutoff radius is introduced (connected to a cutoff energy). If at the cutoff radius the gravitational potential of an isolated star cluster would be $\Phi_0$, then a relative potential $\Psi$ and a relative energy $\varepsilon$ are defined by
\begin{equation}
\Psi = \Phi - \Phi_0 \qquad ; \qquad \varepsilon = E - \Phi_0.
\end{equation}
In that way the cluster extends from the center out to $\varepsilon = 0$ (and $\Psi =0$), and lowered isothermal or Plummer distribution is defined as follows:
\begin{equation}
f(\varepsilon) = 
\begin{cases}  f_\infty \exp(-\varepsilon/\sigma^2) 
  & \varepsilon < 0 (E < \Phi_0) \\
               0  & \varepsilon \ge 0 (E \ge \Phi_0), \\
\end{cases}
\label{King}
\end{equation}
\begin{equation}
f(\varepsilon) = 
\begin{cases}  f_5 \epsilon^{7/2}  
  & \varepsilon < 0 (E < \Phi_0) \\
               0  & \varepsilon \ge 0 (E \ge \Phi_0), \\
\end{cases}
\end{equation}
Again $f_5$ and $f_\infty$ have to be properly chosen normalization factors. The model Eq.~\ref{King} is the widely used King model~\cite{King1966a} (``lowered isothermal''). Note that some papers and books also prefer to change the sign of $E$ (or $\varepsilon$), such that bound objects have a positive value. We do not follow this here to avoid confusion.\\
Even in spherical symmetry the distribution function could be 2D, since we have $E$ and $|L|$ as constants of motion; it corresponds to the possibility that in spherical star clusters still at any given radius $r$ the radial and tangential velocity dispersion may be different. So, a more general approach for the distribution function in case of an isothermal is
\begin{equation}
f(\varepsilon) = 
\begin{cases}  f_\infty \exp(-L^2/L_0^2)\exp(-\varepsilon/\sigma^2) 
  & \varepsilon < 0 (E < \Phi_0) \\
               0  & \varepsilon \ge 0 (E \ge \Phi_0), \\
\end{cases}
\label{Michie}
\end{equation}
which is also known as Michie~\cite{Michie1963b} distribution. Numerical solutions of the Fokker-Planck equations in 2D are based on such 2D distribution functions and Michie models could serve as potential initial models (see Sect.~\ref{OrbitAverage}). It is interesting to note that the most well known 1D King distribution is actually based on the older, even more general (since 2D) Michie model; Ivan King himself gives an account of this\cite{King1981}. \\
1D King models are extensively used for initialising star cluster simulations (e.g~\cite{Rizzutoetal2021,Rizzutoetal2022,Kamlahetal2022a}). While the Plummer model needs two parameters (mass $M$ and scale radius $r_{\mathrm{h}}\simeq 1.305 r_{\mathrm{pl}}$), the King model needs three parameters (mass $M$, scale radius $r_{\mathrm{pl}}$ and dimensionless central potential $W_0$). For intermediate King models ($2.5 \leq W_0 \leq 7.5$), the Plummer models are very similar ($r_{\mathrm{h,Plummer}}=0.366 r_{\mathrm{h,King}}$)~\cite{King2008}. We note that~\cite{GielesZocchi2015} developed a new family of lowered isothermal models called the \textsc{limepy} models. Based on the 1D models of King a generalization in 2D for rotating star clusters is now being used and often described as rotating King models (see Sect.~\ref{Rotation} and citations there).\\
Note that from $f(E)$ or $f(E,L)$ directly a numerical star cluster cannot be constructed, because $E$ depends explicitly on $r$ and implicitly through the gravitational potential (Eq.~\ref{radialvelocity}). Therefore, in order to be self-consistent, the gravitational potential has to be determined by a velocity space integration over the distribution function and then Poisson's equation solved to obtain the stellar density as function of radius (see e.g. the textbook~\cite{BinneyTremaine2008} for examples). In a final step a random procedure has to be used to obtain stellar positions and velocities. If density or gravitational potential are analytically known functions (like in case of a Plummer model) the entire self-consistent model can be constructed in one loop using random numbers~\cite{Aarsethetal1974}.

\subsubsection{Initial stellar mass function}
\label{Initial stellar mass function}
In order to initialise star cluster simulations, we need to draw the ZAMS masses from an assumed distribution. For this purpose, we use an initial stellar mass function (IMF), a ``Hilfskonstrukt"~\cite{Kroupaetal2013,KroupaJerabkova2018}, as a mathematical formulation of an idealised stellar population that has formed from a singular star formation event. We will discuss in Sect.~\ref{Multiple Populations} that this is not the case in nature. An excellent review on the IMF and its construction has been provided by~\cite{Hopkins2018} (and sources therein) and it also explores the universality of the IMF (``unchanging distribution regardless of environment and over the entirety of cosmic
history'') and concludes that general the IMF is not universal. This has consequences for the initial conditions of star cluster simulations across cosmic time and we need to model the IMF of different stellar populations individually.\\
The IMF was established as a concept in a pioneering work by~\cite{Salpeter1955} as a quantisation of stellar masses in the Universe~\cite{KroupaJerabkove2019}. In general, the number of stars in the IMF is given by 
\begin{equation}
    \xi_{*}(m)=\frac{\mathrm{d}N}{\mathrm{d}m},
\end{equation}
where $\mathrm{d}N$ is the number of stars formed in a small region, i.e. an embedded-cluster-forming molecular cloud core, in the mass interval $m$ to $m+\mathrm{d}m$~\cite{Jerabkovaetal2018}. Typically, we express the IMF as a (multi-)power law (powers are typically denoted by ``$\alpha$'') depending on the stellar population that we want to model. For example, for Population I (Pop-I) stars we typically choose an IMF from~\cite{Kroupa2001,Kroupa2002},\cite{Chabrier2003} or~\cite{Maschberger2013}; they are quite similar. The standard \textsc{Nbody}-codes, such as \textsc{Nbody6++GPU} or \textsc{Nbody7} provide tools to initialize a star cluster model with generalized Salpeter or Kroupa~\cite{Kroupa2001} IMF's, in a mass range from 0.08 to 100 or 150 $\mathrm{M}_\odot$. Note that also the initialization of lower mass objects has been prepared in the codes by Pavel Kroupa. Despite many observations over the last decades, the IMF for Pop-I stars is still quite uncertain, see \cite{Hopkins2018} and sources therein. \\
For Pop-III stars, the IMF is very different. It becomes increasingly top-heavy for decreasing metallicity~\cite{Brommetal2002,BrommLarson2004,Marksetal2012,Bromm2013,Stacyetal2016,Jerabkovaetal2018,Kroupaetal2020}. However, we do not have observations of Pop-III stars (yet; although even with the JWST it will be a difficult or impossible undertaking~\cite{Rydbergetal2013}. On the other hand,~\cite{deSouzaetal2013} claim that some hundred SNe detections by JWST may be enough to constrain the IMF of Pop-III stars. See also~\cite{Schaueretal2020} for a further discussion) and therefore, we do not have statistics from which to conclude an IMF. A flat IMF with $\alpha\simeq -1.0$ between 8~$\mathrm{M}_\odot$ and 300~$\mathrm{M}_\odot$ for Pop-III stars has been proposed by~\cite{LazarBromm2021}. However,~\cite{Fraseretal2017} use a Salpeter IMF~\cite{Salpeter1955} of slope $\alpha\simeq -2.35$ with a maximum mass of around 87~$\mathrm{M}_\odot$ instead. We will have to wait for observations of Pop-III stars or their remnants before we can reliably constrain their IMF.

\subsubsection{Initial binary population}
\label{Binary population}
Almost all stars form in binary systems and some in higher order multiple systems~\cite{GoodwinKroupa2005,Kroupa2008,Miloneetal2012a}. As with the IMF in Sect.~\ref{Initial stellar mass function}, there is some debate on the universality of an initial binary population (IBP). In other words, it is contentious that the IBP is independent of environment, in which binaries form~\cite{Marksetal2015,Belloni2018a}. This is typically quietly assumed in the initialisation of star cluster simulations, at least in simulations of clusters made up of Pop-I stars (e.g.~\cite{Askaretal2017a,Kamlahetal2022a}). Although, we would expect this to vary for decreasing metallicity and higher redshift, because the environments and also the primordial gas from which the stars form have very different properties from Pop-I stars (e.g.~\cite{Stacyetal2012,Stacyetal2013}). \\
In any way, the IBP evolves on a cluster crossing timescale $t_{\mathrm{cr}}$. The widest binaries that form are dynamically disrupted, while in star clusters the hardest binaries harden further~\cite{Heggie1975,Hills1975b}. This leads to a pronounced SN-Ia rate in star clusters~\cite{SharaHurley2002}. Binaries generally dominate the global, dynamical evolution of the star cluster by close Newtonian few-body interactions (binary-single and binary-binary encounters)\cite{HeggieHut2003,Mapelli2018a}. \\
A distinction is made following~\cite{Kroupa1995b,Kroupa2008,Belloni2018a} between a birth stellar population, where all protostars are embedded in circum-protostellar material, and an initial stellar population, which consists of pre-main-sequence stars and which are not embedded in circum-protostellar material.


The process from an initial to a fully formed main-sequence binary star population is called pre-main-sequence eigenevolution.
\textcolor{red}{Eigenevolution is the sum of all dissipative physical processes that transfer mass, energy and angular momentum between the companions when they are still very young and accreting (sentence directly cited from \cite{Kroupa2008}). Most obvious is the process of tidal circularization~(e.g. \cite{MardlingAarseth2001}) of tight and initially highly eccentric binaries, which leads to a depletion of high eccentricities for small semi-major axes. In \cite{Kroupa2008}
there is a comprehensive description of eigenevolution and the relation between binary parameters at birth and after the pre-main sequence evolution, when typically $N$-body simulations start 
 (see also further citations in \cite{Kroupa2008} about observational binary data therein, and also \cite{Kroupaetal2013,Kuepperetal2011a,Railtonetal2014,Belloni2018a}). We present this as an example here, but it is not clear whether under all astrophysical circumstances such universal binary parameters are realized.
}\\
Dynamically, a binary star depends on four parameters: its system mass $m_{\mathrm{sys}}=m_{\mathrm{1}}+m_{\mathrm{2}}$, its period $P$ and correspondingly its semi-major axis $a$ (via Kepler's third law), its mass ratio $q=m_{\mathrm{2}}/m_{\mathrm{1}}\leq 1$ and its eccentricity $e$=$(r_{\mathrm{apo}}-r_{\mathrm{peri}})/(r_{\mathrm{apo}}+r_{\mathrm{peri}})$ ~\cite{Kroupa2008}, where $r_{\mathrm{apo}}$ and $r_{\mathrm{peri}}$ are the apocentric and pericentric distances, respectively. Thus, a complete initial binary population in a star cluster depends on the stellar IMF $\xi_{*}(m)$ (see Sect.~\ref{Initial stellar mass function}), the period distribution $f_{\mathrm{P}}(\mathrm{log}P)$, the mass ratio distribution $f_{\mathrm{q}}(q)$ and the eccentricity distribution $f_{\mathrm{e}}(e)$~\cite{Kroupa2008,MoeDiStefano2017,Belloni2018a}:
\begin{enumerate}
	\item $f_{\mathrm{P}}(\mathrm{log}P)$:~\cite{Kroupa1995b} show that 
	\begin{equation}
	    f_{\mathrm{P}}(\mathrm{log}P)=\eta\,\cdot\, \frac{\mathrm{log}(P)-\mathrm{log}(P_{\mathrm{min}})}{\delta + \bigl(\mathrm{log}(P)\!-\!\mathrm{log}(P_{\mathrm{min}})\bigr)^2}\ ,
	\end{equation}
	    where $P_{\mathrm{min}}=10$~days, $\delta=45$, $\eta=2.5$ and $P_{\mathrm{max}}=2.188\times 10^8$~days, because the initial binary fractions $f_{\mathrm{b}}$ is 100\%~\cite{GoodwinKroupa2005}. 
	Adjustments to this distribution were later made for high mass stars with $m> 5$~$\mathrm{M}_{\odot}$ following~\cite{Sanaetal2012a,Ohetal2015}, where for these stars $f_{\mathrm{P}}\propto (\mathrm{log}(P))^{-0.55}$ with $P_{\mathrm{min}}=1.412$~days and $P_{\mathrm{max}}=3.162\times 10^3$~days.
	\item $f_{\mathrm{e}}$: typically, we distribute the binaries thermally, meaning angular momenta are distributed equally $f_{\mathrm{e}}=2e$~\cite{Kroupa1995b}, although this might greatly overpredict observed merger rates according to~\cite{Gelleretal2019}.
	\item $f_{\mathrm{q}}$: the binary stars with members below $5$~$\mathrm{M}_{\odot}$ are distributed randomly and for masses above $5$~$\mathrm{M}_{\odot}$, the binary mass ratios are distributed uniformly ($0.1<q<1.0$)~\cite{Kiminkietal2012,SanaEvans2011,Sanaetal2012a,Kobulnickyetal2014}. 
\end{enumerate}
\textcolor{red}{Many direct $N$-body simulations usually start at a time when the star cluster is gas-free, when eigenevolution has terminated. Also dynamical interactions of binaries with each other and with single stars cause a further dynamical evolution of binary properties during the cluster formation phase. The ``Kroupa''~\cite{Kroupa1995b} period distribution includes many wide binaries. They are dynamically ionized (disrupted) in a time scale very short compared to the relaxation time scale. Monte Carlo simulations (some of them start with binary fractions close to unity~\cite{Leighetal2015}) and direct $N$-body models, starting with lower binary fraction (but most of them tightly bound) converge quickly, as can be seen in~\cite{Kamlahetal2022a}. Therefore one uses typically rather small binary fractions in initial models for long-term evolution of star clusters (order of 5\% - 20\%; \cite{Wangetal2016,ArcaSeddaetal2023a,ArcaSeddaetal2023b,ArcaSeddaetal2023c}.
Also for the same reason simpler power laws for the initial distributions of $\mathrm{log}P$, $e$, and $q$ are used (see e.g. \cite{Wangetal2022,Torniamentietal2022}). Note that in case of rotating star clusters the initial inclination of binaries with respect to the rotation axis of the cluster is a new possibly important parameter, which has not yet been examined.}\\
\textcolor{red}{It is sometimes quoted that 
small binary fractions are chosen in direct $N$-body simulations, because 
tightly bound binaries are computationally expensive; this may not be a problem in the near future. Already the \textsc{PeTar} 
code~\cite{Wangetal2020a,Wangetal2020c} can use efficiently a very large number of binaries in the simulations (see Fig.~\ref{petar-scaling}). Also
\textsc{Nbody6++GPU} potentially allows a good KS binary parallelization,
which is work in progress. Since} the treatment of interacting and relativistic binaries in \textsc{PeTar} is not equivalent to the one used in \textsc{Nbody6++GPU}, tests and comparisons of both codes with respect to binary evolution are ongoing.\\
In summary of what we discussed so far in Sect.~\ref{Global star cluster initial conditions} to Sect.~\ref{Binary population}, during the initialisation of $N$-body simulations we need to distribute the stars in 6D phase space, draw their masses from some IMF, and distribute primordial binaries according to IBP~\cite{Kroupa2008,Kuepperetal2011a,Kuepperetal2011b}. Furthermore, we generally need to make a decision if our star cluster is mass segregated at the beginning of the simulation~\cite{Fregeauetal2002,Subretal2008,Baumgardtetal2008b}, shows fractality~\cite{GoodwinWhitworth2004} and if it is or is not in virial equilibrium. Below we highlight two more areas of active research when it comes to simulations of star clusters and their initialisation: multiple stellar populations in Sect.~\ref{Multiple Populations} and initial bulk rotation of the star cluster in Sect.~\ref{Rotation}, respectively. 

\subsubsection{Multiple stellar populations}
\label{Multiple Populations}
Modern observational methods have made it possible to resolve multiple stellar populations (MSPs) in globular clusters, which can mostly be inferred from photometric diagrams such as colour-magnitude diagrams (CMDs) from multi-band HST photometry (e.g.~\cite{AndersonKing2000,Andersonetal2008,Grattonetal2012,Miloneetal2012a,Miloneetal2013,Milone2020,MiloneMarino2022}),
\textcolor{red}{as well as from abundance variations in integrated light \cite{Bastianetal2019,Bastianetal2020}}. Nowadays, MSPs have been confirmed in around
70 Galactic and extragalactic clusters~\cite{Miloneetal2017a,Miloneetal2017b,Miloneetal2018a,Miloneetal2018b,Miloneetal2020a,Miloneetal2020b,Miloneetal2020c,Milone2020,MiloneMarino2022}. \\
\textcolor{red}{MSPs can have very diverse origins, see e.g. the review \cite{BastianLardo2018}. Recently, kinematic differences between multiple populations have been found in Galactic globular clusters~\cite{Martensetal2023}. Dynamical simulations of different populations can be useful to unravel the origin of MSP and the relation between initial states and currently observed dynamics.} Our direct \textsc{Nbody6++GPU} code allows for the distinction between different populations by defining a corresponding label for each star; this has been used by \cite{Hongetal2017a} to constrain the dynamical origin of multiple populations in intermediate-age clusters in the Magellanic Clouds. Another rarely used feature of \textsc{Nbody6++GPU} (but see \cite{Bialasetal2015}) is that it can start with individual population data for every star (age, metallicity, population index). Also \textsc{MOCCA} simulations have been published \cite{Hypkietal2022} hosting two generations of stars as above in tidally filling and underfilling star clusters. They are able to reproduce the observed fractions and properties of second generation stars in the GCs of the Milky Way. For the time being there is no good way known to handle encounters and mixing of material from different populations in stellar collisions and binary mass overflow interactions in any of the codes.\\
We are still at the beginning of deciphering the complex picture of MSPs and their dynamical evolution using computational methods for collisional stellar systems. Nevertheless, the studies above are encouraging. 

\subsubsection{Rotation}
\label{Rotation}
\begin{center}
\begin{figure}
\includegraphics[width=1.0\textwidth]{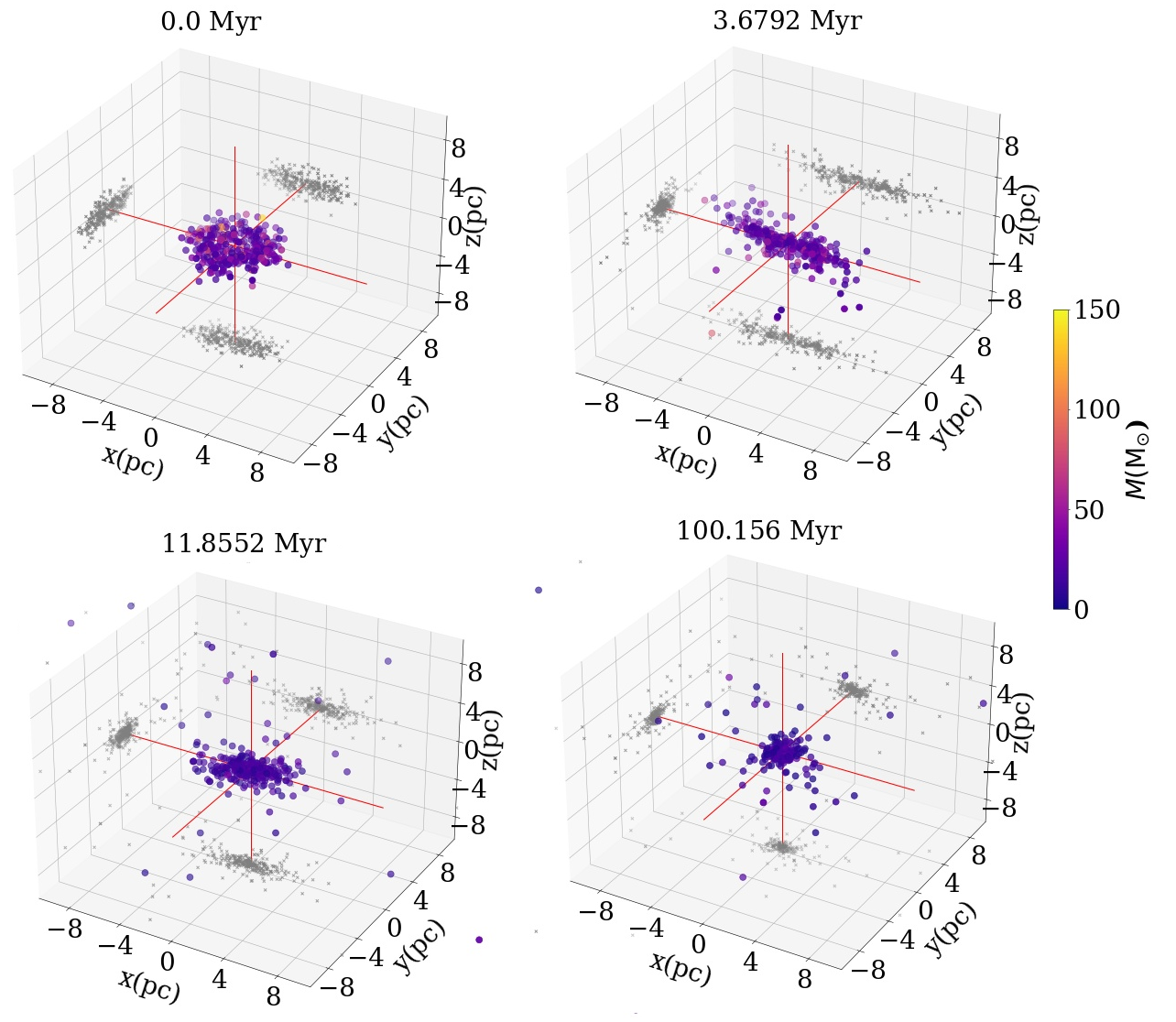}
 \caption{3D scatter plots in showing the spatial distribution of the ZAMS of high mass stars (and the BHs they quickly form) from the simulations that rotate extremely quickly initially ($(W_0=6.0,\omega_0 = 1.8)$, i.e. rotating King models by~\cite{EinselSpurzem1999}) and have stellar evolution (\texttt{level C} by~\cite{Kamlahetal2022a}) at around 0.0~Myr, 3.68~Myr, 11.86~Myr and 100.00~Myr, respectively; The stars and compact objects are color-coded by their mass between 0.0~$\mathrm{M}_{\odot}$ and 150.0~$\mathrm{M}_{\odot}$. The stars and BHs are also projected onto the three dimensional axes, which can be seen from the light-grey dots. We can clearly see the (rotating) triaxial structure, a bar of the BHs and their progenitor stars, at t=3.68~Myr and the spatial reconfiguration of the high mass objects (mostly BHs) to axisymmetric structures already after 11.86~Myr. At 100.00~Myr a practically spherical system of BHs remains that is much more concentrated than the system of their progenitor stars at 0.0~Myr. (Figure adapted from simulations by~\cite{Kamlahetal2022b}).
 }
\label{BHBar_Kamlah}
\end{figure}
\end{center}
The inclusion of initial bulk rotation in direct \textsc{Nbody} simulations of collisional stellar systems is still unusual (e.g.~\cite{Hongetal2013,Wangetal2016,Askaretal2017a,DiCarloetal2019,DiCarloetal2020a,DiCarloetal2020b,DiCarloetal2021,Rizzutoetal2021,Rizzutoetal2022,ArcaSeddaetal2021a,Kamlahetal2022a}), although it has been known for over a century that star clusters even today show significant imprints of rotation, which can, for example, be observed in deviations in the shapes of star clusters from sphericity~\cite{PeaseShapley1917,ShapleySawyer1927,Shapley1930,KopalSlouka1936,King1961,FrenkFall1982,Harris1976,Harris1996,Kormendy1985,WhiteShawl1987,Luptonetal1987,ChenChen2010,Bianchinietal2013}. Moreover, present-day detectors and data processing methods have made it possible to resolve the photometry and kinematics of individual stars all the way down to the cluster centre revealing 
\textcolor{red}{that rotational kinematic features vary between} multiple stellar
populations~\cite{Bianchinietal2016,Bianchinietal2018,Bianchinietal2019,Ferraroetal2018,Giesersetal2018,Giesersetal2019,Lanzonietal2018a,Lanzonietal2018b,Kamannetal2016,Kamannetal2018a,Kamannetal2018b,Kamannetal2019,Sollimaetal2019,Tiongcoetal2019,Tiongcoetal2021}. Additionally, both observations and simulations support these results and find that star clusters show significant fractality~\cite{Balloneetal2020,Pangetal2021a}, and internal rotation at birth in general~\cite{Lahenetal2020b,Balloneetal2021}. Velocity anisotropy has been observed in star clusters with detected elongated structures~\cite{Pangetal2020,Pangetal2021a}, which might be induced by rotation.\\
So, how do we initialise rotating collisional stellar systems? Sometimes it is assumed that there exists a kind of ``Maxwell's demon" that simply switches the direction of initial particle velocities to induce angular momentum to a $N$-body system and assuming the preservation of the spherical distribution function (e.g.~\cite{Plummer1911,King1962,Wilson1975}) and angular momentum in the process (e.g.~\cite{LyndenBell1960,Lingam2018}). This procedure is not physical. We instead need distribution functions that at least depend on two integrals of motion, such as the total energy $E$ and and the total angular momentum in the $z$-direction $L_{\mathrm{z}}$ (assuming that the system rotates around the $z$-axis initially and ignoring a third integral, which in some cases can be approximated by the total angular momentum of a star $L^2$~\cite{LuptonGunn1987}, because a third integral is generally not analytically known). Such rotating equilibrium models were developed by~\cite{Goodman1983a,EinselSpurzem1999,LongarettiLagoute1996,VarriBertin2012}. They can be considered as generalizations of standard King models~\cite{King1966a}, because their energy dependence is a lowered isothermal, and the additional term for the second independent variable is $\exp(-L_z/L_{z0})$ (analogous to \cite{Michie1962}). $L_{z0}$ is a scaling constant; usually a dimensionless rotation parameter $\omega_0$ is used (derived from $L_{z0}$, see e.g.\cite{EinselSpurzem1999}). The models are axisymmetric, with a rigid rotation of the inner parts of the cluster, a maximum of the rotation curve close to the half-mass radius, and a differentially decreasing rotation curve outside in the halo. Rotation supports only a fraction of the total kinetic energy (see Table 1 in \cite{EinselSpurzem1999}, and note that the 2nd column is erroneously labeled, it contains the percentage contained in rotational energy; i.e. for $\omega_0=0.6$ we have 20\% of the total kinetic energy in form of ordered rotational motion). Evolved star clusters obtained from these initial models agree quite well with observed clusters~\cite{Fiestasetal2006}.\\
Due to the 2D velocity distribution function an anisotropy is possible between the velocity dispersions in radial direction (in cylindrical coordinates) and in rotational $\varphi-$ direction; the models are isotropic between radial and vertical direction \textcolor{red}{(parallel to the rotation axis). Such models have been used as initial models} in 2D Fokker-Planck (FP) modelling and direct $N$-body simulation~\cite{Kimetal2002,Kimetal2004,Kimetal2008,Fiestasetal2006,FiestasSpurzem2010,Fiestasetal2012,Ernstetal2007,Hongetal2013,Kamlahetal2022b}. Note that the models by \cite{VarriBertin2012} are using a different form of the distribution function based on the Jacobian of a cluster rotating around the galaxy, but they are as well generalizations of King models for rotation with similar properties. They also have been used as initial models for direct \textsc{Nbody} models~\cite{Tiongcoetal2016a,Tiongcoetal2016b,Tiongcoetal2017,Tiongcoetal2018,Tiongcoetal2019,Tiongcoetal2021,Tiongcoetal2022,Livernoisetal2022}. Furthermore, semi-analytic models exist that~\cite{Szolgyenetal2018,Szolgyenetal2019,Szolgyenetal2021,PanamarevKocsis2022} used to study the formation and evolution of rotating stellar or black hole disks in nuclear star clusters. Most of the aforementioned studies find evidence for the gravogyro catastrophe and its coupling to the gravothermal catastrophe~\cite{InagakiHachisu1978,Hachisu1979,Hachisu1982,AkiyamaSugimoto1989}. It is important to note that both of these effects are entirely gravitational and disappear in the absence of gravity~\cite{LyndenBell1999}. The gravogyro catastrophe happens on all astrophysical scales from rotating stars, to the formation of the Solar system and the dynamics of spiral galaxies. A recent study on the impact of stellar evolution on rotating star clusters, finds that the initial bulk rotation leads to a rotating triaxial structure, a bar, of black holes and their progenitor stars in early star cluster evolution that then takes the shape of an axisymmetric structure, a disk, over time~\cite{Kamlahetal2022b}. For one of the simulations from~\cite{Kamlahetal2022b} the results are shown in Fig.~\ref{BHBar_Kamlah}. This bar formation and dissolution was already found by~\cite{AkiyamaSugimoto1989} in their pioneer low particle number $N$-body simulations ($N=1024$) of rotating star clusters. However, the simulation set-up was very different and naturally less advanced. Since rotation is fundamental to star cluster dynamics and since these environments are typically very dense~\cite{Krumholzetal2019}, it will be important to understand how this process can influence the growth of intermediate mass black holes (see Sect.~\ref{Intermediate mass black holes}) and their progenitor stars going forward. Recently, a study on extremely massive and rotating star Pop-III star clusters does indicate a trend that the larger the initial bulk rotation, the more stellar and BH-BH mergers there are~\cite{Kamlahetal2023}. This might be due to the extremely efficient angular momentum transport in early star cluster evolution (0-2~Myr) and the resulting contraction of of the central region before stellar evolution begins to dominate the pre-core collapse evolution. However, more simulations are needed here to reach a conclusion.   \\
Lastly, we note that due to the assumption of spherical symmetry most of the current Monte Carlo methods for collisional dynamics are currently unable to evolve initially rotation star cluster models, such as those described above (e.g.~\cite{Henon1975,Cohn1979,Stodolkiewicz1982,Stodolkiewicz1986,Giersz1998,Gierszetal2015,Merritt2015,Askaretal2017a,Kremeretal2020a,Kremeretal2021}). A restricted Monte Carlo method for rotating, axisymmetric star clusters (usable even for general geometry, but see problem below) has been presented by~\cite{Vasiliev2015}. It uses Spitzer's Monte Carlo method (see Sect.~\ref{HenonSpitzer}), which distinguishes it from currently common Monte Carlo codes. But it has some more serious approximations, because the random relaxation scatterings are applied only in $v_\parallel$ and $v_\perp$ obtained from a fully isotropic spherically symmetric background. In 2D Fokker-Planck models of axisymmetric rotating star clusters~\cite{EinselSpurzem1999} the background distribution function used for the computation of diffusion coefficients is fully self-consistent and there are five different diffusion coefficients obtained (instead of only two).

\subsection{Formation of massive objects}
\label{Intermediate mass black holes}
\begin{center}
\begin{figure}
\includegraphics[width=0.9\textwidth]{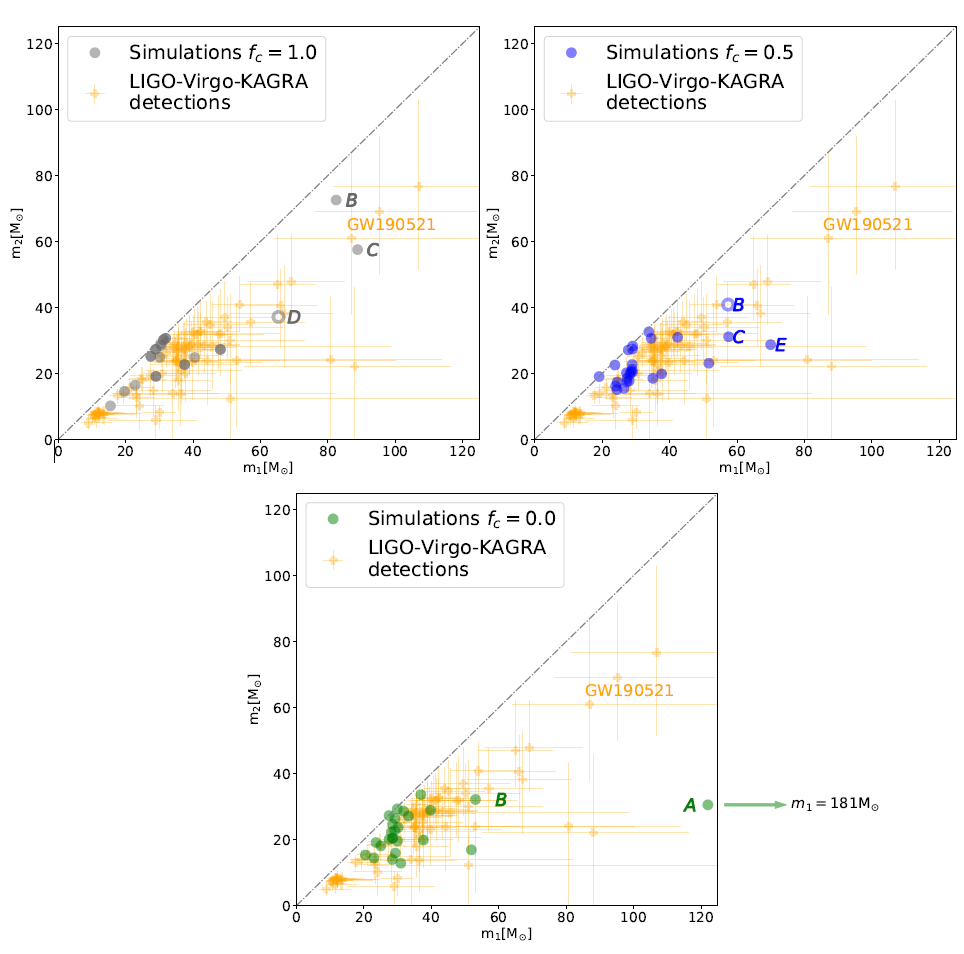}
 \caption{The panels show the primary ($m_1$) and secondary ($m_2$) masses of all BH mergers in the simulations
for an accretion fraction of $f_c$ = 1.0 (left, grey circles), $f_c$ = 0.5, (center, blue circles) and $f_c$ = 0.0, (bottom,
green circles). The currently available LIGO-Virgo-KAGRA gravitational wave detections including error bars
are indicated in orange. BH merger events that might be excluded due to gravitational recoil kicks are indicated
with open circles. In general, the simulated events cover a similar parameter space as all currently available
observations. The $f_c$ = 1.0, simulations provide two possible formation paths for GW190521. Path C is a
second-generation event and has a low probability due to a first-generation BH merger. Path B is more likely as
the event itself is a first-generation BH merger. One of the $f_c$ = 0.0, realizations generated an intermediate-mass
ratio in-spiral of two black holes with 31 and 181 $\mathrm{M}_{\odot}$, respectively, as shown in the bottom panel (Figure from \cite{Rizzutoetal2021}).
 }
\label{LIGOVIRGOKAGRA_RIzzuto}
\end{figure}
\end{center}
\textcolor{red}{The formation of massive objects from collisions of massive stars and mergers involving black holes is a subject with long history and in full detail beyond the scope of this review. The first detection of a massive black hole merger from LIGO-Virgo~(coalescence of two black holes with 85 and 66 $\mathrm{M}_{\odot}$\cite{Abbottetal2020b} has led to new interest into dynamical paths of formation of massive objects (intermediate mass black holes, IMBH) in star clusters. Numerical and theoretical works on IMBH formation in dense star clusters suggests that IMBH formation may occur through differet channels (multiple stellar mergers, accretion of stellar matter onto a stellar mass black hole, or multiple generations of relativistic black hole coalesciences (and mixtures of these processes). The idea that this may happen is not new (cf. e.g. 
\cite{PhinneySigurdsson1991,Gerhard2001,PortegiesZwartMcMillan2002,PortegiesZwartetal2004}).
On the observational side IMBHs remain elusive, a recent review about the issue can be found in~\cite{Greeneetal2020}.} \\
\textcolor{red}{Here we just want to highlight a few recent results obtained with $N$-body and MOCCA codes, showing the feasibility to grow an IMBH within a suitably initialized star cluster
(e.g.~\cite{ArcaSedda2016,ArcaSeddaetal2021a,ArcaSeddaetal2021b,ArcaSedda2019,ArcaSeddaCapuzzoDolcetta2019,Morawskietal2018,Morawskietal2019,DiCarloetal2021,Rizzutoetal2021,Rizzutoetal2022,Kamlahetal2023,ArcaSeddaetal2023b}).} Star cluster environments (particularly young massive or Pop-III clusters) can precipitate the birth and growth of IMBHs in two of the main proposed mechanisms as listed in~\cite{Greeneetal2020}: firstly, stellar mergers in star cluster evolution and subsequent core collapse into an IMBH of the progenitor star or secondly, gravitational runaway mergers of (IM)BHs to form (even more) massive IMBHs. \\
The mergers of (IM)BHs are associated with the emission of a GWs (see Sect.~\ref{More slow non-conservative processes: gravitational radiation and magnetic braking} and sources therein) and GR merger recoil kicks (see Sect.~\ref{Coalescence and collisions} and sources therein). The emitted gravitational radiation can be detected~\cite{ArcaSeddaetal2020b,ArcaSeddaetal2021b}. (a)LIGO~\cite{Aasietal2015,Abbottetal2018d,Abbottetal2019b}, (a)Virgo~\cite{Acerneseetal2015,Abbottetal2018d,Abbottetal2019b} and KAGRA (e.g.~\cite{Abbottetal2018d,Abbottetal2020e,Akutsuetal2019}) 
\textcolor{red}{currently detect stellar mass black hole inspirals and mergers during their last seconds or fractions of seconds. With improved ground based detectors (Einstein Telescope\cite{Branchesietal2023} or Cosmic Explorer\cite{Reitzeetal2019} or space-based instruments (LISA or its Chinese counterparts, see below) inspiralling stellar mass black hole binaries could be detected months and years before their final coalescence (cf. e.g. \cite{Janietal2020}), because they are sensitive at lower frequencies. Planned space based gravitational wave detectors are the European-American LISA\cite{AmaroSeoaneetal2017,Bayleetal2022} project and the Chinese projects TianQin\cite{Luoetal2016} and Taiji\cite{Ruanetal2018,Chenetal2021}. Also these instruments will be more suitable for detection of other black holes inspiralling and merging with IMBHs. Here we present an example from our $N$-body simulations, which shows that from them one can very well predict GW signal patterns and the abundances of GW events related to IMBHs (in the lower mass regime as discussed above) originating from star clusters (e.g.~\cite{ArcaSeddaGualandris2018,ArcaSeddaBeancquista2019}). Fig.~\ref{LIGOVIRGOKAGRA_RIzzuto} shows data from direct $N$-body simulations of~\cite{Rizzutoetal2022}; many BH-BH merger events found in the models fit in almost perfectly with the LIGO-Virgo-KAGRA detections. In these models a new parameter $f_c$ is used, describing how much mass is lost in the collision of a black hole with a main sequence star, $f_c =0$ means no black hole growth in the process, $f_c=1$ means that all the mass of the main sequence star is added to the black hole; this parameter is now present in current \textsc{Nbody6++GPU}, and was introduced by \cite{Rizzutoetal2022}.
}\\


\section{Simulations of nuclear star clusters}
\label{nuclear}
\subsection{Star-accreting supermassive black holes}
As it became clear that the energy emission of quasars and active galactic nuclei originates from a supermassive central black hole (SMBH) the question became interesting, what the equilibrium stellar distribution around it is. First results by~\cite{FrankRees1976,ShapiroLightman1976,LightmanShapiro1977} found the characteristic cusp, where the stellar density would vary as $\rho\propto r^{-7/4}$. Their analysis is comparable to what we would call later a stationary gaseous or momentum model of a stellar cluster. Also already by~\cite{FrankRees1976,Hills1975a} it was noticed that in such a system stars would be tidally disrupted near the SMBH, leading to its growth by mass accretion. Tidally disrupted stars would be preferentially on elongated radial orbits, with low angular momentum. The accretion process would take these stars out of the stellar distribution and create a ``loss cone'' in phase space. In a spherically symmetric cluster the loss cone would be refilled by diffusion of angular momenta due to two-body relaxation~\cite{AmaroSeoaneetal2001,AmaroSeoaneetal2004}. In axisymmetric or triaxial nuclear star clusters angular momentum diffusion can be much faster and lead to higher star accretion rates~\cite{MagorrianTremaine1999,MerrittPoon2004,PoonMerritt2004,WangMerritt2004,MerrittWang2005}.
Also orbit-averaged FP models have been used for this problem~\cite{CohnKulsrud1978,BahcallWolf1976,BahcallWolf1977,Davidetal1987a,Davidetal1987b}. More recently a first direct $N$-body model using \textsc{Nbody6++GPU} has been published~\cite{Panamarevetal2019} (\textsc{DRAGON} simulation of the Galactic Center - it led to the conclusion that the central cusp areas near the SMBH are dominated by stellar mass black holes, which will be accreted to the SMBH under gravitational wave emission.

\subsection{Tidal disruption events}
If a star is tidally disrupted by the SMBH on a parabolic orbit only half of its mass should be accreted while the other half has enough energy to escape from the SMBH (Rees' conjecture,~\cite{Rees1988}). The characteristic light curve of such events decays with $t^{-3/5}$; under realistic conditions, however, tidally disrupted stars will arrive on bound (eccentric) and unbound (hyperbolic) orbits, not just parabolic ones; and that will cause deviations from the standard light curve as well as change the accreted mass fraction. In~\cite{Hayasakietal2018,Zhongetal2022} large direct $N$-body simulations were done to find out the distribution of accreted stars. Furthermore a partial tidal disruption of giant stars was implemented, where the envelope is removed early and a core similar to a white dwarf survives and gets tidally disrupted later~\cite{Zhongetal2022}. Detailed stellar models were also used to determine the mass fraction of a star which is accreted to the SMBH after tidal disruption~\cite{LawSmithetal2020}. In future work this will be used in conjunction with the detailed mass function, single and binary stellar evolution as in the globular cluster simulations.

\section{Practical tools}
\label{Practical Tools}
Here we want to highlight some software tools that have been developed and successfully applied in the context of collisional dynamics.
\subsection{Multiscale and multiphysics simulation with \textsc{AMUSE}}
\label{Multiscale and multiphysics simulation through AMUSE}
The Astrophysical Multipurpose Software Environment (\textsc{AMUSE}) aims to provide a framework by which to simulate multiscale and multiphysics in a hierarchical fashion~\cite{PortegiesZwart2011,PortegiesZwartetal2013,PortegiesZwartetal2018b,PortegiesZwartMcMillan2018}. It does so by constructing new applications from the combination of known codes (solvers). For example, for gravitational dynamics the user may choose from a selection of 18 $N$-body codes~\cite{PortegiesZwartetal2018b}. The \textsc{AMUSE} framework successfully connects gravitational dynamics, radiative transport, stellar evolution and hydrodynamics~\cite{PortegiesZwartMcMillan2018}, which results in incredibly diverse research: gas expulsion in early star cluster evolution~\cite{Levequeetal2022a,Lewisetal2021}, star cluster collisions and massive star cluster formation~\cite{Beilisetal2021}, the interaction of binary stars in gaseous filaments~\cite{MoralesFellhauer2020}, the evolution of star clusters in a cosmological tidal field~\cite{Riederetal2013}, evolution of triple stellar systems with Roche-lobe filling binaries~\cite{deVriesetal2014} and many more. \\
Particle data and relevant quantities are shared between the constituent codes through the \textsc{AMUSE} framework and unit conversion and other data manipulation can be done in this process~\cite{PortegiesZwartetal2018b}. \textsc{AMUSE} is capable of taking care of all sorts of technical problems, such as communication between codes, boundary conditions etc.. However, \textsc{AMUSE} is naturally limited by algorithmic complexity of the software it combines and the astrophysical research objective itself. For example, if we wanted to simulate a globular cluster of realistic size with \textsc{Nbody6++}~\cite{Stiefel1965,AhmadCohen1973,Aarseth1985b,Spurzem1999,Aarseth1999a,Aarseth1999b,Aarseth2003b,Aarseth2008,McMillan1986,Hutetal1995,Makino1991a,Makino1999,Spurzem1999,NitadoriAarseth2012} and evolve the stars in the cluster with \textsc{MESA}~\cite{Paxtonetal2011,Paxtonetal2013,Paxtonetal2015,Paxtonetal2016,Paxtonetal2018,Paxtonetal2019}, it would take an enormous amount of time. This is not a problem of \textsc{AMUSE}, but rather a statement on the plausibility of certain simulations themselves. However, we note that the operations by \textsc{AMUSE} and communication between codes also costs computing time. 

\subsection{Simulation data processing and analysis}
\label{Simulation data processing}
\textsc{AMUSE} (introduced in Sect.~\ref{Multiscale and multiphysics simulation through AMUSE}) provides a large number of in-built data analysis tools simulation data, such as the \textsc{HOP}~\cite{EisensteinHut1998} or the \textsc{Kepler} packages~\cite{PortegiesZwartetal2018b}. These packages can be used by \textsc{AMUSE} users to comfortably analyse data from their simulations without having to write their own scripts. \\
Many $N$-body codes are now supported by built-in simulation data post-processing analysis packages mostly written in \texttt{python}. The direct $N$-body code \textsc{PeTar}~\cite{Wangetal2020a,Wangetal2020c} also provides built-in data analysis tools, such as movie generators from particle data and HR diagrams. Likewise, the \textsc{CMC} code~\cite{Joshietal2000,Pattabiramanetal2013,Breiviketal2020a,Rodriguezetal2022} uses the \texttt{cmctoolkit} package~\cite{Ruietal2021} for converting the simulation output into, e.g., velocity dispersion and surface brightness profiles.\\
Another extremely useful tool for distributed data analysis is provided by \textsc{BEANS}~\cite{Hypki2018}, which can in principle be used for simulation output data format (e.g. \texttt{csv}, \texttt{HDF5}, or \texttt{FITS}~\cite{Hypki2018,Hypkietal2022}). Therefore, this tool can be used for the data analysis of output from all previously mentioned codes. It uses the industry-standard, the Apache Hadoop platform~\cite{DeanGhemawat2004}, for data analysis. This platform is highly optimized for processing huge amounts of (continuously generated) data and is therefore ideal not only in the ``Internet of Things" and the live communication between machines, but also on-the-fly data processing and analysis from $N$-body simulations. \textsc{BEANS} is currently heavily used in data processing and analysis of \textsc{MOCCA} simulations (e.g.~\cite{Hypkietal2022}). Recently, \textsc{BEANS} has also been configured to easily process the output data from \textsc{Nbody6++GPU} simulations. 

\subsection{Photometric mock observations from star cluster simulations}
\label{Mock observations from star cluster simulations}
\begin{center}
\begin{figure}
\includegraphics[width=0.9\textwidth]{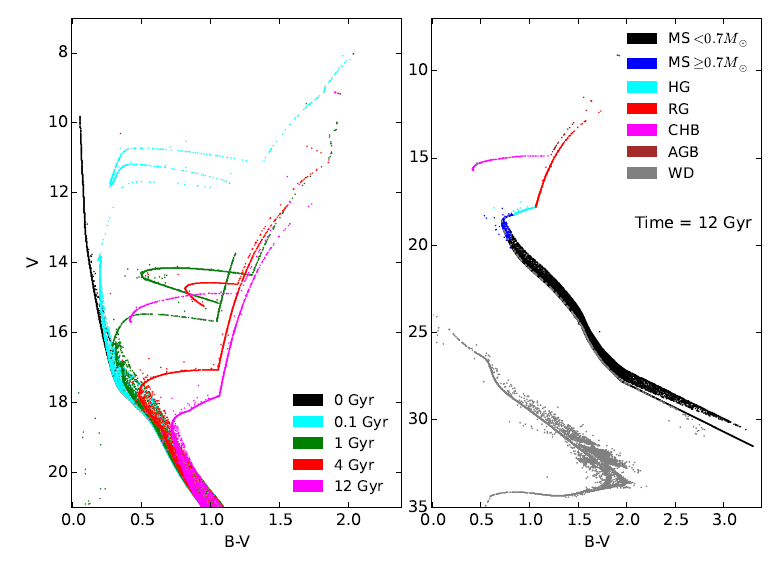}
 \caption{Color (B-V) - magnitude (apparent V) diagram (CMD)
of one of the \texttt{Dragon} simulations: D1-R7-IMF93. The distance modulus 13.82 (5800 pc) is used
here. Left: the evolution of CMD for luminous parts. Black dots
show the initial distribution. After 100 Myr the MS turnoff becomes visible already (cyan). The horizontal branch is populated
after 1 Gyr (green). The MS turnoff moves down to $V\sim 17.5$ mag
after 4 Gyr (red) and $V\sim 18.5$ mag after 12 Gyr (purple). Right:
the full CMD after 12 Gyr. Different colors show different stellar
types. HG: sub-giant branch (Hertzsprung gap); RG: red giant;
CHB: horizontal branch (core helium burning); AGB: asymptotic
giant branch; WD: white dwarf. (Figure from~\cite{Wangetal2016}).}
\label{CMD_Dragon}
\end{figure}
\end{center}
It is oftentimes desirable and useful to create photometric mock observations (CMDs or spectra of star clusters) from star cluster simulations. For this purpose various codes have been written. They mostly rely on the same principle: stellar masses, temperatures, luminosities and metallicities from the $N$-body output data are converted into observational magnitudes in various filter systems (e.g. \textsc{HST, SDSS, 2MASS} and it is typically easy to systems of new observing instruments) and into spectra in a certain wavelength range (e.g.~\cite{Pangetal2016}). Therefore, these mock observations are heavily influenced by the stellar evolution models that the respective simulations evolve the stars with. In the following paragraph, we summarise a number of codes for the creation of the observations outlined above. \\
The Cluster simulatiOn Comparison with ObservAtions (\textsc{COCOA}) code by~\cite{Askaretal2017b,Askaretal2018a} has been used extensively in the creation of mock observations from $N$-body and Monte Carlo simulations (e.g.~\cite{Askaretal2017a,Wangetal2016,Askaretal2017c,Belloni2018a}). \textsc{COCOA} has also been extensively used in \textsc{MOCCA} studies concerning the evolution and detectability of IMBHs in star clusters~\cite{deVita2017,Arosetal2020,Arosetal2021}. In Fig.~\ref{CMD_Dragon} a CMD constructed with \textsc{COCOA} is shown taken from~\cite{Wangetal2016}. They then compared this CMD with HST data of NGC 4372 (the target cluster of this study) from~\cite{Piottoetal2002} and found that the CMDs are mostly very similar. Differences between actual and mock observations were attributed to the intrinsic observational photometry error or the presence of MSPs (see Sect.~\ref{Multiple Populations}) in NGC 4372 that were not resolved by the simulation. Inaccurate stellar evolution modelling will have further influenced the results (in~\cite{Wangetal2016}, the so-called \texttt{level A} was used, see~\cite{Kamlahetal2022a}). This is a practical example of how photometric mock observations and the codes that produce these can be extremely useful in constraining both observation and theory. \\
The GAlaxy EVolutionary synthesis models (\textsc{Galev})~\cite{Kotullaetal2009} was adapted for the use creating mock observations from $N$-body simulations (mostly intended for \textsc{Nbody6++GPU} and codes from that family, but it can be used for any other code that provides stellar masses, temperatures, luminosities and metallicities of the stars) by~\cite{Pangetal2016}. The resulting code \textsc{GalevNB} was applied to study the dynamical origin of MSPs (see Sect.~\ref{Multiple Populations})~\cite{Hongetal2017b}, the dynamical evolution of planetary systems in star clusters~\cite{Kouwenhovenetal2020}, the long-term evolution of binaries in the \texttt{Dragon} simulations~\cite{Wangetal2016,Shuetal2021} and \textsc{PeTar} simulations of open clusters and exploring their UV-excess~\cite{Pangetal2022b}. \\
More codes that are worth mentioning and fulfil the same underlying purpose as \textsc{COCOA} and \textsc{GalevNB} are: Flexible Stellar Population Synthesis (\textsc{FSPS})~\cite{ConroyWhite2009,ConroyGunn2010,Conroyetal2010}, Simulating IFU Star Cluster Observations (\textsc{SISCO}) for integral field unit (IFU) observations of globular clusters~\cite{Bianchinietal2015}, Make Your Own Synthetic ObservaTIonS (\textsc{MYOSOTIS}) for creating generic photometric observations~\cite{Khorramietal2019} and Massive Cluster Evolution and Analysis Package (\textsc{MASSCLEAN}), which serves the same purposes~\cite{PopescuHanson2009}. 

\newpage
\section{Summary and Conclusion}
The gravitational $N$-body problem remains to be one of the oldest and most exciting problems in astrophysics. In this review, we have focused on a branch of this incredibly diverse field: the computation of \textit{collisional} stellar $N$-body systems. The term collisional refers to the cumulative effect of distant elastic two-body encounters here. In nature these systems are commonly represented young and massive, globular, or nuclear star clusters. These clusters are also dense and gravothermal - stellar density are high enough that direct collisions and mergers between stars occur in certain phases, and gravothermal in the sense that two-body relaxation, through distant elastic encounters, provides heat transfer and viscous angular momentum transfer. \\
Despite exhaustive efforts to accurately model these systems, there is still no self-consistent theory available. This is partly because $N$-body systems are dynamically chaotic for a significant fraction of the phase space of initial conditions if $N>2$. For small $N\le 3$, there do exist some stable solutions~\cite{ChencinerMontgomery2000,Montgomery2001}, but these special configurations are extremely unlikely to occur in nature~\cite{Heggie2000}. For all $N>3$, there exist no analytical solutions and we must solve the equation of motions numerically, thus with the help of computers. For increasing $N$, certain configurations of such systems exhibit truly remarkable physical properties. They might exhibit negative heat capacities that result in gravothermal contraction (sometimes also coined the ``gravothermal catastrophe"~(e.g.~\cite{LyndenBellWood1968}). If the star cluster rotates, it could exhibit a negative moment of inertia leading angular momentum transport from high mass particles to low mass particles in the system, while at the same time increasing the rotational velocity of the high mass particles, which is known as the gravogyro catastrophe~(e.g.~\cite{Hachisu1979,Kimetal2008}). In a realistic star cluster simulations, both of these processes will happen at different timescales (convective angular momentum transport is more efficient than conductive heat transport in star clusters), but they are coupled and ultimately reinforce one another~\cite{Hachisu1982}. Importantly, these processes are entirely gravitational in nature and disappear in the absence of self-gravity. \\
The two astrophysical methods that have been most successfully tackling collisional gravitational $N$-body systems are either direct $N$-body simulations (using for gravitational force calculations at least at some times the summation over all $N-1$ other stars) or approximate methods, based on statistical mechanics, using the Fokker-Planck modelling with Monte Carlo methods. While the former method appears physically more accurate, less affected by approximations, and resolves any astrophysical object of interest (star clusters) better, the latter is much faster and computationally less expensive. And recent star-by-star Monte Carlo modelling delivers detailed data comparable to direct $N$-body simulations. It has been demonstrated repeatedly that for global quantities such as the time evolution of binary fractions both methods yield very similar results. Therefore, these methods ultimately complement each other and many works have employed both side by side. \\
All fields in astrophysics have benefited from the revolutionary invention of computers and the ingenious hardware architectures that have been developed since. By the same token, new programming languages and parallelization techniques have had a significant impact on accelerating solving astrophysical problems on hardware and made certain problems possible to begin with. The collisional gravitational $N$-body problem is no exception here. When the astrophysics, the hardware and software domains are mastered, research in this field can culminate in ground-breaking results: the first million-body simulations of globular clusters across cosmic time, the \texttt{Dragon} simulations performed with \textsc{Nbody6++GPU}, are a product of this but more direct $N$-body models of Galactic and extragalactic globular and nuclear star clusters are needed to compare with both observational data and the wealth of data from Monte Carlo surveys. It follows that nowadays the gravitational $N$-body problem is in as much an astrophysics as it is a computer science problem. Examples here are the advent of novel simulation software such as the direct $N$-body codes \textsc{PeTar} and \textsc{Frost}, and new hybrid code approaches such as \textsc{ETICS}, which will make it possible to break into the $10^6-10^9$ particle domain. Massive globular clusters such as 47 Tuc or $\omega$ Cen as well as most nuclear star clusters belong to this parameter range. Will a traditional workhorse such as \textsc{Nbody6++GPU} again keep up and compete here? The first Exaflop/s computer has been inaugurated in the US this year\footnote{\url{https://www.top500.org/}} - for the next steps we need to again adopt our codes to the new developing hardware. It is now 32 years ago that D. Sugimoto announced the million body problem as a challenge for direct $N$-body simulations - here we want to raise the question of when the billion body problem can be tackled and what codes and what hardware do we need? \\
Another challenge lies in the integration of astrophysics into gravitational $N$-body simulations. This involves classical single and binary stellar evolution as well as relativistic dynamics of compact objects, or proper modelling of external tidal fields. The central densities of collisional stellar systems are typically so high, that stars frequently interact with each other; they exchange mass and angular momentum or even collide and merge. In other words, the stars in the simulations do not only evolve on their own, they also interact with nearby stars (and compact objects) through diverse astrophysical processes, such as tidal interaction, dynamical mass transfer, gravitational wave in-spiral and many more. For this purpose there are two main methods that have been thoroughly tested: interpolation between tables or approximation of stellar evolution data by some interpolation (fitting) formulae as functions of mass, age and metallicity. These two approaches are also not in competition, but rather complement one another. The former has traditionally been limited by memory on hardware, since the tables are necessarily rather large. Due to advances in hardware, this problem is diminishing. Unlike the latter approach, fitting formulae, stellar parameters from the given set of detailed tracks are calculated in real time with this method. Therefore, this approach is probably the most flexible, robust and efficient today when combining detailed stellar evolution with stellar dynamics. Furthermore, fitting formulae typically take much more care and time to set up. But the effort is rewarded with rapid, robust and analytic formulae, which can be easily modified and integrated into an $N$-body code and that quickly give stellar luminosity, radius and core mass of the stars as functions of mass, metallicity and age for all stellar evolutionary phases. \\
When full stellar evolution is combined with collisional dynamics and the resulting code is used to study the dynamical evolution of dense star clusters of various make-ups, exciting research can be done. From the formation and evolution of exotic stellar and compact binaries, such as blue stragglers, Cataclysmic Variables, Algols, X-ray binaries, and many others to double-degenerate binaries, such as BH-BH and the elusive BH-NS binaries, to the abundances of compact objects and their dynamical properties an extremely diverse array of astrophysically fascinating populations can be modelled for large metallicity and mass ranges. Therefore, the simulations of collisional stellar systems with modern production codes are the ideal laboratory to study stellar evolution in dense stellar environments. From the gravitational wave in-spiral of compact objects that can be modelled nowadays up to order PN(3.5) as well as the associated general relativistic merger recoil kick with the use of fitting formulae to numerical relativity, the full compact object merger phase can be modelled. As a result, simulations can yield predictions for theoreticians and observers alike on the properties, abundances and dynamics of gravitational wave sources from star clusters (and field) across cosmic time. Additionally, with the use of photometric mock observations from star cluster simulations, simulation data can be translated into magnitudes and fluxes for most mainstream filter systems, which has been done extensively already. In summary, simulations of collisional stellar systems are important and useful tools for unravelling our cosmic history in the age of multi-messenger astronomy. They are at the crossroads of many seemingly disparate astrophysical research fields, much like the simulation target, star clusters, are a fundamental building block in a hierarchy of cosmological structure formation~\cite{Krumholzetal2019}.

\section*{Conflict of interest}
The authors declare that they have no conflict of interest.

\begin{acknowledgements}

In alphabetical order of family name we thank Manuel Arca Sedda, Abbas Askar, Sambaran Banerjee, Peter Berczik, Maxwell Cai, Roberto Capuzzo-Dolcetta, Gaia Fabj, Francesco Flammini-Dotti, Mirek Giersz, Jarrod Hurley, Arek Hypki, Thijs Kouwenhoven, Shuo Li, Michela Mapelli, Thorsten Naab, Keigo Nitadori, Bastian Reinoso, Francesco Rizzuto, Xiaoying Pang, Qi Shu, Long Wang, Kai Wu, and Shiyan Zhong for helpful comments, discussions, and collaboration. AK wants to extend his deep gratitude to Nadine Neumayer for the support and mentorship in difficult times. The authors gratefully acknowledge project support by German Science Foundation (DFG), National Science Foundation of China (NSFC), Chinese Academy of Sciences (CAS), Volkswagen Foundation, and the Gauss Centre for Supercomputing e.V. (www.Gauss-centre.eu) for providing computing time through the John von Neumann Institute for Computing (NIC) on the GCS Supercomputer JUWELS and JUWELS-Booster (J\"ulich Supercomputing Centre\cite{JUWELS2021}) at J\"ulich Supercomputing Centre in Germany (JSC). AK is a fellow of the International Max Planck Research School for Astronomy and Cosmic Physics at the University of Heidelberg (IMPRS-HD).\\
RS wants to thank Mirek Giersz for collaboration and extraordinary hospitality over the past decades during many visits at the Nicolaus Copernicus Astronomical Center in Warsaw, Poland. RS has been Alexander von Humboldt Polish Honorary Research Fellow by the
Foundation for Polish Science. RS is grateful for experiencing many years of excellent, friendly, and open work environment at National Astronomical Observatories of Chinese Academy of Sciences in Beijing, China and at Kavli Institute for Astronomy and Astrophysics at Peking University. European collaborators - if you see some familiar fragments in the introduction area - I have been inspired by some of our old network application texts (RS). RS also wants to thank all team members and collaborators, who are not explicitly mentioned here, including all students, former and current, in Kiel, Heidelberg, Beijing, and across the Silk Road, for their hard work. You provided essential contributions to get to the current level described in this paper. The GRAPE team at Tokyo University, represented first by Daiichiro Sugimoto and later by Junichiro Makino deserves sincere acknowledgment for making GRAPE hardware available early for international collaborators, including our team in Kiel and Heidelberg. RS expresses thanks and gratitude to Sverre Aarseth, Douglas Heggie, Gerhard Hensler, Roland Wielen, and Suijian Xue - without your support and encouragement I would not be here at this point writing this review. The editorial committee of LRCA has been very patient, thank you for this!
\end{acknowledgements}
%

 \bibliographystyle{spphys}       
\bibliography{name.bib}   

\begin{thebibliography}{100}
\providecommand{\url}[1]{{#1}}
\providecommand{\urlprefix}{URL }
\expandafter\ifx\csname urlstyle\endcsname\relax
  \providecommand{\doi}[1]{DOI \discretionary{}{}{}#1}\else
  \providecommand{\doi}{DOI \discretionary{}{}{}\begingroup
  \urlstyle{rm}\Url}\fi

\bibitem{Krumholzetal2019}
M.R. {Krumholz}, C.F. {McKee}, J.~{Bland-Hawthorn}, ARA\&A \textbf{57}, 227
  (2019).
\newblock \doi{10.1146/annurev-astro-091918-104430}

\bibitem{Neumayeretal2020}
N.~{Neumayer}, A.~{Seth}, T.~{B{\"o}ker}, A\&ARv \textbf{28}(1), 4 (2020).
\newblock \doi{10.1007/s00159-020-00125-0}

\bibitem{Harris1996}
W.E. {Harris}, AJ \textbf{112}, 1487 (1996).
\newblock \doi{10.1086/118116}

\bibitem{Bianchinietal2013}
P.~{Bianchini}, A.L. {Varri}, G.~{Bertin}, A.~{Zocchi}, ApJ \textbf{772}(1), 67
  (2013).
\newblock \doi{10.1088/0004-637X/772/1/67}

\bibitem{Harrisetal2017}
W.E. {Harris}, J.P. {Blakeslee}, G.L.H. {Harris}, \apj \textbf{836}(1), 67
  (2017).
\newblock \doi{10.3847/1538-4357/836/1/67}

\bibitem{Reina-Camposetal2019}
M.~{Reina-Campos}, J.M.D. {Kruijssen}, J.L. {Pfeffer}, N.~{Bastian}, R.A.
  {Crain}, MNRAS \textbf{486}(4), 5838 (2019).
\newblock \doi{10.1093/mnras/stz1236}

\bibitem{Boylan-Kolchin2018}
M.~{Boylan-Kolchin}, \mnras \textbf{479}(1), 332 (2018).
\newblock \doi{10.1093/mnras/sty1490}

\bibitem{Ramos-Almendaresetal2020}
F.~{Ramos-Almendares}, L.V. {Sales}, M.G. {Abadi}, J.E. {Doppel}, H.~{Muriel},
  E.W. {Peng}, \mnras \textbf{493}(4), 5357 (2020).
\newblock \doi{10.1093/mnras/staa551}

\bibitem{Vanzellaetal2017}
E.~{Vanzella}, F.~{Calura}, M.~{Meneghetti}, A.~{Mercurio}, M.~{Castellano},
  G.B. {Caminha}, I.~{Balestra}, P.~{Rosati}, P.~{Tozzi}, S.~{De Barros},
  A.~{Grazian}, A.~{D'Ercole}, L.~{Ciotti}, K.~{Caputi}, C.~{Grillo},
  E.~{Merlin}, L.~{Pentericci}, A.~{Fontana}, S.~{Cristiani}, D.~{Coe}, \mnras
  \textbf{467}(4), 4304 (2017).
\newblock \doi{10.1093/mnras/stx351}

\bibitem{Vanzellaetal2019}
E.~{Vanzella}, F.~{Calura}, M.~{Meneghetti}, M.~{Castellano}, G.B. {Caminha},
  A.~{Mercurio}, G.~{Cupani}, P.~{Rosati}, C.~{Grillo}, R.~{Gilli},
  M.~{Mignoli}, G.~{Fiorentino}, C.~{Arcidiacono}, M.~{Lombini},
  F.~{Cortecchia}, \mnras \textbf{483}(3), 3618 (2019).
\newblock \doi{10.1093/mnras/sty3311}

\bibitem{Vanzellaetal2020}
E.~{Vanzella}, M.~{Meneghetti}, G.B. {Caminha}, M.~{Castellano}, F.~{Calura},
  P.~{Rosati}, C.~{Grillo}, M.~{Dijkstra}, M.~{Gronke}, E.~{Sani},
  A.~{Mercurio}, P.~{Tozzi}, M.~{Nonino}, S.~{Cristiani}, M.~{Mignoli},
  L.~{Pentericci}, R.~{Gilli}, T.~{Treu}, K.~{Caputi}, G.~{Cupani},
  A.~{Fontana}, A.~{Grazian}, I.~{Balestra}, \mnras \textbf{494}(1), L81
  (2020).
\newblock \doi{10.1093/mnrasl/slaa041}

\bibitem{Vanzellaetal2021}
E.~{Vanzella}, A.~{Adamo}, F.~{Annibali}, M.~{Annunziatella}, G.~{Bartosch
  Caminha}, P.~{Bergamini}, F.~{Calura}, K.~{Caputi}, M.~{Castellano},
  A.~{Comastri}, M.~{Dickinson}, R.~{Gilli}, C.~{Grillo}, M.~{Gronke},
  C.~{Gruppioni}, M.~{Meneghetti}, A.~{Mercurio}, M.~{Mignoli}, M.~{Nonino},
  S.~{Ravindranath}, M.~{Ricotti}, P.~{Rosati}, E.~{Sani}.
\newblock {Constraining the nature of the first stellar complexes: globular
  cluster precursors and Population III stellar clusters at z 6-7}.
\newblock JWST Proposal. Cycle 1 (2021)

\bibitem{Claeyssensetal2023}
A.~{Claeyssens}, A.~{Adamo}, J.~{Richard}, G.~{Mahler}, M.~{Messa},
  M.~{Dessauges-Zavadsky}, \mnras \textbf{520}(2), 2180 (2023).
\newblock \doi{10.1093/mnras/stac3791}

\bibitem{Charbonneletal2023}
C.~{Charbonnel}, D.~{Schaerer}, N.~{Prantzos}, L.~{Ram{\'\i}rez-Galeano},
  T.~{Fragos}, A.~{Kuruvandothi}, R.~{Marques-Chaves}, M.~{Gieles}, arXiv
  e-prints arXiv:2303.07955 (2023).
\newblock \doi{10.48550/arXiv.2303.07955}

\bibitem{LyndenBellWood1968}
D.~{Lynden-Bell}, R.~{Wood}, MNRAS \textbf{138}, 495 (1968).
\newblock \doi{10.1093/mnras/138.4.495}

\bibitem{Hachisuetal1978}
I.~{Hachisu}, Y.~{Nakada}, K.~{Nomoto}, D.~{Sugimoto}, Progress of Theoretical
  Physics \textbf{60}(2), 393 (1978).
\newblock \doi{10.1143/PTP.60.393}

\bibitem{Hachisu1979}
I.~{Hachisu}, PASJ \textbf{31}, 523 (1979)

\bibitem{Spurzem1991}
R.~{Spurzem}, \mnras \textbf{252}, 177 (1991).
\newblock \doi{10.1093/mnras/252.2.177}

\bibitem{LyndenBellEggleton1980}
D.~{Lynden-Bell}, P.P. {Eggleton}, MNRAS \textbf{191}, 483 (1980).
\newblock \doi{10.1093/mnras/191.3.483}

\bibitem{LouisSpurzem1991}
P.D. {Louis}, R.~{Spurzem}, \mnras \textbf{251}, 408 (1991).
\newblock \doi{10.1093/mnras/251.3.408}

\bibitem{Giesersetal2018}
B.~{Giesers}, S.~{Dreizler}, T.O. {Husser}, S.~{Kamann}, G.~{Anglada
  Escud{\'e}}, J.~{Brinchmann}, C.M. {Carollo}, M.M. {Roth}, P.M. {Weilbacher},
  L.~{Wisotzki}, MNRAS \textbf{475}(1), L15 (2018).
\newblock \doi{10.1093/mnrasl/slx203}

\bibitem{Giesersetal2019}
B.~{Giesers}, S.~{Kamann}, S.~{Dreizler}, T.O. {Husser}, A.~{Askar},
  F.~{G{\"o}ttgens}, J.~{Brinchmann}, M.~{Latour}, P.M. {Weilbacher},
  M.~{Wendt}, M.M. {Roth}, AAP \textbf{632}, A3 (2019).
\newblock \doi{10.1051/0004-6361/201936203}

\bibitem{Kamannetal2020b}
S.~{Kamann}, B.~{Giesers}, N.~{Bastian}, J.~{Brinchmann}, S.~{Dreizler},
  et~al., A\&A \textbf{635}, A65 (2020).
\newblock \doi{10.1051/0004-6361/201936843}

\bibitem{Henon1961}
M.~{H{\'e}non}, AnAp \textbf{24}, 369 (1961)

\bibitem{Heggie1975}
D.C. {Heggie}, MNRAS \textbf{173}, 729 (1975).
\newblock \doi{10.1093/mnras/173.3.729}

\bibitem{Elsonetal1987a}
R.~{Elson}, P.~{Hut}, S.~{Inagaki}, ARA\&A \textbf{25}, 565 (1987).
\newblock \doi{10.1146/annurev.aa.25.090187.003025}

\bibitem{Aasietal2015}
J.~{Aasi}, J.~{Abadie}, B.P. {Abbott}, R.~{Abbott}, T.~{Abbott}, M.R.e.a.
  {Abernathy}, Classical and Quantum Gravity \textbf{32}(11), 115012 (2015).
\newblock \doi{10.1088/0264-9381/32/11/115012}

\bibitem{Abbottetal2018d}
B.P. {Abbott}, R.~{Abbott}, T.D. {Abbott}, M.R. {Abernathy}, F.~{Acernese},
  et~al., LRR \textbf{21}(1), 3 (2018).
\newblock \doi{10.1007/s41114-018-0012-9}

\bibitem{Abbottetal2019b}
B.P. {Abbott}, R.~{Abbott}, T.D. {Abbott}, S.~{Abraham}, F.~{Acernese}, et~al.,
  ApJL \textbf{882}(2), L24 (2019).
\newblock \doi{10.3847/2041-8213/ab3800}

\bibitem{Acerneseetal2015}
F.~{Acernese}, M.~{Agathos}, K.~{Agatsuma}, D.~{Aisa}, N.~{Allemandou}, A.e.a.
  {Allocca}, Classical and Quantum Gravity \textbf{32}(2), 024001 (2015).
\newblock \doi{10.1088/0264-9381/32/2/024001}

\bibitem{Abbottetal2020e}
B.P. {Abbott}, R.~{Abbott}, T.D. {Abbott}, S.~{Abraham}, F.~{Acernese}, et~al.,
  LRR \textbf{23}(1), 3 (2020).
\newblock \doi{10.1007/s41114-020-00026-9}

\bibitem{Akutsuetal2019}
{Kagra Collaboration}, T.~{Akutsu}, M.~{Ando}, K.~{Arai}, Y.~{Arai}, et~al.,
  NatAs \textbf{3}, 35 (2019).
\newblock \doi{10.1038/s41550-018-0658-y}

\bibitem{Branchesietal2023}
M.~{Branchesi}, M.~{Maggiore}, D.~{Alonso}, C.~{Badger}, B.~{Banerjee},
  F.~{Beirnaert}, S.~{Bhagwat}, G.~{Boileau}, S.~{Borhanian}, D.D. {Brown},
  M.L. {Chan}, G.~{Cusin}, S.L. {Danilishin}, J.~{Degallaix}, V.~{De Luca},
  A.~{Dhani}, T.~{Dietrich}, U.~{Dupletsa}, S.~{Foffa}, G.~{Franciolini},
  A.~{Freise}, G.~{Gemme}, B.~{Goncharov}, A.~{Ghosh}, F.~{Gulminelli},
  I.~{Gupta}, P.K. {Gupta}, J.~{Harms}, N.~{Hazra}, S.~{Hild}, T.~{Hinderer},
  I.S. {Heng}, F.~{Iacovelli}, J.~{Janquart}, K.~{Janssens}, A.C. {Jenkins},
  C.~{Kalaghatgi}, X.~{Koroveshi}, T.G.F. {Li}, Y.~{Li}, E.~{Loffredo},
  E.~{Maggio}, M.~{Mancarella}, M.~{Mapelli}, K.~{Martinovic}, A.~{Maselli},
  P.~{Meyers}, A.L. {Miller}, C.~{Mondal}, N.~{Muttoni}, H.~{Narola},
  M.~{Oertel}, G.~{Oganesyan}, C.~{Pacilio}, C.~{Palomba}, P.~{Pani},
  A.~{Pasqualetti}, A.~{Perego}, C.~{P{\`e}rigois}, M.~{Pieroni}, O.~{Juliana
  Piccinni}, A.~{Puecher}, P.~{Puppo}, A.~{Ricciardone}, A.~{Riotto},
  S.~{Ronchini}, M.~{Sakellariadou}, A.~{Samajdar}, F.~{Santoliquido}, B.S.
  {Sathyaprakash}, J.~{Steinlechner}, S.~{Steinlechner}, A.~{Utina}, C.~{Van
  Den Broeck}, T.~{Zhang}, arXiv e-prints arXiv:2303.15923 (2023).
\newblock \doi{10.48550/arXiv.2303.15923}

\bibitem{Reitzeetal2019}
D.~Reitze, R.X. Adhikari, S.~Ballmer, B.~Barish, L.~Barsotti, G.~Billingsley,
  D.A. Brown, Y.~Chen, D.~Coyne, R.~Eisenstein, M.~Evans, P.~Fritschel, E.D.
  Hall, A.~Lazzarini, G.~Lovelace, J.~Read, B.S. Sathyaprakash, D.~Shoemaker,
  J.~Smith, C.~Torrie, S.~Vitale, R.~Weiss, C.~Wipf, M.~Zucker.
\newblock Cosmic explorer: The u.s. contribution to gravitational-wave
  astronomy beyond ligo (2019)

\bibitem{Chandrasekhar1942}
S.~{Chandrasekhar}, \emph{{Principles of stellar dynamics}} (The University of
  Chicago Press, 1942)

\bibitem{Larson1970a}
R.B. {Larson}, MNRAS \textbf{147}, 323 (1970).
\newblock \doi{10.1093/mnras/147.4.323}

\bibitem{BettwieserSpurzem1986}
E.~{Bettwieser}, R.~{Spurzem}, AAP \textbf{161}(1), 102 (1986)

\bibitem{SpurzemTakahashi1995}
R.~{Spurzem}, K.~{Takahashi}, MNRAS \textbf{272}(4), 772 (1995).
\newblock \doi{10.1093/mnras/272.4.772}

\bibitem{Spitzer1987}
L.~Spitzer, \emph{{Dynamical evolution of globular clusters}} (Princeton
  University Press, 1987)

\bibitem{GierszHeggie1994a}
M.~{Giersz}, D.C. {Heggie}, MNRAS \textbf{268}, 257 (1994).
\newblock \doi{10.1093/mnras/268.1.257}

\bibitem{BettwieserSugimoto1984}
E.~{Bettwieser}, D.~{Sugimoto}, MNRAS \textbf{208}, 493 (1984).
\newblock \doi{10.1093/mnras/208.3.493}

\bibitem{Szolgyenetal2018}
A.~Szölgyen, B.~{Kocsis}, PRL \textbf{121}(10), 101101 (2018).
\newblock \doi{10.1103/PhysRevLett.121.101101}

\bibitem{Szolgyenetal2019}
A.~Szölgyen, Y.~{Meiron}, B.~{Kocsis}, ApJ \textbf{887}(2), 123 (2019).
\newblock \doi{10.3847/1538-4357/ab50bb}

\bibitem{Szolgyenetal2021}
A.~Szölgyen, G.~{M{\'a}th{\'e}}, B.~{Kocsis}, ApJ \textbf{919}(2), 140 (2021).
\newblock \doi{10.3847/1538-4357/ac13ab}

\bibitem{Torniamentietal2019}
S.~{Torniamenti}, G.~{Bertin}, P.~{Bianchini}, \aap \textbf{632}, A67 (2019).
\newblock \doi{10.1051/0004-6361/201935878}

\bibitem{Kamlahetal2022b}
A.W.H. {Kamlah}, R.~{Spurzem}, P.~{Berczik}, M.~{Arca Sedda}, F.~{Flammini
  Dotti}, N.~{Neumayer}, X.~{Pang}, Q.~{Shu}, A.~{Tanikawa}, M.~{Giersz},
  \mnras \textbf{516}(3), 3266 (2022).
\newblock \doi{10.1093/mnras/stac2281}

\bibitem{Wangetal2015}
L.~{Wang}, R.~{Spurzem}, S.~{Aarseth}, K.~{Nitadori}, P.~{Berczik}, et~al.,
  MNRAS \textbf{450}(4), 4070 (2015).
\newblock \doi{10.1093/mnras/stv817}

\bibitem{Wangetal2016}
L.~{Wang}, R.~{Spurzem}, S.~{Aarseth}, M.~{Giersz}, A.~{Askar}, et~al., MNRAS
  \textbf{458}(2), 1450 (2016).
\newblock \doi{10.1093/mnras/stw274}

\bibitem{InagakiLyndenBell1983}
S.~{Inagaki}, D.~{Lynden-Bell}, MNRAS \textbf{205}, 913 (1983).
\newblock \doi{10.1093/mnras/205.4.913}

\bibitem{SugimotoBettwieser1983}
D.~{Sugimoto}, E.~{Bettwieser}, MNRAS \textbf{204}, 19P (1983).
\newblock \doi{10.1093/mnras/204.1.19P}

\bibitem{CohnHutWise1989}
H.~{Cohn}, P.~{Hut}, M.~{Wise}, \apj \textbf{342}, 814 (1989).
\newblock \doi{10.1086/167638}

\bibitem{Aarsethetal1974}
S.J. {Aarseth}, M.~{Henon}, R.~{Wielen}, \aap \textbf{37}(1), 183 (1974)

\bibitem{GierszHeggie1994b}
M.~{Giersz}, D.C. {Heggie}, MNRAS \textbf{270}, 298 (1994).
\newblock \doi{10.1093/mnras/270.2.298}

\bibitem{GierszSpurzem1994}
M.~{Giersz}, R.~{Spurzem}, MNRAS \textbf{269}, 241 (1994).
\newblock \doi{10.1093/mnras/269.2.241}

\bibitem{SpurzemAarseth1996}
R.~{Spurzem}, S.J. {Aarseth}, MNRAS \textbf{282}, 19 (1996).
\newblock \doi{10.1093/mnras/282.1.19}

\bibitem{Makino1996}
J.~{Makino}, ApJ \textbf{471}, 796 (1996).
\newblock \doi{10.1086/178007}

\bibitem{Heggie1984}
D.C. {Heggie}, MNRAS \textbf{206}, 179 (1984).
\newblock \doi{10.1093/mnras/206.1.179}

\bibitem{BinneyTremaine2008}
J.~{Binney}, S.~{Tremaine}, \emph{{Galactic Dynamics: Second Edition}}
  (Princeton University Press, 2008)

\bibitem{Sundman1907}
K.~Sundman, Acta Soc. Sci. Fennicae \textbf{34}, 1 (1907).
\newblock
  \urlprefix\url{https://www.scopus.com/record/display.uri?eid=2-s2.0-0011571192&origin=inward&txGid=093af269c50d6af350defc67d56fe2dd}

\bibitem{Sundman1909}
K.~Sundman, Acta Soc. Sci. Fennicae \textbf{35}, 1 (1909).
\newblock
  \urlprefix\url{https://www.scopus.com/record/display.uri?eid=2-s2.0-0011566575&origin=inward&txGid=c54f8aa32a21b6b4faca3197fc17df99}

\bibitem{Szebehely1967}
V.~{Szebehely}, C.F. {Peters}, \aj \textbf{72}, 876 (1967).
\newblock \doi{10.1086/110355}

\bibitem{Tanikawaetal2012}
A.~{Tanikawa}, P.~{Hut}, J.~{Makino}, \na \textbf{17}(3), 272 (2012).
\newblock \doi{10.1016/j.newast.2011.09.001}

\bibitem{Rosenbluthetal1957}
M.N. {Rosenbluth}, W.M. {MacDonald}, D.L. {Judd}, Physical Review
  \textbf{107}(1), 1 (1957).
\newblock \doi{10.1103/PhysRev.107.1}

\bibitem{Schneideretal2011}
J.~{Schneider}, P.~{Amaro-Seoane}, R.~{Spurzem}, \mnras \textbf{410}(1), 432
  (2011).
\newblock \doi{10.1111/j.1365-2966.2010.17454.x}

\bibitem{Bettwieser1983}
E.~{Bettwieser}, MNRAS \textbf{203}, 811 (1983).
\newblock \doi{10.1093/mnras/203.3.811}

\bibitem{Cohn1980}
H.~{Cohn}, ApJ \textbf{242}, 765 (1980).
\newblock \doi{10.1086/158511}

\bibitem{EinselSpurzem1999}
C.~{Einsel}, R.~{Spurzem}, MNRAS \textbf{302}(1), 81 (1999).
\newblock \doi{10.1046/j.1365-8711.1999.02083.x}

\bibitem{Cohn1979}
H.~{Cohn}, ApJ \textbf{234}, 1036 (1979).
\newblock \doi{10.1086/157587}

\bibitem{Takahashi1996a}
K.~{Takahashi}, PASJ \textbf{47}, 561 (1995)

\bibitem{Takahashi1996b}
K.~{Takahashi}, PASJ \textbf{48}, 691 (1996).
\newblock \doi{10.1093/pasj/48.5.691}

\bibitem{Takahashi1997}
K.~{Takahashi}, PASJ \textbf{49}, 547 (1997).
\newblock \doi{10.1093/pasj/49.5.547}

\bibitem{Goodman1983}
J.~{Goodman}, ApJ \textbf{270}, 700 (1983).
\newblock \doi{10.1086/161161}

\bibitem{LuptonGunn1987}
R.H. {Lupton}, J.E. {Gunn}, AJ \textbf{93}, 1106 (1987).
\newblock \doi{10.1086/114394}

\bibitem{LongarettiLagoute1996}
P.Y. {Longaretti}, C.~{Lagoute}, AAP \textbf{308}, 453 (1996)

\bibitem{King1966a}
I.R. {King}, AJ \textbf{71}, 64 (1966).
\newblock \doi{10.1086/109857}

\bibitem{Kimetal2002}
E.~{Kim}, C.~{Einsel}, H.M. {Lee}, R.~{Spurzem}, M.G. {Lee}, MNRAS
  \textbf{334}(2), 310 (2002).
\newblock \doi{10.1046/j.1365-8711.2002.05420.x}

\bibitem{Kimetal2004}
E.~{Kim}, H.M. {Lee}, R.~{Spurzem}, MNRAS \textbf{351}(1), 220 (2004).
\newblock \doi{10.1111/j.1365-2966.2004.07776.x}

\bibitem{Kimetal2008}
E.~{Kim}, I.~{Yoon}, H.M. {Lee}, R.~{Spurzem}, MNRAS \textbf{383}(1), 2 (2008).
\newblock \doi{10.1111/j.1365-2966.2007.12524.x}

\bibitem{AkiyamaSugimoto1989}
K.~{Akiyama}, D.~{Sugimoto}, PASJ \textbf{41}(5), 991 (1989)

\bibitem{Tiongcoetal2022}
M.A. {Tiongco}, E.~{Vesperini}, A.L. {Varri}, MNRAS \textbf{512}(2), 1584
  (2022).
\newblock \doi{10.1093/mnras/stac643}

\bibitem{Livernoisetal2022}
A.R. {Livernois}, E.~{Vesperini}, A.L. {Varri}, J.~{Hong}, M.~{Tiongco}, MNRAS
  \textbf{512}(2), 2584 (2022).
\newblock \doi{10.1093/mnras/stac651}

\bibitem{Kamlahetal2023}
A.W.H. {Kamlah}, A.~{Tanikawa}, M.~{Arca Sedda}, M.~{Giersz}, N.~{Neumayer},
  R.~{Spurzem}, \mnras  (2023).
\newblock In prep.

\bibitem{Hongetal2013}
J.~{Hong}, E.~{Kim}, H.M. {Lee}, R.~{Spurzem}, MNRAS \textbf{430}(4), 2960
  (2013).
\newblock \doi{10.1093/mnras/stt099}

\bibitem{Tiongcoetal2017}
M.A. {Tiongco}, E.~{Vesperini}, A.L. {Varri}, MNRAS \textbf{469}(1), 683
  (2017).
\newblock \doi{10.1093/mnras/stx853}

\bibitem{VarriBertin2012}
A.L. {Varri}, G.~{Bertin}, AAP \textbf{540}, A94 (2012).
\newblock \doi{10.1051/0004-6361/201118300}

\bibitem{Varrietal2018}
A.L. {Varri}, M.X. {Cai}, F.~{Concha-Ram{\'\i}rez}, F.~{Dinnbier},
  N.~{L{\"u}tzgendorf}, V.~{Pavl{\'\i}k}, S.~{Rastello}, A.~{Sollima},
  L.~{Wang}, A.~{Zocchi}, Computational Astrophysics and Cosmology
  \textbf{5}(1), 2 (2018).
\newblock \doi{10.1186/s40668-018-0024-6}

\bibitem{Spitzeretal1971a}
J.~{Spitzer}, Lyman, M.H. {Hart}, \apj \textbf{164}, 399 (1971).
\newblock \doi{10.1086/150855}

\bibitem{Spitzeretal1971b}
J.~{Spitzer}, Lyman, M.H. {Hart}, \apj \textbf{166}, 483 (1971).
\newblock \doi{10.1086/150977}

\bibitem{Spitzeretal1972a}
J.~{Spitzer}, Lyman, S.L. {Shapiro}, \apj \textbf{173}, 529 (1972).
\newblock \doi{10.1086/151442}

\bibitem{Spitzeretal1972b}
J.~{Spitzer}, Lyman, T.X. {Thuan}, \apj \textbf{175}, 31 (1972).
\newblock \doi{10.1086/151537}

\bibitem{Spitzeretal1973}
J.~{Spitzer}, Lyman, R.A. {Chevalier}, \apj \textbf{183}, 565 (1973).
\newblock \doi{10.1086/152247}

\bibitem{Spitzeretal1975a}
J.~{Spitzer}, L., J.M. {Shull}, \apj \textbf{200}, 339 (1975).
\newblock \doi{10.1086/153793}

\bibitem{Spitzeretal1975b}
J.~{Spitzer}, L., J.M. {Shull}, \apj \textbf{201}, 773 (1975).
\newblock \doi{10.1086/153943}

\bibitem{Spitzeretal1980}
J.~{Spitzer}, L., R.D. {Mathieu}, \apj \textbf{241}, 618 (1980).
\newblock \doi{10.1086/158376}

\bibitem{Henon1971}
M.~{H{\'e}non}, \apss \textbf{13}(2), 284 (1971).
\newblock \doi{10.1007/BF00649159}

\bibitem{Stodolkiewicz1982}
J.S. {Stodo\l kiewicz}, \actaa \textbf{32}(1-2), 63 (1982)

\bibitem{Stodolkiewicz1986}
J.S. {Stodo\l kiewicz}, \actaa \textbf{36}(1), 19 (1986)

\bibitem{Giersz1998}
M.~{Giersz}, MNRAS \textbf{298}(4), 1239 (1998).
\newblock \doi{10.1046/j.1365-8711.1998.01734.x}

\bibitem{Joshietal2000}
K.J. {Joshi}, F.A. {Rasio}, S.~{Portegies Zwart}, \apj \textbf{540}(2), 969
  (2000).
\newblock \doi{10.1086/309350}

\bibitem{Gierszetal2013}
M.~{Giersz}, D.C. {Heggie}, J.R. {Hurley}, A.~{Hypki}, MNRAS \textbf{431}(3),
  2184 (2013).
\newblock \doi{10.1093/mnras/stt307}

\bibitem{Gierszetal2015}
M.~{Giersz}, N.~{Leigh}, A.~{Hypki}, N.~{L{\"u}tzgendorf}, A.~{Askar}, MNRAS
  \textbf{454}(3), 3150 (2015).
\newblock \doi{10.1093/mnras/stv2162}

\bibitem{Hongetal2020b}
J.~{Hong}, A.~{Askar}, M.~{Giersz}, A.~{Hypki}, S.J. {Yoon}, MNRAS
  \textbf{498}(3), 4287 (2020).
\newblock \doi{10.1093/mnras/staa2677}

\bibitem{Levequeetal2021}
A.~{Leveque}, M.~{Giersz}, M.~{Paolillo}, MNRAS \textbf{501}(4), 5212 (2021).
\newblock \doi{10.1093/mnras/staa4027}

\bibitem{Levequeetal2022b}
A.~{Leveque}, M.~{Giersz}, M.~{Arca-Sedda}, A.~{Askar}, \mnras \textbf{514}(4),
  5751 (2022).
\newblock \doi{10.1093/mnras/stac1694}

\bibitem{Levequeetal2023}
A.~{Leveque}, M.~{Giersz}, A.~{Askar}, M.~{Arca-Sedda}, A.~{Olejak}, \mnras
  \textbf{520}(2), 2593 (2023).
\newblock \doi{10.1093/mnras/stad240}

\bibitem{Rodriguezetal2021b}
C.L. {Rodriguez}, N.C. {Weatherford}, S.C. {Coughlin}, P.A. {Seoane},
  K.~{Breivik}, S.~{Chatterjee}, G.~{Fragione}, F.~{K{\"A}{\"A}lu},
  K.~{Kremer}, N.Z. {Rui}, C.S. {Ye}, M.~{Zevin}, F.A. {Rasio}.
\newblock {CMC-COSMIC: Cluster Monte Carlo code}.
\newblock Astrophysics Source Code Library, record ascl:2108.023 (2021)

\bibitem{Ruietal2021}
N.Z. {Rui}, K.~{Kremer}, N.C. {Weatherford}, S.~{Chatterjee}, F.A. {Rasio},
  C.L. {Rodriguez}, C.S. {Ye}, \apj \textbf{912}(2), 102 (2021).
\newblock \doi{10.3847/1538-4357/abed49}

\bibitem{Rodriguezetal2022}
C.L. {Rodriguez}, N.C. {Weatherford}, S.C. {Coughlin}, P.~{Amaro-Seoane},
  K.~{Breivik}, et~al., ApJS \textbf{258}(2), 22 (2022).
\newblock \doi{10.3847/1538-4365/ac2edf}

\bibitem{Rodriguezetal2021a}
C.L. {Rodriguez}, K.~{Kremer}, S.~{Chatterjee}, G.~{Fragione}, A.~{Loeb},
  et~al., RNAAS \textbf{5}(1), 19 (2021).
\newblock \doi{10.3847/2515-5172/abdf54}

\bibitem{Kremeretal2021}
K.~{Kremer}, N.Z. {Rui}, N.C. {Weatherford}, S.~{Chatterjee}, G.~{Fragione},
  F.A. {Rasio}, C.L. {Rodriguez}, C.S. {Ye}, \apj \textbf{917}(1), 28 (2021).
\newblock \doi{10.3847/1538-4357/ac06d4}

\bibitem{Yeetal2022}
C.S. {Ye}, K.~{Kremer}, C.L. {Rodriguez}, N.Z. {Rui}, N.C. {Weatherford},
  S.~{Chatterjee}, G.~{Fragione}, F.A. {Rasio}, \apj \textbf{931}(2), 84
  (2022).
\newblock \doi{10.3847/1538-4357/ac5b0b}

\bibitem{Vasiliev2015}
E.~{Vasiliev}, MNRAS \textbf{446}(3), 3150 (2015).
\newblock \doi{10.1093/mnras/stu2360}

\bibitem{Aarseth1963}
S.J. {Aarseth}, \mnras \textbf{126}, 223 (1963).
\newblock \doi{10.1093/mnras/126.3.223}

\bibitem{Aarseth1967}
S.~{Aarseth}, in \emph{Les Nouvelles M\&eacute;thodes de la Dynamique
  Stellaire} (1967), p.~47

\bibitem{KustaanheimoStiefel1965}
P.~{Kustaanheimo}, E.~{Stiefel}, J. Reine Angew. Math. \textbf{218}, 204 (1965)

\bibitem{AarsethZare1974}
S.J. {Aarseth}, K.~{Zare}, Celestial Mechanics \textbf{10}(2), 185 (1974).
\newblock \doi{10.1007/BF01227619}

\bibitem{Heggie1974}
D.C. {Heggie}, CeMec \textbf{10}(2), 217 (1974).
\newblock \doi{10.1007/BF01227621}

\bibitem{AhmadCohen1973}
A.~{Ahmad}, L.~{Cohen}, Journal of Computational Physics \textbf{12}, 389
  (1973).
\newblock \doi{10.1016/0021-9991(73)90160-5}

\bibitem{Aarseth1979}
S.J. {Aarseth}, in \emph{Instabilities in Dynamical Systems. Applications to
  Celestial Mechanics}, \emph{NATO Advanced Study Institute (ASI) Series C},
  vol.~47, ed. by V.G. {Szebehely} (1979), \emph{NATO Advanced Study Institute
  (ASI) Series C}, vol.~47, pp. 69--80

\bibitem{Aarseth1985a}
S.J. {Aarseth}, in \emph{Multiple time scales} (1985), pp. 377--418

\bibitem{LecarAarseth1986}
M.~{Lecar}, S.J. {Aarseth}, ApJ \textbf{305}, 564 (1986).
\newblock \doi{10.1086/164269}

\bibitem{McMillan1986}
P.~{Hut}, S.L.W. {McMillan}, \emph{{The Use of Supercomputers in Stellar
  Dynamics}}, vol. 267 (Springer-Verlag Berlin Heidelberg New York, 1986).
\newblock \doi{10.1007/BFb0116387}

\bibitem{MikkolaAarseth1990}
S.~{Mikkola}, S.J. {Aarseth}, Celestial Mechanics and Dynamical Astronomy
  \textbf{47}(4), 375 (1990)

\bibitem{Aarsethetal1993}
S.J. {Aarseth}, D.N.C. {Lin}, P.L. {Palmer}, ApJ \textbf{403}, 351 (1993).
\newblock \doi{10.1086/172208}

\bibitem{McMillanAarseth1993}
S.L.W. {McMillan}, S.J. {Aarseth}, \apj \textbf{414}, 200 (1993).
\newblock \doi{10.1086/173068}

\bibitem{Aarseth1994}
S.J. {Aarseth}, in \emph{Galactic Dynamics and N-Body Simulations}, vol. 433,
  ed. by G.~{Contopoulos}, N.K. {Spyrou}, L.~{Vlahos} (Springer-Verlag Berlin
  Heidelberg New York, 1994), pp. 277--312

\bibitem{Makino1991a}
J.~{Makino}, ApJ \textbf{369}, 200 (1991).
\newblock \doi{10.1086/169751}

\bibitem{MakinoAarseth1992}
J.~{Makino}, S.J. {Aarseth}, PASJ \textbf{44}, 141 (1992)

\bibitem{Toutetal1997}
C.A. {Tout}, S.J. {Aarseth}, O.R. {Pols}, P.P. {Eggleton}, MNRAS
  \textbf{291}(4), 732 (1997).
\newblock \doi{10.1093/mnras/291.4.732}

\bibitem{Mardlign1995a}
R.A. {Mardling}, ApJ \textbf{450}, 722 (1995).
\newblock \doi{10.1086/176178}

\bibitem{Mardling1995b}
R.A. {Mardling}, ApJ \textbf{450}, 732 (1995).
\newblock \doi{10.1086/176179}

\bibitem{MikkolaAarseth1998}
S.~{Mikkola}, S.J. {Aarseth}, NA \textbf{3}(5), 309 (1998).
\newblock \doi{10.1016/S1384-1076(98)00018-9}

\bibitem{MardlingAarseth1999}
R.~{Mardling}, S.~{Aarseth}, in \emph{The Dynamics of Small Bodies in the Solar
  System, A Major Key to Solar System Studies}, \emph{NATO Advanced Study
  Institute (ASI) Series C}, vol. 522, ed. by B.A. {Steves}, A.E. {Roy} (1999),
  \emph{NATO Advanced Study Institute (ASI) Series C}, vol. 522, p. 385

\bibitem{Aarseth1999b}
S.J. {Aarseth}, PASP \textbf{111}(765), 1333 (1999).
\newblock \doi{10.1086/316455}

\bibitem{MikkolaTanikawa1999a}
S.~{Mikkola}, K.~{Tanikawa}, Celestial Mechanics and Dynamical Astronomy
  \textbf{74}(4), 287 (1999).
\newblock \doi{10.1023/A:1008368322547}

\bibitem{MikkolaTanikawa1999b}
S.~{Mikkola}, K.~{Tanikawa}, MNRAS \textbf{310}(3), 745 (1999).
\newblock \doi{10.1046/j.1365-8711.1999.02982.x}

\bibitem{Spurzem1999}
R.~{Spurzem}, Journal of Computational and Applied Mathematics \textbf{109},
  407 (1999)

\bibitem{Hurleyetal2000}
J.R. {Hurley}, O.R. {Pols}, C.A. {Tout}, MNRAS \textbf{315}(3), 543 (2000).
\newblock \doi{10.1046/j.1365-8711.2000.03426.x}

\bibitem{Hurleyetal2002b}
J.R. {Hurley}, C.A. {Tout}, O.R. {Pols}, MNRAS \textbf{329}(4), 897 (2002).
\newblock \doi{10.1046/j.1365-8711.2002.05038.x}

\bibitem{Makinoetal2003}
J.~{Makino}, T.~{Fukushige}, M.~{Koga}, K.~{Namura}, \pasj \textbf{55}, 1163
  (2003).
\newblock \doi{10.1093/pasj/55.6.1163}

\bibitem{Kupietal2006}
G.~{Kupi}, P.~{Amaro-Seoane}, R.~{Spurzem}, MNRAS \textbf{371}(1), L45 (2006).
\newblock \doi{10.1111/j.1745-3933.2006.00205.x}

\bibitem{PortegiesZwartetal2007}
S.F. {Portegies Zwart}, R.G. {Belleman}, P.M. {Geldof}, \na \textbf{12}(8), 641
  (2007).
\newblock \doi{10.1016/j.newast.2007.05.004}

\bibitem{MikkolaMerritt2008}
S.~{Mikkola}, D.~{Merritt}, AJ \textbf{135}(6), 2398 (2008).
\newblock \doi{10.1088/0004-6256/135/6/2398}

\bibitem{HellstroemMikkola2010}
C.~{Hellstr{\"o}m}, S.~{Mikkola}, Celestial Mechanics and Dynamical Astronomy
  \textbf{106}(2), 143 (2010).
\newblock \doi{10.1007/s10569-009-9248-8}

\bibitem{NitadoriAarseth2012}
K.~{Nitadori}, S.J. {Aarseth}, MNRAS \textbf{424}(1), 545 (2012).
\newblock \doi{10.1111/j.1365-2966.2012.21227.x}

\bibitem{Bercziketal2013}
P.~{Berczik}, R.~{Spurzem}, S.~{Zhong}, L.~{Wang}, K.~Nitadori, T.~{Hamada},
  A.~{Veles}, in \emph{Supercomputing}, \emph{Lecture Notes in Computer
  Science}, vol. 7905, ed. by J.M. {Kunkel}, T.~{Ludwig}, H.~{Meuer} (2013),
  \emph{Lecture Notes in Computer Science}, vol. 7905, pp. 13--25.
\newblock Procs. of 28th Intl. Supercomputing Conf. ISC 2013, Leipzig, Germany

\bibitem{Bremetal2013}
P.~{Brem}, P.~{Amaro-Seoane}, R.~{Spurzem}, MNRAS \textbf{434}(4), 2999 (2013).
\newblock \doi{10.1093/mnras/stt1220}

\bibitem{DehnenHernandez2017}
W.~{Dehnen}, D.M. {Hernandez}, \mnras \textbf{465}(1), 1201 (2017).
\newblock \doi{10.1093/mnras/stw2758}

\bibitem{Wangetal2020c}
L.~{Wang}, K.~{Nitadori}, J.~{Makino}, MNRAS \textbf{493}(3), 3398 (2020).
\newblock \doi{10.1093/mnras/staa480}

\bibitem{Wangetal2020d}
L.~{Wang}, M.~{Iwasawa}, K.~{Nitadori}, J.~{Makino}, MNRAS \textbf{497}(1), 536
  (2020).
\newblock \doi{10.1093/mnras/staa1915}

\bibitem{Rantalaetal2021}
A.~{Rantala}, T.~{Naab}, V.~{Springel}, MNRAS \textbf{502}(4), 5546 (2021).
\newblock \doi{10.1093/mnras/stab057}

\bibitem{vHoerner1960}
S.~{von Hoerner}, \zap \textbf{50}, 184 (1960)

\bibitem{vHoerner1963}
S.~{von Hoerner}, \zap \textbf{57}, 47 (1963)

\bibitem{Aarseth1971}
S.J. {Aarseth}, \apss \textbf{14}(1), 118 (1971).
\newblock \doi{10.1007/BF00649199}

\bibitem{vHoerner2001}
S.~{von Hoerner}, in \emph{Dynamics of Star Clusters and the Milky Way},
  \emph{Astronomical Society of the Pacific Conference Series}, vol. 228, ed.
  by S.~{Deiters}, B.~{Fuchs}, A.~{Just}, R.~{Spurzem}, R.~{Wielen} (2001),
  \emph{Astronomical Society of the Pacific Conference Series}, vol. 228, p.~11

\bibitem{MikkolaAarseth1996}
S.~{Mikkola}, S.J. {Aarseth}, Celestial Mechanics and Dynamical Astronomy
  \textbf{64}(3), 197 (1996).
\newblock \doi{10.1007/BF00728347}

\bibitem{Aarseth1999a}
S.J. {Aarseth}, CeMDA \textbf{73}, 127 (1999).
\newblock \doi{10.1023/A:1008390828807}

\bibitem{Aarseth2003b}
S.J. {Aarseth}, \emph{{Gravitational N-Body Simulations}} (Cambridge University
  Press, 2003)

\bibitem{Sweatman1994}
W.L. {Sweatman}, Journal of Computational Physics \textbf{111}(1), 110 (1994).
\newblock \doi{10.1006/jcph.1994.1048}

\bibitem{MakinoHut1988}
J.~{Makino}, P.~{Hut}, ApJS \textbf{68}, 833 (1988).
\newblock \doi{10.1086/191306}

\bibitem{Makino1991b}
J.~{Makino}, \pasj \textbf{43}, 859 (1991)

\bibitem{LeviCivita1916}
T.~{Levi-Civita}, C. R. Hebd. Acad. Sci. Paris \textbf{162}, 625 (1916).
\newblock Reprinted in Levi-Civita, 1954–1973, vol. III, pp. 589–593

\bibitem{NeutschScherer1992}
W.~{Neutsch}, K.~{Scherer}, \emph{{Celestial Mechanics}} (Bibliographisches
  Institut, 1992).
\newblock 484 pages

\bibitem{Mikkola1997a}
S.~{Mikkola}, Celestial Mechanics and Dynamical Astronomy \textbf{68}(1), 87
  (1997).
\newblock \doi{10.1023/A:1008291715719}

\bibitem{MardlingAarseth2001}
R.A. {Mardling}, S.J. {Aarseth}, MNRAS \textbf{321}(3), 398 (2001).
\newblock \doi{10.1046/j.1365-8711.2001.03974.x}

\bibitem{Sugimotoetal1990}
D.~{Sugimoto}, Y.~{Chikada}, J.~{Makino}, T.~{Ito}, T.~{Ebisuzaki},
  M.~{Umemura}, \nat \textbf{345}(6270), 33 (1990).
\newblock \doi{10.1038/345033a0}

\bibitem{Makinoetal1993}
J.~{Makino}, E.~{Kokubo}, M.~{Taiji}, \pasj \textbf{45}, 349 (1993)

\bibitem{Makinoetal1997}
J.~{Makino}, M.~{Taiji}, T.~{Ebisuzaki}, D.~{Sugimoto}, \apj \textbf{480}(1),
  432 (1997).
\newblock \doi{10.1086/303972}

\bibitem{MakinoTaiji1998}
J.~{Makino}, M.~{Taiji}, \emph{{Scientific Simulations with Special-Purpose
  Computers--the GRAPE Systems}} (Wiley-VCH, 1998)

\bibitem{Harfstetal2007}
S.~{Harfst}, A.~{Gualandris}, D.~{Merritt}, R.~{Spurzem}, S.~{Portegies Zwart},
  P.~{Berczik}, \na \textbf{12}(5), 357 (2007).
\newblock \doi{10.1016/j.newast.2006.11.003}

\bibitem{Schiveetal2008}
H.Y. {Schive}, C.H. {Chien}, S.K. {Wong}, Y.C. {Tsai}, T.~{Chiueh}, \na
  \textbf{13}(6), 418 (2008).
\newblock \doi{10.1016/j.newast.2007.12.005}

\bibitem{NitadoriMakino2008}
K.~{Nitadori}, J.~{Makino}, NewA \textbf{13}(7), 498 (2008).
\newblock \doi{10.1016/j.newast.2008.01.010}

\bibitem{Gaburovetal2009}
E.~{Gaburov}, S.~{Harfst}, S.~{Portegies Zwart}, \na \textbf{14}(7), 630
  (2009).
\newblock \doi{10.1016/j.newast.2009.03.002}

\bibitem{Bellemanetal2008}
R.G. {Belleman}, J.~{B{\'e}dorf}, S.F. {Portegies Zwart}, \na \textbf{13}(2),
  103 (2008).
\newblock \doi{10.1016/j.newast.2007.07.004}

\bibitem{Bellemanetal2014}
R.~{Belleman}, J.~{B{\'e}dorf}, S.F. {Portegies Zwart}.
\newblock {Kirin: N-body simulation library for GPUs}.
\newblock Astrophysics Source Code Library, record ascl:1401.001 (2014)

\bibitem{Makino2002}
J.~{Makino}, \na \textbf{7}(7), 373 (2002).
\newblock \doi{10.1016/S1384-1076(02)00143-4}

\bibitem{Dorbandetal2003}
E.N. {Dorband}, M.~{Hemsendorf}, D.~{Merritt}, Journal of Computational Physics
  \textbf{185}(2), 484 (2003).
\newblock \doi{10.1016/S0021-9991(02)00067-0}

\bibitem{Hemsendorfetal2002}
M.~{Hemsendorf}, S.~{Sigurdsson}, R.~{Spurzem}, \apj \textbf{581}(2), 1256
  (2002).
\newblock \doi{10.1086/344255}

\bibitem{Lippertetal1996}
T.~{Lippert}, G.~{Ritzenh{\"o}fer}, U.~{Glaessner}, H.~{Hoeber}, A.~{Seyfried},
  K.~{Schilling}, International Journal of Modern Physics C \textbf{7}(4), 485
  (1996).
\newblock \doi{10.1142/S0129183196000430}

\bibitem{Lippertetal1998}
T.~{Lippert}, N.~{Petkov}, P.~{Palazzari}, K.~{Schilling}, arXiv e-prints
  cs/9809105 (1998).
\newblock \doi{10.48550/arXiv.cs/9809105}

\bibitem{Abbottetal2016a}
B.P. {Abbott}, R.~{Abbott}, T.D. {Abbott}, M.R. {Abernathy}, F.~{Acernese},
  et~al., PhRvL \textbf{116}(6), 061102 (2016).
\newblock \doi{10.1103/PhysRevLett.116.061102}

\bibitem{Rizzutoetal2021}
F.P. {Rizzuto}, T.~{Naab}, R.~{Spurzem}, M.~{Giersz}, J.P. {Ostriker}, et~al.,
  MNRAS \textbf{501}(4), 5257 (2021).
\newblock \doi{10.1093/mnras/staa3634}

\bibitem{Rizzutoetal2022}
F.P. {Rizzuto}, T.~{Naab}, R.~{Spurzem}, M.~{Arca-Sedda}, M.~{Giersz}, et~al.,
  MNRAS \textbf{512}(1), 884 (2022).
\newblock \doi{10.1093/mnras/stac231}

\bibitem{ArcaSeddaetal2021a}
M.~{Arca-Sedda}, F.P. {Rizzuto}, T.~{Naab}, J.~{Ostriker}, M.~{Giersz},
  R.~{Spurzem}, \apj \textbf{920}(2), 128 (2021).
\newblock \doi{10.3847/1538-4357/ac1419}

\bibitem{DiCarloetal2021}
U.N. {Di Carlo}, M.~{Mapelli}, M.~{Pasquato}, S.~{Rastello}, A.~{Ballone},
  M.~{Dall'Amico}, N.~{Giacobbo}, G.~{Iorio}, M.~{Spera}, S.~{Torniamenti},
  F.~{Haardt}, \mnras \textbf{507}(4), 5132 (2021).
\newblock \doi{10.1093/mnras/stab2390}

\bibitem{Aarseth2012}
S.J. {Aarseth}, MNRAS \textbf{422}(1), 841 (2012).
\newblock \doi{10.1111/j.1365-2966.2012.20666.x}

\bibitem{Banerjeeetal2020}
S.~{Banerjee}, K.~{Belczynski}, C.L. {Fryer}, P.~{Berczik}, J.R. {Hurley},
  et~al., A\&A \textbf{639}, A41 (2020).
\newblock \doi{10.1051/0004-6361/201935332}

\bibitem{Morawskietal2018}
J.~{Morawski}, M.~{Giersz}, A.~{Askar}, K.~{Belczynski}, MNRAS \textbf{481}(2),
  2168 (2018).
\newblock \doi{10.1093/mnras/sty2401}

\bibitem{ArcaSeddaetal2022}
M.e.a. {Arca Sedda}, MNRAS  (2022).
\newblock In prep.

\bibitem{Huangetal2016}
S.Y. {Huang}, R.~{Spurzem}, P.~{Berczik}, Research in Astronomy and
  Astrophysics \textbf{16}(1), 11 (2016).
\newblock \doi{10.1088/1674-4527/16/1/011}

\bibitem{CapuzzoDolcettaetal2008}
S.~{Mikkola}, D.~{Merritt}, AJ \textbf{135}(6), 2398 (2008).
\newblock \doi{10.1088/0004-6256/135/6/2398}

\bibitem{Spera2014}
M.~{Spera}, arXiv arXiv:1411.5234 (2014)

\bibitem{Zhongetal2014}
S.~{Zhong}, P.~{Berczik}, R.~{Spurzem}, \apj \textbf{792}(2), 137 (2014).
\newblock \doi{10.1088/0004-637X/792/2/137}

\bibitem{Lietal2017}
S.~{Li}, F.K. {Liu}, P.~{Berczik}, R.~{Spurzem}, \apj \textbf{834}(2), 195
  (2017).
\newblock \doi{10.3847/1538-4357/834/2/195}

\bibitem{Lietal2019}
S.~{Li}, P.~{Berczik}, X.~{Chen}, F.K. {Liu}, R.~{Spurzem}, Y.~{Qiu}, \apj
  \textbf{883}(2), 132 (2019).
\newblock \doi{10.3847/1538-4357/ab3e4a}

\bibitem{Bortolasetal2018}
E.~{Bortolas}, M.~{Mapelli}, M.~{Spera}, \mnras \textbf{474}(1), 1054 (2018).
\newblock \doi{10.1093/mnras/stx2795}

\bibitem{Miller1964}
R.H. {Miller}, \apj \textbf{140}, 250 (1964).
\newblock \doi{10.1086/147911}

\bibitem{Goodmanetal1993}
J.~{Goodman}, D.C. {Heggie}, P.~{Hut}, \apj \textbf{415}, 715 (1993).
\newblock \doi{10.1086/173196}

\bibitem{Kandrupetal1994}
H.E. {Kandrup}, M.E. {Mahon}, J.~{Smith}, Haywood, \apj \textbf{428}, 458
  (1994).
\newblock \doi{10.1086/174259}

\bibitem{GierszHeggie1996}
M.~{Giersz}, D.C. {Heggie}, MNRAS \textbf{279}(3), 1037 (1996).
\newblock \doi{10.1093/mnras/279.3.1037}

\bibitem{GierszHeggie1997}
M.~{Giersz}, D.C. {Heggie}, MNRAS \textbf{286}(3), 709 (1997).
\newblock \doi{10.1093/mnras/286.3.709}

\bibitem{WisdomHolman1991}
J.~{Wisdom}, M.~{Holman}, \aj \textbf{102}, 1528 (1991).
\newblock \doi{10.1086/115978}

\bibitem{Mikkola1997b}
S.~{Mikkola}, Celestial Mechanics and Dynamical Astronomy \textbf{67}(2), 145
  (1997).
\newblock \doi{10.1023/A:1008217427749}

\bibitem{Hutetal1995}
P.~{Hut}, J.~{Makino}, S.~{McMillan}, ApJL \textbf{443}, L93 (1995).
\newblock \doi{10.1086/187844}

\bibitem{Funatoetal1996}
Y.~{Funato}, P.~{Hut}, S.~{McMillan}, J.~{Makino}, \aj \textbf{112}, 1697
  (1996).
\newblock \doi{10.1086/118136}

\bibitem{Kokuboetal1998}
E.~{Kokubo}, K.~{Yoshinaga}, J.~{Makino}, MNRAS \textbf{297}(4), 1067 (1998).
\newblock \doi{10.1046/j.1365-8711.1998.01581.x}

\bibitem{Glaschkeetal2014}
P.~{Glaschke}, P.~{Amaro-Seoane}, R.~{Spurzem}, \mnras \textbf{445}(4), 3620
  (2014).
\newblock \doi{10.1093/mnras/stu1558}

\bibitem{AmaroSeoaneetal2014}
P.~{Amaro-Seoane}, P.~{Glaschke}, R.~{Spurzem}, \mnras \textbf{445}(4), 3755
  (2014).
\newblock \doi{10.1093/mnras/stu1734}

\bibitem{QuinlanTremaine1992}
G.D. {Quinlan}, S.~{Tremaine}, \mnras \textbf{259}(3), 505 (1992).
\newblock \doi{10.1093/mnras/259.3.505}

\bibitem{WangHernandez2021}
L.~{Wang}, D.M. {Hernandez}, arXiv e-prints arXiv:2104.10843 (2021)

\bibitem{Oshinoetal2011}
S.~{Oshino}, Y.~{Funato}, J.~{Makino}, \pasj \textbf{63}, 881 (2011).
\newblock \doi{10.1093/pasj/63.4.881}

\bibitem{Iwasawaetal2015}
M.~{Iwasawa}, S.~{Portegies Zwart}, J.~{Makino}, Computational Astrophysics and
  Cosmology \textbf{2}, 6 (2015).
\newblock \doi{10.1186/s40668-015-0010-1}

\bibitem{Iwasawaetal2016}
M.~{Iwasawa}, A.~{Tanikawa}, N.~{Hosono}, K.~{Nitadori}, T.~{Muranushi},
  J.~{Makino}, \pasj \textbf{68}(4), 54 (2016).
\newblock \doi{10.1093/pasj/psw053}

\bibitem{Iwasawaetal2017}
M.~{Iwasawa}, S.~{Oshino}, M.S. {Fujii}, Y.~{Hori}, \pasj \textbf{69}(5), 81
  (2017).
\newblock \doi{10.1093/pasj/psx073}

\bibitem{PretoTremaine1999}
M.~{Preto}, S.~{Tremaine}, \aj \textbf{118}(5), 2532 (1999).
\newblock \doi{10.1086/301102}

\bibitem{Iwasawaetal2020}
M.~{Iwasawa}, D.~{Namekata}, K.~{Nitadori}, K.~{Nomura}, L.~{Wang},
  M.~{Tsubouchi}, J.~{Makino}, \pasj \textbf{72}(1), 13 (2020).
\newblock \doi{10.1093/pasj/psz133}

\bibitem{Hurleyetal2005}
J.R. {Hurley}, O.R. {Pols}, S.J. {Aarseth}, C.A. {Tout}, MNRAS \textbf{363}(1),
  293 (2005).
\newblock \doi{10.1111/j.1365-2966.2005.09448.x}

\bibitem{JUWELS2021}
{J\"{u}lich Supercomputing Centre}, Journal of large-scale research facilities
  \textbf{7}(A138) (2021).
\newblock \doi{10.17815/jlsrf-7-183}.
\newblock \urlprefix\url{http://dx.doi.org/10.17815/jlsrf-7-183}

\bibitem{Spurzemetal2022}
R.~{Spurzem}, F.~{Rizzuto}, M.~{Arca Sedda}, A.~{Kamlah}, P.~{Berczik},
  Q.~{Shu}, A.~{Tanikawa}, T.~{Naab}, in \emph{NIC Symposium 2022},
  \emph{Publication Series of the John von Neumann Institute for Computing
  (NIC), NIC Series}, vol.~51, ed. by M.~{M\"uller}, C.~{Peter}, A.~{Trautmann}
  (2022), \emph{Publication Series of the John von Neumann Institute for
  Computing (NIC), NIC Series}, vol.~51, pp. 159--172

\bibitem{Bedorfetal2012b}
J.~{B{\'e}dorf}, E.~{Gaburov}, S.~{Portegies Zwart}.
\newblock {Bonsai: N-body GPU tree-code}.
\newblock Astrophysics Source Code Library, record ascl:1212.001 (2012)

\bibitem{Chin1997}
S.A. {Chin}, Physics Letters A \textbf{226}(6), 344 (1997).
\newblock \doi{10.1016/S0375-9601(97)00003-0}

\bibitem{ChinChen2005}
S.A. {Chin}, C.R. {Chen}, Celestial Mechanics and Dynamical Astronomy
  \textbf{91}(3-4), 301 (2005).
\newblock \doi{10.1007/s10569-004-4622-z}

\bibitem{Chin2007}
S.A. {Chin}, \pre \textbf{75}(3), 036701 (2007).
\newblock \doi{10.1103/PhysRevE.75.036701}

\bibitem{HernquistOstriker1992}
L.~{Hernquist}, J.P. {Ostriker}, \apj \textbf{386}, 375 (1992).
\newblock \doi{10.1086/171025}

\bibitem{Meironetal2014}
Y.~{Meiron}, B.~{Li}, K.~{Holley-Bockelmann}, R.~{Spurzem}, \apj
  \textbf{792}(2), 98 (2014).
\newblock \doi{10.1088/0004-637X/792/2/98}

\bibitem{Avramovetal2021}
B.~{Avramov}, P.~{Berczik}, Y.~{Meiron}, A.~{Acharya}, A.~{Just}, \aap
  \textbf{649}, A41 (2021).
\newblock \doi{10.1051/0004-6361/202039698}

\bibitem{Churchetal2009}
R.P. {Church}, C.A. {Tout}, J.R. {Hurley}, PASA \textbf{26}(1), 92 (2009).
\newblock \doi{10.1071/AS08062}

\bibitem{Kippenhahnetal2012}
R.~{Kippenhahn}, A.~{Weigert}, A.~{Weiss}, \emph{{Stellar Structure and
  Evolution}} (Springer Berlin Heidelberg, 2012).
\newblock \doi{10.1007/978-3-642-30304-3}

\bibitem{SalarisCassini2017}
M.~{Salaris}, S.~{Cassisi}, Royal Society Open Science \textbf{4}(8), 170192
  (2017).
\newblock \doi{10.1098/rsos.170192}

\bibitem{Paxtonetal2011}
B.~{Paxton}, L.~{Bildsten}, A.~{Dotter}, F.~{Herwig}, P.~{Lesaffre}, et~al.,
  ApJS \textbf{192}(1), 3 (2011).
\newblock \doi{10.1088/0067-0049/192/1/3}

\bibitem{Paxtonetal2013}
B.~{Paxton}, M.~{Cantiello}, P.~{Arras}, L.~{Bildsten}, E.F. {Brown}, et~al.,
  ApJS \textbf{208}(1), 4 (2013).
\newblock \doi{10.1088/0067-0049/208/1/4}

\bibitem{Paxtonetal2015}
B.~{Paxton}, P.~{Marchant}, J.~{Schwab}, E.B. {Bauer}, L.~{Bildsten}, et~al.,
  ApJS \textbf{220}(1), 15 (2015).
\newblock \doi{10.1088/0067-0049/220/1/15}

\bibitem{Paxtonetal2016}
B.~{Paxton}, P.~{Marchant}, J.~{Schwab}, E.B. {Bauer}, L.~{Bildsten}, et~al.,
  ApJS \textbf{223}(1), 18 (2016).
\newblock \doi{10.3847/0067-0049/223/1/18}

\bibitem{Paxtonetal2018}
B.~{Paxton}, J.~{Schwab}, E.B. {Bauer}, L.~{Bildsten}, S.~{Blinnikov}, et~al.,
  ApJS \textbf{234}(2), 34 (2018).
\newblock \doi{10.3847/1538-4365/aaa5a8}

\bibitem{Paxtonetal2019}
B.~{Paxton}, R.~{Smolec}, J.~{Schwab}, A.~{Gautschy}, L.~{Bildsten}, et~al.,
  ApJS \textbf{243}(1), 10 (2019).
\newblock \doi{10.3847/1538-4365/ab2241}

\bibitem{Takahashietal2016}
K.~{Takahashi}, T.~{Yoshida}, H.~{Umeda}, K.~{Sumiyoshi}, S.~{Yamada}, MNRAS
  \textbf{456}(2), 1320 (2016).
\newblock \doi{10.1093/mnras/stv2649}

\bibitem{Takahashietal2018}
K.~{Takahashi}, T.~{Yoshida}, H.~{Umeda}, ApJ \textbf{857}(2), 111 (2018).
\newblock \doi{10.3847/1538-4357/aab95f}

\bibitem{Takahashietal2019}
K.~{Takahashi}, K.~{Sumiyoshi}, S.~{Yamada}, H.~{Umeda}, T.~{Yoshida}, ApJ
  \textbf{871}(2), 153 (2019).
\newblock \doi{10.3847/1538-4357/aaf8a8}

\bibitem{Yoshidaetal2019}
T.~{Yoshida}, T.~{Takiwaki}, K.~{Kotake}, K.~{Takahashi}, K.~{Nakamura},
  et~al., ApJ \textbf{881}(1), 16 (2019).
\newblock \doi{10.3847/1538-4357/ab2b9d}

\bibitem{Decin2020}
L.~{Decin}, arXiv e-prints arXiv:2011.13472 (2020)

\bibitem{Vink2021}
J.S. {Vink}, arXiv e-prints arXiv:2109.08164 (2021)

\bibitem{SanderVink2020}
A.A.C. {Sander}, J.S. {Vink}, MNRAS \textbf{499}(1), 873 (2020).
\newblock \doi{10.1093/mnras/staa2712}

\bibitem{Trabucchietal2019}
M.~{Trabucchi}, P.R. {Wood}, J.~{Montalb{\'a}n}, P.~{Marigo}, G.~{Pastorelli},
  L.~{Girardi}, MNRAS \textbf{482}(1), 929 (2019).
\newblock \doi{10.1093/mnras/sty2745}

\bibitem{Kamlahetal2022a}
A.W.H. {Kamlah}, A.~{Leveque}, R.~{Spurzem}, M.~{Arca Sedda}, A.~{Askar},
  S.~{Banerjee}, P.~{Berczik}, M.~{Giersz}, J.~{Hurley}, D.~{Belloni},
  L.~{K{\"u}hmichel}, L.~{Wang}, MNRAS \textbf{511}(3), 4060 (2022).
\newblock \doi{10.1093/mnras/stab3748}

\bibitem{Belczynskietal2016}
K.~{Belczynski}, A.~{Heger}, W.~{Gladysz}, A.J. {Ruiter}, S.~{Woosley}, et~al.,
  A\&A \textbf{594}, A97 (2016).
\newblock \doi{10.1051/0004-6361/201628980}

\bibitem{Leungetal2019c}
S.C. {Leung}, K.~{Nomoto}, S.~{Blinnikov}, ApJ \textbf{887}(1), 72 (2019).
\newblock \doi{10.3847/1538-4357/ab4fe5}

\bibitem{Leungetal2020b}
S.C. {Leung}, S.~{Blinnikov}, K.~{Ishidoshiro}, A.~{Kozlov}, K.~{Nomoto}, ApJ
  \textbf{889}(2), 75 (2020).
\newblock \doi{10.3847/1538-4357/ab6211}

\bibitem{Belczynskietal2010}
K.~{Belczynski}, T.~{Bulik}, C.L. {Fryer}, A.~{Ruiter}, F.~{Valsecchi}, et~al.,
  ApJ \textbf{714}(2), 1217 (2010).
\newblock \doi{10.1088/0004-637X/714/2/1217}

\bibitem{Fryeretal2012}
C.L. {Fryer}, K.~{Belczynski}, G.~{Wiktorowicz}, M.~{Dominik}, V.~{Kalogera},
  et~al., ApJ \textbf{749}(1), 91 (2012).
\newblock \doi{10.1088/0004-637X/749/1/91}

\bibitem{Willemsetal2005}
B.~{Willems}, M.~{Henninger}, T.~{Levin}, N.~{Ivanova}, V.~{Kalogera},
  K.~{McGhee}, F.X. {Timmes}, C.L. {Fryer}, ApJ \textbf{625}(1), 324 (2005).
\newblock \doi{10.1086/429557}

\bibitem{Fragosetal2009}
T.~{Fragos}, B.~{Willems}, V.~{Kalogera}, N.~{Ivanova}, G.~{Rockefeller}, C.L.
  {Fryer}, P.A. {Young}, ApJ \textbf{697}(2), 1057 (2009).
\newblock \doi{10.1088/0004-637X/697/2/1057}

\bibitem{Wongetal2012}
T.W. {Wong}, F.~{Valsecchi}, T.~{Fragos}, V.~{Kalogera}, ApJ \textbf{747}(2),
  111 (2012).
\newblock \doi{10.1088/0004-637X/747/2/111}

\bibitem{Wongetal2014}
T.W. {Wong}, F.~{Valsecchi}, A.~{Ansari}, T.~{Fragos}, E.~{Glebbeek},
  V.~{Kalogera}, J.~{McClintock}, ApJ \textbf{790}(2), 119 (2014).
\newblock \doi{10.1088/0004-637X/790/2/119}

\bibitem{HurleyShara2003}
J.R. {Hurley}, M.M. {Shara}, ApJ \textbf{589}(1), 179 (2003).
\newblock \doi{10.1086/374637}

\bibitem{Linares2018}
M.~{Linares}, in \emph{42nd COSPAR Scientific Assembly}, vol.~42 (2018),
  vol.~42, pp. E1.3--27--18

\bibitem{Linares2020}
M.~{Linares}, in \emph{Multifrequency Behaviour of High Energy Cosmic Sources -
  XIII. 3-8 June 2019. Palermo} (2020), p.~23

\bibitem{LattimerPrakash2004}
J.M. {Lattimer}, M.~{Prakash}, Sci \textbf{304}(5670), 536 (2004).
\newblock \doi{10.1126/science.1090720}

\bibitem{Lattimer2012}
J.M. {Lattimer}, ARNPS \textbf{62}(1), 485 (2012).
\newblock \doi{10.1146/annurev-nucl-102711-095018}

\bibitem{Nomoto1984}
K.~{Nomoto}, ApJ \textbf{277}, 791 (1984).
\newblock \doi{10.1086/161749}

\bibitem{Nomoto1987}
K.~{Nomoto}, ApJ \textbf{322}, 206 (1987).
\newblock \doi{10.1086/165716}

\bibitem{Podsiadlowskietal2004b}
P.~{Podsiadlowski}, N.~{Langer}, A.J.T. {Poelarends}, S.~{Rappaport},
  A.~{Heger}, et~al., ApJ \textbf{612}(2), 1044 (2004).
\newblock \doi{10.1086/421713}

\bibitem{Kieletal2008a}
P.D. {Kiel}, J.R. {Hurley}, M.~{Bailes}, J.R. {Murray}, MNRAS \textbf{388}(1),
  393 (2008).
\newblock \doi{10.1111/j.1365-2966.2008.13402.x}

\bibitem{Ivanovaetal2008}
N.~{Ivanova}, C.O. {Heinke}, F.A. {Rasio}, K.~{Belczynski}, J.M. {Fregeau},
  MNRAS \textbf{386}(1), 553 (2008).
\newblock \doi{10.1111/j.1365-2966.2008.13064.x}

\bibitem{Leungetal2020a}
S.C. {Leung}, K.~{Nomoto}, T.~{Suzuki}, ApJ \textbf{889}(1), 34 (2020).
\newblock \doi{10.3847/1538-4357/ab5d2f}

\bibitem{Belczynskietal2008}
K.~{Belczynski}, V.~{Kalogera}, F.A. {Rasio}, R.E. {Taam}, A.~{Zezas}, et~al.,
  ApJS \textbf{174}(1), 223 (2008).
\newblock \doi{10.1086/521026}

\bibitem{EldridgeTout2004b}
J.J. {Eldridge}, C.A. {Tout}, MNRAS \textbf{353}(1), 87 (2004).
\newblock \doi{10.1111/j.1365-2966.2004.08041.x}

\bibitem{Fryeretal2001}
C.L. {Fryer}, S.E. {Woosley}, A.~{Heger}, ApJ \textbf{550}(1), 372 (2001).
\newblock \doi{10.1086/319719}

\bibitem{Yoshidaetal2016a}
T.~{Yoshida}, H.~{Umeda}, K.~{Maeda}, T.~{Ishii}, MNRAS \textbf{457}(1), 351
  (2016).
\newblock \doi{10.1093/mnras/stv3002}

\bibitem{SperaMapelli2017}
M.~{Spera}, M.~{Mapelli}, MNRAS \textbf{470}(4), 4739 (2017).
\newblock \doi{10.1093/mnras/stx1576}

\bibitem{Woosley2017}
S.E. {Woosley}, ApJ \textbf{836}(2), 244 (2017).
\newblock \doi{10.3847/1538-4357/836/2/244}

\bibitem{WoosleyHeger2021}
S.E. {Woosley}, A.~{Heger}, ApJL \textbf{912}(2), L31 (2021).
\newblock \doi{10.3847/2041-8213/abf2c4}

\bibitem{Kuepperetal2011a}
A.H.W. {Kuepper}, T.~{Maschberger}, P.~{Kroupa}, H.~{Baumgardt}.
\newblock {McLuster: A Tool to Make a Star Cluster} (2011)

\bibitem{Levequeetal2022a}
A.~{Leveque}, M.~{Giersz}, S.~{Banerjee}, E.~{Vesperini}, J.~{Hong},
  S.~{Portegies Zwart}, \mnras \textbf{514}(4), 5739 (2022).
\newblock \doi{10.1093/mnras/stac1690}

\bibitem{Wangetal2022}
L.~{Wang}, A.~{Tanikawa}, M.S. {Fujii}, \mnras \textbf{509}(4), 4713 (2022).
\newblock \doi{10.1093/mnras/stab3255}

\bibitem{Hobbsetal2005}
G.~{Hobbs}, D.R. {Lorimer}, A.G. {Lyne}, M.~{Kramer}, MNRAS \textbf{360}(3),
  974 (2005).
\newblock \doi{10.1111/j.1365-2966.2005.09087.x}

\bibitem{HansenPhinney1997}
B.M.S. {Hansen}, E.S. {Phinney}, MNRAS \textbf{291}(3), 569 (1997).
\newblock \doi{10.1093/mnras/291.3.569}

\bibitem{Fellhaueretal2003}
M.~{Fellhauer}, D.N.C. {Lin}, M.~{Bolte}, S.J. {Aarseth}, K.A. {Williams}, ApJL
  \textbf{595}(1), L53 (2003).
\newblock \doi{10.1086/379005}

\bibitem{GessnerJanka2018}
A.~{Gessner}, H.T. {Janka}, ApJ \textbf{865}(1), 61 (2018).
\newblock \doi{10.3847/1538-4357/aadbae}

\bibitem{Clark1975}
G.W. {Clark}, ApJL \textbf{199}, L143 (1975).
\newblock \doi{10.1086/181869}

\bibitem{Abbottetal2017a}
B.P. {Abbott}, R.~{Abbott}, T.D. {Abbott}, F.~{Acernese}, K.~{Ackley}, et~al.,
  PhRvL \textbf{119}(16), 161101 (2017).
\newblock \doi{10.1103/PhysRevLett.119.161101}

\bibitem{Abbottetal2020a}
R.~{Abbott}, T.D. {Abbott}, S.~{Abraham}, F.~{Acernese}, K.~{Ackley}, et~al.,
  ApJL \textbf{896}(2), L44 (2020).
\newblock \doi{10.3847/2041-8213/ab960f}

\bibitem{Manchesteretal2005}
R.N. {Manchester}, G.B. {Hobbs}, A.~{Teoh}, M.~{Hobbs}, AJ \textbf{129}(4),
  1993 (2005).
\newblock \doi{10.1086/428488}

\bibitem{Banerjee2021a}
S.~{Banerjee}, MNRAS \textbf{500}(3), 3002 (2021).
\newblock \doi{10.1093/mnras/staa2392}

\bibitem{Schecketal2004}
L.~{Scheck}, T.~{Plewa}, H.T. {Janka}, K.~{Kifonidis}, E.~{M{\"u}ller}, PhRvL
  \textbf{92}(1), 011103 (2004).
\newblock \doi{10.1103/PhysRevLett.92.011103}

\bibitem{FryerYoung2007}
C.L. {Fryer}, P.A. {Young}, ApJ \textbf{659}(2), 1438 (2007).
\newblock \doi{10.1086/513003}

\bibitem{Schecketal2008}
L.~{Scheck}, H.T. {Janka}, T.~{Foglizzo}, K.~{Kifonidis}, A\&A \textbf{477}(3),
  931 (2008).
\newblock \doi{10.1051/0004-6361:20077701}

\bibitem{BurrowsHayes1996}
A.~{Burrows}, J.~{Hayes}, PhRvL \textbf{76}(3), 352 (1996).
\newblock \doi{10.1103/PhysRevLett.76.352}

\bibitem{Fryer2004}
C.L. {Fryer}, ApJ \textbf{601}(2), L175 (2004).
\newblock \doi{10.1086/382044}

\bibitem{MeakinArnett2006}
C.A. {Meakin}, D.~{Arnett}, ApJL \textbf{637}(1), L53 (2006).
\newblock \doi{10.1086/500544}

\bibitem{MeakinArnett2007}
C.A. {Meakin}, D.~{Arnett}, ApJ \textbf{665}(1), 690 (2007).
\newblock \doi{10.1086/519372}

\bibitem{Fulleretal2003}
G.M. {Fuller}, A.~{Kusenko}, I.~{Mocioiu}, S.~{Pascoli}, PhRvD \textbf{68}(10),
  103002 (2003).
\newblock \doi{10.1103/PhysRevD.68.103002}

\bibitem{FryerKusenko2006}
C.L. {Fryer}, A.~{Kusenko}, ApJS \textbf{163}(2), 335 (2006).
\newblock \doi{10.1086/500933}

\bibitem{Morawskietal2019}
J.~{Morawski}, M.~{Giersz}, A.~{Askar}, K.~{Belczynski}, MNRAS \textbf{486}(3),
  3402 (2019).
\newblock \doi{10.1093/mnras/stz1074}

\bibitem{Kerr1963}
R.P. {Kerr}, PhRvL \textbf{11}(5), 237 (1963).
\newblock \doi{10.1103/PhysRevLett.11.237}

\bibitem{Belczynskietal2020}
K.~{Belczynski}, J.~{Klencki}, C.E. {Fields}, A.~{Olejak}, E.~{Berti}, et~al.,
  A\&A \textbf{636}, A104 (2020).
\newblock \doi{10.1051/0004-6361/201936528}

\bibitem{Kroupa2001}
P.~{Kroupa}, MNRAS \textbf{322}(2), 231 (2001).
\newblock \doi{10.1046/j.1365-8711.2001.04022.x}

\bibitem{Spruit2002}
H.C. {Spruit}, AAP \textbf{381}, 923 (2002).
\newblock \doi{10.1051/0004-6361:20011465}

\bibitem{FullerMa2019}
J.~{Fuller}, L.~{Ma}, ApJL \textbf{881}(1), L1 (2019).
\newblock \doi{10.3847/2041-8213/ab339b}

\bibitem{Fulleretal2019}
J.~{Fuller}, A.L. {Piro}, A.S. {Jermyn}, MNRAS \textbf{485}(3), 3661 (2019).
\newblock \doi{10.1093/mnras/stz514}

\bibitem{Eggenbergeretal2008}
P.~{Eggenberger}, G.~{Meynet}, A.~{Maeder}, R.~{Hirschi}, C.~{Charbonnel},
  et~al., Ap\&SS \textbf{316}(1-4), 43 (2008).
\newblock \doi{10.1007/s10509-007-9511-y}

\bibitem{Ekstroem2012}
S.~{Ekstr{\"o}m}, C.~{Georgy}, P.~{Eggenberger}, G.~{Meynet}, N.~{Mowlavi},
  et~al., A\&A \textbf{537}, A146 (2012).
\newblock \doi{10.1051/0004-6361/201117751}

\bibitem{Eggleton1996}
P.P. {Eggleton}, in \emph{Dynamical Evolution of Star Clusters: Confrontation
  of Theory and Observations}, vol. 174, ed. by P.~{Hut}, J.~{Makino} (1996),
  vol. 174, p. 213

\bibitem{Eggleton2006}
P.~{Eggleton}, \emph{{Evolutionary Processes in Binary and Multiple Stars}}
  (Cambridge University Press, 2006)

\bibitem{Tout2008b}
C.A. {Tout}, in \emph{The Cambridge N-Body Lectures}, vol. 760, ed. by S.J.
  {Aarseth}, C.A. {Tout}, R.A. {Mardling} (Springer-Verlag Berlin Heidelberg,
  2008), p. 297.
\newblock \doi{10.1007/978-1-4020-8431-7_11}

\bibitem{BondiHoyle1944}
H.~{Bondi}, F.~{Hoyle}, MNRAS \textbf{104}, 273 (1944).
\newblock \doi{10.1093/mnras/104.5.273}

\bibitem{Lurieetal2017}
J.C. {Lurie}, K.~{Vyhmeister}, S.L. {Hawley}, J.~{Adilia}, A.~{Chen}, J.R.A.
  {Davenport}, M.~{Juri{\'c}}, M.~{Puig-Holzman}, K.L. {Weisenburger}, AJ
  \textbf{154}(6), 250 (2017).
\newblock \doi{10.3847/1538-3881/aa974d}

\bibitem{Mazeh2008}
T.~{Mazeh}, in \emph{EAS Publications Series}, \emph{EAS Publications Series},
  vol.~29, ed. by M.J. {Goupil}, J.P. {Zahn} (2008), \emph{EAS Publications
  Series}, vol.~29, pp. 1--65.
\newblock \doi{10.1051/eas:0829001}

\bibitem{MeibomMathieu2005}
S.~{Meibom}, R.D. {Mathieu}, ApJ \textbf{620}(2), 970 (2005).
\newblock \doi{10.1086/427082}

\bibitem{Zahn1977}
J.P. {Zahn}, A\&A \textbf{500}, 121 (1977)

\bibitem{Hut1981}
P.~{Hut}, A\&A \textbf{99}, 126 (1981)

\bibitem{Hut1980}
P.~{Hut}, A\&A \textbf{92}(1-2), 167 (1980)

\bibitem{Rasioetal1996}
F.A. {Rasio}, C.A. {Tout}, S.H. {Lubow}, M.~{Livio}, ApJ \textbf{470}, 1187
  (1996).
\newblock \doi{10.1086/177941}

\bibitem{Eggletonetal1998}
P.P. {Eggleton}, L.G. {Kiseleva}, P.~{Hut}, ApJ \textbf{499}(2), 853 (1998).
\newblock \doi{10.1086/305670}

\bibitem{Zahn1970}
J.P. {Zahn}, A\&A \textbf{4}, 452 (1970)

\bibitem{Zahn1974}
J.P. {Zahn}, in \emph{Stellar Instability and Evolution}, vol.~59, ed. by
  P.~{Ledoux}, A.~{Noels}, A.W. {Rodgers} (1974), vol.~59, p. 185

\bibitem{Zahn1975}
J.P. {Zahn}, A\&A \textbf{41}, 329 (1975)

\bibitem{Siessetal2013}
L.~{Siess}, R.G. {Izzard}, P.J. {Davis}, R.~{Deschamps}, A\&A \textbf{550},
  A100 (2013).
\newblock \doi{10.1051/0004-6361/201220327}

\bibitem{Toutetal2008}
C.A. {Tout}, D.T. {Wickramasinghe}, J.~{Liebert}, L.~{Ferrario}, J.E.
  {Pringle}, MNRAS \textbf{387}(2), 897 (2008).
\newblock \doi{10.1111/j.1365-2966.2008.13291.x}

\bibitem{Kopal1978}
Z.~{Kopal}, \emph{{Dynamics of close binary systems}} (Dordrecht: Reidel,
  1978).
\newblock \doi{10.1007/978-94-009-9780-6}

\bibitem{Zahn1989}
J.P. {Zahn}, A\&A \textbf{220}(1-2), 112 (1989)

\bibitem{Zahn1991}
J.P. {Zahn}, A\&A \textbf{252}, 179 (1991)

\bibitem{Zahn1992}
J.P. {Zahn}, A\&A \textbf{265}, 115 (1992)

\bibitem{Campbell1984}
C.G. {Campbell}, MNRAS \textbf{207}, 433 (1984).
\newblock \doi{10.1093/mnras/207.3.433}

\bibitem{Eggleton1983a}
P.P. {Eggleton}, ApJ \textbf{268}, 368 (1983).
\newblock \doi{10.1086/160960}

\bibitem{Webbink1985b}
R.F. {Webbink}, in \emph{Interacting Binary Stars}, ed. by J.E. {Pringle}, R.A.
  {Wade} (Cambridge University Press, 1985), p.~39

\bibitem{Webbink2003}
R.F. {Webbink}, in \emph{3D Stellar Evolution}, \emph{Astronomical Society of
  the Pacific Conference Series}, vol. 293, ed. by S.~{Turcotte}, S.C.
  {Keller}, R.M. {Cavallo} (2003), \emph{Astronomical Society of the Pacific
  Conference Series}, vol. 293, p.~76

\bibitem{Paczynski1976}
B.~{Paczynski}, in \emph{Structure and Evolution of Close Binary Systems},
  vol.~73, ed. by P.~{Eggleton}, S.~{Mitton}, J.~{Whelan} (1976), vol.~73,
  p.~75

\bibitem{Ivanovaetal2013}
N.~{Ivanova}, S.~{Justham}, X.~{Chen}, O.~{De Marco}, C.L. {Fryer}, et~al.,
  A\&ARv \textbf{21}, 59 (2013).
\newblock \doi{10.1007/s00159-013-0059-2}

\bibitem{Ivanova2019}
N.~{Ivanova}, in \emph{KITP Conference: Merging Visions: Exploring
  Compact-Object Binaries with Gravity and Light} (2019), p.~7

\bibitem{Ivanova2016}
N.~{Ivanova}, in \emph{Star Clusters and Black Holes in Galaxies across Cosmic
  Time}, vol. 312, ed. by Y.~{Meiron}, S.~{Li}, F.K. {Liu}, R.~{Spurzem}
  (2016), vol. 312, pp. 203--212.
\newblock \doi{10.1017/S1743921315007826}

\bibitem{Ivanova2018}
N.~{Ivanova}, ApJL \textbf{858}(2), L24 (2018).
\newblock \doi{10.3847/2041-8213/aac101}

\bibitem{Olejaketal2021}
A.~{Olejak}, K.~{Belczynski}, N.~{Ivanova}, \aap \textbf{651}, A100 (2021).
\newblock \doi{10.1051/0004-6361/202140520}

\bibitem{Webbink1984}
R.F. {Webbink}, ApJ \textbf{277}, 355 (1984).
\newblock \doi{10.1086/161701}

\bibitem{Fragos2019}
T.~Fragos, J.J. Andrews, E.~Ramirez-Ruiz, G.~Meynet, V.~Kalogera, R.E. Taam,
  A.~Zezas, The Astrophysical Journal \textbf{883}(2), L45 (2019).
\newblock \doi{10.3847/2041-8213/ab40d1}.
\newblock \urlprefix\url{http://dx.doi.org/10.3847/2041-8213/ab40d1}

\bibitem{DewiTauris2000}
J.D.M. {Dewi}, T.M. {Tauris}, A\&A \textbf{360}, 1043 (2000)

\bibitem{Claeysetal2014}
J.S.W. {Claeys}, O.R. {Pols}, R.G. {Izzard}, J.~{Vink}, F.W.M. {Verbunt}, A\&A
  \textbf{563}, A83 (2014).
\newblock \doi{10.1051/0004-6361/201322714}

\bibitem{IvanovaNandez2016}
N.~{Ivanova}, J.L.A. {Nandez}, MNRAS \textbf{462}(1), 362 (2016).
\newblock \doi{10.1093/mnras/stw1676}

\bibitem{Ivanovaetal2020}
N.~Ivanova, S.~Justham, P.~Ricker, \emph{Common Envelope Evolution}.
\newblock 2514-3433 (IOP Publishing, 2020).
\newblock \doi{10.1088/2514-3433/abb6f0}.
\newblock \urlprefix\url{http://dx.doi.org/10.1088/2514-3433/abb6f0}

\bibitem{Eldrigeetal2017}
J.J. {Eldridge}, E.R. {Stanway}, L.~{Xiao}, L.A.S. {McClelland}, G.~{Taylor},
  M.~{Ng}, S.M.L. {Greis}, J.C. {Bray}, PASA \textbf{34}, e058 (2017).
\newblock \doi{10.1017/pasa.2017.51}

\bibitem{Mapelli2018b}
M.~{Mapelli}, arXiv arXiv:1809.09130 (2018)

\bibitem{Breiviketal2020a}
K.~{Breivik}, S.~{Coughlin}, M.~{Zevin}, C.L. {Rodriguez}, K.~{Kremer}, et~al.,
  ApJ \textbf{898}(1), 71 (2020).
\newblock \doi{10.3847/1538-4357/ab9d85}

\bibitem{Tranietal2022}
A.A. {Trani}, S.~{Rieder}, A.~{Tanikawa}, G.~{Iorio}, R.~{Martini},
  G.~{Karelin}, H.~{Glanz}, S.~{Portegies Zwart}, \prd \textbf{106}(4), 043014
  (2022).
\newblock \doi{10.1103/PhysRevD.106.043014}

\bibitem{Kruckowetal2021}
M.U. {Kruckow}, P.G. {Neunteufel}, R.~{Di Stefano}, Y.~{Gao}, C.~{Kobayashi},
  ApJ \textbf{920}(2), 86 (2021).
\newblock \doi{10.3847/1538-4357/ac13ac}

\bibitem{Taurisetal2013b}
T.M. {Tauris}, N.~{Langer}, T.J. {Moriya}, P.~{Podsiadlowski}, S.C. {Yoon},
  et~al., ApJL \textbf{778}(2), L23 (2013).
\newblock \doi{10.1088/2041-8205/778/2/L23}

\bibitem{Tauris2015}
T.M. {Tauris}, MNRAS \textbf{448}, L6 (2015).
\newblock \doi{10.1093/mnrasl/slu189}

\bibitem{Taurisetal2017}
T.M. {Tauris}, M.~{Kramer}, P.C.C. {Freire}, N.~{Wex}, H.T. {Janka}, et~al.,
  ApJ \textbf{846}(2), 170 (2017).
\newblock \doi{10.3847/1538-4357/aa7e89}

\bibitem{Schneideretal2021}
F.R.N. {Schneider}, P.~{Podsiadlowski}, B.~{M{\"u}ller}, A\&A \textbf{645}, A5
  (2021).
\newblock \doi{10.1051/0004-6361/202039219}

\bibitem{Belczynskietal2017}
K.~{Belczynski}, T.~{Ryu}, R.~{Perna}, E.~{Berti}, T.L. {Tanaka}, et~al., MNRAS
  \textbf{471}(4), 4702 (2017).
\newblock \doi{10.1093/mnras/stx1759}

\bibitem{ThorneZytkow1977}
K.S. {Thorne}, A.N. {{\.Z}ytkow}, ApJ \textbf{212}, 832 (1977).
\newblock \doi{10.1086/155109}

\bibitem{Olejaketal2020a}
A.~{Olejak}, K.~{Belczynski}, T.~{Bulik}, M.~{Sobolewska}, A\&A \textbf{638},
  A94 (2020).
\newblock \doi{10.1051/0004-6361/201936557}

\bibitem{Costaetal2022}
G.~{Costa}, A.~{Ballone}, M.~{Mapelli}, A.~{Bressan}, \mnras \textbf{516}(1),
  1072 (2022).
\newblock \doi{10.1093/mnras/stac2222}

\bibitem{Balloneetal2023}
A.~{Ballone}, G.~{Costa}, M.~{Mapelli}, M.~{MacLeod}, S.~{Torniamenti}, J.M.
  {Pacheco-Arias}, \mnras \textbf{519}(4), 5191 (2023).
\newblock \doi{10.1093/mnras/stac3752}

\bibitem{Loustoetal2012}
C.O. {Lousto}, Y.~{Zlochower}, M.~{Dotti}, M.~{Volonteri}, PhRvD
  \textbf{85}(8), 084015 (2012).
\newblock \doi{10.1103/PhysRevD.85.084015}

\bibitem{Schoedel2014b}
R.~{Sch{\"o}del}, A.~{Feldmeier}, N.~{Neumayer}, L.~{Meyer}, S.~{Yelda}, CQGra
  \textbf{31}(24), 244007 (2014).
\newblock \doi{10.1088/0264-9381/31/24/244007}

\bibitem{Bakeretal2007b}
J.G. {Baker}, W.D. {Boggs}, J.~{Centrella}, B.J. {Kelly}, S.T. {McWilliams},
  et~al., ApJ \textbf{668}(2), 1140 (2007).
\newblock \doi{10.1086/521330}

\bibitem{Bakeretal2008}
J.G. {Baker}, W.D. {Boggs}, J.~{Centrella}, B.J. {Kelly}, S.T. {McWilliams},
  et~al., PhRvD \textbf{78}(4), 044046 (2008).
\newblock \doi{10.1103/PhysRevD.78.044046}

\bibitem{PortegiesZwartetal2010}
S.F. {Portegies Zwart}, S.L.W. {McMillan}, M.~{Gieles}, ARA\&A \textbf{48}, 431
  (2010).
\newblock \doi{10.1146/annurev-astro-081309-130834}

\bibitem{Baumgardtetal2018}
H.~{Baumgardt}, M.~{Hilker}, MNRAS \textbf{478}(2), 1520 (2018).
\newblock \doi{10.1093/mnras/sty1057}

\bibitem{vanMeteretal2010b}
J.R. {van Meter}, M.C. {Miller}, J.G. {Baker}, W.D. {Boggs}, B.J. {Kelly}, ApJ
  \textbf{719}(2), 1427 (2010).
\newblock \doi{10.1088/0004-637X/719/2/1427}

\bibitem{HoffmanLoeb2007}
L.~{Hoffman}, A.~{Loeb}, MNRAS \textbf{377}(3), 957 (2007).
\newblock \doi{10.1111/j.1365-2966.2007.11694.x}

\bibitem{JimenezFortezaetal2017}
X.~{Jim{\'e}nez-Forteza}, D.~{Keitel}, S.~{Husa}, M.~{Hannam}, S.~{Khan},
  et~al., PhRvD \textbf{95}(6), 064024 (2017).
\newblock \doi{10.1103/PhysRevD.95.064024}

\bibitem{ArcaSedda2020}
M.~{Arca Sedda}, CmPhy \textbf{3}(1), 43 (2020).
\newblock \doi{10.1038/s42005-020-0310-x}

\bibitem{Chattopadhyayetal2021}
D.~{Chattopadhyay}, S.~{Stevenson}, J.R. {Hurley}, M.~{Bailes},
  F.~{Broekgaarden}, MNRAS \textbf{504}(3), 3682 (2021).
\newblock \doi{10.1093/mnras/stab973}

\bibitem{Banerjee2022}
S.~{Banerjee}, \aap \textbf{665}, A20 (2022).
\newblock \doi{10.1051/0004-6361/202142331}

\bibitem{DiCarloetal2019}
U.N. {Di Carlo}, N.~{Giacobbo}, M.~{Mapelli}, M.~{Pasquato}, M.~{Spera},
  et~al., MNRAS \textbf{487}(2), 2947 (2019).
\newblock \doi{10.1093/mnras/stz1453}

\bibitem{DiCarloetal2020a}
U.N. {Di Carlo}, M.~{Mapelli}, Y.~{Bouffanais}, N.~{Giacobbo},
  F.~{Santoliquido}, et~al., MNRAS \textbf{497}(1), 1043 (2020).
\newblock \doi{10.1093/mnras/staa1997}

\bibitem{DiCarloetal2020b}
U.N. {Di Carlo}, M.~{Mapelli}, N.~{Giacobbo}, M.~{Spera}, Y.~{Bouffanais},
  et~al., MNRAS \textbf{498}(1), 495 (2020).
\newblock \doi{10.1093/mnras/staa2286}

\bibitem{Campanellietal2007}
M.~{Campanelli}, C.~{Lousto}, Y.~{Zlochower}, D.~{Merritt}, ApJL
  \textbf{659}(1), L5 (2007).
\newblock \doi{10.1086/516712}

\bibitem{Rezollaetal2008}
L.~{Rezzolla}, E.~{Barausse}, E.N. {Dorband}, D.~{Pollney}, C.~{Reisswig},
  et~al., PhRvD \textbf{78}(4), 044002 (2008).
\newblock \doi{10.1103/PhysRevD.78.044002}

\bibitem{Hughes2009}
S.A. {Hughes}, ARA\&A \textbf{47}(1), 107 (2009).
\newblock \doi{10.1146/annurev-astro-082708-101711}

\bibitem{BelczynskiBanerjee2020}
K.~{Belczynski}, S.~{Banerjee}, A\&A \textbf{640}, L20 (2020).
\newblock \doi{10.1051/0004-6361/202038427}

\bibitem{SaioNomoto2004}
H.~{Saio}, K.~{Nomoto}, ApJ \textbf{615}(1), 444 (2004).
\newblock \doi{10.1086/423976}

\bibitem{SaioNomoto1985}
H.~{Saio}, K.~{Nomoto}, A\&A \textbf{150}(1), L21 (1985)

\bibitem{SaioNomoto1998}
H.~{Saio}, K.~{Nomoto}, ApJ \textbf{500}(1), 388 (1998).
\newblock \doi{10.1086/305696}

\bibitem{Ruiteretal2019}
A.J. Ruiter, L.~Ferrario, K.~Belczynski, I.R. Seitenzahl, R.M. Crocker, A.I.
  Karakas, Monthly Notices of the Royal Astronomical Society \textbf{484}(1),
  698–711 (2019).
\newblock \doi{10.1093/mnras/stz001}.
\newblock \urlprefix\url{http://dx.doi.org/10.1093/mnras/stz001}

\bibitem{PetersMathews1963}
P.C. {Peters}, J.~{Mathews}, PhRv \textbf{131}(1), 435 (1963).
\newblock \doi{10.1103/PhysRev.131.435}

\bibitem{Peters1964}
P.C. {Peters}, PhRv \textbf{136}(4B), 1224 (1964).
\newblock \doi{10.1103/PhysRev.136.B1224}

\bibitem{Mestel1968a}
L.~{Mestel}, MNRAS \textbf{138}, 359 (1968).
\newblock \doi{10.1093/mnras/138.3.359}

\bibitem{Mestel1968b}
L.~{Mestel}, MNRAS \textbf{140}, 177 (1968).
\newblock \doi{10.1093/mnras/140.2.177}

\bibitem{MestelSpruit1987}
L.~{Mestel}, H.C. {Spruit}, MNRAS \textbf{226}, 57 (1987).
\newblock \doi{10.1093/mnras/226.1.57}

\bibitem{Schreiberetal2016}
M.R. {Schreiber}, M.~{Zorotovic}, T.P.G. {Wijnen}, MNRAS \textbf{455}(1), L16
  (2016).
\newblock \doi{10.1093/mnrasl/slv144}

\bibitem{Zorotovicetal2016}
M.~{Zorotovic}, M.R. {Schreiber}, S.G. {Parsons}, B.T. {G{\"a}nsicke},
  A.~{Hardy}, et~al., MNRAS \textbf{457}(4), 3867 (2016).
\newblock \doi{10.1093/mnras/stw246}

\bibitem{Bellonietal2018b}
D.~{Belloni}, P.~{Kroupa}, H.J. {Rocha-Pinto}, M.~{Giersz}, MNRAS
  \textbf{474}(3), 3740 (2018).
\newblock \doi{10.1093/mnras/stx3034}

\bibitem{KielHurley2006}
P.D. {Kiel}, J.R. {Hurley}, MNRAS \textbf{369}(3), 1152 (2006).
\newblock \doi{10.1111/j.1365-2966.2006.10400.x}

\bibitem{KielHurley2009}
P.D. {Kiel}, J.R. {Hurley}, MNRAS \textbf{395}(4), 2326 (2009).
\newblock \doi{10.1111/j.1365-2966.2009.14711.x}

\bibitem{Polsetal1998}
O.R. {Pols}, K.P. {Schr{\"o}der}, J.R. {Hurley}, C.A. {Tout}, P.P. {Eggleton},
  MNRAS \textbf{298}(2), 525 (1998).
\newblock \doi{10.1046/j.1365-8711.1998.01658.x}

\bibitem{Agrawaletal2020}
P.~{Agrawal}, J.~{Hurley}, S.~{Stevenson}, D.~{Sz{\'e}csi}, C.~{Flynn}, MNRAS
  \textbf{497}(4), 4549 (2020).
\newblock \doi{10.1093/mnras/staa2264}

\bibitem{MaederMeynet1989}
A.~{Maeder}, G.~{Meynet}, AAP \textbf{210}, 155 (1989)

\bibitem{Schalleretal1992}
G.~{Schaller}, D.~{Schaerer}, G.~{Meynet}, A.~{Maeder}, A\&AS \textbf{96}, 269
  (1992)

\bibitem{Alongietal1993}
M.~{Alongi}, G.~{Bertelli}, A.~{Bressan}, C.~{Chiosi}, F.~{Fagotto},
  L.~{Greggio}, E.~{Nasi}, A\&AS \textbf{97}, 851 (1993)

\bibitem{Bressanetal1993}
A.~{Bressan}, F.~{Fagotto}, G.~{Bertelli}, C.~{Chiosi}, A\&AS \textbf{100}, 647
  (1993)

\bibitem{Fagottoetal1994a}
F.~{Fagotto}, A.~{Bressan}, G.~{Bertelli}, C.~{Chiosi}, A\&AS \textbf{104}, 365
  (1994)

\bibitem{Fagottoetal1994b}
F.~{Fagotto}, A.~{Bressan}, G.~{Bertelli}, C.~{Chiosi}, A\&AS \textbf{105}, 29
  (1994)

\bibitem{Claret1995}
A.~{Claret}, A\&AS \textbf{109}, 441 (1995)

\bibitem{ClaretGimenez1995}
A.~{Claret}, A.~{Gimenez}, A\&AS \textbf{114}, 549 (1995)

\bibitem{Biermann1932}
L.~{Biermann}, ZA \textbf{5}, 117 (1932)

\bibitem{Gabrieletal2014}
M.~{Gabriel}, A.~{Noels}, J.~{Montalb{\'a}n}, A.~{Miglio}, A\&A \textbf{569},
  A63 (2014).
\newblock \doi{10.1051/0004-6361/201423442}

\bibitem{JoyceChaboyer2018}
M.~{Joyce}, B.~{Chaboyer}, ApJ \textbf{856}(1), 10 (2018).
\newblock \doi{10.3847/1538-4357/aab200}

\bibitem{Schroederetal1997}
K.P. {Schroder}, O.R. {Pols}, P.P. {Eggleton}, MNRAS \textbf{285}(4), 696
  (1997).
\newblock \doi{10.1093/mnras/285.4.696}

\bibitem{Polsetal1997}
O.R. {Pols}, C.A. {Tout}, K.P. {Schroder}, P.P. {Eggleton}, J.~{Manners}, MNRAS
  \textbf{289}(4), 869 (1997).
\newblock \doi{10.1093/mnras/289.4.869}

\bibitem{BoehmVitense1958}
E.~{B{\"o}hm-Vitense}, ZA \textbf{46}, 108 (1958)

\bibitem{Pasettoetal2018}
S.~Pasetto, C.~Chiosi, M.~Cropper, E.~Grebel, Journal of Physics: Conference
  Series \textbf{940}, 012020 (2018).
\newblock \doi{10.1088/1742-6596/940/1/012020}

\bibitem{Speraetal2015}
M.~{Spera}, M.~{Mapelli}, A.~{Bressan}, MNRAS \textbf{451}(4), 4086 (2015).
\newblock \doi{10.1093/mnras/stv1161}

\bibitem{Speraetal2019}
M.~{Spera}, M.~{Mapelli}, N.~{Giacobbo}, A.A. {Trani}, A.~{Bressan}, et~al.,
  MNRAS \textbf{485}(1), 889 (2019).
\newblock \doi{10.1093/mnras/stz359}

\bibitem{Siciliaetal2022}
A.~{Sicilia}, A.~{Lapi}, L.~{Boco}, M.~{Spera}, U.N. {Di Carlo}, M.~{Mapelli},
  F.~{Shankar}, D.M. {Alexander}, A.~{Bressan}, L.~{Danese}, \apj
  \textbf{924}(2), 56 (2022).
\newblock \doi{10.3847/1538-4357/ac34fb}

\bibitem{Kruckowetal2018}
M.U. {Kruckow}, T.M. {Tauris}, N.~{Langer}, M.~{Kramer}, R.G. {Izzard}, MNRAS
  \textbf{481}(2), 1908 (2018).
\newblock \doi{10.1093/mnras/sty2190}

\bibitem{Kruckow2020}
M.U. {Kruckow}, A\&A \textbf{639}, A123 (2020).
\newblock \doi{10.1051/0004-6361/202037519}

\bibitem{Eggleton1971}
P.P. {Eggleton}, MNRAS \textbf{151}, 351 (1971).
\newblock \doi{10.1093/mnras/151.3.351}

\bibitem{Eggleton1972}
P.P. {Eggleton}, MNRAS \textbf{156}, 361 (1972).
\newblock \doi{10.1093/mnras/w156.3.361}

\bibitem{Eggletonetal1973a}
P.P. {Eggleton}, MNRAS \textbf{163}, 279 (1973).
\newblock \doi{10.1093/mnras/163.3.279}

\bibitem{Eggletonetal1973b}
P.P. {Eggleton}, J.~{Faulkner}, B.P. {Flannery}, A\&A \textbf{23}, 325 (1973)

\bibitem{Pols1995}
O.R. {Pols}, C.A. {Tout}, P.P. {Eggleton}, Z.~{Han}, MNRAS \textbf{274}(3), 964
  (1995).
\newblock \doi{10.1093/mnras/274.3.964}

\bibitem{Yoonetal2006}
S.C. {Yoon}, N.~{Langer}, C.~{Norman}, Nuovo Cimento B Serie \textbf{121}(12),
  1631 (2006).
\newblock \doi{10.1393/ncb/i2007-10344-4}

\bibitem{Yoonetal2012}
S.C. {Yoon}, A.~{Dierks}, N.~{Langer}, AAP \textbf{542}, A113 (2012).
\newblock \doi{10.1051/0004-6361/201117769}

\bibitem{Brottetal2011}
I.~{Brott}, C.J. {Evans}, I.~{Hunter}, A.~{de Koter}, N.~{Langer}, P.L.
  {Dufton}, M.~{Cantiello}, C.~{Trundle}, D.J. {Lennon}, S.E. {de Mink}, S.C.
  {Yoon}, P.~{Anders}, AAP \textbf{530}, A116 (2011).
\newblock \doi{10.1051/0004-6361/201016114}

\bibitem{Koehleretal2015}
K.~{K{\"o}hler}, N.~{Langer}, A.~{de Koter}, S.E. {de Mink}, P.A. {Crowther},
  C.J. {Evans}, G.~{Gr{\"a}fener}, H.~{Sana}, D.~{Sanyal}, F.R.N. {Schneider},
  J.S. {Vink}, AAP \textbf{573}, A71 (2015).
\newblock \doi{10.1051/0004-6361/201424356}

\bibitem{Szecsietal2015}
D.~{Sz{\'e}csi}, N.~{Langer}, S.C. {Yoon}, D.~{Sanyal}, S.~{de Mink}, C.J.
  {Evans}, T.~{Dermine}, AAP \textbf{581}, A15 (2015).
\newblock \doi{10.1051/0004-6361/201526617}

\bibitem{Szecsietal2022}
D.~{Sz{\'e}csi}, P.~{Agrawal}, R.~{W{\"u}nsch}, N.~{Langer}, \aap \textbf{658},
  A125 (2022).
\newblock \doi{10.1051/0004-6361/202141536}

\bibitem{Aarseth1996}
S.J. {Aarseth}, in \emph{The Origins, Evolution, and Destinies of Binary Stars
  in Clusters}, \emph{Astronomical Society of the Pacific Conference Series},
  vol.~90, ed. by E.F. {Milone}, J.C. {Mermilliod} (1996), \emph{Astronomical
  Society of the Pacific Conference Series}, vol.~90, p. 423

\bibitem{Eggletonetal1989}
P.P. {Eggleton}, M.J. {Fitchett}, C.A. {Tout}, ApJ \textbf{347}, 998 (1989).
\newblock \doi{10.1086/168190}

\bibitem{Hurley2008a}
J.R. {Hurley}, in \emph{The Cambridge N-Body Lectures}, vol. 760, ed. by S.J.
  {Aarseth}, C.A. {Tout}, R.A. {Mardling} (Springer-Verlag Berlin Heidelberg,
  2008), p. 283.
\newblock \doi{10.1007/978-1-4020-8431-7_10}

\bibitem{Hurleyetal2013a}
J.R. {Hurley}, O.R. {Pols}, C.A. {Tout}.
\newblock {SSE: Single Star Evolution}.
\newblock Astrophysics Source Code Library, record ascl:1303.015 (2013)

\bibitem{Railtonetal2014}
A.D. {Railton}, C.A. {Tout}, S.J. {Aarseth}, PASA \textbf{31}, e017 (2014).
\newblock \doi{10.1017/pasa.2014.10}

\bibitem{TangJoyce2021}
J.~{Tang}, M.~{Joyce}, RNAAS \textbf{5}(5), 117 (2021).
\newblock \doi{10.3847/2515-5172/ac01ca}

\bibitem{Agrawaletal2022}
P.~{Agrawal}, S.~{Stevenson}, D.~{Sz{\'e}csi}, J.~{Hurley}, \aap \textbf{668},
  A90 (2022).
\newblock \doi{10.1051/0004-6361/202244044}

\bibitem{Hurley2008b}
J.R. {Hurley}, in \emph{The Cambridge N-Body Lectures}, vol. 760, ed. by S.J.
  {Aarseth}, C.A. {Tout}, R.A. {Mardling} (Springer-Verlag Berlin Heidelberg,
  2008), p. 321.
\newblock \doi{10.1007/978-1-4020-8431-7_12}

\bibitem{Hurleyetal2013b}
J.R. {Hurley}, C.A. {Tout}, O.R. {Pols}.
\newblock {BSE: Binary Star Evolution}.
\newblock Astrophysics Source Code Library, record ascl:1303.014 (2013)

\bibitem{Askaretal2017a}
A.~{Askar}, M.~{Szkudlarek}, D.~{Gondek-Rosi{\'n}ska}, M.~{Giersz}, T.~{Bulik},
  MNRAS \textbf{464}(1), L36 (2017).
\newblock \doi{10.1093/mnrasl/slw177}

\bibitem{TeamCompas2021}
{Team COMPAS}, {:}, J.~{Riley}, P.~{Agrawal}, J.W. {Barrett}, K.N.K. {Boyett},
  F.S. {Broekgaarden}, D.~{Chattopadhyay}, S.M. {Gaebel}, F.~{Gittins},
  R.~{Hirai}, G.~{Howitt}, S.~{Justham}, L.~{Khandelwal}, F.~{Kummer}, M.Y.M.
  {Lau}, I.~{Mandel}, S.E. {de Mink}, C.~{Neijssel}, T.~{Riley}, L.~{van Son},
  S.~{Stevenson}, A.~{Vigna-Gomez}, S.~{Vinciguerra}, T.~{Wagg}, R.~{Willcox},
  arXiv e-prints arXiv:2109.10352 (2021)

\bibitem{Hamersetal2020}
A.S. {Hamers}, M.~{Safarzadeh}, ApJ \textbf{898}(2), 99 (2020).
\newblock \doi{10.3847/1538-4357/ab9b27}

\bibitem{GiacobboMapelli2018}
N.~{Giacobbo}, M.~{Mapelli}, MNRAS \textbf{480}(2), 2011 (2018).
\newblock \doi{10.1093/mnras/sty1999}

\bibitem{GiacobboMapelli2019b}
N.~{Giacobbo}, M.~{Mapelli}, MNRAS \textbf{486}(2), 2494 (2019).
\newblock \doi{10.1093/mnras/stz892}

\bibitem{Mapellietal2020a}
M.~{Mapelli}, M.~{Spera}, E.~{Montanari}, M.~{Limongi}, A.~{Chieffi}, et~al.,
  ApJ \textbf{888}(2), 76 (2020).
\newblock \doi{10.3847/1538-4357/ab584d}

\bibitem{Lietal2023}
Z.M. {Li}, B.~{Kayastha}, A.~{Kamlah}, P.~{Berczik}, Y.Y. {Deng}, R.~{Spurzem},
  Research in Astronomy and Astrophysics \textbf{23}(2), 025019 (2023).
\newblock \doi{10.1088/1674-4527/aca94f}

\bibitem{Belczynskietal2002}
K.~{Belczynski}, V.~{Kalogera}, T.~{Bulik}, ApJ \textbf{572}(1), 407 (2002).
\newblock \doi{10.1086/340304}

\bibitem{Kremeretal2019}
K.~{Kremer}, C.L. {Rodriguez}, P.~{Amaro-Seoane}, K.~{Breivik},
  S.~{Chatterjee}, et~al., PhRvD \textbf{99}(6), 063003 (2019).
\newblock \doi{10.1103/PhysRevD.99.063003}

\bibitem{Izzardetal2004}
R.G. {Izzard}, C.A. {Tout}, A.I. {Karakas}, O.R. {Pols}, MNRAS \textbf{350}(2),
  407 (2004).
\newblock \doi{10.1111/j.1365-2966.2004.07446.x}

\bibitem{Izzardetal2006}
R.G. {Izzard}, L.M. {Dray}, A.I. {Karakas}, M.~{Lugaro}, C.A. {Tout}, AAP
  \textbf{460}(2), 565 (2006).
\newblock \doi{10.1051/0004-6361:20066129}

\bibitem{Izzardetal2009}
R.G. {Izzard}, E.~{Glebbeek}, R.J. {Stancliffe}, O.R. {Pols}, AAP
  \textbf{508}(3), 1359 (2009).
\newblock \doi{10.1051/0004-6361/200912827}

\bibitem{Tanikawaetal2020}
A.~{Tanikawa}, T.~{Yoshida}, T.~{Kinugawa}, K.~{Takahashi}, H.~{Umeda}, MNRAS
  \textbf{495}(4), 4170 (2020).
\newblock \doi{10.1093/mnras/staa1417}

\bibitem{Tanikawaetal2021a}
A.~{Tanikawa}, H.~{Susa}, T.~{Yoshida}, A.A. {Trani}, T.~{Kinugawa}, ApJ
  \textbf{910}(1), 30 (2021).
\newblock \doi{10.3847/1538-4357/abe40d}

\bibitem{Tanikawaetal2021c}
A.~{Tanikawa}, T.~{Kinugawa}, T.~{Yoshida}, K.~{Hijikawa}, H.~{Umeda},
  MNRAS.tmp  (2021).
\newblock \doi{10.1093/mnras/stab1421}

\bibitem{Hijikawaetal2021}
K.~{Hijikawa}, A.~{Tanikawa}, T.~{Kinugawa}, T.~{Yoshida}, H.~{Umeda},
  MNRAS.tmp  (2021).
\newblock \doi{10.1093/mnrasl/slab052}

\bibitem{Plummer1915}
H.C. {Plummer}, \mnras \textbf{76}, 107 (1915).
\newblock \doi{10.1093/mnras/76.2.107}

\bibitem{BahcallWolf1976}
J.N. {Bahcall}, R.A. {Wolf}, \apj \textbf{209}, 214 (1976).
\newblock \doi{10.1086/154711}

\bibitem{FrankRees1976}
J.~{Frank}, M.J. {Rees}, \mnras \textbf{176}, 633 (1976).
\newblock \doi{10.1093/mnras/176.3.633}

\bibitem{Pretoetal2004}
M.~{Preto}, D.~{Merritt}, R.~{Spurzem}, \apjl \textbf{613}(2), L109 (2004).
\newblock \doi{10.1086/425139}

\bibitem{Chandrasekhar1939}
S.~{Chandrasekhar}, \emph{{An introduction to the study of stellar structure}}
  (The University of Chicago Press, 1939)

\bibitem{Michie1963b}
R.W. {Michie}, \mnras \textbf{126}, 331 (1963).
\newblock \doi{10.1093/mnras/126.4.331}

\bibitem{King1981}
I.R. {King}, \qjras \textbf{22}, 227 (1981)

\bibitem{King2008}
I.R. {King}, in \emph{Dynamical Evolution of Dense Stellar Systems}, vol. 246,
  ed. by E.~{Vesperini}, M.~{Giersz}, A.~{Sills} (2008), vol. 246, pp.
  131--140.
\newblock \doi{10.1017/S1743921308015470}

\bibitem{GielesZocchi2015}
M.~{Gieles}, A.~{Zocchi}, MNRAS \textbf{454}(1), 576 (2015).
\newblock \doi{10.1093/mnras/stv1848}

\bibitem{Kroupaetal2013}
P.~{Kroupa}, C.~{Weidner}, J.~{Pflamm-Altenburg}, I.~{Thies},
  J.~{Dabringhausen}, M.~{Marks}, T.~{Maschberger}, in \emph{Planets, Stars and
  Stellar Systems. Volume 5: Galactic Structure and Stellar Populations},
  vol.~5, ed. by T.D. {Oswalt}, G.~{Gilmore} (Springer Dordrecht Heidelberg New
  York London, 2013), p. 115.
\newblock \doi{10.1007/978-94-007-5612-0_4}

\bibitem{KroupaJerabkova2018}
P.~{Kroupa}, T.~{Jerabkova}, arXiv e-prints arXiv:1806.10605 (2018)

\bibitem{Hopkins2018}
A.M. {Hopkins}, PASA \textbf{35}, e039 (2018).
\newblock \doi{10.1017/pasa.2018.29}

\bibitem{Salpeter1955}
E.E. {Salpeter}, ApJ \textbf{121}, 161 (1955).
\newblock \doi{10.1086/145971}

\bibitem{KroupaJerabkove2019}
P.~{Kroupa}, T.~{Jerabkova}, NatAs \textbf{3}, 482 (2019).
\newblock \doi{10.1038/s41550-019-0793-0}

\bibitem{Jerabkovaetal2018}
T.~{Je{\v{r}}{\'a}bkov{\'a}}, A.~{Hasani Zonoozi}, P.~{Kroupa}, G.~{Beccari},
  Z.~{Yan}, A.~{Vazdekis}, Z.Y. {Zhang}, AAP \textbf{620}, A39 (2018).
\newblock \doi{10.1051/0004-6361/201833055}

\bibitem{Kroupa2002}
P.~{Kroupa}, Sci \textbf{295}(5552), 82 (2002).
\newblock \doi{10.1126/science.1067524}

\bibitem{Chabrier2003}
G.~{Chabrier}, PASP \textbf{115}(809), 763 (2003).
\newblock \doi{10.1086/376392}

\bibitem{Maschberger2013}
T.~{Maschberger}, MNRAS \textbf{429}(2), 1725 (2013).
\newblock \doi{10.1093/mnras/sts479}

\bibitem{Brommetal2002}
V.~{Bromm}, P.S. {Coppi}, R.B. {Larson}, ApJ \textbf{564}(1), 23 (2002).
\newblock \doi{10.1086/323947}

\bibitem{BrommLarson2004}
V.~{Bromm}, R.B. {Larson}, ARA\&A \textbf{42}(1), 79 (2004).
\newblock \doi{10.1146/annurev.astro.42.053102.134034}

\bibitem{Marksetal2012}
M.~{Marks}, P.~{Kroupa}, J.~{Dabringhausen}, M.S. {Pawlowski}, MNRAS
  \textbf{422}(3), 2246 (2012).
\newblock \doi{10.1111/j.1365-2966.2012.20767.x}

\bibitem{Bromm2013}
V.~{Bromm}, Reports on Progress in Physics \textbf{76}(11), 112901 (2013).
\newblock \doi{10.1088/0034-4885/76/11/112901}

\bibitem{Stacyetal2016}
A.~{Stacy}, V.~{Bromm}, A.T. {Lee}, MNRAS \textbf{462}(2), 1307 (2016).
\newblock \doi{10.1093/mnras/stw1728}

\bibitem{Kroupaetal2020}
P.~{Kroupa}, L.~{Subr}, T.~{Jerabkova}, L.~{Wang}, MNRAS \textbf{498}(4), 5652
  (2020).
\newblock \doi{10.1093/mnras/staa2276}

\bibitem{Rydbergetal2013}
C.E. {Rydberg}, E.~{Zackrisson}, P.~{Lundqvist}, P.~{Scott}, MNRAS
  \textbf{429}(4), 3658 (2013).
\newblock \doi{10.1093/mnras/sts653}

\bibitem{deSouzaetal2013}
R.S. {de Souza}, B.~{Ciardi}, U.~{Maio}, A.~{Ferrara}, MNRAS \textbf{428}(3),
  2109 (2013).
\newblock \doi{10.1093/mnras/sts181}

\bibitem{Schaueretal2020}
A.T.P. {Schauer}, N.~{Drory}, V.~{Bromm}, ApJ \textbf{904}(2), 145 (2020).
\newblock \doi{10.3847/1538-4357/abbc0b}

\bibitem{LazarBromm2021}
A.~{Lazar}, V.~{Bromm}, arXiv e-prints arXiv:2110.11956 (2021)

\bibitem{Fraseretal2017}
M.~{Fraser}, A.R. {Casey}, G.~{Gilmore}, A.~{Heger}, C.~{Chan}, MNRAS
  \textbf{468}(1), 418 (2017).
\newblock \doi{10.1093/mnras/stx480}

\bibitem{GoodwinKroupa2005}
S.P. {Goodwin}, P.~{Kroupa}, AAP \textbf{439}(2), 565 (2005).
\newblock \doi{10.1051/0004-6361:20052654}

\bibitem{Kroupa2008}
P.~{Kroupa}, in \emph{The Cambridge N-Body Lectures}, vol. 760, ed. by S.J.
  {Aarseth}, C.A. {Tout}, R.A. {Mardling} (Springer-Verlag Berlin Heidelberg,
  2008), p. 181.
\newblock \doi{10.1007/978-1-4020-8431-7_8}

\bibitem{Miloneetal2012a}
A.P. {Milone}, G.~{Piotto}, L.R. {Bedin}, S.~{Cassisi}, J.~{Anderson}, et~al.,
  A\&A \textbf{537}, A77 (2012).
\newblock \doi{10.1051/0004-6361/201116539}

\bibitem{Marksetal2015}
M.~{Marks}, M.~{Janson}, P.~{Kroupa}, N.~{Leigh}, I.~{Thies}, MNRAS
  \textbf{452}(1), 1014 (2015).
\newblock \doi{10.1093/mnras/stv1361}

\bibitem{Belloni2018a}
D.~{Belloni}, A.~{Askar}, M.~{Giersz}, P.~{Kroupa}, H.J. {Rocha-Pinto}, MNRAS
  \textbf{471}(3), 2812 (2017).
\newblock \doi{10.1093/mnras/stx1763}

\bibitem{Stacyetal2012}
A.~{Stacy}, A.H. {Pawlik}, V.~{Bromm}, A.~{Loeb}, MNRAS \textbf{421}(1), 894
  (2012).
\newblock \doi{10.1111/j.1365-2966.2011.20373.x}

\bibitem{Stacyetal2013}
A.~{Stacy}, V.~{Bromm}, MNRAS \textbf{433}(2), 1094 (2013).
\newblock \doi{10.1093/mnras/stt789}

\bibitem{Hills1975b}
J.G. {Hills}, \nat \textbf{254}(5498), 295 (1975).
\newblock \doi{10.1038/254295a0}

\bibitem{SharaHurley2002}
M.M. {Shara}, J.R. {Hurley}, ApJ \textbf{571}(2), 830 (2002).
\newblock \doi{10.1086/340062}

\bibitem{HeggieHut2003}
D.~{Heggie}, P.~{Hut}, \emph{{The Gravitational Million-Body Problem: A
  Multidisciplinary Approach to Star Cluster Dynamics}} (Cambridge University
  Press, 2003)

\bibitem{Mapelli2018a}
M.~{Mapelli}, arXiv arXiv:1807.07944 (2018)

\bibitem{Kroupa1995b}
P.~{Kroupa}, MNRAS \textbf{277}, 1507 (1995).
\newblock \doi{10.1093/mnras/277.4.1507}

\bibitem{MoeDiStefano2017}
M.~{Moe}, R.~{Di Stefano}, ApJS \textbf{230}(2), 15 (2017).
\newblock \doi{10.3847/1538-4365/aa6fb6}

\bibitem{Sanaetal2012a}
H.~{Sana}, S.E. {de Mink}, A.~{de Koter}, N.~{Langer}, C.J. {Evans}, et~al.,
  Sci \textbf{337}(6093), 444 (2012).
\newblock \doi{10.1126/science.1223344}

\bibitem{Ohetal2015}
S.~{Oh}, P.~{Kroupa}, J.~{Pflamm-Altenburg}, ApJ \textbf{805}(2), 92 (2015).
\newblock \doi{10.1088/0004-637X/805/2/92}

\bibitem{Gelleretal2019}
A.M. {Geller}, N.W.C. {Leigh}, M.~{Giersz}, K.~{Kremer}, F.A. {Rasio}, ApJ
  \textbf{872}(2), 165 (2019).
\newblock \doi{10.3847/1538-4357/ab0214}

\bibitem{Kiminkietal2012}
D.C. {Kiminki}, H.A. {Kobulnicky}, I.~{Ewing}, M.M. {Bagley Kiminki},
  M.~{Lundquist}, et~al., ApJ \textbf{747}(1), 41 (2012).
\newblock \doi{10.1088/0004-637X/747/1/41}

\bibitem{SanaEvans2011}
H.~{Sana}, C.J. {Evans}, in \emph{Active OB Stars: Structure, Evolution, Mass
  Loss, and Critical Limits}, vol. 272, ed. by C.~{Neiner}, G.~{Wade},
  G.~{Meynet}, G.~{Peters} (2011), vol. 272, pp. 474--485.
\newblock \doi{10.1017/S1743921311011124}

\bibitem{Kobulnickyetal2014}
H.A. {Kobulnicky}, D.C. {Kiminki}, M.J. {Lundquist}, J.~{Burke}, J.~{Chapman},
  et~al., ApJS \textbf{213}(2), 34 (2014).
\newblock \doi{10.1088/0067-0049/213/2/34}

\bibitem{Leighetal2015}
N.W.C. {Leigh}, M.~{Giersz}, M.~{Marks}, J.J. {Webb}, A.~{Hypki}, et~al., MNRAS
  \textbf{446}(1), 226 (2015).
\newblock \doi{10.1093/mnras/stu2110}

\bibitem{ArcaSeddaetal2023a}
M.~{Arca Sedda}, A.~{Kamlah}, R.~{Spurzem}, M.~{Giersz}, P.~{Berczik},
  S.~{Rastello}, G.. {Iorio}, M.~{Mapelli}, M.~{Gatto}, E.~{Grebel}, \mnras
  (2023).
\newblock Subm. to MNRAS

\bibitem{ArcaSeddaetal2023b}
M.~{Arca Sedda}, A.~{Kamlah}, R.~{Spurzem}, F.~{Rizzuto}, M.~{Giersz},
  T.~{Naab}, P.~{Berczik}, \mnras  (2023).
\newblock Subm. to MNRAS

\bibitem{ArcaSeddaetal2023c}
M.~{Arca Sedda}, A.~{Kamlah}, R.~{Spurzem}, F.~{Rizzuto}, M.~{Giersz},
  T.~{Naab}, P.~{Berczik}, \mnras  (2023).
\newblock Subm. to MNRAS

\bibitem{Torniamentietal2022}
S.~{Torniamenti}, S.~{Rastello}, M.~{Mapelli}, U.N. {Di Carlo}, A.~{Ballone},
  M.~{Pasquato}, \mnras \textbf{517}(2), 2953 (2022).
\newblock \doi{10.1093/mnras/stac2841}

\bibitem{Wangetal2020a}
L.~{Wang}, P.~{Kroupa}, K.~{Takahashi}, T.~{Jerabkova}, MNRAS \textbf{491}(1),
  440 (2020).
\newblock \doi{10.1093/mnras/stz3033}

\bibitem{Kuepperetal2011b}
A.H.W. {K{\"u}pper}, T.~{Maschberger}, P.~{Kroupa}, H.~{Baumgardt}, MNRAS
  \textbf{417}(3), 2300 (2011).
\newblock \doi{10.1111/j.1365-2966.2011.19412.x}

\bibitem{Fregeauetal2002}
J.M. {Fregeau}, K.J. {Joshi}, S.F. {Portegies Zwart}, F.A. {Rasio}, ApJ
  \textbf{570}(1), 171 (2002).
\newblock \doi{10.1086/339576}

\bibitem{Subretal2008}
L.~{{\v{S}}ubr}, P.~{Kroupa}, H.~{Baumgardt}, MNRAS \textbf{385}(3), 1673
  (2008).
\newblock \doi{10.1111/j.1365-2966.2008.12993.x}

\bibitem{Baumgardtetal2008b}
H.~{Baumgardt}, G.~{De Marchi}, P.~{Kroupa}, ApJ \textbf{685}(1), 247 (2008).
\newblock \doi{10.1086/590488}

\bibitem{GoodwinWhitworth2004}
S.P. {Goodwin}, A.P. {Whitworth}, A\&A \textbf{413}, 929 (2004).
\newblock \doi{10.1051/0004-6361:20031529}

\bibitem{AndersonKing2000}
J.~{Anderson}, I.R. {King}, PASP \textbf{112}(776), 1360 (2000).
\newblock \doi{10.1086/316632}

\bibitem{Andersonetal2008}
J.~{Anderson}, A.~{Sarajedini}, L.R. {Bedin}, I.R. {King}, G.~{Piotto}, et~al.,
  AJ \textbf{135}(6), 2055 (2008).
\newblock \doi{10.1088/0004-6256/135/6/2055}

\bibitem{Grattonetal2012}
R.G. {Gratton}, E.~{Carretta}, A.~{Bragaglia}, AAPR \textbf{20}, 50 (2012).
\newblock \doi{10.1007/s00159-012-0050-3}

\bibitem{Miloneetal2013}
A.P. {Milone}, A.F. {Marino}, G.~{Piotto}, L.R. {Bedin}, J.~{Anderson}, et~al.,
  ApJ \textbf{767}(2), 120 (2013).
\newblock \doi{10.1088/0004-637X/767/2/120}

\bibitem{Milone2020}
A.P. {Milone}, in \emph{Star Clusters: From the Milky Way to the Early
  Universe}, vol. 351, ed. by A.~{Bragaglia}, M.~{Davies}, A.~{Sills},
  E.~{Vesperini} (2020), vol. 351, pp. 251--260.
\newblock \doi{10.1017/S1743921319010044}

\bibitem{MiloneMarino2022}
A.P. {Milone}, A.F. {Marino}, arXiv arXiv:2206.10564 (2022)

\bibitem{Bastianetal2019}
N.~{Bastian}, C.~{Usher}, S.~{Kamann}, C.~{Lardo}, S.S. {Larsen},
  I.~{Cabrera-Ziri}, W.~{Chantereau}, S.~{Martocchia}, M.~{Salaris}, R.P.
  {Schiavon}, R.~{Asa'd}, M.~{Hilker}, \mnras \textbf{489}(1), L80 (2019).
\newblock \doi{10.1093/mnrasl/slz130}

\bibitem{Bastianetal2020}
N.~{Bastian}, C.~{Lardo}, C.~{Usher}, S.~{Kamann}, S.S. {Larsen},
  I.~{Cabrera-Ziri}, W.~{Chantereau}, S.~{Martocchia}, M.~{Salaris},
  R.~{Asa'd}, M.~{Hilker}, \mnras \textbf{494}(1), 332 (2020).
\newblock \doi{10.1093/mnras/staa716}

\bibitem{Miloneetal2017a}
A.P. {Milone}, G.~{Piotto}, A.~{Renzini}, A.F. {Marino}, L.R. {Bedin}, et~al.,
  MNRAS \textbf{464}(3), 3636 (2017).
\newblock \doi{10.1093/mnras/stw2531}

\bibitem{Miloneetal2017b}
A.P. {Milone}, A.F. {Marino}, L.R. {Bedin}, J.~{Anderson}, D.~{Apai}, et~al.,
  MNRAS \textbf{469}(1), 800 (2017).
\newblock \doi{10.1093/mnras/stx836}

\bibitem{Miloneetal2018a}
A.P. {Milone}, A.F. {Marino}, A.~{Mastrobuono-Battisti}, E.P. {Lagioia}, MNRAS
  \textbf{479}(4), 5005 (2018).
\newblock \doi{10.1093/mnras/sty1873}

\bibitem{Miloneetal2018b}
A.P. {Milone}, A.F. {Marino}, A.~{Renzini}, F.~{D'Antona}, J.~{Anderson},
  et~al., MNRAS \textbf{481}(4), 5098 (2018).
\newblock \doi{10.1093/mnras/sty2573}

\bibitem{Miloneetal2020a}
A.P. {Milone}, A.F. {Marino}, G.S. {Da Costa}, E.P. {Lagioia}, F.~{D'Antona},
  et~al., MNRAS \textbf{491}(1), 515 (2020).
\newblock \doi{10.1093/mnras/stz2999}

\bibitem{Miloneetal2020b}
A.P. {Milone}, E.~{Vesperini}, A.F. {Marino}, J.~{Hong}, R.~{van der Marel},
  et~al., MNRAS \textbf{492}(4), 5457 (2020).
\newblock \doi{10.1093/mnras/stz3629}

\bibitem{Miloneetal2020c}
A.P. {Milone}, A.F. {Marino}, A.~{Renzini}, C.~{Li}, S.~{Jang}, et~al., MNRAS
  \textbf{497}(3), 3846 (2020).
\newblock \doi{10.1093/mnras/staa2119}

\bibitem{BastianLardo2018}
N.~{Bastian}, C.~{Lardo}, \araa \textbf{56}, 83 (2018).
\newblock \doi{10.1146/annurev-astro-081817-051839}

\bibitem{Martensetal2023}
S.~{Martens}, S.~{Kamann}, S.~{Dreizler}, F.~{G{\"o}ttgens}, T.O. {Husser},
  M.~{Latour}, E.~{Balakina}, D.~{Krajnovi{\'c}}, R.~{Pechetti}, P.M.
  {Weilbacher}, \aap \textbf{671}, A106 (2023).
\newblock \doi{10.1051/0004-6361/202244787}

\bibitem{Hongetal2017a}
J.~{Hong}, E.~{Vesperini}, D.~{Belloni}, M.~{Giersz}, MNRAS \textbf{464}(2),
  2511 (2017).
\newblock \doi{10.1093/mnras/stw2595}

\bibitem{Bialasetal2015}
D.~{Bialas}, T.~{Lisker}, C.~{Olczak}, R.~{Spurzem}, R.~{Kotulla}, \aap
  \textbf{576}, A103 (2015).
\newblock \doi{10.1051/0004-6361/201425235}

\bibitem{Hypkietal2022}
A.~{Hypki}, M.~{Giersz}, J.~{Hong}, A.~{Leveque}, A.~{Askar}, D.~{Belloni},
  M.~{Otulakowska-Hypka}, \mnras \textbf{517}(4), 4768 (2022).
\newblock \doi{10.1093/mnras/stac2815}

\bibitem{PeaseShapley1917}
F.G. {Pease}, H.~{Shapley}, ApJ \textbf{45}, 225 (1917).
\newblock \doi{10.1086/142324}

\bibitem{ShapleySawyer1927}
H.~{Shapley}, H.B. {Sawyer}, Harvard College Observatory Bulletin \textbf{852},
  22 (1927)

\bibitem{Shapley1930}
H.~{Shapley}, \emph{{Star Clusters}}, vol.~2 (McGraw-Hill Book Co., 1930)

\bibitem{KopalSlouka1936}
Z.~{Kopal}, H.~{Slouka}, Natur \textbf{137}(3467), 621 (1936).
\newblock \doi{10.1038/137621a0}

\bibitem{King1961}
I.~{King}, AJ \textbf{66}, 68 (1961).
\newblock \doi{10.1086/108376}

\bibitem{FrenkFall1982}
C.S. {Frenk}, S.M. {Fall}, MNRAS \textbf{199}, 565 (1982).
\newblock \doi{10.1093/mnras/199.3.565}

\bibitem{Harris1976}
W.E. {Harris}, AJ \textbf{81}, 1095 (1976).
\newblock \doi{10.1086/111991}

\bibitem{Kormendy1985}
J.~{Kormendy}, ApJ \textbf{295}, 73 (1985).
\newblock \doi{10.1086/163350}

\bibitem{WhiteShawl1987}
R.E. {White}, S.J. {Shawl}, ApJ \textbf{317}, 246 (1987).
\newblock \doi{10.1086/165273}

\bibitem{Luptonetal1987}
R.H. {Lupton}, J.E. {Gunn}, R.F. {Griffin}, AJ \textbf{93}, 1114 (1987).
\newblock \doi{10.1086/114395}

\bibitem{ChenChen2010}
C.W. {Chen}, W.P. {Chen}, ApJ \textbf{721}(2), 1790 (2010).
\newblock \doi{10.1088/0004-637X/721/2/1790}

\bibitem{Bianchinietal2016}
P.~{Bianchini}, G.~{van de Ven}, M.A. {Norris}, E.~{Schinnerer}, A.L. {Varri},
  MNRAS \textbf{458}(4), 3644 (2016).
\newblock \doi{10.1093/mnras/stw552}

\bibitem{Bianchinietal2018}
P.~{Bianchini}, R.P. {van der Marel}, A.~{del Pino}, L.L. {Watkins},
  A.~{Bellini}, M.A. {Fardal}, M.~{Libralato}, A.~{Sills}, \mnras
  \textbf{481}(2), 2125 (2018).
\newblock \doi{10.1093/mnras/sty2365}

\bibitem{Bianchinietal2019}
P.~{Bianchini}, R.~{Ibata}, B.~{Famaey}, ApJL \textbf{887}(1), L12 (2019).
\newblock \doi{10.3847/2041-8213/ab58d1}

\bibitem{Ferraroetal2018}
F.R. {Ferraro}, A.~{Mucciarelli}, B.~{Lanzoni}, C.~{Pallanca}, E.~{Lapenna},
  L.~{Origlia}, E.~{Dalessandro}, E.~{Valenti}, G.~{Beccari}, M.~{Bellazzini},
  E.~{Vesperini}, A.~{Varri}, A.~{Sollima}, ApJ \textbf{860}(1), 50 (2018).
\newblock \doi{10.3847/1538-4357/aabe2f}

\bibitem{Lanzonietal2018a}
B.~{Lanzoni}, F.R. {Ferraro}, A.~{Mucciarelli}, C.~{Pallanca}, E.~{Lapenna},
  L.~{Origlia}, E.~{Dalessandro}, E.~{Valenti}, M.~{Bellazzini}, M.A.
  {Tiongco}, A.L. {Varri}, E.~{Vesperini}, G.~{Beccari}, ApJ \textbf{861}(1),
  16 (2018).
\newblock \doi{10.3847/1538-4357/aac26a}

\bibitem{Lanzonietal2018b}
B.~{Lanzoni}, F.R. {Ferraro}, A.~{Mucciarelli}, C.~{Pallanca}, M.A. {Tiongco},
  A.~{Varri}, E.~{Vesperini}, M.~{Bellazzini}, E.~{Dalessandro}, L.~{Origlia},
  E.~{Valenti}, A.~{Sollima}, E.~{Lapenna}, G.~{Beccari}, ApJ \textbf{865}(1),
  11 (2018).
\newblock \doi{10.3847/1538-4357/aad810}

\bibitem{Kamannetal2016}
S.~{Kamann}, T.O. {Husser}, J.~{Brinchmann}, E.~{Emsellem}, P.M. {Weilbacher},
  et~al., A\&A \textbf{588}, A149 (2016).
\newblock \doi{10.1051/0004-6361/201527065}

\bibitem{Kamannetal2018a}
S.~{Kamann}, T.O. {Husser}, S.~{Dreizler}, E.~{Emsellem}, P.M. {Weilbacher},
  et~al., MNRAS \textbf{473}(4), 5591 (2018).
\newblock \doi{10.1093/mnras/stx2719}

\bibitem{Kamannetal2018b}
S.~{Kamann}, N.~{Bastian}, T.O. {Husser}, S.~{Martocchia}, C.~{Usher}, et~al.,
  MNRAS \textbf{480}(2), 1689 (2018).
\newblock \doi{10.1093/mnras/sty1958}

\bibitem{Kamannetal2019}
S.~{Kamann}, N.J. {Bastian}, M.~{Gieles}, E.~{Balbinot},
  V.~{H{\'e}nault-Brunet}, MNRAS \textbf{483}(2), 2197 (2019).
\newblock \doi{10.1093/mnras/sty3144}

\bibitem{Sollimaetal2019}
A.~{Sollima}, H.~{Baumgardt}, M.~{Hilker}, MNRAS \textbf{485}(1), 1460 (2019).
\newblock \doi{10.1093/mnras/stz505}

\bibitem{Tiongcoetal2019}
M.A. {Tiongco}, E.~{Vesperini}, A.L. {Varri}, MNRAS \textbf{487}(4), 5535
  (2019).
\newblock \doi{10.1093/mnras/stz1595}

\bibitem{Tiongcoetal2021}
M.~{Tiongco}, A.~{Collier}, A.L. {Varri}, MNRAS \textbf{506}(3), 4488 (2021).
\newblock \doi{10.1093/mnras/stab1968}

\bibitem{Balloneetal2020}
A.~{Ballone}, M.~{Mapelli}, U.N. {Di Carlo}, S.~{Torniamenti}, M.~{Spera},
  et~al., MNRAS \textbf{496}(1), 49 (2020).
\newblock \doi{10.1093/mnras/staa1383}

\bibitem{Pangetal2021a}
X.~{Pang}, Y.~{Li}, Z.~{Yu}, S.Y. {Tang}, F.~{Dinnbier}, et~al., ApJ
  \textbf{912}(2), 162 (2021).
\newblock \doi{10.3847/1538-4357/abeaac}

\bibitem{Lahenetal2020b}
N.~{Lah{\'e}n}, T.~{Naab}, P.H. {Johansson}, B.~{Elmegreen}, C.Y. {Hu}, et~al.,
  ApJ \textbf{904}(1), 71 (2020).
\newblock \doi{10.3847/1538-4357/abc001}

\bibitem{Balloneetal2021}
A.~{Ballone}, S.~{Torniamenti}, M.~{Mapelli}, U.N. {Di Carlo}, M.~{Spera},
  et~al., MNRAS \textbf{501}(2), 2920 (2021).
\newblock \doi{10.1093/mnras/staa3763}

\bibitem{Pangetal2020}
X.~{Pang}, Y.~{Li}, S.Y. {Tang}, M.~{Pasquato}, M.B.N. {Kouwenhoven}, ApJL
  \textbf{900}(1), L4 (2020).
\newblock \doi{10.3847/2041-8213/abad28}

\bibitem{Plummer1911}
H.C. {Plummer}, MNRAS \textbf{71}, 460 (1911).
\newblock \doi{10.1093/mnras/71.5.460}

\bibitem{King1962}
I.~{King}, AJ \textbf{67}, 471 (1962).
\newblock \doi{10.1086/108756}

\bibitem{Wilson1975}
C.P. {Wilson}, AJ \textbf{80}, 175 (1975).
\newblock \doi{10.1086/111729}

\bibitem{LyndenBell1960}
D.~{Lynden-Bell}, MNRAS \textbf{120}, 204 (1960).
\newblock \doi{10.1093/mnras/120.3.204}

\bibitem{Lingam2018}
M.~{Lingam}, MNRAS \textbf{473}(2), 1719 (2018).
\newblock \doi{10.1093/mnras/stx2531}

\bibitem{Goodman1983a}
J.J. {Goodman}, {Dynamical Relaxation in Stellar Systems.}
\newblock Ph.D. thesis, Princeton Univ., NJ. (1983)

\bibitem{Michie1962}
R.W. {Michie}, \aj \textbf{67}, 582 (1962).
\newblock \doi{10.1086/108880}

\bibitem{Fiestasetal2006}
J.~{Fiestas}, R.~{Spurzem}, E.~{Kim}, MNRAS \textbf{373}(2), 677 (2006).
\newblock \doi{10.1111/j.1365-2966.2006.11036.x}

\bibitem{FiestasSpurzem2010}
J.~{Fiestas}, R.~{Spurzem}, MNRAS \textbf{405}(1), 194 (2010).
\newblock \doi{10.1111/j.1365-2966.2010.16479.x}

\bibitem{Fiestasetal2012}
J.~{Fiestas}, O.~{Porth}, P.~{Berczik}, R.~{Spurzem}, MNRAS \textbf{419}(1), 57
  (2012).
\newblock \doi{10.1111/j.1365-2966.2011.19670.x}

\bibitem{Ernstetal2007}
A.~{Ernst}, P.~{Glaschke}, J.~{Fiestas}, A.~{Just}, R.~{Spurzem}, MNRAS
  \textbf{377}(2), 465 (2007).
\newblock \doi{10.1111/j.1365-2966.2007.11602.x}

\bibitem{Tiongcoetal2016a}
M.A. {Tiongco}, E.~{Vesperini}, A.L. {Varri}, MNRAS \textbf{455}(4), 3693
  (2016).
\newblock \doi{10.1093/mnras/stv2574}

\bibitem{Tiongcoetal2016b}
M.A. {Tiongco}, E.~{Vesperini}, A.L. {Varri}, MNRAS \textbf{461}(1), 402
  (2016).
\newblock \doi{10.1093/mnras/stw1341}

\bibitem{Tiongcoetal2018}
M.A. {Tiongco}, E.~{Vesperini}, A.L. {Varri}, MNRAS \textbf{475}(1), L86
  (2018).
\newblock \doi{10.1093/mnrasl/sly009}

\bibitem{PanamarevKocsis2022}
T.~{Panamarev}, B.~{Kocsis}, \mnras \textbf{517}(4), 6205 (2022).
\newblock \doi{10.1093/mnras/stac3050}

\bibitem{InagakiHachisu1978}
S.~{Inagaki}, I.~{Hachisu}, PASJ \textbf{30}, 39 (1978)

\bibitem{Hachisu1982}
I.~{Hachisu}, PASJ \textbf{34}, 313 (1982)

\bibitem{LyndenBell1999}
D.~{Lynden-Bell}, Physica A Statistical Mechanics and its Applications
  \textbf{263}(1-4), 293 (1999).
\newblock \doi{10.1016/S0378-4371(98)00518-4}

\bibitem{Henon1975}
M.~{H{\'e}non}, in \emph{Dynamics of the Solar Systems}, vol.~69, ed. by
  A.~{Hayli} (1975), vol.~69, p. 133

\bibitem{Merritt2015}
D.~{Merritt}, ApJ \textbf{804}(1), 52 (2015).
\newblock \doi{10.1088/0004-637X/804/1/52}

\bibitem{Kremeretal2020a}
K.~{Kremer}, C.S. {Ye}, N.Z. {Rui}, N.C. {Weatherford}, S.~{Chatterjee},
  et~al., ApJS \textbf{247}(2), 48 (2020).
\newblock \doi{10.3847/1538-4365/ab7919}

\bibitem{Abbottetal2020b}
R.~{Abbott}, T.D. {Abbott}, S.~{Abraham}, F.~{Acernese}, K.~{Ackley}, et~al.,
  PhRvL \textbf{125}(10), 101102 (2020).
\newblock \doi{10.1103/PhysRevLett.125.101102}

\bibitem{PhinneySigurdsson1991}
E.S. {Phinney}, S.~{Sigurdsson}, Natur \textbf{349}(6306), 220 (1991).
\newblock \doi{10.1038/349220a0}

\bibitem{Gerhard2001}
O.~{Gerhard}, ApJL \textbf{546}(1), L39 (2001).
\newblock \doi{10.1086/318054}

\bibitem{PortegiesZwartMcMillan2002}
S.F. {Portegies Zwart}, S.L.W. {McMillan}, ApJ \textbf{576}(2), 899 (2002).
\newblock \doi{10.1086/341798}

\bibitem{PortegiesZwartetal2004}
S.F. {Portegies Zwart}, H.~{Baumgardt}, P.~{Hut}, J.~{Makino}, S.L.W.
  {McMillan}, Natur \textbf{428}(6984), 724 (2004).
\newblock \doi{10.1038/nature02448}

\bibitem{Greeneetal2020}
J.E. {Greene}, J.~{Strader}, L.C. {Ho}, ARA\&A \textbf{58}, 257 (2020).
\newblock \doi{10.1146/annurev-astro-032620-021835}

\bibitem{ArcaSedda2016}
M.~{Arca-Sedda}, MNRAS \textbf{455}(1), 35 (2016).
\newblock \doi{10.1093/mnras/stv2265}

\bibitem{ArcaSeddaetal2021b}
M.~{Arca Sedda}, P.~{Amaro Seoane}, X.~{Chen}, A\&A \textbf{652}, A54 (2021).
\newblock \doi{10.1051/0004-6361/202037785}

\bibitem{ArcaSedda2019}
M.~{Arca Sedda}, A.~{Askar}, M.~{Giersz}, arXiv arXiv:1905.00902 (2019)

\bibitem{ArcaSeddaCapuzzoDolcetta2019}
M.~{Arca-Sedda}, R.~{Capuzzo-Dolcetta}, MNRAS \textbf{483}(1), 152 (2019).
\newblock \doi{10.1093/mnras/sty3096}

\bibitem{ArcaSeddaetal2020b}
M.~{Arca Sedda}, C.P.L. {Berry}, K.~{Jani}, P.~{Amaro-Seoane}, P.~{Auclair},
  et~al., CQGra \textbf{37}(21), 215011 (2020).
\newblock \doi{10.1088/1361-6382/abb5c1}

\bibitem{Janietal2020}
K.~{Jani}, D.~{Shoemaker}, C.~{Cutler}, Nature Astronomy \textbf{4}, 260
  (2020).
\newblock \doi{10.1038/s41550-019-0932-7}

\bibitem{AmaroSeoaneetal2017}
P.~{Amaro-Seoane}, H.~{Audley}, S.~{Babak}, J.~{Baker}, E.~{Barausse}, et~al.,
  arXiv arXiv:1702.00786 (2017)

\bibitem{Bayleetal2022}
J.B. {Bayle}, B.~{Bonga}, C.~{Caprini}, D.~{Doneva}, M.~{Muratore},
  A.~{Petiteau}, E.~{Rossi}, L.~{Shao}, Nature Astronomy \textbf{6}, 1334
  (2022).
\newblock \doi{10.1038/s41550-022-01847-0}

\bibitem{Luoetal2016}
J.~{Luo}, L.S. {Chen}, H.Z. {Duan}, Y.G. {Gong}, S.~{Hu}, J.~{Ji}, Q.~{Liu},
  J.~{Mei}, V.~{Milyukov}, M.~{Sazhin}, C.G. {Shao}, V.T. {Toth}, H.B. {Tu},
  Y.~{Wang}, Y.~{Wang}, H.C. {Yeh}, M.S. {Zhan}, Y.~{Zhang}, V.~{Zharov}, Z.B.
  {Zhou}, Classical and Quantum Gravity \textbf{33}(3), 035010 (2016).
\newblock \doi{10.1088/0264-9381/33/3/035010}

\bibitem{Ruanetal2018}
W.H. {Ruan}, Z.K. {Guo}, R.G. {Cai}, Y.Z. {Zhang}, arXiv e-prints
  arXiv:1807.09495 (2018).
\newblock \doi{10.48550/arXiv.1807.09495}

\bibitem{Chenetal2021}
J.~{Chen}, C.S. {Yan}, Y.J. {Lu}, Y.T. {Zhao}, J.Q. {Ge}, Research in Astronomy
  and Astrophysics \textbf{21}(11), 285 (2021).
\newblock \doi{10.1088/1674-4527/21/11/285}

\bibitem{ArcaSeddaGualandris2018}
M.~{Arca-Sedda}, A.~{Gualandris}, MNRAS \textbf{477}(4), 4423 (2018).
\newblock \doi{10.1093/mnras/sty922}

\bibitem{ArcaSeddaBeancquista2019}
M.~{Arca Sedda}, M.~{Benacquista}, MNRAS \textbf{482}(3), 2991 (2019).
\newblock \doi{10.1093/mnras/sty2764}

\bibitem{ShapiroLightman1976}
S.L. {Shapiro}, A.P. {Lightman}, \nat \textbf{262}(5571), 743 (1976).
\newblock \doi{10.1038/262743a0}

\bibitem{LightmanShapiro1977}
A.P. {Lightman}, S.L. {Shapiro}, \apj \textbf{211}, 244 (1977).
\newblock \doi{10.1086/154925}

\bibitem{Hills1975a}
J.G. {Hills}, AJ \textbf{80}, 809 (1975).
\newblock \doi{10.1086/111815}

\bibitem{AmaroSeoaneetal2001}
P.~{Amaro-Seoane}, R.~{Spurzem}, \mnras \textbf{327}(3), 995 (2001).
\newblock \doi{10.1046/j.1365-8711.2001.04799.x}

\bibitem{AmaroSeoaneetal2004}
P.~{Amaro-Seoane}, M.~{Freitag}, R.~{Spurzem}, \mnras \textbf{352}(2), 655
  (2004).
\newblock \doi{10.1111/j.1365-2966.2004.07956.x}

\bibitem{MagorrianTremaine1999}
J.~{Magorrian}, S.~{Tremaine}, \mnras \textbf{309}(2), 447 (1999).
\newblock \doi{10.1046/j.1365-8711.1999.02853.x}

\bibitem{MerrittPoon2004}
D.~{Merritt}, M.Y. {Poon}, \apj \textbf{606}(2), 788 (2004).
\newblock \doi{10.1086/382497}

\bibitem{PoonMerritt2004}
M.Y. {Poon}, D.~{Merritt}, \apj \textbf{606}(2), 774 (2004).
\newblock \doi{10.1086/383190}

\bibitem{WangMerritt2004}
J.~{Wang}, D.~{Merritt}, \apj \textbf{600}(1), 149 (2004).
\newblock \doi{10.1086/379767}

\bibitem{MerrittWang2005}
D.~{Merritt}, J.~{Wang}, \apjl \textbf{621}(2), L101 (2005).
\newblock \doi{10.1086/429272}

\bibitem{CohnKulsrud1978}
H.~{Cohn}, R.M. {Kulsrud}, \apj \textbf{226}, 1087 (1978).
\newblock \doi{10.1086/156685}

\bibitem{BahcallWolf1977}
J.N. {Bahcall}, R.A. {Wolf}, \apj \textbf{216}, 883 (1977).
\newblock \doi{10.1086/155534}

\bibitem{Davidetal1987a}
L.P. {David}, R.H. {Durisen}, H.N. {Cohn}, \apj \textbf{313}, 556 (1987).
\newblock \doi{10.1086/164997}

\bibitem{Davidetal1987b}
L.P. {David}, R.H. {Durisen}, H.N. {Cohn}, \apj \textbf{316}, 505 (1987).
\newblock \doi{10.1086/165222}

\bibitem{Panamarevetal2019}
T.~{Panamarev}, A.~{Just}, R.~{Spurzem}, P.~{Berczik}, L.~{Wang}, et~al., MNRAS
  \textbf{484}(3), 3279 (2019).
\newblock \doi{10.1093/mnras/stz208}

\bibitem{Rees1988}
M.J. {Rees}, \nat \textbf{333}(6173), 523 (1988).
\newblock \doi{10.1038/333523a0}

\bibitem{Hayasakietal2018}
K.~{Hayasaki}, S.~{Zhong}, S.~{Li}, P.~{Berczik}, R.~{Spurzem}, \apj
  \textbf{855}(2), 129 (2018).
\newblock \doi{10.3847/1538-4357/aab0a5}

\bibitem{Zhongetal2022}
S.~{Zhong}, S.~{Li}, P.~{Berczik}, R.~{Spurzem}, \apj \textbf{933}(1), 96
  (2022).
\newblock \doi{10.3847/1538-4357/ac71ad}

\bibitem{LawSmithetal2020}
J.A.P. {Law-Smith}, D.A. {Coulter}, J.~{Guillochon}, B.~{Mockler},
  E.~{Ramirez-Ruiz}, \apj \textbf{905}(2), 141 (2020).
\newblock \doi{10.3847/1538-4357/abc489}

\bibitem{PortegiesZwart2011}
S.~{Portegies Zwart}.
\newblock {AMUSE: Astrophysical Multipurpose Software Environment}.
\newblock Astrophysics Source Code Library, record ascl:1107.007 (2011)

\bibitem{PortegiesZwartetal2013}
S.~{Portegies Zwart}, S.L.W. {McMillan}, E.~{van Elteren}, I.~{Pelupessy},
  N.~{de Vries}, CoPhC \textbf{184}(3), 456 (2013).
\newblock \doi{10.1016/j.cpc.2012.09.024}

\bibitem{PortegiesZwartetal2018b}
S.~{Portegies Zwart}, A.~{van Elteren}, I.~{Pelupessy}, S.~{McMillan},
  S.~{Rieder}, et~al.
\newblock {Amuse: The Astrophysical Multipurpose Software Environment} (2018).
\newblock \doi{10.5281/zenodo.1443252}

\bibitem{PortegiesZwartMcMillan2018}
S.~{Portegies Zwart}, S.~{McMillan}, \emph{{Astrophysical Recipes; The art of
  AMUSE}} (IOP Publishing, 2018).
\newblock \doi{10.1088/978-0-7503-1320-9}

\bibitem{Lewisetal2021}
S.~{Lewis}, C.~{Cournoyer-Cloutier}, A.~{Tran}, W.~{Farner}, S.~{McMillan},
  et~al., in \emph{American Astronomical Society Meeting Abstracts},
  \emph{American Astronomical Society Meeting Abstracts}, vol.~53 (2021),
  \emph{American Astronomical Society Meeting Abstracts}, vol.~53, p. 232.08

\bibitem{Beilisetal2021}
D.~{Beilis}, S.~{Beck}, J.~{Lacy}, in \emph{American Astronomical Society
  Meeting Abstracts}, \emph{American Astronomical Society Meeting Abstracts},
  vol.~53 (2021), \emph{American Astronomical Society Meeting Abstracts},
  vol.~53, p. 150.01

\bibitem{MoralesFellhauer2020}
M.C. {Morales}, M.~{Fellhauer}, BAAA \textbf{61C}, 94 (2020)

\bibitem{Riederetal2013}
S.~{Rieder}, T.~{Ishiyama}, P.~{Langelaan}, J.~{Makino}, S.L.W. {McMillan},
  et~al., MNRAS \textbf{436}(4), 3695 (2013).
\newblock \doi{10.1093/mnras/stt1848}

\bibitem{deVriesetal2014}
N.~{de Vries}, S.~{Portegies Zwart}, J.~{Figueira}, MNRAS \textbf{438}(3), 1909
  (2014).
\newblock \doi{10.1093/mnras/stt1688}

\bibitem{Stiefel1965}
E.~Stiefel, P.~Kustaanheimo, Journal für die reine und angewandte Mathematik
  \textbf{218}, 204 (1965).
\newblock \urlprefix\url{http://eudml.org/doc/150684}

\bibitem{Aarseth1985b}
S.J. {Aarseth}, in \emph{Dynamics of Star Clusters}, vol. 113, ed. by
  J.~{Goodman}, P.~{Hut} (1985), vol. 113, pp. 251--258

\bibitem{Aarseth2008}
S.J. {Aarseth}, in \emph{The Cambridge N-Body Lectures}, vol. 760, ed. by S.J.
  {Aarseth}, C.A. {Tout}, R.A. {Mardling} (Springer-Verlag Berlin Heidelberg,
  2008), p.~1.
\newblock \doi{10.1007/978-1-4020-8431-7_1}

\bibitem{Makino1999}
J.~{Makino}, Journal of Computational Physics \textbf{151}(2), 910 (1999).
\newblock \doi{10.1006/jcph.1999.6226}

\bibitem{EisensteinHut1998}
D.J. {Eisenstein}, P.~{Hut}, ApJ \textbf{498}(1), 137 (1998).
\newblock \doi{10.1086/305535}

\bibitem{Pattabiramanetal2013}
B.~{Pattabiraman}, S.~{Umbreit}, W.k. {Liao}, A.~{Choudhary}, V.~{Kalogera},
  et~al., ApJS \textbf{204}(2), 15 (2013).
\newblock \doi{10.1088/0067-0049/204/2/15}

\bibitem{Hypki2018}
A.~{Hypki}, MNRAS \textbf{477}(3), 3076 (2018).
\newblock \doi{10.1093/mnras/sty803}

\bibitem{DeanGhemawat2004}
J.~Dean, S.~Ghemawat, Communications of the ACM \textbf{51}, 137 (2004).
\newblock \doi{10.1145/1327452.1327492}

\bibitem{Pangetal2016}
X.Y. {Pang}, C.~{Olczak}, D.F. {Guo}, R.~{Spurzem}, R.~{Kotulla}, RAA
  \textbf{16}(3), 37 (2016).
\newblock \doi{10.1088/1674-4527/16/3/037}

\bibitem{Askaretal2017b}
A.~{Askar}, M.~{Giersz}, W.~{Pych}, E.~{Dalessandro}.
\newblock {COCOA: Simulating Observations of Star Cluster Simulations} (2017)

\bibitem{Askaretal2018a}
A.~{Askar}, M.~{Giersz}, W.~{Pych}, E.~{Dalessandro}, MNRAS \textbf{475}(3),
  4170 (2018).
\newblock \doi{10.1093/mnras/sty101}

\bibitem{Askaretal2017c}
A.~{Askar}, P.~{Bianchini}, R.~{de Vita}, M.~{Giersz}, A.~{Hypki}, et~al.,
  MNRAS \textbf{464}(3), 3090 (2017).
\newblock \doi{10.1093/mnras/stw2573}

\bibitem{deVita2017}
R.~{de Vita}, M.~{Trenti}, P.~{Bianchini}, A.~{Askar}, M.~{Giersz}, et~al.,
  MNRAS \textbf{467}(4), 4057 (2017).
\newblock \doi{10.1093/mnras/stx325}

\bibitem{Arosetal2020}
F.I. {Aros}, A.C. {Sippel}, A.~{Mastrobuono-Battisti}, A.~{Askar},
  P.~{Bianchini}, et~al., MNRAS \textbf{499}(4), 4646 (2020).
\newblock \doi{10.1093/mnras/staa2821}

\bibitem{Arosetal2021}
F.I. {Aros}, A.C. {Sippel}, A.~{Mastrobuono-Battisti}, P.~{Bianchini},
  A.~{Askar}, et~al., MNRAS \textbf{508}(3), 4385 (2021).
\newblock \doi{10.1093/mnras/stab2872}

\bibitem{Piottoetal2002}
G.~{Piotto}, I.R. {King}, S.G. {Djorgovski}, C.~{Sosin}, M.~{Zoccali}, et~al.,
  A$\&$A \textbf{391}, 945 (2002).
\newblock \doi{10.1051/0004-6361:20020820}

\bibitem{Kotullaetal2009}
R.~{Kotulla}, U.~{Fritze}, P.~{Weilbacher}, P.~{Anders}, MNRAS \textbf{396}(1),
  462 (2009).
\newblock \doi{10.1111/j.1365-2966.2009.14717.x}

\bibitem{Hongetal2017b}
J.~{Hong}, R.~{de Grijs}, A.~{Askar}, P.~{Berczik}, C.~{Li}, et~al., MNRAS
  \textbf{472}(1), 67 (2017).
\newblock \doi{10.1093/mnras/stx1954}

\bibitem{Kouwenhovenetal2020}
M.B.N. {Kouwenhoven}, F.~{Flammini Dotti}, Q.~{Shu}, X.~{Xu}, K.~{Wu}, et~al.,
  in \emph{Journal of Physics Conference Series}, \emph{Journal of Physics
  Conference Series}, vol. 1523 (2020), \emph{Journal of Physics Conference
  Series}, vol. 1523, p. 012011.
\newblock \doi{10.1088/1742-6596/1523/1/012011}

\bibitem{Shuetal2021}
Q.~{Shu}, X.~{Pang}, F.~{Flammini Dotti}, M.B.N. {Kouwenhoven}, M.~{Arca
  Sedda}, R.~{Spurzem}, \apjs \textbf{253}(1), 14 (2021).
\newblock \doi{10.3847/1538-4365/abcfb8}

\bibitem{Pangetal2022b}
X.~{Pang}, Q.~{Shu}, L.~{Wang}, M.B.N. {Kouwenhoven}, Research in Astronomy and
  Astrophysics \textbf{22}(9), 095015 (2022).
\newblock \doi{10.1088/1674-4527/ac7f0f}

\bibitem{ConroyWhite2009}
C.~{Conroy}, J.E. {Gunn}, M.~{White}, ApJ \textbf{699}(1), 486 (2009).
\newblock \doi{10.1088/0004-637X/699/1/486}

\bibitem{ConroyGunn2010}
C.~{Conroy}, J.E. {Gunn}, ApJ \textbf{712}(2), 833 (2010).
\newblock \doi{10.1088/0004-637X/712/2/833}

\bibitem{Conroyetal2010}
C.~{Conroy}, M.~{White}, J.E. {Gunn}, ApJ \textbf{708}(1), 58 (2010).
\newblock \doi{10.1088/0004-637X/708/1/58}

\bibitem{Bianchinietal2015}
P.~{Bianchini}, M.A. {Norris}, G.~{van de Ven}, E.~{Schinnerer}, MNRAS
  \textbf{453}(1), 365 (2015).
\newblock \doi{10.1093/mnras/stv1651}

\bibitem{Khorramietal2019}
Z.~{Khorrami}, P.~{Khalaj}, A.S.M. {Buckner}, P.C. {Clark}, E.~{Moraux},
  et~al., MNRAS \textbf{485}(3), 3124 (2019).
\newblock \doi{10.1093/mnras/stz490}

\bibitem{PopescuHanson2009}
B.~{Popescu}, M.M. {Hanson}, AJ \textbf{138}(6), 1724 (2009).
\newblock \doi{10.1088/0004-6256/138/6/1724}

\bibitem{ChencinerMontgomery2000}
A.~{Chenciner}, R.~{Montgomery}, arXiv Mathematics e-prints math/0011268 (2000)

\bibitem{Montgomery2001}
R.~{Montgomery}, Notices of the AMS \textbf{48}(5), 471 (2001)

\bibitem{Heggie2000}
D.C. {Heggie}, \mnras \textbf{318}(4), L61 (2000).
\newblock \doi{10.1046/j.1365-8711.2000.04027.x}

\end{thebibliography}



\glsaddall
\renewcommand{\glossarypreamble}{\glsfindwidesttoplevelname[\currentglossary]}
\setglossarystyle{alttree}
\printnoidxglossary[sort=word,title=Acronyms]
\end{document}